\documentclass[12pt]{iopart}

\usepackage{graphicx,enumerate}

\eqnobysec

\def\changed#1{{#1}}

\begin{document}
\title{Solitons in the Higgs phase 
{\small 
-- the moduli matrix approach --
}  
}

\author{Minoru Eto, Youichi Isozumi, Muneto Nitta, Keisuke Ohashi 
and Norisuke Sakai}

\address{Department of Physics, Tokyo Institute of 
Technology \\
Tokyo 152-8551, JAPAN  }
\ead{ 
meto,
isozumi,
nitta,
keisuke,
nsakai@th.phys.titech.ac.jp
}
\begin{abstract} 
We review our recent work on solitons in the Higgs phase. 
We use $U(N_{\rm C})$ gauge theory with $N_{\rm F}$ 
Higgs scalar fields in the fundamental representation,  
which can be extended to possess eight supercharges. 
We propose the moduli matrix as a fundamental tool to 
exhaust all BPS solutions, 
and to characterize all possible 
moduli parameters. 
Moduli spaces of domain walls (kinks) 
and vortices, which are the only elementary solitons in 
the Higgs phase, are found in terms of the moduli matrix.
Stable monopoles and instantons 
can exist in the Higgs phase 
if they are attached by vortices to form composite 
solitons. 
The moduli spaces of these composite solitons are also 
worked out in terms of the moduli matrix. 
Webs of walls can also be formed with characteristic 
difference between Abelian and non-Abelian gauge theories.
\changed{
Instanton-vortex systems, 
monopole-vortex-wall systems, 
and webs of walls in 
Abelian gauge theories are found to 
admit negative energy objects 
with the instanton charge (called intersectons), 
the monopole charge (called boojums) 
and the Hitchin charge, respectively. 
}
We characterize the total moduli space of these elementary 
as well as composite solitons. 
In particular 
the total moduli space of walls is given by the complex Grassmann 
manifold $
SU(N_{\rm F}) / 
[SU(N_{\rm C})\times SU(N_{\rm F}-{N}_{\rm C}) \times U(1)]$ 
and is decomposed into various topological 
sectors corresponding to boundary condition specified by 
particular vacua. 
The moduli space of $k$ vortices is also completely determined and 
is reformulated as the half ADHM construction. 
Effective Lagrangians are constructed on walls and 
vortices in a compact form. 
We also present several new results on interactions of 
various solitons, such as monopoles, vortices, and walls. 
Review parts contain our works on 
domain walls 
  \cite{Isozumi:2004jc} (hep-th/0404198)  
  \cite{Isozumi:2004va} (hep-th/0405194)
  \cite{Eto:2004vy} (hep-th/0412024)
  \cite{Eto:2005wf} (hep-th/0503033) 
  \cite{Sakai:2005kz} (hep-th/0505136), 
vortices 
  \cite{Eto:2005yh} (hep-th/0511088) 
  \cite{Eto:2006mz} (hep-th/0601181), 
domain wall webs  
  \cite{Eto:2005cp} (hep-th/0506135)
  \cite{Eto:2005fm} (hep-th/0508241)   
  \cite{Eto:2005mx} (hep-th/0509127),
monopole-vortex-wall systems 
  \cite{Isozumi:2004vg} (hep-th/0405129) 
  \cite{Sakai:2005sp} (hep-th/0501207),  
instanton-vortex systems 
  \cite{Eto:2004rz} (hep-th/0412048), 
effective Lagrangian on walls and vortices 
  \cite{Eto:2006uw} (hep-th/0602289), 
classification of BPS equations
  \cite{Eto:2005sw} (hep-th/0506257), 
and 
Skyrmions 
  \cite{Eto:2005cc} (hep-th/0508130).


\end{abstract}

\section{Introduction}
\label{int}
Topological solitons play very important roles 
in broad area of physics 
\cite{MantonSutcliffe,VilenkinShellard,Mermin,Rajaraman,Tong:2005un}. 
They appear various situations in condensed matter physics, 
cosmology, nuclear physics and high energy physics including 
string theory. 
In field theory it is useful to classify solitons by 
co-dimensions on which solitons depend. 
Kinks (domain walls), vortices, 
monopoles and instantons are well-known
typical solitons with co-dimensions one, two, three and four, 
respectively.\footnote{
In this paper we keep terminology of ``instantons" 
for Yang-Mills instantons in four Euclidean space. 
They become particles in 4+1 dimensions.
}   
They carry topological charges classified by certain homotopy groups 
according to their co-dimensions. 
If the spatial dimension of spacetime is larger 
than the co-dimensions, 
solitons are extended objects having world volume 
and are sometimes called ``{\it branes}".
D-branes are solitons in string theory whereas 
topological solitons in higher dimensional field theory 
are models of branes. 
D-branes and field theory solitons are closely related or 
sometimes are identified in various situations.  
Recently the {\it brane-world scenario} 
\cite{Horava:1996ma,Arkani-Hamed:1998rs,Randall:1999ee}
are also realized on topological solitons in field theory or 
D-branes in string theory.

When solitons/branes saturate a lower energy bound, 
called the Bogomol'nyi bound, 
they are the most stable among solitons with the same topological charge,   
and are called {\it Bogomol'nyi-Prasad-Sommerfield (BPS)} 
solitons~\cite{Bogomolny:1975de}.  
BPS solitons can be naturally realized in 
supersymmetric (SUSY) field theories and 
preserve some fraction of the original SUSY~\cite{Witten:1978mh}. 
From the discussion of SUSY representation,  
they are non-perturbatively stable  and therefore 
play crucial roles in 
non-perturbative study of SUSY gauge theories and string 
theory \cite{Seiberg:1994rs}.

Since there exist no force between BPS solitons 
the most general solutions of solitons contain parameters corresponding to 
positions of solitons. 
Combined with parameters in the internal space, they are called 
the {\it moduli parameters}. 
A space parametrized by the moduli parameters is 
no longer a flat space but a curved space called the {\it moduli space},  
possibly containing singularities. 
The moduli space is the most important tool to study BPS solitons. 
When solitons can be regarded as particles, 
say for instantons in $d=4+1$, 
monopoles in $d=3+1$, vortices in $d=2+1$, kinks in $d=1+1$ and so on, 
geodesics in their moduli space describe classical 
scattering of solitons~\cite{Manton:1981mp}. 
In quantum theory, for instance, 
the instanton calculus is reduced to
the integration over the instanton moduli space \cite{Dorey:2002ik}. 
The same discussion should hold for a ``monopole calculus" in $d=2+1$,  
a ``vortex calculus" in $d=1+1$ and so on. 
On the other hand, when solitons have world volume, 
for instance vortex-string in $d=3+1$, 
moduli are promoted to massless moduli fields 
in the effective field theory on the world volume of solitons. 
Therefore moduli space is crucial to consider the brane world scenario,  
solitons in higher dimensions or string theory.
The moduli fields describe local deformations along 
the world volume of solitons. 
This fact is useful when we consider composite solitons 
made of solitons with different co-dimensions. 
Namely, composite solitons may sometimes be regarded as 
solitons in the effective field theory of 
the other (host) solitons \cite{Townsend:1999hi}. 
For instance a D/fundamental string ending on a D-brane 
can be realized as a soliton called the BIon \cite{Callan:1997kz} 
in the Dirac-Born-Infeld theory on the D-brane.


Construction of solutions and the moduli spaces 
of instantons and monopoles 
were established long time ago and are well known as 
the ADHM \cite{Atiyah:1978ri,Corrigan:1983sv} 
and the Nahm \cite{Nahm:1979yw,Corrigan:1983sv} constructions, respectively. 
Instantons and monopoles are naturally realized as 1/2 BPS solitons 
in SUSY gauge theory with sixteen supercharges. 
The effective theories on them are nonlinear sigma models 
with eight supercharges, 
whose target spaces must be hyper-K\"ahler 
\cite{Alvarez-Gaume:1981hm}--\cite{Eguchi:1978xp}, 
and therefore the moduli spaces of instantons and monopoles 
are hyper-K\"ahler.

Vacua outside monopoles and instantons 
are in the Coulomb phase and in
the unbroken phase of the gauge symmetry, respectively. 
Contrary to this fact, 
vacua outside kinks or vortices are in the Higgs phase 
where gauge symmetry is completely broken. 
These solitons can be constructed
as 1/2 BPS solitons in SUSY gauge theory 
with eight supercharges, 
where the so-called Fayet-Iliopoulos (FI) term~\cite{Fayet:1974jb} 
should be contained in the Lagrangian to realize the Higgs vacua.  
The moduli space of kinks and vortices are K\"ahler \cite{Zumino:1979et}
because they preserve four supercharges.  
Kinks (domain walls) in SUSY $U(1)$ gauge theory with 
eight supercharges were firstly found in \cite{Abraham:1992vb} 
in the strong gauge coupling (sigma model) limit 
and have been developed recently 
\cite{Lambert:1999ix}--\cite{Tong:2003ik}, 
\cite{Eto:2005wf,Sakai:2005kz,Hanany:2005bq}. 
Domain walls in non-Abelian gauge theory 
have been firstly discussed 
in \cite{Shifman:2003uh,Isozumi:2004jc,Isozumi:2004va} 
and have been further studied 
\cite{Eto:2004vy,Eto:2005cc,Tong:2005nf,Eto:2006uw}.  
In particular their moduli space has been 
determined to be complex Grassmann manifold \cite{Eto:2004vy}. 
On the other hand, 
vortices were found earlier 
by Abrikosov, Nielsen and Olesen \cite{Abrikosov:1956sx}
in $U(1)$ gauge theory coupled with one complex Higgs field,   
and are now referred as the ANO vortices. 
Their moduli space was constructed \cite{Taubes:1979tm}--\cite{Chen:2004xu}. 
When the number of Higgs fields are large enough 
vortices are called semi-local vortices \cite{Vachaspati:1991dz}, 
and their moduli space contains size moduli similarly to lumps
 \cite{Ward:1985ij,Stokoe:1986ic,Leese:1991hr} 
or sigma model instantons \cite{Polyakov:1975yp}.
Study of vortices in non-Abelian gauge theory, 
called {\it non-Abelian vortices}, 
was initiated in \cite{Hanany:2003hp,Auzzi:2003fs} 
and has been extensively discussed 
\cite{Hanany:2003hp}--\cite{Gorsky:2006if}.\footnote{
Another type of non-Abelian vortices were discussed 
earlier~\cite{Alford:1990mk}.
} 
Especially 
their moduli space has been determined in the framework of 
field theory \cite{Eto:2005yh} as well as 
string theory \cite{Hanany:2003hp}.

One aim of this paper is to give a comprehensive understanding 
of the moduli spaces of 1/2 BPS kinks and vortices.
The other aim is to study the moduli spaces of 
various 1/4 BPS composite solitons as discussed below.\footnote{
Composite solitons were also studied in non-supersymmetric 
field theories \cite{Carroll:1997pz, network, Matsuda:2004bg}.
}

Domain walls can make a junction as a 1/4 BPS state \cite{Gibbons:1999np}  
and these wall junctions in SUSY theories with four supercharges 
were further studied in 
\cite{Oda:1999az,Ito:2000zf,Naganuma:2001br,Shifman:1999ri} 
(see \cite{Eto:2005cp} for more complete references). 
Domain wall junction in SUSY $U(1)$ gauge theory 
with eight supercharges 
was constructed \cite{Kakimoto:2003zu} by embedding 
an exact solution in \cite{Oda:1999az,Ito:2000zf,Naganuma:2001br}. 
Finally in \cite{Eto:2005cp,Eto:2005fm} 
the full solutions of domain wall junction, 
called {\it domain wall webs}, have been constructed 
in SUSY non-Abelian gauge theory with eight supercharges. 
The Hitchin charge is found to be 
localized around junction points 
which is always negative in Abelian gauge theory \cite{Eto:2005cp}
and can be either negative or positive in non-Abelian gauge theory 
\cite{Eto:2005fm}. 
This configuration shares the many properties with 
the $(p,q)$ 5-brane webs \cite{Aharony:1997ju}.

As noted above, monopoles and instantons do not live 
in the Higgs phase. 
Question is what happens if monopoles or instantons 
are put into the Higgs phase. 
This situation can be realized 
by considering SUSY gauge theory with the FI term. 
In the Higgs phase, magnetic flux from a monopole is squeezed by 
the Meissner effect into a vortex, 
and the configuration becomes 
a confined monopole with vortices attached 
\cite{Hindmarsh:1985xc}--\cite{Ferretti:2006jg}. 
This configuration is interesting 
because it gives 
a dual picture of color confinement \cite{Auzzi:2003em}. 
The confined monopole can be regarded as a kink 
in the effective field theory on a vortex \cite{Tong:2003pz}. 
In SUSY theory the configuration preserves 
a quarter of eight supercharges and 
is a 1/4 BPS state \cite{Shifman:2004dr}.  
Moreover it was found \cite{Gauntlett:2000de} 
in the strong gauge coupling limit 
that vortices can end on a domain wall  
to form a 1/4 BPS state, like strings ending on a D-brane. 
This configuration was further studied 
in gauge theory without taking the strong coupling limit \cite{Shifman:2002jm}.
Finally it was found \cite{Isozumi:2004vg} that 
all monopoles, vortices and domain walls can coexist 
as a 1/4 BPS state. 
Full solutions constructed in \cite{Isozumi:2004vg} 
resemble with the Hanany-Witten type brane 
configuration \cite{Hanany:1996ie}.
The negative monopole charge (energy) has also been found 
\cite{Isozumi:2004vg} in $U(1)$ gauge theory 
and has been later called {\it boojum} \cite{Sakai:2005sp,Auzzi:2005yw}.

1/4 BPS composite configurations made of instantons and vortices 
have also been found as solutions of self-dual Yang-Mills equation 
coupled with Higgs fields 
(the SDYM-Higgs equation) 
in $d=5,6$ SUSY gauge theory 
with eight supercharges~\cite{Hanany:2004ea,Eto:2004rz}. 
Monopoles in the Higgs phase can be obtained 
by putting a periodic array of these instantons 
along one space direction inside the vortex world volume 
\cite{Eto:2004rz}, while  
the BPS equation of monopoles is obtained by 
the Scherk-Schwarz dimensional reduction \cite{Scherk:1979zr} 
of the SDYM-Higgs equation. 
All other BPS equations introduced above can be obtained from 
the SDYM-Higgs equation by the Scherk-Schwarz 
and/or ordinary dimensional reductions. 
The negative instanton charge (energy) has been also 
found \cite{Eto:2004rz} 
at intersection of vortices in Abelian gauge theory, 
and is called {\it intersecton}.

Surprisingly enough this SDYM-Higgs equation was independently found by 
mathematicians \cite{Bradlow:1990ir, MundetiRiera:1999fd,CRGS} earlier than physicists 
\cite{Hanany:2004ea,Eto:2004rz}. 
Moreover they consider it in a more general setting, 
namely with a K\"ahler manifold in any dimension as a base space 
where solitons live and 
with a general target manifold of scalar fields,  
unlike ordinary Higgs fields in linear representation.
They call their equation simply as a {\it vortex equation}. 
If we take a base space as ${\bf C}^2$ and a target space 
as a vector space, 
the vortex equation reduces to our SDYM-Higgs equation. 
Whereas if we take a base space as ${\bf C}$, 
the vortex equation reduces 
to the BPS equation of vortices \cite{Baptista:2004rk}.
Some integration over the moduli space of the vortex equation 
defines a new topological invariant called 
the Hamiltonian Gromov-Witten invariant~\cite{CRGS,sympleticGW} 
which generalizes 
the Gromov-Witten invariant and the Donaldson invariant. 
Therefore studying the moduli space of the SDYM-Higgs equation 
is very important in mathematics as well as physics.

\medskip
In this paper we focus on the solitons in the Higgs phase; 
domain walls, vortices, and
composite solitons of monopoles/instantons. 
We solve the half (the hypermultiplet part) of BPS equations 
by introducing the {\it moduli matrix}.
The rest (the vector multiplet part) of BPS equations is 
difficult to solve in general. 
When the number of Higgs fields is larger then the number 
of colors,  
they can be solved analytically in the strong gauge coupling limit 
in which the gauge theories reduce to nonlinear sigma models 
with hyper-K\"ahler target spaces. 
In general cases, we assume that 
the vector multiplet part of BPS equations 
produces no additional moduli parameters. 
This assumption was rigorously proved in certain situations,  
for instance in the case of the ANO vortices \cite{Taubes:1979tm} 
and in the case of compact K\"ahler base 
spaces \cite{MundetiRiera:1999fd,CRGS},  
and is now called the Hitchin-Kobayashi correspondence in 
the mathematical literature. 
In cases of odd co-dimensions it is a rather difficult problem 
but it was proved for domain walls in $U(1)$ gauge theory \cite{Sakai:2005kz}
and the index theorem \cite{Lee:2002gv,Sakai:2005sp} supports it 
for the case of domain walls in non-Abelian gauge theory.
Therefore this assumption is correct for the most cases, 
and we consider that all moduli parameters in the BPS equations 
are contained in the moduli matrix. 
We concretely discuss the correspondence between the moduli parameters 
in the moduli matrix and actual soliton configurations  
in various cases, 1) domain walls \cite{Isozumi:2004jc,Isozumi:2004va}, 
2) vortices \cite{Eto:2005yh,Eto:2006mz}, 
3) domain wall junctions or webs \cite{Eto:2005cp,Eto:2005fm},
4) composites of monopoles (boojums), vortices and walls \cite{Isozumi:2004vg},
5) composites of instantons and vortices \cite{Eto:2004rz}.  
We will see that composite solitons in non-Abelian gauge theory have 
much more variety than those in Abelian gauge theory.
One interesting property which all systems commonly share  
is the presence of a negative/positive charge 
localized around junction points of composite solitons;
The junction charge is always negative in Abelian gauge theory while  
it can be either negative or positive in non-Abelian gauge theory.

This paper contains many new results; 
We extend analysis of non-Abelian vortices in \cite{Eto:2005yh} 
to semi-local non-Abelian vortices which contain 
non-normalizable zero modes. 
Relation to K\"ahler quotient construction \cite{Hanany:2003hp} 
of the vortex moduli space is completely clarified.
The half ADHM construction of vortices is found. 
We construct effective Lagrangian on non-Abelian 
(semi-local) vortices in a compact form, 
which generalizes the Abelian cases 
\cite{Taubes:1979tm}--\cite{Chen:2004xu}. 
Relation between moduli parameters in 
1/2 BPS states in massless theory and 
1/4 BPS states in massive theory is found; 
for instance 
orientational moduli of a non-Abelian vortex 
are translated to position moduli of a monopole.
We give a complete answer to the question addressed in 
\cite{Sakai:2005sp,Lee:2005sv} whether 
a confined monopole attached by a vortex ending on a domain wall 
can pass through that domain wall by 
changing moduli or not. 
Namely we find that a monopole can pass through a domain wall 
if and only if positions of vortices attached to the wall 
from both sides coincide. 
If they do not coincide, no monopole exists as 
a BPS state, suggesting repulsive force between 
a monopole and a boojum on a junction point of the vortex and the wall.

\bigskip
\bigskip
This paper is organized as follows. 
In section \ref{mdl} we present the model and 
investigate its vacua.
In \ref{sc:mdl:model} we give the Lagrangian of 
$U(N_{\rm C})$ gauge theory with 
$N_{\rm F}$ Higgs fields 
in the fundamental representation 
in spacetime dimensions $d= 1+1,\cdots,5+1$.   
In section \ref{sc:mdl:vacua-UN} we analyze 
the vacuum structure of our model 
with the massless or massive Higgs fields. 
In section \ref{sc:mdl:BPSWSIGC} 
we discuss the strong gauge coupling limit 
of the model with large number of Higgs fields 
($N_{\rm F}>N_{\rm C}$),  
in which the model reduces to a nonlinear sigma model 
whose target space is a hyper-K\"ahler manifold. 
In section \ref{hlf} we discuss 1/2 BPS solitons in the Higgs phase, 
namely domain walls in section \ref{wll} and vortices in section \ref{vtx}. 
In section \ref{efa} we construct 
the effective action on these solitons.
In section \ref{qrt} we discuss 1/4 BPS composite solitons. 
First in section \ref{4eq} 
we present sets of 1/4 BPS equations which we consider in this paper.
In section \ref{juc} we work out solutions of domain wall webs, 
or junction made of domain walls.  
In section \ref{mon} we work out composite states 
of monopoles (boojums), vortices and domain walls. 
In section \ref{ist} we work out composite states 
of instantons and (intersecting) vortices. 
In section \ref{sos} we interpret some of 
these 1/4 BPS composite solitons 
as 1/2 BPS solitons on host 1/2 BPS solitons. 
Finally section \ref{cad} is devoted to a discussion. 

\section{Model and vacua}
\label{mdl}
\subsection{$U(N_{\rm C})$ gauge theory with 
$N_{\rm F}$ flavors
}
\label{sc:mdl:model}
We are mostly interested in $U(N_{\rm C})$ gauge theory in 
$(d-1)+1$ dimensions with a number of adjoint scalar 
fields $\Sigma_p$ and $N_{\rm F}$ flavors of scalar 
fields in the fundamental representation as an 
$N_{\rm C} \times N_{\rm F}$ matrix $H$ 
\begin{eqnarray}
{\cal L} &=& {\cal L}_{\rm kin} - V, 
\label{eq:mdl:total_lagrangian}
\\ 
{\cal L}_{\rm kin} &=& 
{\rm Tr}\left(- {1\over 2g^2}F_{\mu\nu}F^{\mu\nu} 
+\frac{1}{g^2}{\cal D}_\mu\Sigma_p{\cal D}^\mu\Sigma_p 
+{\cal D}^\mu H \left({\cal D}_\mu H\right)^\dagger \right), 
\label{eq:mdl:lagrangian}
\end{eqnarray}
where the covariant derivatives and field strengths are 
defined as 
${\cal D}_\mu \Sigma_p=\partial_\mu\Sigma_p + i[W_\mu, \Sigma_p]$, 
${\cal D}_\mu H=(\partial_\mu + iW_\mu)H$, 
$F_{\mu\nu}=-i[{\cal D}_\mu,\,{\cal D}_\nu]$. 
Our convention for the metric is 
$\eta_{\mu\nu} = {\rm diag}(+,-,\cdots,-)$. 
The scalar potential $V$ is given in terms of 
diagonal mass matrices $M_p$ and a real parameter $c$ as 
\begin{eqnarray}
V&=& 
{\rm Tr}
\Big[
\frac{g^2}{4}
\left(c\mathbf{1}
-H  H^{\dagger} 
\right)^2 
+ (\Sigma_p H - H M_p) 
 (\Sigma_p H - H M_p)^\dagger 
\Big] . 
\label{eq:mdl:scalar_pot}
\end{eqnarray}
This Lagrangian is obtained as the bosonic part of the 
Lagrangian with eight supercharges by ignoring one of the 
scalars in the fundamental representation: 
$H^1\equiv H$, $H^2=0$. 
Although the gauge couplings for $U(1)$ and $SU(N_{\rm C})$ 
are independent, we have chosen these to be identical to 
obtain simple solutions classically. 
The real positive parameter $c$ 
is called the Fayet-Iliopoulos (FI) parameter, 
which can appear in supersymmetric $U(1)$ gauge theories 
\cite{Fayet:1974jb}. 
Since we are interested in the Higgs phase, it is crucial 
to have this parameter $c$. 
We use a matrix notation for 
these component fields, such as $W_\mu=W_\mu^I T_I$, 
where $T_I \ (I = 0, 1,2,\cdots, N_{\rm C}^2-1)$ 
are matrix generators of the gauge 
group $G$ in the fundamental representation 
satisfying 
${\rm Tr}(T_I T_J) = \frac12\delta_{IJ}$, 
$[ T_I , T_J]=i f_{IJ}{}^{K} T_K$ with $T^0$ as the 
$U(1)$ generator. 
In order to embed this Lagrangian into 
a supersymmetric gauge theory with eight supercharges, 
space-time dimensions are restricted as $d \leq 6$ and
the number of adjoint scalars and mass matrices are 
given by $6-d$ ($p=1, \cdots, 6-d$), 
since these theories can be obtained by dimensional 
reductions with possible twisted boundary conditions 
(the Scherk-Schwarz dimensional reduction \cite{Scherk:1979zr}) 
as described below.

Let us note that a common mass $M_{p}=m_p{\bf 1}$ 
for all flavors 
can be absorbed into a shift of the adjoint scalar 
field $\Sigma_p$, and has no physical significance. 
In this paper, we assume either massless hypermultiplets, 
or fully non-degenerate mass parameters 
$m_{pA}\not=m_{pB}$, for $A\not=B$ unless 
stated otherwise. 
Then the flavor symmetry $SU(N_{\rm F})$ for the massless 
case reduces in the massive case to 
\begin{eqnarray}
 G_{\rm F} = U(1)_{\rm F}^{N_{\rm F}-1},
 \label{eq:mdl:break-flavor}
\end{eqnarray} 
where 
$U(1)_{\rm F}$ corresponding to common phase is
gauged by $U(1)_{\rm G}$ local gauge symmetry.

Let us discuss supersymmetric extension of the Lagrangian 
given by equations (\ref{eq:mdl:total_lagrangian})--(\ref{eq:mdl:scalar_pot}).
(Those who are unfamiliar with supersymmetry can skip 
the rest of this subsection and can go to 
section \ref{sc:mdl:vacua-UN}.)
Gauge theories with eight supercharges are most 
conveniently constructed first in $5+1$ dimensions 
and theories in lower dimensions follow from dimensional 
reductions. 
The gamma matrices satisfy 
$\{\Gamma^M,\Gamma^N\} = 2\eta^{MN}$, 
and the totally antisymmetric product of the gamma matrices 
$\Gamma^M, \cdots, \Gamma^N$ are denoted by 
$\Gamma^{M\cdots N}$. 
The charge conjugation matrix $C$ is defined by 
$C^{-1}\Gamma_MC = \Gamma_M^T$ and satisfy $C^T = -C$. 
The building blocks for gauge theories with eight 
supercharges are vector multiplets and hypermultiplets. 
The vector multiplet in $5+1$ dimensions consists of 
a gauge field $W_M^I$ ($M=0,1,2,3,4,5$) for generators 
of gauge group $I$, an $SU(2)_R$ triplet of real 
auxiliary fields $Y_{a}^I$, and an $SU(2)_R$ doublet of 
gauginos $\lambda^{iI}$ ($i=1,2$) which are an 
$SU(2)$-Majorana Weyl spinor, namely 
$\Gamma_7\lambda^i=\lambda^i$ and 
$\lambda^i=C\varepsilon^{ij}(\bar\lambda_j)^T$. 
Here $\Gamma_7$ is defined by $\Gamma_7 = \Gamma^{012345}$ 
and $C$ is the charge conjugation matrix in $5+1$ 
dimensions. 
All these fields are in the adjoint representation of $G$.

We have hypermultiplets as matter fields, 
consisting of an $SU(2)_R$ doublet of complex scalar 
fields $H^{irA}$ and Dirac field $\psi^{rA}$ 
(hyperino) whose chirality is $\Gamma_7\psi^{rA}=-\psi^{rA}$. 
Color (flavor) indices are denoted as $r,s,\cdots$ 
($A,B, \cdots$). 
The hypermultiplet in $5+1$ dimensions 
does not allow (finite numbers of) auxiliary fields and 
superalgebra closes only on-shell, although the vector 
multiplet has auxiliary fields.

We shall consider a model with minimal 
kinetic terms for vector and hypermultiplets. 
In $5+1$ dimensions, the model allows only two types of 
parameters, gauge couplings $g_I$ and a triplet of 
the Fayet-Iliopoulos (FI) parameters $\zeta_a$ with $a=1,2,3$. 
There exist the triplets of the FI parameters 
as many as $U(1)$ factors of gauge group in general.
To distinguish different gauge couplings for different 
factor groups, we retained suffix $I$ for $g_I$. 
The bosonic part of the Lagrangian is given by 
\begin{eqnarray}
{\cal L}_{6} &=&-{1\over 4g_I^2}F^I_{MN}F^{IMN }
+\left({\cal D}_M H^{irA}\right)^*{\cal D}^M H^{irA}
+{\cal L}_{\rm aux}, 
\label{eq:mdl:lag6_kin}
\\
{\cal L}_{\rm aux}&=&
{1\over 2g_I^2}(Y^{I}_a)^2 
-\zeta_aY^{0}_a +(H^{irA})^*  (\sigma^a)^i{}_j 
(Y_a)^r{}_s H^{jsA}, 
\label{eq:mdl:lag6}
\end{eqnarray}
The equation of motion for auxiliary fields $Y^{I}_a$ gives 
\begin{eqnarray}
Y^{I}_a ={1 \over g^2_I}\left[\zeta_a\delta^I_0 
-(H^{irA})^*(\sigma _a)^i{}_j(T_I)^r{}_sH^{jsA}\right] . 
\label{eq:mdl:auxiliary_fields}
\end{eqnarray}

The supersymmetry transformation for the spinor fields 
in $5+1$ dimensions are given in terms of 
an $SU(2)$-Majorana Weyl spinor parameter $\varepsilon ^i$ 
satisfying 
$\varepsilon ^i
=C\epsilon ^{ij}(\bar \varepsilon _j)^{\rm T},\ 
\Gamma _7\varepsilon ^i=+\varepsilon ^i $ 
\begin{eqnarray}
 \delta _\varepsilon \lambda ^i=\frac12 \Gamma ^{M N}F_{M N}
 \varepsilon ^i+Y_a(i\sigma _a)^i{}_j\varepsilon ^j,\quad
\delta _\varepsilon \psi^{rA} =
-\sqrt{2}i\Gamma ^M {\cal D}_M H^{irA}
\epsilon _{ij}\varepsilon ^j .
\label{eq:mdl:susy_transf}
\end{eqnarray}

We can obtain the $(d-1)+1$-dimensional ($d < 6$) 
supersymmetric gauge theory with 8 supercharges, 
by performing the Scherk-Schwarz (SS) \cite{Scherk:1979zr} 
and/or 
the trivial dimensional reductions 
($6-d$)-times from the $5+1$ dimensional theory 
(\ref{eq:mdl:lag6}), 
after compactifying the $p$-th ($p=5,4,\cdots,d$) 
direction to $S^1$ with radius $R_p$. 
The twisted boundary condition for the SS 
dimensional reduction along the $x^p$-direction  
is given by 
\begin{eqnarray}
 H^{iA}(x^\mu,x^p+2\pi R_p) = 
 H^{iA}(x^\mu, x^p)e^{i\alpha _{pA}},
\quad \left(|\alpha _{pA}| \ll 2\pi\right),
  \label{eq:mdl:SS-red.}
\end{eqnarray}
where $\mu$ is spacetime index in $(d-1)+1$ dimensions. 
We have used the flavor symmetry 
(\ref{eq:mdl:break-flavor}) commuting with supersymmetry 
for this twisting and so supersymmetry is preserved,  
unlike twisting by symmetry not commuting with supersymmetry 
often used in the context in which case supersymmetry is broken. 
If we consider the effective Lagrangian at sufficiently 
low energies, we can discard 
an infinite tower of the Kaluza-Klein modes and 
retain only the lightest mass field as a function 
of the $(d-1)+1$ dimensional spacetime coordinates 
\begin{eqnarray}
W_\mu(x^\mu,x^p) \rightarrow W_\mu(x^\mu),\quad 
W_p(x^\mu,x^p) \rightarrow -\Sigma_p (x^\mu),
\label{eq:mdl:red_gauge}
\end{eqnarray}
\begin{eqnarray}
\hspace{-1cm}
H^{iA}(x^\mu,x^p)\rightarrow {1\over \prod_p 
\sqrt{2\pi R_p}}
H^{iA}(x^\mu)\exp\left({i\sum_pm_{pA}x^p}\right),\quad 
m_{pA}\equiv {\alpha _{pA}\over 2\pi R_p}. 
\label{eq:mdl:red_hyper}
\end{eqnarray}
Integrating the $5+1$ dimensional Lagrangian in 
equation (\ref{eq:mdl:lag6}) over the $x^p$-coordinates 
and introducing 
the auxiliary fields $F_i^{rA}$ for hypermultiplets, 
we obtain the $(d-1)+1$ dimensional effective Lagrangian 
\begin{eqnarray}
{\cal L}_{d} &=& 
- {1\over 4g_I^2}F^I_{\mu\nu}F^{I\mu\nu} + 
\frac{1}{2g_I^2}{\cal D}_\mu\Sigma_p^I
{\cal D}^\mu\Sigma_p^I + 
\left({\cal D}_\mu H^{irA}\right)^*
{\cal D}^\mu H^{irA} 
\nonumber \\
&&
-(H^{irA})^*[(
\Sigma_p-m_{pA} )^2]^r{}_s H^{isA}
+{\cal L}_{\rm aux}, 
\label{eq:mdl:lag_d}
\end{eqnarray}
\begin{eqnarray}
\hspace{-1cm} 
{\cal L}_{\rm aux}
=
{1\over 2g_I^2}(Y^{I}_a)^2 
-\zeta_a Y^{0}_a 
+(H^{irA})^*  (\sigma^a)^i{}_j 
(Y_a)^r{}_s H^{jsA}
+(F_i^{rA})^* F_i^{rA},
\label{eq:mdl:lag_d_aux}
\end{eqnarray}
where we have redefined 
the gauge couplings and the FI parameters in $(d-1)+1$ 
dimensions from $5+1$ dimensions as 
$g_I^2\rightarrow \left(\prod_p2\pi R_p\right)g_I^2$,
$\zeta_a\rightarrow \zeta_a/ \left(\prod_p2\pi R_p\right)$.
We obtain $(6-d)$ adjoint real scalar fields $\Sigma_p$ 
and $(6-d)$ real mass parameters for hypermultiplets 
in $(d-1)+1$ dimensions. 
The $SU(2)_R$ symmetry allows us to 
choose the FI parameters to lie in the third direction 
without loss of generality 
$\zeta_a = (0, \ 0, \ c\sqrt{N_{\rm C}/2}),  \ \ c>0$, 
although we cannot reduce all the FI parameters 
to the third direction if there are more FI parameters. 
Since the equations of motion for auxiliary fields are 
given by (\ref{eq:mdl:auxiliary_fields}) and $F_i^{rA}=0$, 
we obtain the on-shell version of the bosonic part of the 
Lagrangian with the scalar potential $V$ as given in 
equation (\ref{eq:mdl:lagrangian}). 
However, we ignored in equation (\ref{eq:mdl:lagrangian}) 
one of the hypermultiplet 
scalars $H^2=0$, since $H^2$ vanishes for almost all 
soliton solutions as we see in the following sections. 

\subsection{Vacua 
}
\label{sc:mdl:vacua-UN}

SUSY vacuum is equivalent to the vanishing vacuum 
energy, 
which requires 
both contributions from vector and hypermultiplets 
to $V$ in equation (\ref{eq:mdl:lagrangian}) to vanish. 
The SUSY condition $Y_a=0$ for 
vector multiplets can be rewritten 
as 
\begin{eqnarray} 
H^{1}  H^{1\dagger}  - H^{2} H^{2\dagger} 
=c\mathbf{1}_{N_{\rm C}}, 
\qquad 
H^2 H^{1\dagger}= 0. 
\label{eq:mdl:D-term-cond}
\end{eqnarray}
This condition implies that some of hypermultiplets 
have to be non-vanishing. 
Since the non-vanishing hypermultiplets in the fundamental 
representation breaks gauge symmetry, we call these vacua 
as Higgs branch of vacua. 

In the case of massless theory, the vanishing contribution 
from the hypermultiplet gives 
for each index $A$ 
\begin{eqnarray}
(\Sigma_p )^r{}_s H^{isA}=0,   
\label{eq:mdl:susy-cond-H-massless}
\end{eqnarray}
which requires 
$\Sigma_p=0$ 
for all $p$. 
Therefore we find that 
the Higgs branches for the massless hypermultiplets 
are hyper-K\"ahler quotient~\cite{Lindstrom:1983rt,Hitchin:1986ea} 
given by ${\cal M}_{\rm vac} = \{H^{irA} | Y^I_a=0\}/G$,  
where $G$ denotes the gauge group. 
In our specific case of $U(N_{\rm C})$ gauge group with 
$N_{\rm F} (> N_{\rm C})$ massless hypermultiplets in the 
fundamental representation, the moduli space is given by 
the cotangent bundle over the complex Grassmann 
manifold \cite{Lindstrom:1983rt}
\begin{eqnarray}
 {\cal M}^{M_p=0}_{\rm vac} \simeq
 T^* G_{N_{\rm F},N_{\rm C}} \simeq 
 T^* \left[SU(N_{\rm F}) \over 
 SU(N_{\rm C}) \times SU(N_{\rm F}-N_{\rm C}) \times U(1) 
 \right] \, . \label{eq:mdl:T*Gr}
\end{eqnarray}
The real dimension of the Higgs branch is 
$4N_{\rm C}(N_{\rm F} - N_{\rm C})$.

In the massive theory, 
the vanishing contribution to vacuum energy 
from hypermultiplets 
gives 
\begin{eqnarray}
(\Sigma_p -m_{pA} \mathbf{1})^r{}_s H^{isA}=0 , \  
\label{eq:mdl:susy-cond-H-massive}
\end{eqnarray}
for each index $A$. 
This is satisfied by choosing the adjoint scalar $\Sigma_p$ 
to be diagonal matrices whose $r$-th elements are specified 
by the the non-degenerate mass $m_{pA_r}$ for the 
hypermultiplet with non-vanishing $r$ color and $A_r$ flavor 
\begin{eqnarray}
\hspace{-1cm}
 H^{1rA}=\sqrt{c}\,\delta ^{A_r}{}_A,\quad H^{2rA}=0, 
\quad 
\Sigma_p ={\rm diag.}(m_{pA_1},\,m_{pA_2},\,\cdots,\,
m_{pA_{N_{\rm C}}}).
\label{eq:mdl:color-flavor-locking}
\end{eqnarray}
Therefore we find that 
the Higgs branch of vacua of the massless case 
is lifted by masses 
except for fixed points of the tri-holomorphic $U(1)$ Killing 
vectors~\cite{Alvarez-Gaume:1983ab} 
induced by the $U(1)$ actions in equation (\ref{eq:mdl:SS-red.}) 
or (\ref{eq:mdl:break-flavor}), 
when we introduce masses in lower dimensions 
by the SS dimensional reductions. 
Introducing non-degenerate masses, 
only $N_{\rm F}!/[N_{\rm C}! \times (N_{\rm F}-N_{\rm C})!]$ 
discrete points out of the massless moduli space 
$T^*G_{N_{\rm F},N_{\rm C}}$ 
remain as vacua~\cite{Arai:2003tc}. 
These discrete vacua are often called color-flavor locking 
vacua. 
In the particular case of $N_{\rm F} = N_{\rm C}$, 
we have the unique vacuum 
up to gauge transformations. 
Throughout this paper the vacuum given by
equation (\ref{eq:mdl:color-flavor-locking})
is labeled by 
\begin{eqnarray} 
 \langle A_1,A_2,\cdots,A_{N_{\rm C}}\rangle 
   \label{eq:mdl:vac-label}
\end{eqnarray} 
or briefly by $\langle \{A_r\}\rangle$.
This kind of labels  may also 
be used for defining an 
$N_{\rm C}\times N_{\rm C}$ minor matrix 
$H^{\langle \{A_r\}\rangle}$ from 
the $N_{\rm C}\times N_{\rm F}$ 
matrix $H$ as $(H^{\langle \{A_r\}\rangle})^{qs}=H^{q A_s}$.

\subsection{Infinite gauge coupling and nonlinear sigma models}
\label{sc:mdl:BPSWSIGC}

SUSY gauge theories reduce to 
nonlinear sigma models in the strong gauge coupling limit 
$g^2 \to \infty$. 
With eight supercharges, they become hyper-K\"ahler (HK) 
nonlinear sigma 
models~\cite{Zumino:1979et,Alvarez-Gaume:1981hm,Alvarez-Gaume:1983ab}
on the Higgs branch~\cite{Argyres:1996eh,Antoniadis:1996ra} 
of gauge theories as 
their target spaces. 
This construction of HK manifold is called 
a HK quotient~\cite{Lindstrom:1983rt,Hitchin:1986ea}. 
If hypermultiplets are massless, 
the HK nonlinear sigma models receive no potentials. 
If hypermultiplets have masses, 
the models are called massive HK nonlinear sigma models 
which possess potentials as the square of 
tri-holomorphic Killing vectors on the target 
manifold~\cite{Alvarez-Gaume:1983ab}. 
Most vacua are lifted 
with this potential 
leaving some discrete points as vacua, 
which are characterized by fixed points of the 
Killing vector. 
In our case of $U(N_{\rm C})$ gauge theory with $N_{\rm F}$ 
hypermultiplets in the fundamental representation, the model 
reduces to the massive hyper-K\"ahler nonlinear sigma model on 
$T^* G_{N_{\rm F},N_{\rm C}}$ in equation (\ref{eq:mdl:T*Gr}).  
With our choice of the FI parameters, 
$H^1$ parameterizes the base manifold 
$G_{N_{\rm F},N_{\rm C}}$, 
whereas $H^2$ its cotangent space. 
Thus we obtain the K\"ahler nonlinear sigma model on 
the Grassmann manifold $G_{N_{\rm F},N_{\rm C}}$ 
if we set $H^2 =0$ ~\cite{Higashijima:1999ki}.

Let us give the concrete Lagrangian of the 
nonlinear sigma models. 
Since the gauge kinetic terms for $W_\mu$ and $\Sigma_p$ 
(and their superpartners) disappear in the limit of 
infinite coupling, we obtain the Lagrangian 
\begin{eqnarray}
\mathcal{L}^{g\rightarrow\infty}
=
{\rm Tr}[({\cal D}_\mu H^i)^\dagger {\cal D}^\mu H^i]
+
{\rm Tr}[(H^{i\dagger} \Sigma_p - M_p H^{i\dagger})
(\Sigma_p H^i - H^i M_p)]. 
\label{eq:mdl:reduced-L}
\end{eqnarray}
The auxiliary fields $Y^a$ serve as Lagrange multiplier 
fields to give constraints 
(\ref{eq:mdl:D-term-cond}) as their equations of motion. 
equation (\ref{eq:mdl:reduced-L}) gives equations of 
motion for $W_\mu$ and 
$\Sigma_p$ as auxiliary fields expressible in terms of 
hypermultiplets 
\begin{eqnarray}
 && W^{I}_\mu= i (A^{-1})^{IJ} 
   {\rm Tr}
[(H^i {\partial}_\mu H^{i\dagger}
-{\partial}_\mu H^i H^{i\dagger} )T_J ],  
\label{eq:mdl:constr_gauge}
    \\
 && \Sigma^I_p = 2 (A^{-1})^{IJ} 
  {\rm Tr}(H^{i\dagger} T_J H^i M_p),
 \label{eq:mdl:constraint1}
\end{eqnarray}
where $(A^{-1})^{IJ}$ is an inverse matrix of $A_{IJ}$ 
defined by 
$A_{IJ} = {\rm Tr}(H^{i\dagger} \{ T_I,T_J \} H^i) $. 
As a result the Lagrangian (\ref{eq:mdl:reduced-L}) 
with the constraints (\ref{eq:mdl:D-term-cond}) gives 
the nonlinear sigma model, after eliminating $W_\mu, \Sigma_p$. 
This is the HK nonlinear sigma 
model~\cite{Lindstrom:1983rt,Arai:2003tc} 
on the cotangent bundle over the complex Grassmann manifold 
in equation (\ref{eq:mdl:T*Gr}). 
The isometry of the metric, which is the symmetry 
of the kinetic term, is $SU(N_{\rm F})$, although it 
is broken to its maximal Abelian subgroup 
$U(1)^{N_{\rm F}-1}$ by the potential. 
In the massless limit $M_p=0$, the potential $V$ vanishes 
and the whole manifold becomes vacua, 
the Higgs branch of our gauge theory. 
Turning on the hypermultiplet masses, 
we obtain the potential allowing 
only discrete points as SUSY vacua~\cite{Arai:2003tc}, 
which are fixed points 
of the invariant subgroup $U(1)^{N_{\rm F}-1}$ 
of the potential. 
The number of vacua is 
$N_{\rm F} !/[N_{\rm C} ! (N_{\rm F}-N_{\rm C}) !]$, 
which is the same as the case of the finite gauge coupling. 

In the case of $N_{\rm C} =1$ the target space 
reduces to the cotangent bundle over 
the compact projective space ${\bf C}P^{N_{\rm F}-1}$, 
$T^* {\bf C}P^{N_{\rm F}-1}
=T^* [SU(N_{\rm F})/SU(N_{\rm F} -1) \times U(1)]$ \cite{Curtright:1979yz}. 
This is a toric HK (or hypertoric) manifold and the massive model 
has discrete $N_{\rm F}$ vacua \cite{Gauntlett:2000ib}. 
If $N_{\rm F}=2$ the target space 
$T^*{\bf C}P^1$ is the simplest HK manifold, 
the Eguchi-Hanson space~\cite{Eguchi:1978xp}.

From the target manifold (\ref{eq:mdl:T*Gr}) one can easily 
see that there exists a duality 
between theories with the same number of flavors and 
two different gauge groups 
in the case of the infinite gauge 
coupling~\cite{Argyres:1996eh,Arai:2003tc}: 
\begin{eqnarray}
 U(N_{\rm C}) \leftrightarrow 
 U(N_{\rm F}-N_{\rm C}) .
\label{eq:mdl:duality}
\end{eqnarray}
This duality holds for the entire Lagrangian of the 
nonlinear sigma models. 

\section{1/2 BPS solitons}
\label{hlf}
\subsection{Walls}
\label{wll}
\subsubsection{BPS Equations for Domain Walls} 

Domain walls are static BPS solitons of co-dimension one 
interpolating between different discrete vacua 
like equation (\ref{eq:mdl:color-flavor-locking}). 
In order to obtain domain wall solutions 
we require that all fields should depend on 
one spatial coordinate, say $y \equiv x^2$. 
We also set $H^2=0$ and define $H \equiv H^1$. 
We have shown in 
Appendix B in \cite{Isozumi:2004va} that the condition $H^2=0$ is deduced 
in our model 
(but it is not always the case in general models~\cite{Eto:2005wf}).
The Bogomol'nyi completion of the energy density for 
domain walls can be performed as 
\begin{eqnarray}
{\cal E} 
 &=& {1 \over g^2}{\rm Tr}\left({\cal  D}_y \Sigma -
 {g^2\over 2}\left(c{\bf 1}_{N_{\rm C}} - H H^\dagger 
 \right)\right)^2
\nonumber \\
&&{}
+ {\rm Tr}\left[
({\cal D}_y H + \Sigma H - H M) 
({\cal D}_y H + \Sigma H - H M)^\dagger\right] 
\nonumber \\ && {}
+ c \,\partial_y{\rm Tr}\Sigma 
- \partial_y \left\{{\rm Tr}
\left[ 
\left(\Sigma H - H M\right)H^\dagger
\right]\right\}. 
\label{eq:wll:bogomolnyi-wall}
\end{eqnarray} 
This energy bound is saturated when 
the BPS equations for domain walls are satisfied 
\begin{eqnarray}
D_y H = - \Sigma H + H M, \quad
D_y \Sigma = 
{g^2\over 2}\left(c{\bf 1}_{N_{\rm C}}-H H^\dagger 
\right), 
\label{eq:wll:BPSeq-wall}
\end{eqnarray}
and the energy per unit volume (tension) of domain walls 
interpolating between the vacuum $\langle\{A_r\}\rangle$ at 
$y\rightarrow +\infty$ and the vacuum $\langle\{B_r\}\rangle$ at 
$y\rightarrow -\infty$ 
is obtained as 
\begin{eqnarray}
T_{\rm w} &=& 
\int^{+\infty}_{-\infty}
\hspace{-1.5em}dy \; {\cal E}
=c 
\left[{\rm Tr}\Sigma \right]^{+\infty}_{-\infty}
=c \left(\sum_{k=1}^{N_{\rm C}}m_{A_k}
-\sum_{k=1}^{N_{\rm C}}m_{B_k}\right) .
\label{eq:wll:tension}
\end{eqnarray}
The tension $T_{\rm w}$ depend only on boundary conditions at 
spatial infinities $y \to \pm \infty$ 
and is a topological charge.

\begin{figure}[ht]
\begin{center}
\begin{tabular}{ccc}
\includegraphics[height=6cm]{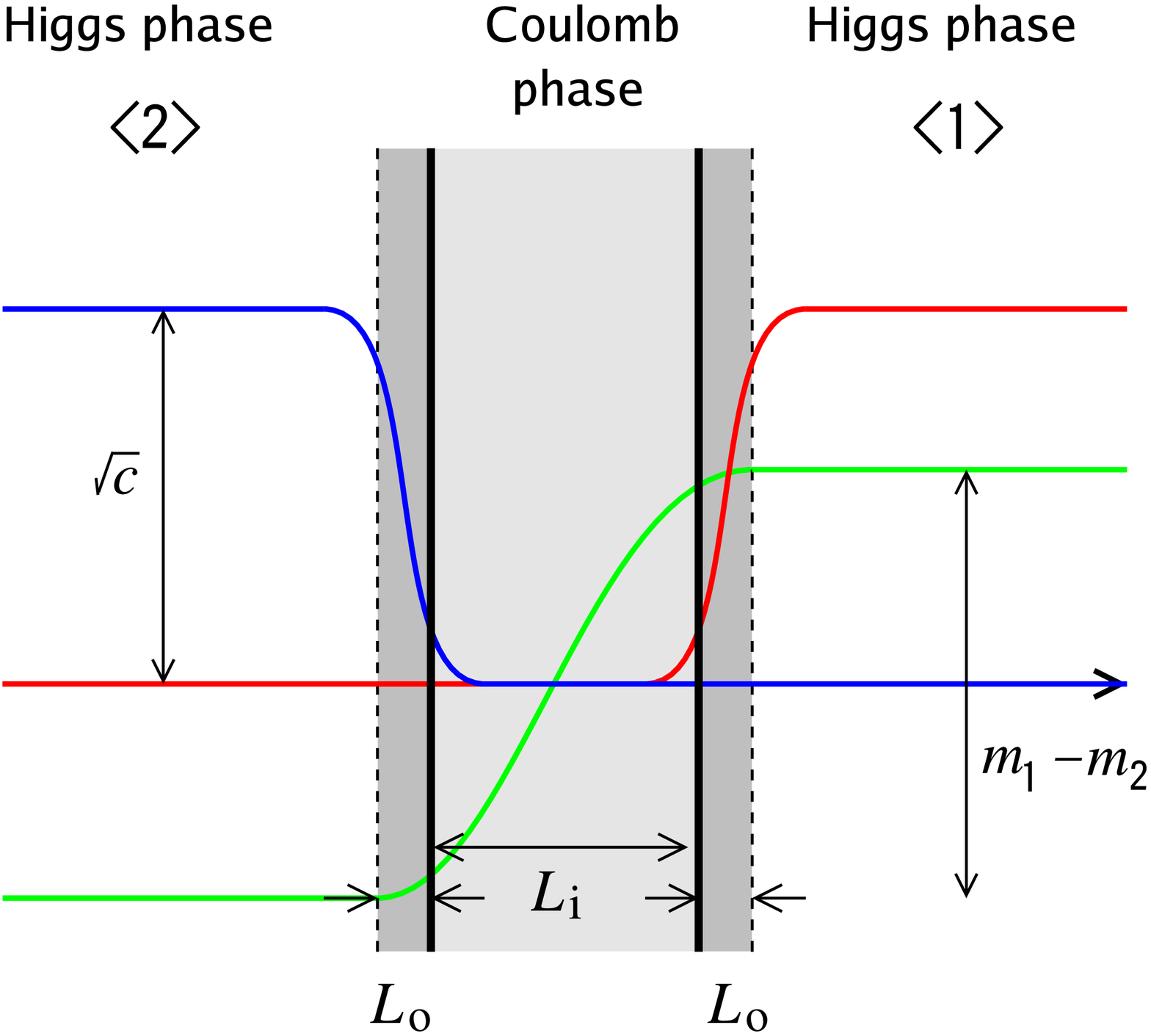}
&\quad&
\includegraphics[height=6cm]{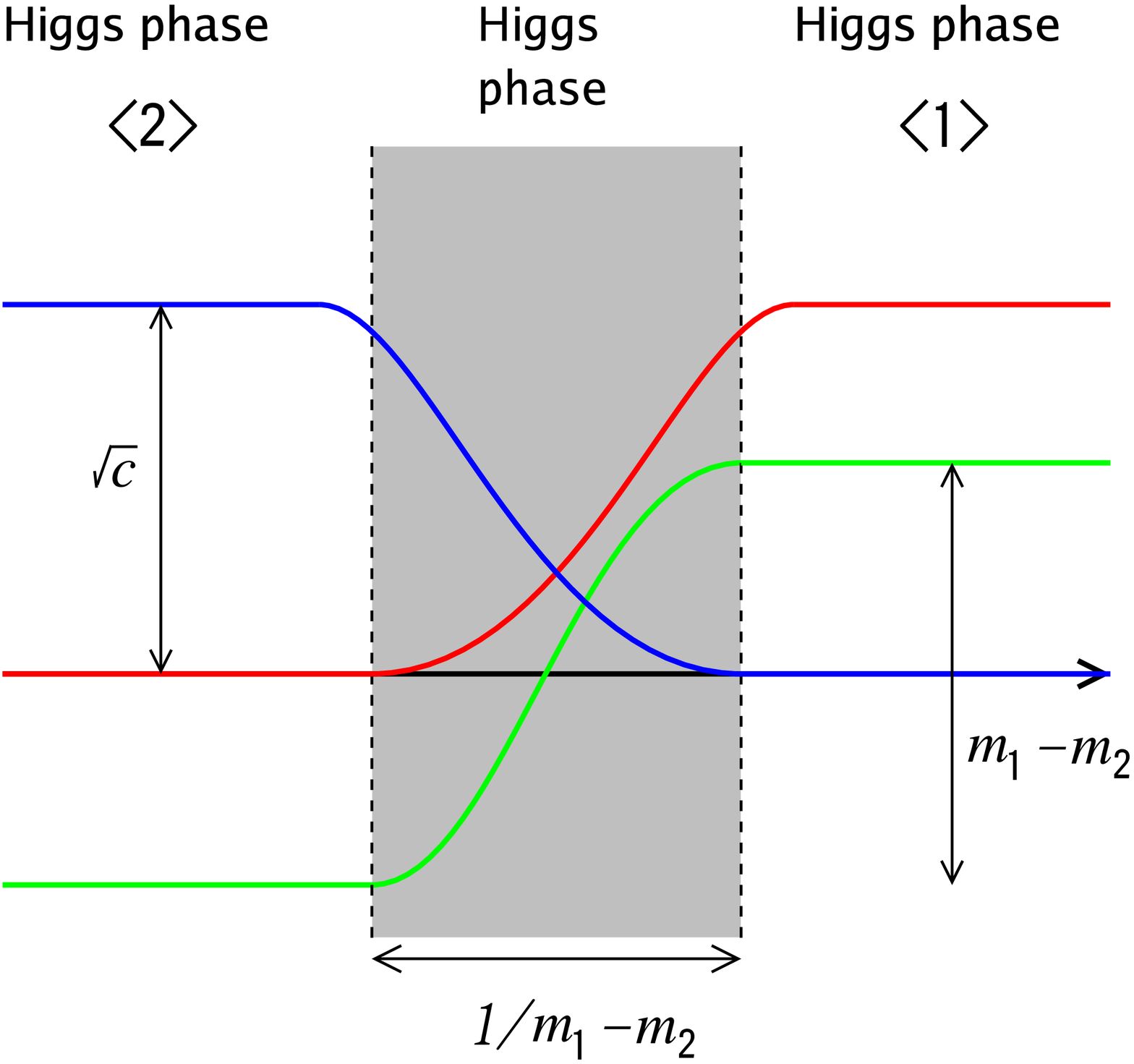}\\
\small{\sf (a) three-layer structure if $g\sqrt c \ll |\Delta m|$} & & 
\small{\sf (b) single-layer structure if $g\sqrt c \gg |\Delta m|$}
\vspace*{-.3cm}
\end{tabular}
\caption{\small{\sf Internal structures of the domain walls.}}
\label{fig:wll:wall}
\vspace*{-.3cm}
\end{center}
\end{figure}

There exists one dimensionless parameter $g\sqrt c/|\Delta m|$ 
in our system.
Depending on whether the gauge coupling constant is 
weak ($g \sqrt c \ll |\Delta m|$ ) 
or strong ($|\Delta m| \ll g \sqrt c$), 
domain walls have different internal structure. 
Let us review  internal structure of 
$U(1)$ gauge theory~\cite{Shifman:2002jm}. 
Walls have a three-layer structure shown in 
Fig.~\ref{fig:wll:wall}(a) 
in weak gauge coupling.
The outer two thin layers have the same width of 
order $L_{\rm o}=1/g\sqrt c$ 
and the internal fat layer has width of order 
$L_{\rm i}=|\Delta m|/g^2c\,(\gg L_{\rm o})$.
The wall in $U(1)$ gauge theory with $N_{\rm F}=2$ interpolating between 
the vacuum 
$\langle 1\rangle\ \left(H=\sqrt{c}(1,0),\ \Sigma = m_1\right)$ 
at $y\to -\infty$ and the vacuum 
$\langle 2\rangle\ \left(H=\sqrt{c}(0,1),\ \Sigma = m_2\right)$ 
at $y\to +\infty$ 
is shown in the Fig.~\ref{fig:wll:wall} (a).  
The first (second) flavor component of the Higgs field 
exponentially decreases 
in the left (right) outer layer 
so that the entire $U(1)$ gauge symmetry is restored in 
the inner core. 

In the strong gauge coupling ($g\sqrt c \gg |\Delta m|$)
the internal structure becomes simpler
for both Abelian and non-Abelian cases. 
The middle layer disappears 
and two outer layers of the Higgs phase grow 
with the total width being of order $1/|\Delta m|$.

Internal structure becomes important 
at finite or weak gauge coupling, 
for instance when we discuss domain wall junction 
\cite{Eto:2005cp, Eto:2005fm} in Sec.~\ref{juc} 
or Skyrmion as instantons inside domain walls \cite{Eto:2005cc}.

\subsubsection{Wall Solutions and Their Moduli Space} 

Let us solve the BPS equations (\ref{eq:wll:BPSeq-wall}). 
Defining an $N$ by $N$ invertible matrix 
$S(y) \in GL(N_{\rm C},{\bf C})$ by the ``Wilson line"  
\begin{eqnarray}
 S(y) \equiv {\bf P} \exp \left(\int dy ( \Sigma + iW_y ) \right)
  \label{eq:wll:Wilson}
\end{eqnarray}
with ${\bf P}$ denoting the path ordering, 
we obtain the relation
\begin{eqnarray}
 \Sigma + iW_y = S^{-1}(y) \partial_y S(y) .
  \label{eq:wll:def-S}
\end{eqnarray} 
Using $S$ the first equation in (\ref{eq:wll:BPSeq-wall}) can be solved as 
\begin{eqnarray}
 H = S^{-1}(y) H_0 e^{My}.
  \label{eq:wll:sol-H} 
\end{eqnarray}
Defining a $U(N_{\rm C})$ gauge invariant 
\begin{eqnarray}
 \Omega \equiv SS^\dagger, 
 \label{eq:wll:omega}
\end{eqnarray} 
the second equation in (\ref{eq:wll:BPSeq-wall}) can be rewritten as 
\begin{eqnarray}
\partial_y\left( \Omega^{-1}\partial_y \Omega\right) = g^2 
  \left(c\, - \Omega^{-1}H_0 \,e^{2My} H_0{}^\dagger \right). 
\label{eq:wll:master-eq-wall}
\end{eqnarray}
We call this the master equation for domain walls. 
This equation is difficult to solve analytically in general. 
\footnote{
Non-integrability of this equation has been addressed 
recently in \cite{Inami:2006wr} by using the 
Painlev\'{e} test.
} 
However it can be solved immediately as 
\begin{eqnarray}
 \Omega_{g \to \infty} 
 \equiv \Omega_0 = c^{-1} H_0 e^{2 M y} H_0{}^\dagger 
  \label{eq:wll:infinite}
\end{eqnarray}
in the strong gauge coupling limit $g^2 \to \infty$,  
in which the model reduces to the HK nonlinear sigma model.
Some exact solutions are also known 
for particular finite gauge coupling 
with restricted moduli parameters~\cite{Isozumi:2003rp}. 
Existence and uniqueness of solutions of 
(\ref{eq:wll:master-eq-wall}) were proved 
for the $U(1)$ gauge group \cite{Sakai:2005kz}. 
One can expect from the index theorem \cite{Sakai:2005sp} 
that it holds for the $U(N_{\rm C})$ gauge group. 

Therefore we conclude that all moduli parameters in wall solutions 
are contained in the moduli matrix $H_0$. 
However one should note that 
two sets $(S, H_0)$ and $(S', H_0{}')$ related by
{\it the V-transformation }
\begin{eqnarray}
 S' = V S,\quad 
 H_0{}'= V H_0 ,\quad 
  V \in GL (N_{\rm C},{\bf C})
  \label{eq:wll:V-equiv-wall}
\end{eqnarray} 
give the same physical 
quantities $W_y$ and $\Sigma$,
where the quantity $\Omega$ transforms as
\begin{eqnarray}
 \Omega'=V\Omega V^\dagger \label{eq:wll:omegaV}
\end{eqnarray}
and the 
equation (\ref{eq:wll:master-eq-wall}) is covariant.    
Thus we need to identify these two as $(S, H_0)\sim$$(S', H_0{}')$,
which we call {\it the V-equivalence relation}. 
The moduli space of the {\it BPS equations} 
(\ref{eq:wll:BPSeq-wall})
is found to be the complex Grassmann manifold 
\begin{eqnarray}
{\cal M}^{\rm total}_{\rm wall}
 &\simeq& \{H_0 | H_0 \sim V H_0, V \in GL(N_{\rm C},{\bf C})\} \nonumber \\ 
 &\simeq& G_{N_{\rm F},N_{\rm C}}
  \simeq {SU(N_{\rm F}) \over 
  SU(N_{\rm C}) \times SU(N_{\rm F} - N_{\rm C}) \times U(1)}\, , 
  \label{eq:wll:Gr}
\end{eqnarray} 
with dimension 
$\dim {\cal M}^{\rm total}_{\rm wall}  
  = 2 N_{\rm C} (N_{\rm F} - N_{\rm C})$. 
We did not put any boundary conditions at $y \to \pm \infty$ 
to get the moduli space (\ref{eq:wll:Gr}).
Therefore it contains configurations with 
all possible boundary conditions, 
and can be decomposed into 
the sum of topological sectors 
\begin{eqnarray}
 {\cal M}^{\rm total}_{\rm wall} 
 = \sum_{\rm BPS} {\cal M}^{\langle A_1,\cdots, A_{N_{\rm C}} \rangle 
                 \leftarrow \langle B_1,\cdots, B_{N_{\rm C}} \rangle}
  \label{eq:wll:decom}
\end{eqnarray}
Here each topological sector  
${\cal M}^{\langle A_1,\cdots, A_{N_{\rm C}} \rangle 
\leftarrow \langle B_1,\cdots, B_{N_{\rm C}} \rangle}$
is specified by the boundary conditions, 
$\langle A_1,\cdots, A_{N_{\rm C}}\rangle$ at $y \to + \infty$ 
and  
$\langle B_1,\cdots, B_{N_{\rm C}}\rangle$ at $y \to - \infty$. 
It is interesting to observe that 
this space also contains vacuum sectors, 
$B_r = A_r$ for all $r$, 
as isolated points because 
these states of course satisfy the BPS equations 
(\ref{eq:wll:BPSeq-wall}). 
More explicit decomposition will be explained 
in the next subsection. 
We often call ${\cal M}^{\rm total}_{\rm wall}$ 
the {\it total moduli space} for domain walls.
One has to note that 
we cannot define the usual Manton's metric 
on the total moduli space 
because it is made by gluing different 
topological sectors. 
The Manton's metric is defined 
in each topological sector. 

We have seen that the total moduli space of domain walls 
is the complex Grassmann manifold (\ref{eq:wll:Gr}).
On the other hand, the moduli space of {\it vacua} for 
the corresponding model with massless hypermultiplet is 
the cotangent bundle over the complex Grassmann manifold 
(\ref{eq:mdl:T*Gr}). 
This is not just a coincidence. 
It has been shown in \cite{Eto:2005wf} that 
the moduli space of domain walls in a massive theory 
is a special Lagrangian submanifold 
the moduli space of vacua in 
the corresponding massless theory.

For any given moduli matrix $H_0$ 
the $V$-equivalence relation (\ref{eq:wll:V-equiv-wall}) 
can be uniquely fixed to 
obtain the following matrix, 
called {\it the standard form}: \\
 \begin{eqnarray}
&&\hspace{3em}_{A_1}
\hspace{1.7em}_{A_2}
\hspace{2.3em}_{A_{N_{\rm C}}} \ {}_{\leftarrow \ }
\hspace{.1em}_{B_1}
\hspace{2.5em}_{B_{N_{\rm C}}}
\hspace{2.5em}_{B_2}\nonumber\\
 H_0&=&\left(
\begin{array}{ccccccccccccc}
 & 1 & * & & \cdots & & * & e^{v_1} & & & & & {\bf O}\\
 & & & 1 & * & & & \cdots & & & * & e^{v_2} &\\
 & & & & & & & \vdots & & & & & \\
{\bf O} & & & & & 1 & * & \cdots & * & e^{v_{N_{\rm C}}} & & &\\
\end{array}\right),
\label{eq:wll:standard-form}
\end{eqnarray}
where $A_r$ is ordered as $A_r < A_{r+1}$ but $B_r$ is not.
Here in the $r$-th row 
the left-most non-zero $(r, A_r)$-elements are fixed to be one,  
the right-most non-zero $(r,B_r)$-elements  
are denoted by 
$e^{v_r} (\in {\bf C}^* \equiv {\bf C} - \{0\} \simeq {\bf R} \times S^1)$.
Some elements between 
them must vanish to fix $V$-transformation (\ref{eq:wll:V-equiv-wall}), 
but some of them denoted by $* (\in {\bf C})$ are 
complex parameters which can vanish. 
(See Appendix B of \cite{Isozumi:2004va} 
for how to fix $V$-transformation completely.)
Substituting the standard form (\ref{eq:wll:standard-form}) into 
the solution (\ref{eq:wll:sol-H}) 
one find that configuration interpolates between 
$\langle A_1,\cdots, A_{N_{\rm C}}\rangle$ at $y \to + \infty$ 
and  
$\langle B_1,\cdots, B_{N_{\rm C}}\rangle$ at $y \to - \infty$. 
In order to obtain the topological sector
${\cal M}^{\langle A_1,\cdots, A_{N_{\rm C}} \rangle
\leftarrow \langle B_1,\cdots, B_{N_{\rm C}} \rangle}$ 
we have to gather matrices in the standard form 
(\ref{eq:wll:standard-form}) with all possible ordering 
of $B_r$. 
We can show that the generic region of 
the topological sector 
${\cal M}^{\langle A_1,\cdots, A_{N_{\rm C}} \rangle 
\leftarrow \langle B_1,\cdots, B_{N_{\rm C}} \rangle}$ 
is covered by the moduli parameters in 
the moduli matrix with ordered $B_r (< B_{r+1})$.  
Therefore its complex dimension is calculated to be 
\begin{eqnarray}
 \dim_{\bf C} {\cal M}^{\left<\left\{A_r\right\}\right> \leftarrow 
\left<\left\{B_r\right\}\right>} 
 =  \sum_{r=1}^{N_{\rm C}} 
    (B_r - A_r).
\label{eq:wll:dim-formula}
\end{eqnarray}
The maximal number of domain walls 
is realized in the maximal topological sector 
${\cal M}^{\langle 1,2,\cdots,N_{\rm C}\rangle 
\leftarrow 
\langle N_{\rm F}-N_{\rm C}+1,
\cdots,N_{\rm F}-1,N_{\rm F}
\rangle}$ 
with complex dimension 
$N_{\rm C} (N_{\rm F} - N_{\rm C})$.

When a moduli matrix contains only one modulus, like 
\begin{eqnarray}
 H_0 = \left( \begin{array}{cccccc}
  1 & 0 & 0 & 0 & 0 & 0\\
  0 & 1 & 0 & 0 & e^{r} & 0\\ 
  0 & 0 & 1 & 0 & 0 & 0
 \end{array}\right) , 
  \label{eq:wll:single}
\end{eqnarray}
we call the configuration generated by this matrix 
as a {\it single wall}.
In particular, we call a single wall generated by
a moduli matrix (\ref{eq:wll:single}) with 
no zeros between $1$ and $e^r$ as 
an {\it  elementary wall}. 
Whereas a single wall 
with some zeros between $1$ and $e^r$ 
is called a {\it composite wall}, 
because it can be broken into 
a set of elementary walls with a moduli deformation.

\subsubsection{Properties} 

In order to clarify the meaning of the moduli parameters  
we explain how to estimate the positions of domain walls from 
the moduli matrix $H_0$ according to Appendix A of \cite{Isozumi:2004va}.
We also show explicit decomposition (\ref{eq:wll:decom}) of 
the total moduli space (\ref{eq:wll:Gr})
by using simple examples in this subsection.

Using $\Omega$ in (\ref{eq:wll:omega}) 
the energy density ${\cal E}$ of domain wall, 
the integrand in equation (\ref{eq:wll:tension}),
can be rewritten as 
\begin{eqnarray}
 {\cal E} 
 = c \partial_y {\rm Tr} \Sigma 
   + {1\over g^2} (\mbox{$\partial_y^4$ term}) 
 = c \partial_y^2 (\log \det \Omega) 
   + {1\over g^2} (\mbox{$\partial_y^4$ term}).
 \label{eq:wll:wall-tension2}
\end{eqnarray}
The $\partial_y^4$ term can be neglected 
when we discuss wall positions.
Apart from the core of domain wall, 
$\Omega$ approach to $\Omega_0 = c^{-1} H_0 e^{2My}H_0^\dagger$ 
in equation (\ref{eq:wll:infinite}).  
There the energy density (\ref{eq:wll:wall-tension2}) 
can be expressed by the moduli matrix as   
\begin{eqnarray}
 {\cal E} &\approx& 
 c \; \partial_y^2 \log 
 \det\left(\frac{1}{c}H_0e^{2My}H_0^\dagger \right) \nonumber\\
&=& 
 c \; \partial_y^2 \log 
 \sum_{\langle\{ A_r\}\rangle}
  \left[ 
  \left| \tau^{\langle\{ A_r\} \rangle} \right|^2
  {\rm exp}
  \left({2\sum_{r=1}^{N_{\rm C}} m_{A_r} y}\right)
  \right]
 . \label{eq:wll:detOmega} 
\end{eqnarray} 
Here the sum is taken over 
all possible vacua
$\langle \{A_r\} \rangle = \langle A_1, A_2, \cdots , A_{N_{\rm C}}\rangle$ 
and $\tau^{\langle \{A_r\} \rangle}$ is defined by 
\begin{eqnarray}
 \tau^{\langle \{A_r\} \rangle}
  \equiv  \exp(a^{\left<\{A_r\}\right>} + i b^{\left<\{A_r\}\right>})
  \equiv \det H_0^{\langle \{A_r\} \rangle}  
  \label{eq:wll:def-tau}
\end{eqnarray}
with $(H_0^{\langle \{A_r\} \rangle})^{st} = H_0^{s A_t}$ 
an $N_{\rm C}$ by $N_{\rm C}$ minor matrix of $H_0$. 
It is useful to define a {\it weight} of 
a vacuum $\langle \{A_r\} \rangle$ 
as $e^{2{\cal W}^{\left<\{A_r\} \right>} }$ with 
\begin{eqnarray}
 {\cal W}^{\left<\{A_r\} \right>} (y) 
  \equiv 
   \sum_{r=1}^{N_{\rm C}}
    m_{A_r}y + a^{\left<\{A_r\}\right>},
      \label{eq:wll:weight} 
\end{eqnarray}
and an {\it average magnitude } of the vacua by 
\begin{eqnarray}
 \exp{2\langle {\cal W}\rangle}\equiv 
\sum_{\left< \{A_r\} \right>} \exp{2 {\cal W}^{\left< \{A_r\} \right>}}
\end{eqnarray}
Then the energy density can be rewritten as 
\begin{eqnarray}
{\cal E} \approx
 c\,\partial_y^2 \langle {\cal W}\rangle= \frac{c}{2} \partial_y^2 
 \log \sum_{\left< \{A_r\} \right>} \exp{2 {\cal W}^{\left< \{A_r\} \right>}}.
 \label{eq:wll:energy} 
\end{eqnarray}
This approximation is valid away from the core of domain walls 
but not good near their core.
This expression holds exactly in the whole region 
at the strong gauge coupling limit. 

It may be useful to order 
$\tau$'s according to 
the sum of masses of hypermultiplets 
corresponding to the labels of flavors, like  
\begin{eqnarray}
 \{ \cdots, \tau^{\langle \{A_r\} \rangle}, \cdots ,
    \tau^{\langle \{B_r\} \rangle}, \cdots\}, 
 \quad \mbox{so that} \quad
 \sum_{r=1}^{N_{\rm C}} m_{A_r} > \sum_{r=1}^{N_{\rm C}} m_{B_r} .
 \label{eq:wll:taus}
\end{eqnarray} 
When only one $\tau$ is nonzero with the rests vanishing as
\begin{eqnarray}
 \{ 0, \cdots, 0, \tau^{\langle \{A_r\} \rangle},0, \cdots ,0 \}, 
 \label{eq:wll:one-tau}
\end{eqnarray} 
only one weight $e^{2{\cal W}^{\left<\{A_r\} \right>}}$ survives 
and the logarithm $\log \det \Omega$ 
inside the $y$-derivative in equation (\ref{eq:wll:energy}) 
becomes linear with respect to $y$. 
Therefore the energy (\ref{eq:wll:energy}) vanishes and 
the configuration is in a SUSY vacuum. 
Next let us consider general situation. 
In a region of $y$ such that 
one ${\cal W}^{\left<\left\{A_r\right\}\right>}$ is larger than the rests, 
$\exp {\cal W}^{\left<\left\{A_r\right\}\right>}$ is dominant in the logarithm 
in equation (\ref{eq:wll:energy}). 
Therefore the logarithm $\log \det \Omega$ 
inside differentiations 
in equation (\ref{eq:wll:energy}) 
is almost linear with respect to $y$, 
the energy (\ref{eq:wll:energy}) vanishes  
and configuration is close to a SUSY vacuum in that region of $y$. 
The energy does not vanish only when 
two or more ${\cal W}^{\left<\left\{A_r\right\}\right>}$'s are comparable. 
If two ${\cal W}^{\left<\left\{A_r\right\}\right>}$'s are comparable 
and are larger than the rests, 
there exists a domain wall.  
This is a key observation throughout this paper.

\medskip
We now discuss the $U(1)$ gauge theory in detail.
There exist $N$ vacua,  
$\left< A \right>$ with $A=1,\cdots,N$.
The moduli matrix and weight are  
\begin{eqnarray}
 H_0 = \sqrt c
 \left(
 \tau^{\left<1\right>},
 \tau^{\left<2\right>},\cdots,
 \tau^{\left<N_{\rm F}\right>}\right) , 
  \label{eq:wll:H0-for-U(1)} \\
 \exp\left(2{\cal W}^{\left< A \right>}(y)\right) 
 = 
 \exp
  2\left(
   m_{A}y + a^{\left<A\right>}
  \right),
 \label{eq:wll:weight_ab}
\end{eqnarray}
respectively, 
with $\tau^{\left<A\right>} \in {\bf C}$ and 
 $a^{\left<A\right>} = {\rm Re}( \log \tau^{\left<A\right>})$.
Here $\tau^{\left<A\right>}$ are regarded as 
the homogeneous coordinates of the total moduli
space ${\bf C}P^{N_{\rm F}-1}$. 
Any single wall is generated by 
a moduli matrix (\ref{eq:wll:weight_ab}) 
with only two non-vanishing $\tau$'s. 
If these two are a nearest pair, 
an elementary wall is generated. 

\begin{figure}[htb]
\begin{center}
\includegraphics[width=6cm,clip]{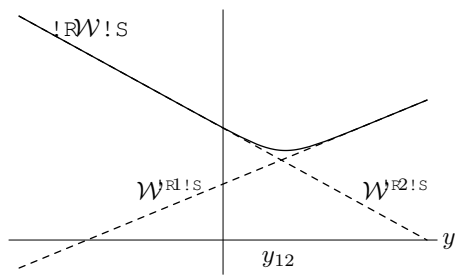}
\caption{ 
Comparison of the profile of 
$\langle {\cal W}\rangle,{\cal W}^{\langle 1 \rangle},
{\cal W}^{\langle 2 \rangle}$ as functions of $y$. 
Linear functions ${\cal W}^{\langle A \rangle}$ are good approximations in their 
respective dominant regions. 
}
\label{fig:wll:position-of-wall0}
\end{center}
\end{figure}

\underline{Example 1: single wall}.
We now restrict ourselves to the simplest case, $N_{\rm F}=2$. 
This model contains two vacua 
$\left< 1 \right>$ and $\left< 2 \right>$ 
allowing one domain wall connecting them. 
The weights of these vacua are 
$e^{2{\cal W}^{\left< 1 \right>}}$ and 
$e^{2{\cal W}^{\left< 2 \right>}}$, respectively. 
When one weight $e^{2{\cal W}^{\left< A \right>}}$ is larger than 
the other, configuration approaches to the vacuum $\langle A \rangle$ 
as seen in figure \ref{fig:wll:position-of-wall0}.
Energy density is concentrated 
around the region where both the weights are comparable. 
Therefore the wall position is determined by equating them, 
\begin{eqnarray}
 y = - 
  {a^{\left< 1 \right>} - a^{\left< 2 \right>}\over
       m_{1} - m_{2} } 
  = - {\log |\tau^{\left< 1\right>}/\tau^{\left< 2\right>}| \over m_{1} - m_{2}} .
  \label{eq:wll:weight_balance_ab}
\end{eqnarray}

We also have the $U(1)$ modulus in the phase of 
$\tau^{\left< 1 \right>}/\tau^{\left< 2 \right>}$ 
which does not affect the shape of the wall.
This is a Nambu-Goldstone mode coming from 
the flavor $U(1)$ symmetry spontaneously broken by 
the wall configuration. 
The moduli space of single wall is a cylinder 
${\cal M}^{k=1} \simeq {\bf R} \times S^1 
\simeq {\bf C}^* \equiv {\bf C} - \{0\}$.
This is non-compact. 
In the limit of $a^{\left< 1 \right>} \to - \infty$
or $a^{\left< 2 \right>} \to - \infty$ 
the configuration becomes a vacuum. 
These limits naturally define how to add two points, 
which correspond to the two vacuum states, 
to ${\cal M}^{k=1}$. 
We thus obtain the total moduli space as a compact space:
\begin{eqnarray}
 {\bf C}P^1 \simeq S^2 
 = {\cal M}^{k=1} + \mbox{two points} 
 = {\bf R} \times S^1 + \mbox{two points}.
  \label{eq:wll:CP1}
\end{eqnarray}
This is an explicit illustration of the 
decomposition (\ref{eq:wll:decom}) 
of the total moduli space.

\begin{figure}
\begin{center}
\includegraphics[width=5cm,clip]{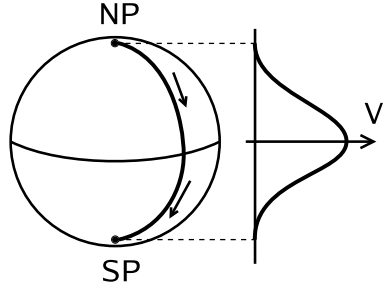}
\caption{ ${\bf C}P^1$ and the potential $V$.
The base space of $T^*{\bf C}P^1$, 
${\bf C}P^1 \simeq S^2$, is displayed.  
This model contains two discrete vacua denoted by 
$N$ and $S$.  
The potential $V$ is also displayed on 
the right of the ${\bf C}P^1$.
It admits a single wall solution 
connecting these two vacua expressed by a curve.
The $U(1)$ isometry around the axis connecting 
$N$ and $S$ is spontaneously broken by the wall configuration.
}
\label{fig:wll:CP1}
\end{center}
\end{figure}

In the strong gauge coupling limit, 
the model reduces to a nonlinear sigma model 
on $T^* {\bf C}P^1$, the Eguchi-Hanson space, 
allowing a single domain wall~\cite{Abraham:1992vb,Arai:2002xa,Arai:2003es}. 
In figure \ref{fig:wll:CP1} we display  
the base space ${\bf C}P^1$ of the target space, 
the potential $V$ on it, two vacua $N$ and $S$, 
and a curve in the target space mapped from 
a domain wall solution connecting these vacua. 

\begin{figure}[htb]
\begin{center}
\begin{tabular}{ccc}
\includegraphics[width=6cm,clip]{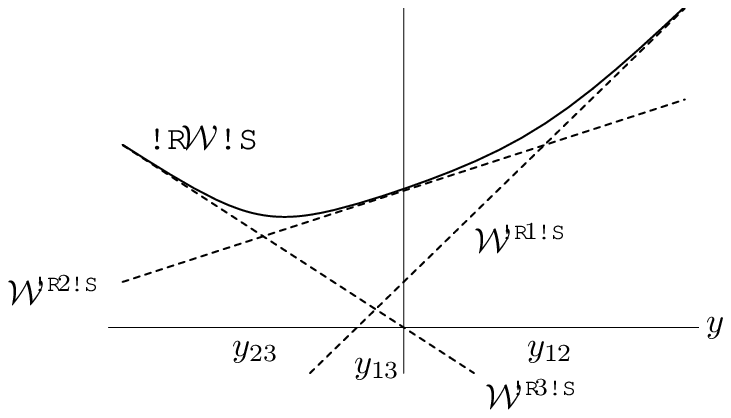}
&\quad&
\includegraphics[width=6cm,clip]{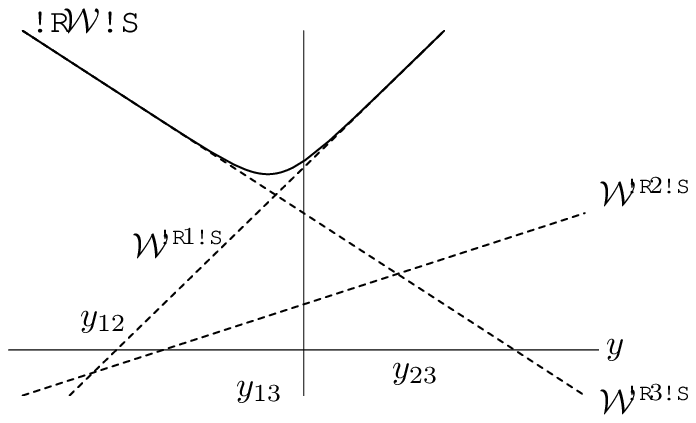}\\
\small{\sf a) $y_{23}<y_{12}$} & & 
\small{\sf b) $y_{23}>y_{12}$}
\vspace*{-.3cm}
\end{tabular}
\caption{
Comparison of the profile of 
$\langle {\cal W}\rangle,{\cal W}^{\langle 1 \rangle},
{\cal W}^{\langle 2 \rangle}$ as functions of $y$. 
$\langle {\cal W}\rangle$ connects smoothly 
dominant linear functions ${\cal W}^{\langle A \rangle}$
in respective regions. 
}
\label{fig:wll:position-of-wall}
\end{center}
\end{figure}

\medskip
\underline{Example 2: double wall}. (Appendix A of \cite{Isozumi:2004va})
Let us switch to the second simplest case, $N_{\rm F}=3$. 
This model contains three vacua 
$\left< 1 \right>$, $\left< 2 \right>$ and 
$\left< 3 \right>$ whose weights are 
$e^{2{\cal W}^{\left< 1 \right>}}$, $e^{2{\cal W}^{\left< 2 \right>}}$ 
and $e^{2{\cal W}^{\left< 3 \right>}}$, respectively. 
This model admits three single walls 
generated by the moduli matrices 
\begin{eqnarray}
 H_0 = \sqrt c
 \left(
 \tau^{\left<1\right>},
 \tau^{\left<2\right>},0\right), \quad 
 \sqrt c
 \left(0,
 \tau^{\left<2\right>},
 \tau^{\left<3\right>}\right), \quad 
 \sqrt c
 \left(
 \tau^{\left<1\right>},0,
 \tau^{\left<3\right>}\right). 
 \label{eq:wll:three-moduli}
\end{eqnarray} 
We now show that 
the first two are elementary wall while 
the last is not elementary but 
a composite of the first two, 
as defined below equation (\ref{eq:wll:single}). 
Let us consider the full moduli matrix 
$H_0 = \sqrt c
 \left(
 \tau^{\left<1\right>},\tau^{\left<2\right>},\tau^{\left<3\right>}
\right)$. 
By equating two of the three weights 
we have three solutions of $y$: 
\begin{eqnarray}
 y_{12}  = - {a^{\left<1\right>} - a^{\left<2\right>} \over m_1-m_2},\quad 
 y_{23}  = - {a^{\left<2\right>} - a^{\left<3\right>} \over m_2-m_3},\quad 
 y_{13}  = - {a^{\left<1\right>} - a^{\left<3\right>} \over m_1-m_3}.
  \label{eq:wll:positions}
\end{eqnarray}
Not all of these correspond to wall positions. 
To see this we draw the three linear functions 
${\cal W}^{\left< A \right>}$ and 
$\langle {\cal W}\rangle =$$1/2\log\left(
   e^{2{\cal W}^{\left< 1 \right>}(y)}
 + e^{2{\cal W}^{\left< 2 \right>}(y)}
 + e^{2{\cal W}^{\left< 3 \right>}(y)}\right)$ 
in figure \ref{fig:wll:position-of-wall} according to 
the two cases a) $y_{23} < y_{12}$ and b) $y_{12} < y_{23}$.  
We observe that there exist two domain walls in 
the case a) $y_{23} < y_{12}$ but 
only one wall in 
the case b) $y_{12} < y_{23}$.  
By taking a limit $a^{\left<1\right>} \to - \infty$ 
($e^{2{\cal W}^{\left< 1 \right>}} \to 0$)
or $a^{\left<3\right>} \to - \infty$ 
($e^{2{\cal W}^{\left< 3 \right>}} \to 0$),  
we obtain a configuration of a single wall 
located at $y_{23}$ or $y_{12}$, respectively. 
They are configurations of elemantary walls 
generated by the first two moduli matrices in 
equation (\ref{eq:wll:three-moduli}). 
The configuration a) in figure \ref{fig:wll:position-of-wall}  
is the case that these two walls appraoch each other with finite distance.   
If these two walls get close further we obtain the configuration b) in figure. 
This looks almost a single wall. 
In the limit 
 $a^{\left<2\right>} \to - \infty$ 
($e^{2{\cal W}^{\left< 2 \right>}} \to 0$), 
the configuration really becomes a single wall generated 
by the last moduli matrix of 
(\ref{eq:wll:three-moduli}). 
Therefore the last one generate a composite wall 
made of two elementary walls compressed.
This is a common feature when 
 Abelian dowain walls interact. 

The moduli space of the double wall 
is ${\cal M}^{k=2} \simeq {\bf C}^* \times {\bf C}$ 
with ${\bf C}^*$ denoting the overall position and phase and 
${\bf C}$ denoting relative ones.  
This is non-compact. 
In the two limits $a^{\left<1\right>} \to - \infty$ 
and $a^{\left<3\right>} \to - \infty$ which we took above, 
the configuration reach two single-wall sectors 
${\cal M}^{k=1}_{(1)}$ and 
${\cal M}^{k=1}_{(2)}$ both of which 
are isomorphic to ${\bf C}^*$. 
These limits naturally define gluing 
the two single-wall sectors to 
the double wall sector ${\cal M}^{k=2}$. 
Finally the three vacuum sectors are 
added to it, 
resulting the total moduli space ${\bf C}P^2$: 
\begin{eqnarray}
 {\bf C}P^2 = {\cal M}^{k=2} + {\cal M}^{k=1}_{(1)} 
 + {\cal M}^{k=1}_{(2)} + \mbox{three points}.
 \label{eq:wll:CP2}
\end{eqnarray}
This is another explicit illustration of 
the decomposition (\ref{eq:wll:decom}) 
of the total moduli space. 

Note that the function 
$\log \sum_{\left<A\right>} e^{2{\cal W}^{\left<A\right>}}$ 
in equation (\ref{eq:wll:energy}) 
can be approximated by piecewise linear function 
obtained by the largest weight ${\cal W}^{\left<A\right>}$ 
in each region of $y$, 
as seen in (\ref{fig:wll:position-of-wall}).   
This is known as tropical geometry in mathematical literature.

\medskip
The $U(1)$ gauge theory ($N_{\rm C}=1$) with $N_{\rm F}$ flavors 
admits the $N_{\rm F}$ vacua and the $N_{\rm F}-1$ walls which are ordered. 
\begin{eqnarray}
 {\bf C}P^{N_{\rm F}-1} 
 = \sum_{k=1}^{N_{\rm F}-1} \sum_{i_k} {\cal M}^{k}_{(i_k)}.
\label{eq:wll:CPNF-1}
\end{eqnarray}
We now make a comment on symmetry properties 
of domain walls. 
In the Abelian case with $N_{\rm F}$, 
the number of walls are $N_{\rm F}-1$. 
Each wall carries 
approximate Nambu-Goldstone modes for 
translational invariance if they are well separated.
Only the overall translation is an exact Nambu-Goldsonte mode. 
They carry Nambu-Goldstone modes 
for spontaneously broken $U(1)^{N_{\rm F}-1}$ flavor symmetry.

\medskip
Next let us turn our attention to non-Abelian gauge theory 
($N_{\rm C} > 1$). 
We have defined single walls, elementary walls and 
composite walls below equation (\ref{eq:wll:single}). 
However these definitions are not covariant 
under the $V$-transformation (\ref{eq:wll:V-equiv-wall}). 
They can be defined covariantly as follows.  
To this end we first should note that 
$\tau$'s defined in equation (\ref{eq:wll:def-tau}) 
are the so-called Pl\"ucker coordinates 
of the complex Grassmann manifold.
These coordinates 
$\left\{\tau^{\left<\{A_r\}\right>}\right\}$ 
are not independent but 
satisfy the so-called Pl\"ucker relations 
\begin{eqnarray}
 \sum_{k=0}^{N_{\rm C}} (-1)^k
 \tau^{\langle A_1\cdots A_{N_{\rm C}-1}B_k\rangle} 
 \tau^{\langle B_0\cdots 
 \underline{B_k}\cdots B_{N_{\rm C}}\rangle} 
 =0
\label{eq:wll:plucker}
\end{eqnarray}
where the bar under $B_k$ denotes removing $B_k$ from 
$\langle B_0\cdots B_k\cdots B_{N_{\rm C}}\rangle$. 
Among these equations,  
only $_{N_{\rm F}}C_{N_{\rm C}} - 1 - N_{\rm C} (N_{\rm F}-N_{\rm C})$ 
equations give independent constraints 
with reducing the number of independent coordinates to 
the complex dimension $N_{\rm C} (N_{\rm F}-N_{\rm C})$ 
of the Grassmann manifold.

Using the Pl\"ucker coordinates,
single walls are defined to be configurations 
generated by two non-vanishing $\tau$'s 
with the rests vanishing. 
These are configurations 
interpolating between two vacua 
$\left<\ \underline{\cdots}\ A \right>$ and
$\left<\ \underline{\cdots}\ B \right>$ $(A\neq B)$,
where
underlined dots denote the same set of labels. 
We can show that 
the Pl\"ucker relations (\ref{eq:wll:plucker}) 
do not forbid these configurations. 
If the labels are different by one the configurations 
are said to be elementary walls, 
whereas if 
they are different by more than one
the configurations are said to be composite walls. 
On the other hand, 
the Pl\"ucker relations (\ref{eq:wll:plucker}) 
forbid configurations to interpolate between 
two vacua whose
labels have two or more different integers, 
like $\left<\ \underline{\cdots}\ 123\right>$ and 
$\left<\ \underline{\cdots}\ 456\right>$.

\underline{Example}. 
Let us consider the simplest model 
of $N_{\rm F}=4$ and $N_{\rm C}=2$ with one nontrivial 
Pl\"ucker relation. 
This model contains the six vacua, 
$\langle 12 \rangle$, 
$\langle 13 \rangle$, 
$\langle 14 \rangle$, 
$\langle 23 \rangle$, 
$\langle 24 \rangle$ and
$\langle 34 \rangle$. 
The Pl\"ucker relation (\ref{eq:wll:plucker}) becomes 
\begin{eqnarray}
 \tau^{\langle 12\rangle} \tau^{\langle 34\rangle} 
 - \tau^{\langle 13\rangle} \tau^{\langle 24\rangle}
 + \tau^{\langle 14\rangle} \tau^{\langle 23\rangle} 
 =0.  
 \label{eq:wll:plucker42}
\end{eqnarray}
This allows, for example,
$\tau^{\left<12\right>}$ and $\tau^{\left<13\right>}$ 
to be non-vanishing with the rests vanishing. 
So we have a single wall connecting 
$\langle 12 \rangle$ and 
$\langle 13 \rangle$. 
However, when all $\tau$'s except for 
$\tau^{\left< 12 \right>}$ and $\tau^{\left<34 \right>}$ vanish, 
the Pl\"ucker relation (\ref{eq:wll:plucker42})
reduces to $\tau^{\langle 12\rangle} \tau^{\langle 34\rangle} =0$, 
which requires one of them also to vanish. 
We thus see that there exits 
no domain wall interpolating between 
two vacua $\left< 12 \right>$ and $\left<34 \right>$.

\medskip
Configurations of the single domain walls can also be estimated by
comparing weights of the two vacua as those in the Abelian gauge theory:
The domain wall interpolating $\left<\ \underline{\cdots}\ A \right>$ 
and $\left<\ \underline{\cdots}\ B \right>$ is
given by
${\cal W}^{\left< \underline{\cdots}A \right>} = 
{\cal W}^{\left< \underline{\cdots}B \right>}$.
Then we again obtain the same transition as 
equation (\ref{eq:wll:weight_balance_ab})
\begin{eqnarray}
 y
 = - {a^{\left<\underline{\cdots}A \right>} 
    - a^{\left<\underline{\cdots}B \right>} 
      \over m_{A} - m_{B}} .
\label{eq:wll:weight_balance}
\end{eqnarray}

Of course the Pl\"ucker relations 
(\ref{eq:wll:plucker42}) can 
forbid a set of three or more than three 
$\tau$'s to be non-vanishing with the rests vanishing.
In other words, if it is allowed by
the Pl\"ucker relations 
(\ref{eq:wll:plucker42}), 
that configuration can be realized.

We make several comments on 
characteristic properties of 
domain walls in non-Abelian gauge theory.

Unlike the case of $U(1)$ gauge theory, 
all of moduli are not 
(approximate) Nambu-Goldstone (NG) modes. 
There exist $N_{\rm C} (N_{\rm F} - N_{\rm C})$ walls. 
They carry 
approximate NG modes 
for translational symmetry with 
the overall being an exact NG mode, 
if they are well separated. 
Only $N_{\rm F}-1$ phases are 
NG modes for spontaneously broken 
$U(1)^{N_{\rm F}-1}$ flavor symmetry. 
However 
the rests $N_{\rm C} (N_{\rm F} - N_{\rm C}) - N_{\rm F}+1$ 
are not related with any symmetry, 
but are required by unbroken SUSY. 
These additional modes are called 
quasi-NG modes in the context 
of spontaneously broken global symmetry 
with keeping SUSY~\cite{Bando:1983ab,Higashijima:1999ki}. 

It may be worth to point out that 
a gauge field $W_y$ in co-dimensional direction 
can exist in wall configurations in non-Abelian gauge theory,
unlike the Abelian cases where it can be eliminated 
by a gauge transformation. 
See reference \cite{Isozumi:2004va} in detail.

In the strong coupling limit 
exact duality relation holds, 
$N_{\rm C} \leftrightarrow N_{\rm F} - N_{\rm C}$ 
in equation (\ref{eq:mdl:duality}). 
This relation can be promoted to wall solutions 
as shown in Appendix D in \cite{Isozumi:2004va}.
Although this duality is not exact for finite coupling 
there still exists a one-to-one dual map 
by the relation 
\begin{eqnarray}
 H_0 \tilde H_0^\dagger = 0 
 \label{eq:wll:duality}
\end{eqnarray}
among the moduli matrix $H_0$ in the original theory 
and the $(N_{\rm F} - N_{\rm C}) \times N_{\rm F}$ 
moduli matrix $\tilde H_0$ 
of the dual theory. 
This relation determines $\tilde H_0$ uniquely 
from $H_0$ up to the $V$-equivalence (\ref{eq:wll:V-equiv-wall}).

\subsubsection{D-Brane Configuration}

We found the ordering rules of non-Abelian domain walls 
in \cite{Isozumi:2004va}. 
In this subsection we show that these rules can be obtained easily from 
D-brane configuration in string theory \cite{Eto:2004vy}. 
This configuration was obtained by generalizing 
the one for the U(1) gauge theory considered in 
\cite{Lambert:1999ix}. 
We restrict dimensionality to $d=3+1$ in this subsection, 
but we can consider from dimension $d=1+1$ to $d=4+1$ 
by taking T-duality.
We realize our theory with 
gauge group $U(N_{\rm C})$ 
on $N_{\rm C}$ D$3$-branes 
with the background of $N_{\rm F}$ D$7$-branes; 
\begin{eqnarray}
 \mbox{$N_{\rm C}$ D$3$:} && 0123     \nonumber\\
 \mbox{$N_{\rm F}$ D$7$:} && 01234567 \nonumber\\
 \mbox{${\bf C}^2/{\bf Z}_2$ ALE:}  && \hspace{8mm} 4567 .
  \label{eq:wll:D3-D7}
\end{eqnarray}
A string connecting D$3$-branes provides the gauge multiplets 
whereas a string connecting D$3$-branes and D$7$-branes 
provides the hypermultiplets in the fundamental representation. 
In order to get rid of adjoint hypermultiplet 
we have divided four spatial directions of their world volume 
by ${\bf Z}_2$ to form 
the orbifold ${\bf C}^2 / {\bf Z}_2$.
The orbifold singularity is blown up to the Eguchi-Hanson space 
by $S^2$ with the area 
\begin{eqnarray}
 A = c g_s l_s^{4} = {c \over \tau_3}  \label{eq:wll:S2area}
\end{eqnarray}
with $g_s$ the string coupling, 
$l_s=\sqrt {\alpha'}$ the string length and 
$\tau_{3} = 1/ g_s l_s^{4}$ 
the D$3$-brane tension.  
Our D$3$-branes are 
fractional D$3$-branes
that is, D$5$-branes wrapping around $S^2$.  
The gauge coupling constant $g$ of the gauge theory realized on the 
D$3$-brane is 
\begin{eqnarray}
 {1 \over g^2} 
   = {b \over \, g_s } \label{eq:wll:g^2}
\end{eqnarray}
with 
$b$ the $B$-field flux integrated over the $S^2$, 
$b \sim A B_{ij}$. 
The positions of the D$7$-branes in the $x^8$-direction 
gives the masses for the fundamental hypermultiplets 
whereas the positions of the D$3$-branes in the $x^8$-direction 
is determined by the VEV of $\Sigma$ 
(when $\Sigma$ can be diagonalized $\Sigma = {\rm diag} \Sigma_{rr}$): 
\begin{eqnarray} 
   x^8|_{A-{\rm th \; D}7} 
       = l_s^2 m_A  , \quad 
   x^8|_{r-{\rm th \; D}3} 
   = l_s^2 \Sigma_{rr} (x^1) . 
\end{eqnarray}

Any D$3$-brane 
must lie in a D$7$-brane 
as vacuum states, but   
at most one D$3$-brane  
can lie in each D$7$-brane 
because of the s-rule \cite{Hori:1997ab}. 
Therefore the vacuum 
$\langle A_1,\cdots, A_{N_{\rm C}}\rangle$ 
is realized with $A_r$ denoting 
positions of D$3$-branes, 
and the number of 
vacua is ${}_{N_{\rm F}} C_{N_{\rm C}}$ 
with reproducing field theory. 

\begin{figure}[thb]
\begin{center}
\includegraphics[width=12cm,clip]{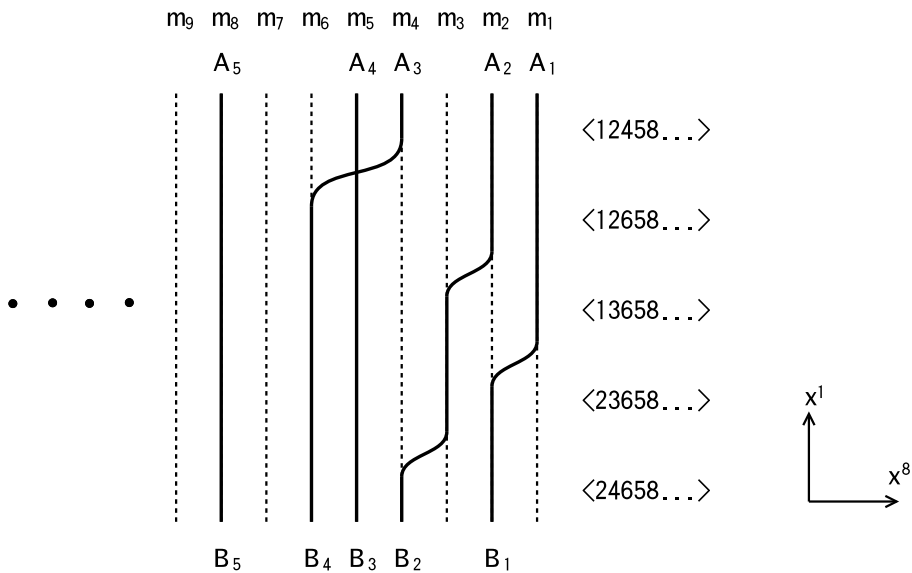}
\caption{\small
Multiple non-Abelian walls as kinky D-branes. 
}
\label{fig:wll:fig9}
\end{center}
\end{figure}

\begin{figure}[thb]
\begin{center}
\begin{tabular}{ccccc}
\includegraphics[width=3.8cm,clip]{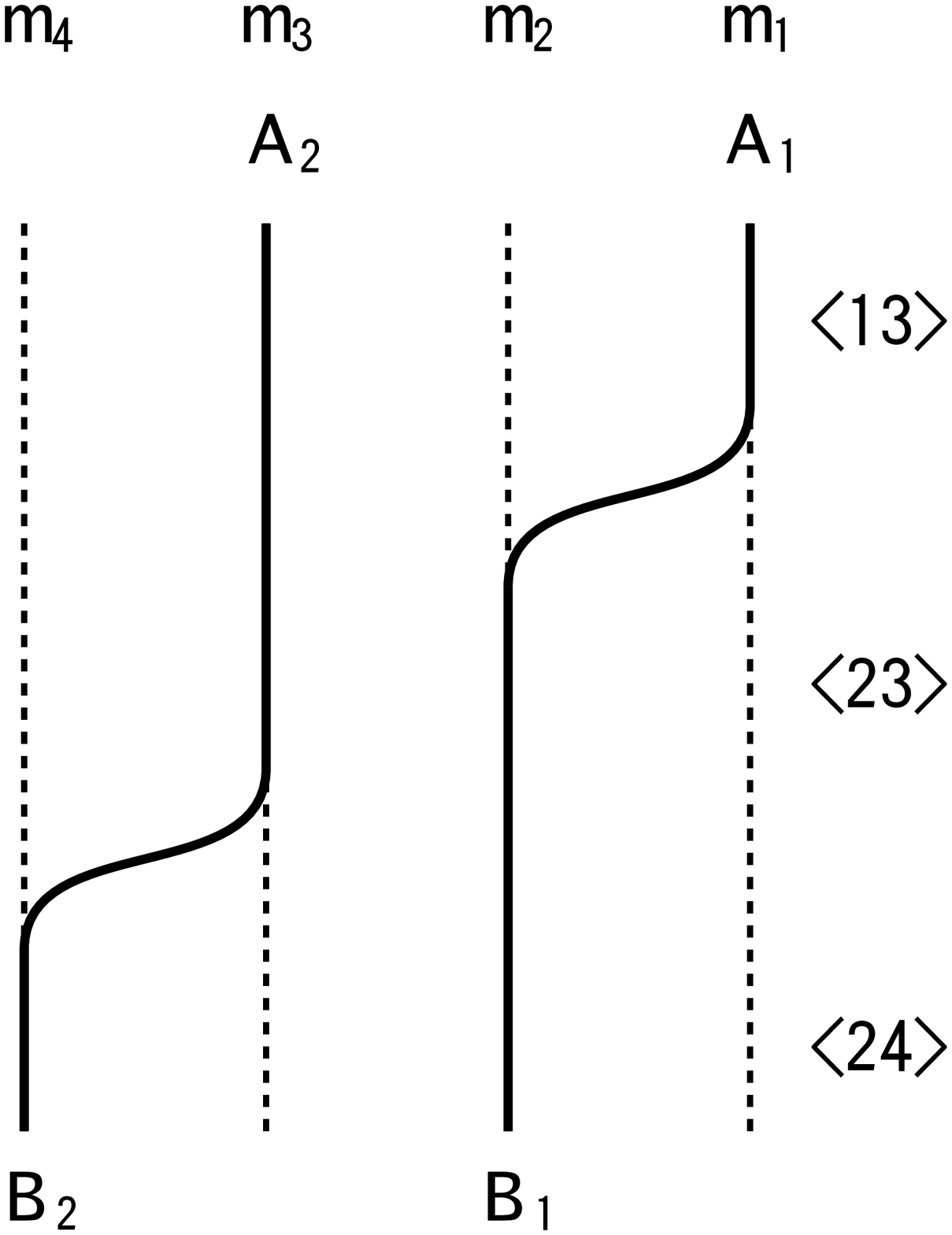}
 & 
\includegraphics[width=3.8cm,clip]{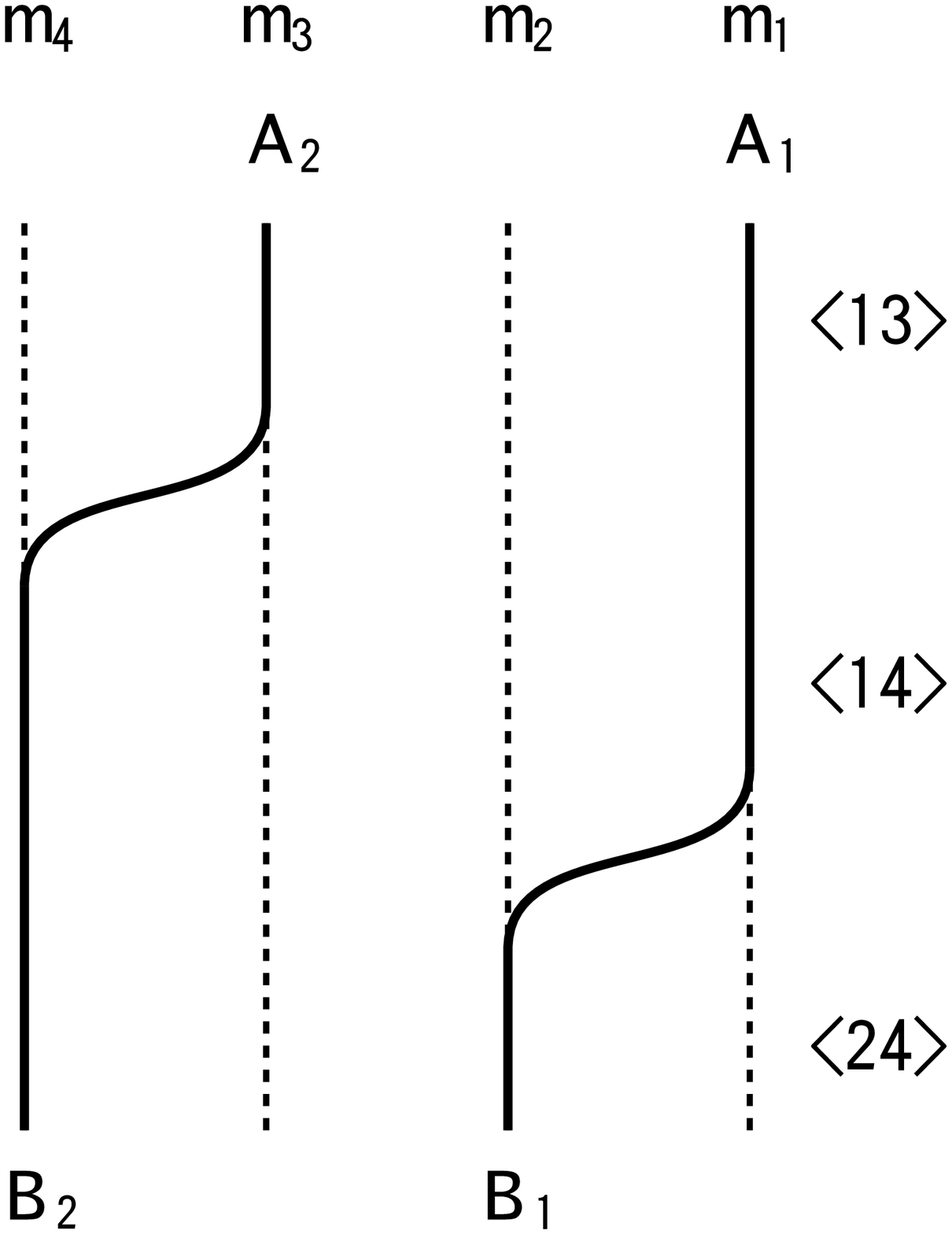} 
& \hspace{1cm} &
\includegraphics[width=2.8cm,clip]{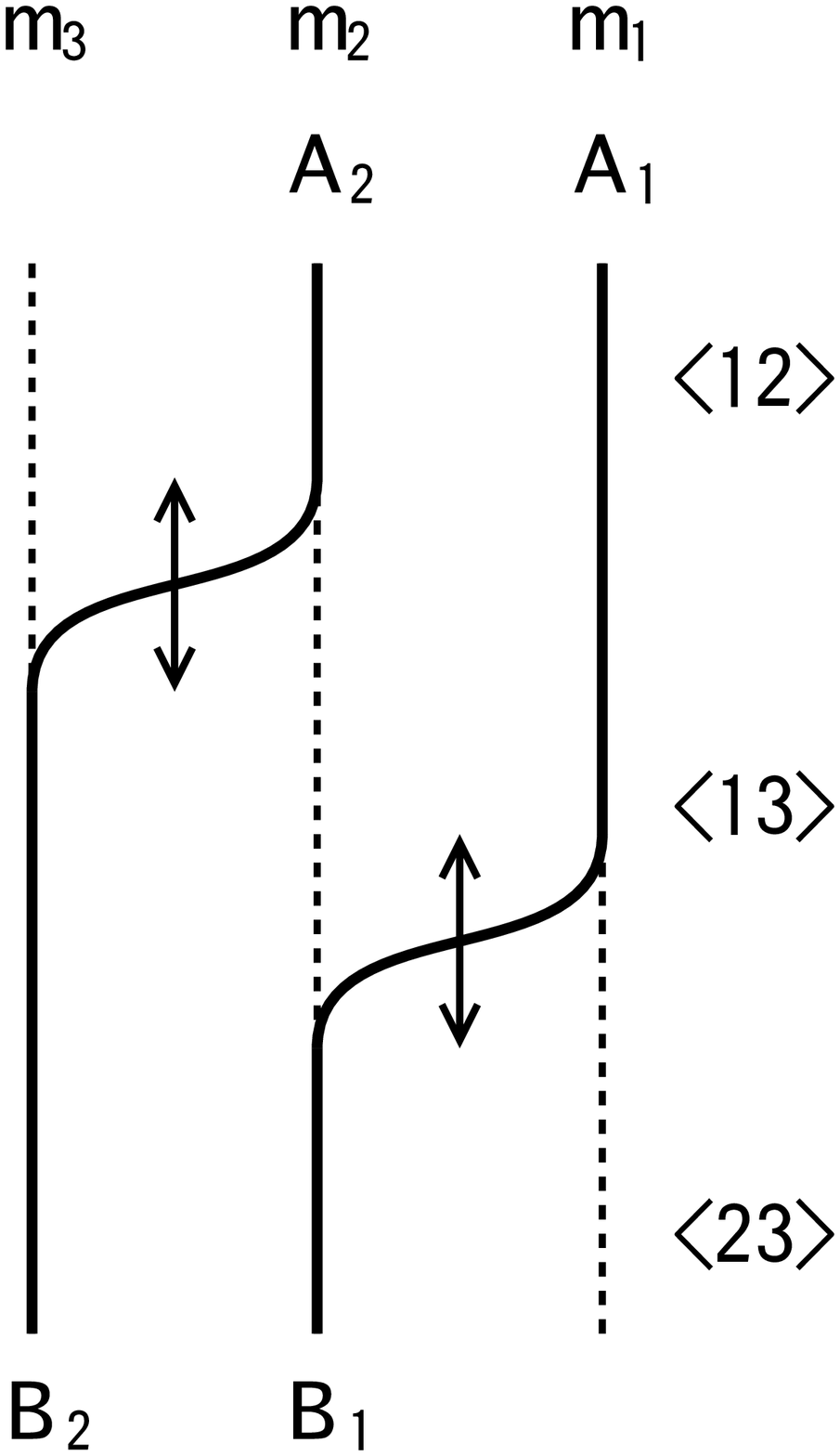}
&
\includegraphics[width=2.8cm,clip]{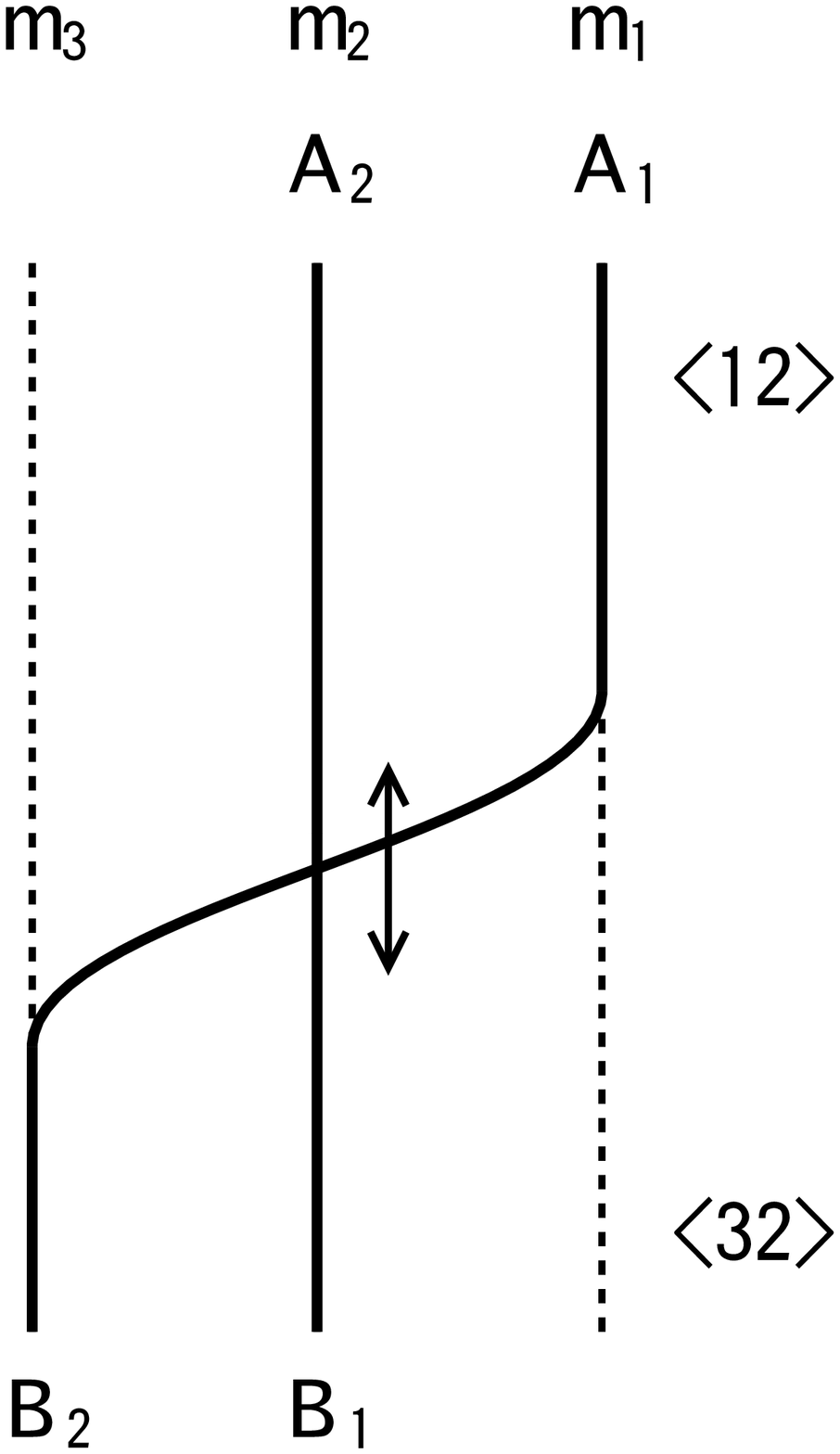}
\\
(a) & & & (b) & 
\end{tabular}
\caption{\small
(a) Penetrable walls in $N_{\rm F}=4$ and $N_{\rm C}=2$ 
and (b) Impenetrable walls $N_{\rm F}=3$ and $N_{\rm C}=2$. 
}
\label{fig:wll:fig13}
\end{center}
\end{figure}

As domain wall states,  
$\Sigma$ depends on 
one coordinate $y \equiv x^1$. 
All D$3$-branes lie in a set of 
$N_{\rm C}$ out of $N_{\rm F}$ 
D$7$-branes in the limit 
$y \to + \infty$, 
giving $\langle A_1,\cdots, A_{N_{\rm C}}\rangle$,
but lie in another set of D$7$-branes 
in the opposite limit $y \to - \infty$,  
giving another vacuum 
$\langle B_1,\cdots, B_{N_{\rm C}}\rangle$. 
The $N_{\rm C}$ D$3$-branes exhibit kinks somewhere 
in the $y$-coordinate  
as illustrated in figure \ref{fig:wll:fig9}. 
Here we labeled $B_r$ such that 
the $A_r$-th brane at $y \to + \infty$ goes to the $B_r$-th brane 
at $y \to - \infty$. 
If we separate adjacent walls far enough 
the configuration between these walls 
approach a vacuum  as illustrated in the right of 
figure \ref{fig:wll:fig9}. 
These configurations clarify 
dynamics of domain walls easily. 
In non-Abelian gauge theory 
two domain walls can penetrate each other 
if they are made of separated 
D$3$-branes like figure \ref{fig:wll:fig13} (a)
but they cannot  
if they are made of adjacent 
D$3$-branes like figure \ref{fig:wll:fig13} (b). 
In the latter case, 
reconnection of D$3$-branes 
occur in the limit that two walls are compressed.

Taking a T-duality along the $x^4$-direction 
in the configuration (\ref{eq:wll:D3-D7}), 
the ALE geometry is mapped to two NS$5$-branes 
separated in the $x^4$-direction. 
The configuration becomes 
the Hanany-Witten type brane configuration \cite{Hanany:1996ie}
\begin{eqnarray}
 \mbox{$N_{\rm C}$ D$4$:} && 01234     \nonumber\\
 \mbox{$N_{\rm F}$ D$6$:} && 0123 \hspace{2mm} 567 \nonumber\\
 \mbox{$2$ NS$5$:}  && 0123 \hspace{8.5mm} 89 . 
  \label{eq:wll:HWset1}
\end{eqnarray}
The relations between the positions of branes and 
physical quantities in field theory on D$4$-branes are 
summarized as  
\begin{eqnarray} 
 && x^8|_{r-{\rm th \; D}4} 
        = l_s^2 \Sigma_{rr} (x^1), \nonumber\\
 && x^8|_{A{\rm -th \; D}6} = l_s^2 m_A, \nonumber \\
 && \Delta x^4|_{{\rm NS}5} = {g_s l_s \over g^2}, \quad
  (\Delta x^5, \Delta x^6, \Delta x^7) |_{{\rm NS}5}
  = g_s l_s^3 (0,0,c).  \label{eq:wll:quantities}
\end{eqnarray}
D-brane configurations of domain walls 
are obtained completely parallel to 
the configuration before taking the T-duality.
However this configuration has some merits. 
First the strong gauge coupling limit corresponds to 
zero separation $\Delta x^4 =0$ of two NS$5$-brane along $x^4$. 
In that limit, the duality (\ref{eq:mdl:duality}) 
becomes exact \cite{Eto:2004vy} due to
the Hanany-Witten effect \cite{Hanany:1996ie}. 
By using this configuration in the strong gauge coupling limit, 
it has been shown in \cite{Hanany:2005bq} 
that the domain wall moduli space 
has half properties of the monopole moduli space 
and that the former can be described by 
the half Nahm construction.

If we put D7(D6)-branes separated along the $x^9$-direction  
in configurations before (after) taking T-duality, 
complex masses of hypermultiplets appear. 
We can consider configuration with 
$\Sigma$ and one more scalar 
depending on $x^2$ as well as $x^1$ 
as a 1/4 BPS state \cite{Eto:2005mx}. 
That is a domain wall junction discussed in section \ref{juc}.

\subsubsection{More General Models}  \quad
We have considered non-degenerate masses of hypermultiplets so far. 
If we consider (partially) degenerate masses more interesting 
physics appear \cite{Shifman:2003uh}. 
Non-Abelian $U(n)$ flavor symmetry arises in the original theory 
instead of $U(1)^{N_{\rm F}-1}$ in equation (\ref{eq:mdl:break-flavor}),  
and some of them are broken and associated Nambu-Goldstone bosons 
can be localized on a wall   
unlike only $U(1)$ localization in the case of non-degenerate masses. 
Nonlinear sigma model on $U(N)$ (called the chiral Lagrangian) 
appears on domain walls 
in the model with $N_{\rm F} = 2 N_{\rm C} \equiv 2N$ 
with masses $m_A = m$ for $A=1,\cdots,N$ and $m_A = -m$ 
for $A=N+1,\cdots,2N$. Including four derivative term,   
the Skyrme model appears on domain walls in that model \cite{Eto:2005cc}.

\begin{figure}[thb]
\begin{center}
\includegraphics[width=7cm,clip]{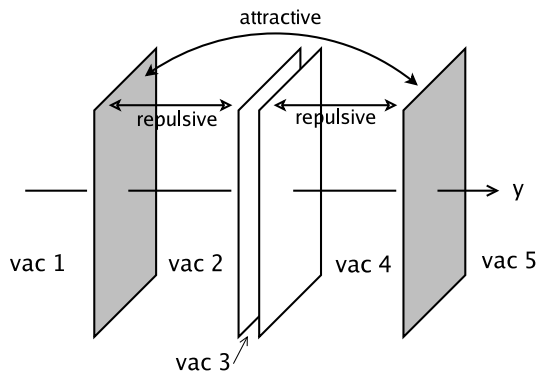} 
\caption{\small
Interactions between four domain walls
in the $T^*F_n$ sigma model. 
}
\label{fig:wll:Fn}
\end{center}
\end{figure}

It has been shown in 
\cite{Eto:2005wf} that 
the moduli space of domain walls 
is always the union of special Lagrangian submanifolds 
of the moduli space of vacua of the corresponding massless theory.
As an example, 
domain walls and their moduli space have been considered 
in the linear sigma model giving the cotangent bundle 
over the Hirzebruch surface $F_n$. 
Interestingly, as special Lagrangian submanifolds, 
this model contains a weighted projective space 
$W{\bf C}P^2_{1,1,n}$ in addition to $F_n$. 
The moduli space of the domain walls 
has been shown to be the union of these special Lagrangian submanifolds,
both of which is connected along 
a lower dimensional submanifold. 
Interesting consequence of this model is as follows. 
This model admits four domain walls which are ordered. 
The inner two walls are always compressed to form a single wall 
in the presence of outer two walls, 
and the position of that single wall is locked 
between the outer two walls.  
However if we take away outer two walls to infinities, 
the compressed walls can be broken into two walls. 
These phenomena can be regarded as an 
evidence for the attractive/repulsive 
force exist between some pairs of domain walls 
as in figure \ref{fig:wll:Fn}. 

\subsection{Vortices}
\label{vtx}
In this section, we consider vortices as $1/2$ BPS states.
There exist various types of vortices.
First we consider the ANO vortices embedded into 
non-Abelian gauge theory with $N_{\rm F}=N_{\rm C}$, 
which are usually called non-Abelian vortices.  
We determine their moduli space completely 
by using the moduli matrix \cite{Eto:2005yh}.   
Then we find a complete relationship between 
our moduli space and 
the moduli space constructed by the K\"ahler quotient, 
which was given in \cite{Hanany:2003hp} 
by using a D-brane configuration in string theory. 
Proving this equivalence has been initiated in \cite{Eto:2005yh} 
and is completed in this paper as a new result.  
Next we extend these results  
to the case of semi-local vortices \cite{Vachaspati:1991dz}, 
which exist in the theory with 
the large number of Higgs fields ($N_{\rm F} > N_{\rm C}$).  
This part is also new. 

\subsubsection{Vortex Solutions and Their Moduli Space}

We consider the case of massless hypermultiplets which gives 
only continuously degenerated and connected vacua. 
The case of hypermultiplets with non-degenerate masses  
will be investigated in section \ref{mon}.
In the following we simply set $H^2=0$ and $H \equiv H^1$ 
because of boundary conditions and BPS equations. 
Although the adjoint scalars $\Sigma_p$ ($p=1, \cdots, 6-d$) 
appear in $d=3,4,5$ from higher dimensional components 
of the gauge field, 
they trivially vanish in the vacua and 
also in vortex solutions. 
Therefore we can consistently 
set $\Sigma_p=0$. 
In the theory with $N_{\rm F}=N_{\rm C}$ in any dimension, 
the vacuum is unique and is in the so-called color-flavor 
locking phase, 
$H^1 = \sqrt c {\bf 1}_{N_{\rm C}}$ and $H^2=0$,   
where symmetry of the Lagrangian is spontaneously 
broken down to $SU(N_{\rm C})_{\rm G+F}$. 
This symmetry will be further broken 
in the presence of vortices, and 
therefore it acts as an isometry on the moduli space.

The Bogomol'nyi completion of energy density for vortices 
in the $x^1$-$x^2$ plane can be
obtained as
\begin{eqnarray}
{\cal E}&=&{\rm Tr}\left[\frac1{g^2}
\left(B_3 + {g^2 \over 2} (c {\bf 1}_N - H H^\dagger)\right)^2
+\left({\cal D}_1 H + i {\cal D}_2 H\right)
\left({\cal D}_1 H + i {\cal D}_2 H\right)^\dagger\right]\nonumber \\
&&+{\rm Tr}\left[-c\,B_3+2i\partial _{[1}H{\cal D}_{2]}H^\dagger \right]
\label{eq:vtx:energy}
\end{eqnarray}
with a magnetic field $B_3 \equiv F_{12}$. 
This leads to the vortex equations 
\begin{eqnarray}
 0 &=& {\cal D}_1 H + i {\cal D}_2 H, 
   \label{eq:vtx:BPSeq1}
\\ 
 0 &=& B_3 + {g^2 \over 2} (c {\bf 1}_N - H H^\dagger) ,
   \label{eq:vtx:BPSeq2}
\end{eqnarray}
and their tension 
\begin{eqnarray}
 T = -c \int d^2 x\ {\rm Tr} B_3 = 2 \pi c k,
  \label{eq:vtx:tension}
\end{eqnarray}
with $k (\in {\bf Z})$ 
measuring the winding number 
of the 
$U(1)$ part of broken $U(N_{\rm C})$ gauge symmetry. 
The integer $k$ is called the vorticity or the vortex number. 

\begin{figure}[ht]
\begin{center}
\begin{tabular}{ccc}
\includegraphics[height=4cm]{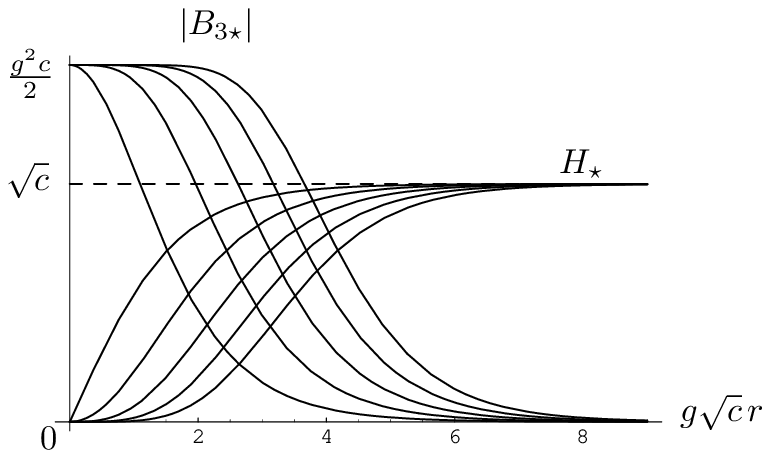}
&\quad&
\includegraphics[height=4cm]{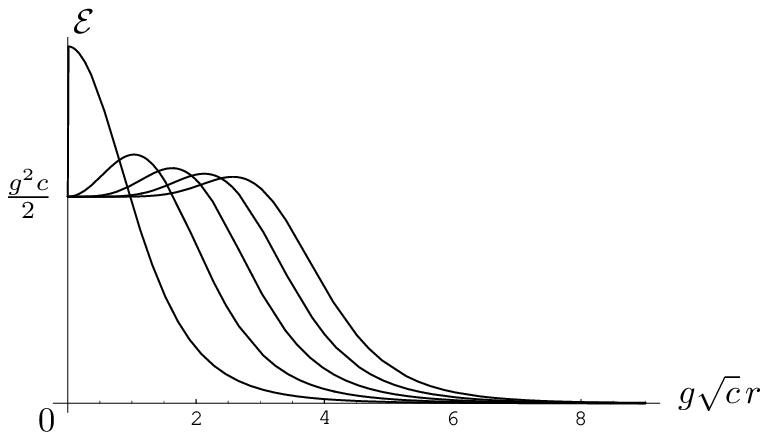}\\
\small{\sf (a) magnetic flux and Higgs field} & & 
\small{\sf (b) energy}
\vspace*{-.3cm}
\end{tabular}
\caption{\small{\sf 
Distributions of 
numerical vortex solutions with 
vorticity $k=1,\cdots,5$ as functions of the radius 
$r$. 
Magnetic flux is centered at $r=0$, whereas the Higgs field 
vanishes at $r=0$ and approaches to the vacuum value 
$\sqrt{c}$ at $r\rightarrow \infty$. 
Energy density has a dip at $r=0$ except for the case of 
unit ($k=1$) vorticity. 
}}
\label{fig:vtx:anovortex}
\vspace*{-.3cm}
\end{center}
\end{figure}

Let us first consider the simplest example 
of the model with $N_{\rm C}=N_{\rm F}=1$ 
in order to extract fundamental properties of vortices.
Vortices in this model are called 
Abrikosov-Nielsen-Olesen (ANO) vortices \cite{Abrikosov:1956sx}.  
A profile function of the ANO vortex has been established numerically 
very well, 
although no analytic solution is known.
We illustrate numerical solutions of 
the profile function with the vortex number $k=1,\cdots,5$ 
in figure \ref{fig:vtx:anovortex}.
One can see that the Higgs field vanishes 
at the center of the ANO vortex, 
and the winding to the Higges vacuum is resolved smoothly. 
Then the magnetic flux emerges there, whose intensity is given by
$B_3=-g^2c/2$ due to the vortex equation (\ref{eq:vtx:BPSeq2}).  
Therefore a characteristic size of ANO vortex  
can be estimated to be of order $1/(g\sqrt{c})$ 
by taking the total flux $2\pi $ into account.

In the non-Abelian case with $N_{\rm F}=N_{\rm C} \equiv N \geq 2$,
a solution for single vortex    
can be constructed by embedding such ANO vortex solution $(B_{3\star}, H_\star)$ in the Abelian
case into those in the non-Abelian case, like 
\begin{eqnarray}
 B_3=U{\rm diag}\,(B_{3\star},0,\cdots,0)\,U^{-1},\quad 
 H=U{\rm diag}\,(H_\star,\sqrt{c},\cdots,\sqrt{c})\,U^{-1}.
 \label{eq:vtx:ANOembedding}
\end{eqnarray}
Here $U$ takes a value in a coset space, 
the projective space 
$SU(N)/[SU(N-1) \times U(1)]
\simeq {\bf C}P^{N-1}$,  
arising from the fact that the $SU(N)_{\rm G+F}$ 
symmetry is spontaneously broken by the existence of 
the vortex. 
It parametrizes the orientation of the non-Abelian vortex 
in the internal space, whose moduli are called 
{\it orientational moduli}.  
Note that  at the center of the ANO vortex 
$x^{1,2}=x^{1,2}_0$, 
the rank of the $N \times N$ matrix $H$ reduces to 
$N - 1$, $({\rm det}(H(x^{1,2}_0))=0)$,
implying the existence of an $N$-column vector 
 $\vec \phi$ 
satisfying 
\begin{eqnarray}
 H(x^{1,2}_0)\vec \phi =0 .\label{eq:vtx:orientation}
\end{eqnarray}
Components of this vector are precisely 
the homogeneous coordinates of
the orientational moduli ${\bf C}P^{N-1}$. 
Its components are actually given by 
$\vec \phi =U(1,0,\cdots)^{\rm T}$.
Roughly speaking, the moduli space of multiple non-Abelian vortices 
is parametrized by a set of the position moduli 
and the orientational moduli,
both of which are attached to each vortex as we will see later.  

%

We now turn back to general cases with arbitrary 
$N_{\rm C}$ and $N_{\rm F} (> N_{\rm C})$.
The vortex equation (\ref{eq:vtx:BPSeq1}) can be solved 
by use of the method similar to that in the case of domain walls.   
Defining a complex coordinate 
$z \equiv x^1+ix^2$, 
the first of the vortex equations (\ref{eq:vtx:BPSeq1}) 
can be solved 
as \cite{Eto:2005yh}
\begin{eqnarray}
 H =S^{-1} H_0(z),
  \quad
 W_1+iW_2 =-i2S^{-1}{\bar \partial}_z S.   
  \label{eq:vtx:solution}
\end{eqnarray}
Here $S=S(z,\bar z) \in GL(N_{\rm C},{\bf C})$ 
is defined in the second of the equations 
(\ref{eq:vtx:solution}), 
and $H_0(z)$ is an arbitrary 
$N_{\rm C}$ by $N_{\rm F}$ matrix whose components 
are holomorphic with respect to $z$. 
We call $H_0$ the {\it moduli matrix} of vortices. 
With a gauge invariant quantity
\begin{eqnarray}  
 \Omega (z,\bar z) \equiv S(z,\bar z) S^\dagger (z,\bar z) 
  \label{eq:vtx:omega} 
\end{eqnarray} 
the second vortex equations (\ref{eq:vtx:BPSeq2})
can be rewritten as
\begin{eqnarray}
 \partial_z (\Omega^{-1} \bar \partial_z \Omega ) 
 = {g^2 \over 4} (c{\bf 1}_{N_{\rm C}} - \Omega^{-1} H_0 H_0^\dagger) .
 \label{eq:vtx:master}
\end{eqnarray}
We call this the {\it master equation} for vortices \footnote{
The master equation reduces to the so-called 
Taubes equation \cite{Taubes:1979tm} 
in the case of ANO vortices ($N_{\rm C} = N_{\rm F} = 1$)
by rewriting 
$c \Omega(z,\bar z)=|H_0|^2 e^{-\xi (z,\bar z)}$ 
with $H_0 = \prod_i(z-z_i)$. 
Note that $\log \Omega$ 
is regular everywhere 
while $\xi$ is singular at vortex positions.
Non-integrability of the master equation has been shown in 
\cite{Inami:2006wr}.
}. 
This equation is expected 
to give no additional moduli parameters. 
It was proved for the ANO vortices ($N_{\rm F} = N_{\rm C}=1$) 
\cite{Taubes:1979tm} 
and is consistent with 
the index theorem \cite{Hanany:2003hp} in general 
$N_{\rm C}$ and $N_{\rm F}$ as seen below. 
 
Therefore we assume that the moduli matrix $H_0$ describes
 thoroughly the moduli space of vortices.   
We should, however, note that 
there exists a redundancy in the solution (\ref{eq:vtx:solution}): 
physical quantities $H$ and $W_{1,2}$ 
are invariant under 
the following $V$-transformations 
\begin{eqnarray}
\hspace{-1cm}
 H_0 (z)\to H_0' (z)= V(z) H_0(z), 
\quad S(z,\bar z) \to S'(z,\bar z)= V(z) S(z,\bar z) , 
  \label{eq:vtx:V-trans}
\end{eqnarray}
with $V(z) \in GL(N_{\rm C},{\bf C})$ for 
${}^\forall z\in {\bf C}$, 
whose elements are holomorphic with respect to $z$.  
Incorporating all possible boundary conditions, 
we find that the total moduli space of vortices 
${\cal M}^{\rm total}_{N_{\rm C},N_{\rm F}}$ is given by 
\begin{eqnarray} 
 {\cal M}^{\rm total}_{N_{\rm C},N_{\rm F}}&=&
 {\left\{H_0(z)|H_0(z)\in M_{N_{\rm C},N_{\rm F}}\right\}\over 
 \left\{V(z)|V(z)\in M_{N_{\rm C},N_{\rm C}}, 
 {\rm det}V(z)\not=0 \right\}}
  \label{eq:vtx:totalquotient}
\end{eqnarray}
where $M_{N,N'}$ denotes a set of 
holomorphic $N\times N'$ matrices. 
This is of course an infinite dimensional space which may 
not be defined well in general. 

The original definition of the total moduli space is 
the space of solutions of two BPS equations in 
(\ref{eq:vtx:BPSeq1}) and (\ref{eq:vtx:BPSeq2}) 
divided by the $U(N_{\rm C})$ 
local gauge equivalence denoted as $G(x)$: 
$\{{\rm eq.(\ref{eq:vtx:BPSeq1}) },\; {\rm eq.(\ref{eq:vtx:BPSeq2})}\}/G(x)$. 
On the other hand, we have solved the first 
vortex equation (\ref{eq:vtx:BPSeq1}), but have to 
assume the existence and the uniqueness of the solution of 
the master equation (\ref{eq:vtx:master}), in order to 
arrive at the total moduli space (\ref{eq:vtx:totalquotient}). 
Let us note that the first vortex equation 
(\ref{eq:vtx:BPSeq1}) is invariant under the complex 
extension 
$G^{\bf C} = U(N_{\rm C})^{\bf C} = GL(N_{\rm C},{\bf C})$ 
of the local gauge group $G = U(N_{\rm C})$. 
Our procedure to obtain the total 
moduli space ${\cal M}^{\rm total}_{N_{\rm C},N_{\rm F}}$ 
in equation (\ref{eq:vtx:totalquotient}) implies that 
it can be rewritten as 
$\{{\rm eq.(\ref{eq:vtx:BPSeq1})}\}/G^{\bf C}(x)$. 
Therefore the uniqueness and existence of solution of 
the master equation (\ref{eq:vtx:master}) is equivalent to 
the isomorphism between these spaces, 
$\{{\rm eq.(\ref{eq:vtx:BPSeq1}) },\; 
{\rm eq.(\ref{eq:vtx:BPSeq2})}\}/G(x)
 \simeq 
\{{\rm eq.(\ref{eq:vtx:BPSeq1})}\}/G^{\bf C}(x)$. 
This isomorphism is rigorously proven at least if we compactify 
the base space (co-dimensions of vortices) ${\bf C}$ 
to ${\bf C}P^1$. 
This is called the Hitchin-Kobayashi correspondence 
\cite{MundetiRiera:1999fd,CRGS,Baptista:2004rk}.\footnote{
Actually it is proved for arbitrary gauge group $G$ 
with arbitrary matter contents and arbitrary {\it compact} base space. 
It may be interesting to note that this isomorphism 
is an infinite dimensional version of the K\"ahler quotient.
} 
We will establish the finite dimensional version of 
this equivalence (for each topological sector) 
directly in another method using moduli matrix 
in section \ref{sec:vtx:kahler}. 

We require that the total energy of configurations must be finite 
in order to obtain physically meaningful vortex configurations. 
This implies that any point at infinity $S^1_{\infty}$ must belong 
to the same gauge equivalence class of vacua. 
Therefore elements of the moduli matrix $H_0$ must be 
polynomial functions of $z$. 
(If exponential factors exist 
they become dominant at the boundary $S^1_{\infty}$ 
and the configuration fails to converge to 
the same gauge equivalence class there.) 
Furthermore the topological sector of the moduli space 
of vortices should be determined under 
a fixed boundary condition 
with a given vorticity $k$. 

The energy density (\ref{eq:vtx:energy}) of BPS states can be rewritten 
in terms of the gauge invariant matrix $\Omega$ in equation 
(\ref{eq:vtx:omega}) as   
\begin{eqnarray}
{\cal E}|_{\rm BPS}&=&{\rm Tr}\left[-c\,B_3+2i\partial _{[1}H{\cal D}_{2]}H^\dagger \right] 
\Big|_{\rm BPS}\nonumber\\
&=&2c\,{\bar \partial _z}{\partial _z}\left(1-{4\over g^2c}{\bar \partial _z}{\partial _z}
\right)\log \det \Omega .
  \label{eq:vtx:energy2}
\end{eqnarray}
The last four-derivative term above does not contribute to the tension
if a configuration approaches to a vacuum on the boundary.
Equation (\ref{eq:vtx:master}) implies asymptotic behavior 
at infinity $z \to \infty$ becomes $\Omega \to \frac{1}{c} H_0 H_0^\dagger$.   
The condition of vorticity $k$ requires 
\begin{eqnarray}
 T=2\pi c\,k 
&=&-{c\over 2}i\oint dz {\partial _z}\log\det(H_0H_0^\dagger )+{\rm c.c.} 
\label{eq:vtx:tension2}
\end{eqnarray}
The total moduli space is decomposed 
into topological sectors ${\cal M}_{N_{\rm F},N_{\rm C},k}$ 
with vorticity $k$. 
 
\subsubsection{The case with $N_{\rm F}=N_{\rm C}$: 
the non-Abelian ANO vortices}
Let us consider the case with $N_{\rm C}=N_{\rm F} \equiv N$.
In this case the vacuum, given by 
$H=\sqrt{c}{\bf 1}_N$, is unique and no flat direction exists.  
The tension formula (\ref{eq:vtx:tension2}) with $N_{\rm F}=N_{\rm C}$ 
implies that the vorticity $k$ can be rewritten as 
\begin{eqnarray}
k = {1\over 2\pi }{\rm Im} \oint dz\ \partial {\rm log}({\rm det}H_0) .
\label{eq:vtx:ANOtension}
\end{eqnarray}
We thus obtain the boundary condition on $S^1_{\infty}$ 
for $H_0$ as
\begin{eqnarray}
 {\rm det}(H_0) \sim z^k  \quad \mbox{for} \quad z \to \infty, 
   \label{eq:vtx:asympt}
\end{eqnarray}
that is, $\det H_0(z)$ has $k$ zeros. 
We denote positions of zeros by $z=z_i$ ($i=1,\cdots,k$). 
These can be recognized as the positions of vortices: 
\begin{eqnarray}
 P (z) \equiv \det H_0 (z) =\prod_{i=1}^k(z-z_i),
\label{eq:vtx:monic}
\end{eqnarray}
and the orientation moduli $\vec \phi _i$ 
of the $i$-th vortex are determined by  
\begin{eqnarray}
 H_0(z_i)\vec \phi _i=0 \quad \leftrightarrow\quad  
H(z=z_i,\bar z=\bar z_i)\vec \phi _i=0.
 \label{eq:vtx:k-orientations}
\end{eqnarray}
The moduli space ${\cal M}_{N,k}$ for 
$k$-vortices in $U(N)$ gauge theory 
can be formally expressed as a quotient 
\begin{eqnarray} 
 {\cal M}_{N,k}={\left\{H_0(z)|H_0(z)\in M_N, 
 {\rm deg}\, ({\rm det}(H_0(z))) = k \right\}\over 
 \left\{V(z)|V(z)\in M_N, 
 {\rm det}V(z) = 1 \right\}}
 \label{eq:vtx:k-vortexmoduli}
\end{eqnarray}
where $M_N$ denotes a set of 
$N\times N$ matrices of polynomial function of $z$, 
and ``deg'' denotes a degree of polynomials. 
The condition ${\rm det}V(z) = 1$ holds because 
we have fixed $P(z)$ as a monic polynomial (coefficient of 
highest power is unity) as 
in equation (\ref{eq:vtx:monic}) by 
using the $V$-transformation (\ref{eq:vtx:V-trans}). 
This is a finite dimensional subspace of the total moduli space 
(\ref{eq:vtx:totalquotient}).

The $V$-transformation (\ref{eq:vtx:V-trans}) allows us to reduce 
degrees of polynomials in $H_0$ by applying the division 
algorithm.  
After fixing the $V$-transformation completely,
any moduli matrix $H_0$ 
can be uniquely transformed to a triangular matrix, 
which we call the {\it standard form} of vortices: 
\begin{eqnarray}
H_0
 =\left(\begin{array}{ccccc}
  P_1(z)&R_{2,1}(z)&R_{3,1}(z)&\cdots    &R_{N,1}(z) \\
       0& P_2(z)   &R_{3,2}(z)&\cdots    &R_{N,2}(z)\\
  \vdots&          & \ddots   &          & \vdots \\
        &          &          &&R_{N,N-1}(z)\\
       0&\cdots    &          &0         & P_{N}(z)
	   \end{array}\right) .
 \label{eq:vtx:standard}
\end{eqnarray}
Here $P_r(z)$ are monic polynomials defined by
$P_r(z) = \prod_{i=1}^{k_r}(z-z_{r,i})$ 
with $z_{r,i} \in {\bf C}$,
and $R_{r,m}(z)\in {\rm Pol}(z;k_r)$  
where ${\rm Pol}(z;n)$ denotes 
a set of polynomial functions 
of order less than $n$. 
We would like to emphasize that 
the standard form (\ref{eq:vtx:standard}) 
has {\it one-to-one correspondence} to 
a point in the moduli space ${\cal M}_{N,k}$. 
Since $\tau (z)=\prod_{r=1}^NP_r(z)\sim z^k$ 
asymptotically 
for $z \to \infty$, 
we obtain the vortex number 
$k = \sum_{r=1}^Nk_r$ from equation (\ref{eq:vtx:tension2}).
The positions of the $k$-vortices are the zeros of $P_r(z)$.
Collecting all matrices with given $k$
in the standard form (\ref{eq:vtx:standard}) 
we obtain the whole moduli space  ${\cal M}_{N,k}$ for $k$-vortices, 
as in the case of domain walls. 
Its generic points are 
parameterized by the matrix 
with $k_N = k$ and $k_r=0$ for 
$r \neq N$, given by 
\begin{eqnarray}
 H_0 (z) = 
 \left(
 \begin{array}{cc}
 {\bf 1}_{N-1} & - \vec R(z) \\
 0 & P(z)
 \end{array}
 \right)
\label{eq:vtx:generic-form}
\end{eqnarray}
where $(\vec R(z))^r = R_r(z) \in {\rm Pol}(z;k)$ 
constitutes an $N-1$ vector and 
$P(z) = \prod_{i=1}^k(z-z_i)$.
This moduli matrix contains 
the maximal number of the moduli parameters in 
${\cal M}_{N,k}$.
Thus the dimension of the moduli space is 
${\rm dim}({\cal M}_{N,k})=2k N$. 
This coincides with the index theorem shown in \cite{Hanany:2003hp} 
implying the uniqueness and existence of a solution of 
the master equation (\ref{eq:vtx:master}).

The standard form (\ref{eq:vtx:standard}) has the merit 
of covering the entire moduli space only once without any 
overlap. 
To clarify the global structure of the moduli space, 
however, it may be more useful to parameterize 
the moduli space with overlapping patches. 
We can parameterize the moduli space  
by a set of ${}_{k+N-1}C_k$ patches defined by 
\begin{eqnarray}
 (H_0)^r{_s} = z^{k_s}\delta^r{_s} - T^r{_s}(z),\quad
 T^r{}_s(z)=\sum_{n=1}^{k_s}(T_n)^r{}_s z^{n-1}\in {\rm Pol}(z;k_s).
\label{eq:vtx:patches} 
\end{eqnarray}
Coefficients $(T_n)^r{}_s$ of monomials in $T^r{}_s(z)$ are 
moduli parameters as 
coordinates in a patch. 
We denote this patch by 
${\cal U}^{(k_1,k_2,\cdots,k_N)}$:
\begin{eqnarray}
 {\cal U}^{(k_1,k_2,\cdots,k_N)} = \{ (T_{n_s})^r{}_s \} \quad 
  n_s = 1,\cdots,k_s, \quad r=1,\cdots,N_{\rm C}
  \label{eq:vtx:patch}
\end{eqnarray}  
We can show that each patch 
fixes the V-transformation (\ref{eq:vtx:V-trans}) completely 
including any discrete subgroup, 
and therefore that 
the isomorphism 
${\cal U}^{(k_1,k_2,\cdots,k_N)} \simeq {\bf C}^{kN}$ holds.
The transition functions between these patches 
are given by the $V$-transformation (\ref{eq:vtx:V-trans}), 
completely defining the moduli space 
as a {\it smooth} manifold,  
\begin{eqnarray} 
 {\cal M}_{N,k} \simeq \bigcup {\cal U}^{(k_1,k_2,\cdots,k_N)}.
  \label{eq:vtx:union}
\end{eqnarray} 

To see this explicitly we show an example of 
single vortex ($k=1$). 
In this case there exist $N$ patches defined in $N$ $H_0$'s given by
\begin{eqnarray}
H_0(z) &\sim& 
\left(\begin{array}{cccc}
 1&      &0&-b_1^{(N)}    \\
  &\ddots& &\vdots  \\
 0&      &1&-b_{N-1}^{(N)}\\
 0&\dots &0&z-z_0     
\end{array}\right)
\! \sim \!
 \left(\begin{array}{cccc}
  1&      &-b_1^{(N-1)}  &0 \\
   &\ddots&\vdots & \\
  0&      &z-z_0  &0\\
  0&\dots &-b_{N}^{(N-1)}&1     
 \end{array}\right)\!\sim \!\cdots  \nonumber \\ 
\! &\sim& \! 
 \left(\begin{array}{cccc}
     z - z_0 &0     &\cdots  &0 \\
  -b_2^{(1)} &1     &        &0 \\
   \vdots    &      &\ddots  & \\
  -b_N^{(1)} &0     &        &1     
 \end{array}\right)\! ,
  \label{eq:vtx:single}
\end{eqnarray}
and transition functions between them are 
summarized by 
\begin{eqnarray}
 \vec \phi \sim 
\left(\begin{array}{c}
b_1^{(N)}\\ \vdots\\ \vdots \\ b_{N-1}^{(N)}\\ 1
\end{array}\right) 
= b_{N-1}^{(N)}
\left(\begin{array}{c}
b_1^{(N-1)}\\\vdots \\
   b_{N-2}^{(N-1)}\\1 \\b_{N}^{(N-1)}
      \end{array}\right) 
= \cdots
= b_1^{(N)}\left(\begin{array}{c}
1\\ b_{2}^{(1)}\\ \vdots \\ b_{N-1}^{(1)} \\ b_N^{(1)}
\end{array}\right) .
  \label{eq:vtx:single2}
\end{eqnarray}
These $b$'s are the standard coordinates for ${\bf C}P^{N-1}$, 
and we identify components of the vector $\vec \phi $ 
as the orientational moduli 
satisfying equation (\ref{eq:vtx:k-orientations}). 
We thus confirm 
${\cal M}_{N,k=1}$ $\simeq$ ${\bf C} \times {\bf C}P^{N-1}$ 
recovering the result \cite{Auzzi:2003fs}
obtained by a symmetry argument. 
To see the procedure more explicitly,  
we show, in the case of $N=2$, that the $V$-transformation connects
sets of coordinates in two patches as
\begin{eqnarray}
\left(\begin{array}{cc}
 z-z_0& 0\\-b' & 1     
\end{array}\right)
\! \sim \!
\left(\begin{array}{cc}
 0& -b\\1/b&z-z_0     
\end{array}\right)
\left(\begin{array}{cc}
 z-z_0& 0\\-1/b & 1     
\end{array}\right)
=\left(\begin{array}{cc}
 1& -b\\0&z-z_0     
\end{array}\right), 
  \label{eq:vtx:N=2}
\end{eqnarray}
where we obtain a transition function $b'=1/b$.

The next example is the case of $N=2$ and $k=2$ which is 
more interesting and cannot be obtained by symmetry 
argument only.    
The moduli space ${\cal M}_{N=2,k=2}$ is 
parameterized by the three patches 
${\cal U}^{(0,2)}$, ${\cal U}^{(1,1)}$, 
${\cal U}^{(2,0)}$ defined in $H_0$'s given by 
\begin{eqnarray}
H_0=\left(
\begin{array}{cc}
1 & -az-b\\
0 & z^2-\alpha z - \beta
\end{array}
\right),
\left(
\begin{array}{cc}
z-\phi & -\varphi\\
-\tilde\varphi & z-\tilde\phi
\end{array}
\right),
\left(
\begin{array}{cc}
z^2 - \alpha z - \beta & 0\\
-a'z- b' & 1
\end{array}
\right), \label{eq:vtx:HofN2k2}
\end{eqnarray}
respectively. 
We find that the transition functions between ${\cal U}^{(0,2)}$ and 
${\cal U}^{(1,1)}$ are given by 
\begin{eqnarray}
 a={1\over \tilde \varphi },\quad b=-{\tilde \phi \over \tilde \varphi },\quad 
\alpha =\phi +\tilde{\phi} ,\quad 
\beta =\varphi \tilde \varphi -\phi \tilde \phi 
\label{eq:vtx:02to11}
\end{eqnarray}
and that those between ${\cal U}^{(0,2)}$ and 
${\cal U}^{(2,0)}$ are given by 
\begin{eqnarray}
 a= {a'\over {a'}^2\beta -a'b'\alpha -{b'}^2},\quad b=-{b'+a'\alpha \over {a'}^2\beta -a'b'\alpha -{b'}^2}
\label{eq:vtx:02to20}
\end{eqnarray}
with common parameters $\alpha ,\beta $. 
Positions of two vortices $z_1, z_2$ 
are given by solving an equation $P(z_i)=0$.  
We find that orientations of the vortices
satisfying equation (\ref{eq:vtx:k-orientations})
are expressed by four kinds of forms:
\begin{eqnarray}
 \vec \phi _i\sim 
\left(\begin{array}{c}
a z_i+b\\ 1  \end{array}\right)
\sim  \left(\begin{array}{c}
z_i-\tilde \phi \\ \tilde \varphi   \end{array}\right)
\sim 
\left(\begin{array}{c}
\varphi \\  z_i-\phi   \end{array}\right)
\sim 
\left(\begin{array}{c}
1\\a' z_i+b'  \end{array}\right)
   \label{eq:vtx:fourkinds}
\end{eqnarray}
with the equivalence relation $\vec \phi \sim \vec \phi '=\lambda \vec \phi ,
(\lambda \in {\bf C}^*)$
of ${\bf C}P^{N-1}$.
The above equivalence relations between 
the various forms for orientation are consistent with 
the transition functions 
(\ref{eq:vtx:02to11}) and (\ref{eq:vtx:02to20}), 
since the orientations are, by definition, independent of 
the patches which we take.   

We now see properties of the three patches 
${\cal U}^{(0,2)}$, ${\cal U}^{(1,1)}$ and 
${\cal U}^{(2,0)}$.
If we set $a=0\, (a'=0)$ in ${\cal U}^{(0,2)} ({\cal U}^{(2,0)})$, 
the orientations of two vortices are parallel 
\begin{eqnarray}
 \vec \phi _1\sim \vec \phi _2\sim (b,1)^{\rm T}\,(\sim(1,b')^{\rm T} ). 
  \label{eq:vtx:parallel}
\end{eqnarray} 
This is in contrast to the patch ${\cal U}^{(1,1)}$ where 
parallel vortices are impossible, 
as long as the two vortices are separated. 
Configurations for parallel multiple vortices can 
be realized by embedding the configuration for multiple 
ANO vortices in the Abelian gauge theory in the same way 
as equation (\ref{eq:vtx:ANOembedding}).  
In contrast we can take the orientations of two vortices 
opposite each other in the patch ${\cal U}^{(1,1)}$, like 
\begin{eqnarray}
 \vec \phi _1=(1,0)^{\rm T},\quad   
 \vec \phi _2=(0,1)^{\rm T}  
  \label{eq:vtx:opposite}
\end{eqnarray}  
by setting 
$\phi =z_1,\tilde \phi =z_2$ and $\varphi =\tilde \varphi =0$.    
In this case, the moduli matrix $H_0(z)$, 
$\Omega $ as a solution of equation (\ref{eq:vtx:master}) and 
physical fields $B_3,\,H$ are all diagonal,
and thus we find that this case is realized by embedding 
two sets of single ANO vortices in the Abelian case
into two different diagonal components of 
the moduli matrices of this non-Abelian case. 
The moduli space for non-Abelian vortices described by patches 
(\ref{eq:vtx:HofN2k2}) are far larger than subspaces which can
be described by embedding the Abelian cases. 
Such subspaces can be interpolated with continuous moduli 
in the whole moduli space. 
Actually, as long as the vortices are separated $z_1\not=z_2$,
the positions $z_1,z_2\in {\bf C}$ and the orientations 
$\vec \phi _1,\vec\phi _2 \in {\bf C}P^1$ are independent of each others
and can parametrize the moduli space, as we discuss  later.

\begin{figure}[ht]
\begin{center}
\begin{tabular}{ccc}
\includegraphics[height=5cm]{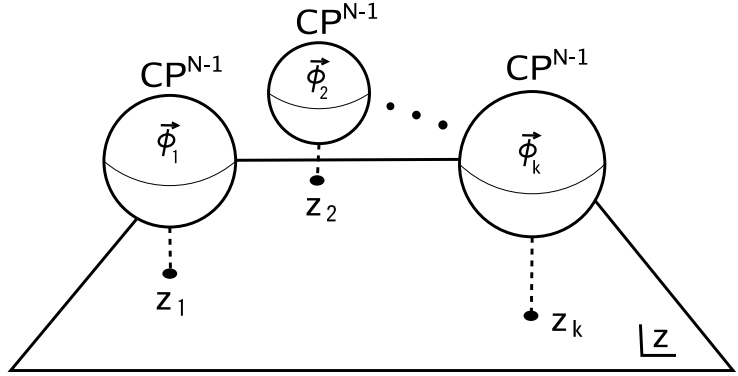}
\end{tabular}
\caption{\small{\sf the moduli space for separated vortices}}
\label{fig:vtx:vortexmoduli}
\end{center}
\end{figure}

In generic cases with arbitrary $N$ and $k$, 
we can find that
orientational moduli $\vec \phi _i\in {\bf C}P^{N-1}$ 
are attached to each vortex at $z=z_i\in {\bf C}$ as 
an independent moduli parameters. 
Thus the asymptotic form (open set) of the moduli space 
for separated vortices are found to be 
\begin{eqnarray}
 {\cal M}_{N,k} \leftarrow  
 \left({\bf C}\times{\bf C}P^{N-1}\right)^k / {\mathcal S}_k
 \label{moduli-asympt}
\end{eqnarray} 
with ${\mathcal S}_k$ permutation group 
exchanging the positions of the vortices.\footnote{
Interestingly 
this is a ``half" of the open set of the moduli space 
of $k$ separated $U(N)$ instantons on non-commutative 
${\bf R}^4$, 
$({\bf C}^2 \times T^* {\bf C}P^{N-1})^k/{\mathcal S}_k$. 
The singularity of the latter is resolved 
in terms of the Hilbert scheme at least for $N=1$ \cite{Na}. 
Also it was pointed out in \cite{Hanany:2003hp} 
that the moduli space of vortices 
is a special Lagrangian submanifold 
of the moduli space of non-commutative instantons.
} 
Here ``$\leftarrow$" denotes a map resolving the 
singularities on the right hand side.
We sketch the structure of separated vortices in figure 
\ref{fig:vtx:vortexmoduli}.
Equation (\ref{moduli-asympt}) 
can be easily expected from physical intuition; 
for instance the $k=2$ case was found 
in \cite{Hashimoto:2005hi}. 
The most important thing is how 
orbifold singularities of 
the right hand side in 
(\ref{moduli-asympt}) are resolved 
by coincident vortices \cite{Eto:2005yh}.  
In the $N=1$ case, 
${\cal M}_{N=1,k} \simeq {\bf C}^k/{\mathcal S}_k \simeq {\bf C}^k$ 
holds instead of (\ref{moduli-asympt}) \cite{Taubes:1979tm}, 
and the problem of singularity does not occur. 

\subsubsection{Equivalence to the K\"ahler quotient 
\label{sec:vtx:kahler}}
In this subsection we rewrite our moduli space of vortices  
in the form of K\"ahler quotient which 
was originally found in \cite{Hanany:2003hp} by using a D-brane configuration.
This form is close to the ADHM construction of instantons  
and so we may call it the {\it half ADHM construction}.

First of all, let us consider a vector whose $N$ components are
elements of Pol$(z;k)$ satisfying an equation
\begin{eqnarray} 
 H_0(z)\vec \phi (z)=\vec J(z)P(z) = 0 \quad {\rm mod~}P(z)
 \label{eq:vtx:Eq_Phi(z)}
\end{eqnarray}
where $P(z)\equiv {\rm det}(H_0(z))$ and $\vec J(z)$ is a certain holomorphic 
polynomial obtained,  
that is, the equation requires that the l.h.s. can
be divided by the polynomial $P(z)$.
We can show there exist $k$ linearly-independent solutions 
$\{\vec \phi _i(z)\}, (i=1,\cdots,k)$
for $\vec \phi (z)$ with
given $H_0(z)$. 
We obtain the $N\times k$ matrices 
${\bf \Phi }(z)$ and 
${\bf J}(z)$, defined by
\begin{eqnarray}
 {\bf \Phi }(z)=(\vec \phi _1(z),\vec \phi _2(z),\cdots,\vec \phi_k(z)), 
 \quad
 {\bf J}(z)=(\vec J_1(z), \vec J_2(z), 
   \cdots,\vec J_k(z)),  \label{eq:vtx:PhiJ}
\end{eqnarray}
with satisfying 
\begin{eqnarray}
 H_0(z){\bf \Phi }(z)={\bf J}(z)P(z) .
 \label{eq:vtx:HtoPhi}
\end{eqnarray}

Let us consider a product $z\,{\bf \Phi }(z)$.
Since components of this product  
are not elements of Pol$(z,k)$ but Pol$(z,k+1)$ generally, 
this matrix leads to an $N\times k$ {\it constant} matrix ${\bf \Psi }$
as a quotient of division by the polynomial $P(z)$.
Moreover a remainder of this division can be written  
as ${\bf \Phi }(z)$ multiplied by a certain $k\times k$ 
{\it constant} matrix ${\bf Z}$ since each column vector of the remainder 
is also a solution of equation (\ref{eq:vtx:Eq_Phi(z)}). 
Therefore we find the product determines the matrices 
$\bf Z$ and $\bf \Psi$ uniquely as 
\begin{eqnarray}
 z\, {\bf \Phi }(z)={\bf \Phi }(z){\bf Z}+{\bf \Psi }P(z). 
\label{eq:vtx:PhitoZPsi}
\end{eqnarray}
Note that when we extract 
the matrix ${\bf \Phi }(z)$ from the moduli matrix  $H_0$, 
there exists a redundancy due to a rearrangement of 
$\vec \phi _i(z)$ which leads to an equivalence relation 
${\bf \Phi }(z)\simeq {\bf \Phi }'(z)
={\bf \Phi }(z)U^{-1}$ with $U\in GL(k,{\bf C})$.
This fact implies that we should also consider 
an equivalence relation for the matrices $\bf Z, \Psi $ as 
\begin{eqnarray}
 ({\bf Z},\,{\bf \Psi })\simeq ({\bf Z}',\,{\bf \Psi }')
=(U{\bf Z}U^{-1},\,{\bf \Psi }U^{-1}),
\label{eq:vtx:GLaction}
\end{eqnarray}
where a $GL(k,{\bf C})$ action on ${\bf \Phi }(z)$ and $\{\bf Z,\,\Psi \}$ is
free: if ${\bf \Phi }(z)X=0$ then $X=0$. 
Since the action is free, the quotient space is smooth. 
This is precisely the complexified 
transformation appearing in 
the K\"ahler quotient construction. 
On the other hand the K\"ahler quotient construction of the moduli space 
in \cite{Hanany:2003hp} is given by $k$ by $k$ 
matrix $Z$ and $N$ by $k$ matrix $\psi$.
The concrete correspondence is obtained by fixing the 
imaginary part of the complexified transformation as 
\begin{eqnarray}
\{{\bf Z},{\bf \Psi}\}/\!/GL(k,{\bf C}) 
\simeq 
\left\{(Z,\psi) | \left[Z^\dagger,Z\right]
+\psi^\dagger \psi \propto {\bf 1}_k\right\}/U(k) 
 \equiv {\cal M}^{\rm HT}.
\label{eq:vtx:Kahler} 
\end{eqnarray}
Therefore, the above procedure defines the mapping from 
our moduli space ${\cal M}_{N,k}$ (\ref{eq:vtx:k-vortexmoduli}) 
into the K\"ahler quotient (\ref{eq:vtx:Kahler}).\footnote{
The choice of the D-term condition $\left[Z^\dagger,Z\right]
+\psi^\dagger \psi \propto {\bf 1}_k$ is not unique. 
There exist many candidates of it but all of them give 
topologically isomorphic manifolds \cite{Eto:2004ii}.
}  
This is a topological sector version of the Hitchin-Kobayahsi 
correspondence as informed below equation (\ref{eq:vtx:totalquotient}). 

By combining equations (\ref{eq:vtx:HtoPhi}) and (\ref{eq:vtx:PhitoZPsi}),
we derive a direct relation between the moduli matrix $H_0(z)$  
and the matrices $\{\bf Z, \Psi \}$ as 
\begin{eqnarray}
\nabla ^\dagger L=0,\quad {\rm with~} L^\dagger \equiv \left(H_0(z),\,{\bf J}(z)\right),\quad 
\nabla \equiv \left(
\begin{array}{c}
-{\bf \Psi } \\ z-{\bf Z}
\end{array}\right).  
  \label{eq:vtx:rel-L}
\end{eqnarray}
By use of this equation, 
we can concretely relate coefficients (coordinates) 
$(T_m)^r{}_s$ in a patch 
${\cal U}^{k_1,\cdots,k_{N_{\rm C}}} $ of $H_0$ (\ref{eq:vtx:patches})
with the matrices $\{\bf Z, \Psi \}$ as
\begin{eqnarray}
({\bf \Psi })^{r}{}_{(s,m)}&=&\left\{
\begin{array}{cl}
\delta ^{r}_s\delta ^1_m & {\rm for~} k_r>0 \\
(T_m)^{r}{}_s& {\rm for~} k_r=0\, ,  
\end{array}
\right.\nonumber\\
 ({\bf Z})^{(r,n)}{}_{(s,m)}&=&\left\{
\begin{array}{cl}
 \delta ^{n+1}_m\delta ^r_s& {\rm for~} 1\leq n<k_r\\
(T_m)^r{}_s                &{\rm for~} n=k_r \, ,
\end{array}\right. 
\label{eq:vtx:PZpatch}
\end{eqnarray}
where the label $(s,m)$ runs from $(s,1)$ to $(s,k_s)$ with
$1\leq s\leq N_{\rm C}$, and 
the equation  
\begin{eqnarray} 
 \det(z-{\bf Z})=\det(H_0(z))=P(z) 
  \label{eq:vtx:z-Z}
\end{eqnarray} 
holds. 
To show this relation, we need ${\bf J}(z)$ 
in the patch ${\cal U}^{k_1,\cdots,k_{N_{\rm C}}} $ as
\begin{eqnarray}
 ({\bf J}(z))^{r}{}_{(s,m)}=\left.{H_0^{rs}(z)\over z^m}\right|_{\rm reg}
=z^{k_s-m}\delta ^r_s-\sum_{l=m+1}^{k_s}(T_l)^r{}_sz^{l-m-1}
  \label{eq:vtx:rel-J}
\end{eqnarray}
where `reg' implies to remove terms with negative power of $z$. 
By substituting equations (\ref{eq:vtx:patches}), 
(\ref{eq:vtx:PZpatch}) and (\ref{eq:vtx:rel-J}), 
we can confirm equation (\ref{eq:vtx:rel-L}). 

Using the whole set of $H_0$'s in equation (\ref{eq:vtx:patches}), 
we obtain the set of $({\bf Z},{\bf \Psi})$'s 
in equation (\ref{eq:vtx:PZpatch}). 
On the other hand, 
the $GL(k,{\bf C})$ action (\ref{eq:vtx:GLaction}) in 
the K\"ahler quotient (\ref{eq:vtx:Kahler}) 
can be fixed to the form (\ref{eq:vtx:PZpatch}). 
Namely the K\"ahler quotient (\ref{eq:vtx:Kahler}) 
is also covered by ${}_{N+k-1}C_k$ patches 
given by (\ref{eq:vtx:PZpatch}) where 
$(T_m)^r{}_s$ are coordinates of the patches. 
Combining this result with 
the argument (\ref{eq:vtx:Eq_Phi(z)})--(\ref{eq:vtx:Kahler}), 
we finally find that 
the moduli space of vortices given by our moduli matrix 
is completely equivalent to that defined by 
the K\"ahler quotient: 
\begin{eqnarray}
 {\cal M}_{N,k}\simeq \{{\bf Z},\,\ {\bf \Psi} \}/\!/ GL(k,{\bf C}).
  \label{eq:vtx:kahlerquo}
\end{eqnarray}
Thus this result confirms the result in \cite{Hanany:2003hp} 
from the field theoretical point of view, 
while they used a method of D-brane construction.

Let us examine the relation with simple examples. 
For the separated vortices $z_i\not=z_j$, we find 
$\vec \phi _i(z)=\vec \phi _i\prod_{j\not=i}(z-z_j)$ 
with orientations $\vec \phi _i$
satisfying equations (\ref{eq:vtx:k-orientations}) gives 
${\bf \Psi}$ for the orientational moduli and 
the diagonal matrix ${\bf Z}$ whose elements 
correspond to the positions of the vortices  
\begin{eqnarray}
 {\bf Z} = {\rm diag}(z_1,z_2,\cdots,z_k), 
\quad
{\bf \Psi }
 = \left(
 \begin{array}{ccc}
    \vec \phi _1,& \cdots &, \vec \phi _k 
 \end{array}\right) , \label{eq:vtx:matrices}
\end{eqnarray}
although the matrix ${\bf Z}$ 
is not always diagonalizable by $GL(k,{\bf C})$ if 
there are coincident vortices.

In what follows we illustrate 
the correspondence in the case of $(N,k)=(2,2)$. 
The moduli data in the patches (\ref{eq:vtx:HofN2k2}) 
can be summarized by  
two matrices ${\bf Z}$ and ${\bf \Psi}$ as follows 
\begin{eqnarray}
\left(\begin{array}{c}
{\bf \Psi} \\{\bf Z}
      \end{array}\right)
= &&
\left(
\begin{array}{cc}
 b & a\\
 1 & 0\\
 0 & 1\\
 \beta & \alpha
\end{array}\right),
\left(
\begin{array}{cc}
 1 & 0\\
 0 & 1\\
\phi & \varphi\\
\tilde\varphi & \tilde\phi
\end{array}
\right),
\left(
\begin{array}{cc}
  1 & 0\\
 b' & a'\\
 0 & 1\\
 \beta & \alpha
\end{array}
\right).
 \label{(k,N)=(2,2)}
\end{eqnarray}
The transition functions (\ref{eq:vtx:02to11}) and (\ref{eq:vtx:02to20})
between these 
three patches can be expressed 
by the complexified gauge transformation  
between moduli data  
as $\left({\bf Z}',{\bf \Psi}'\right) 
= \left(U{\bf Z}U^{-1},{\bf \Psi}U^{-1}\right)$
with appropriate $U\in GL(2,{\bf C})$.

\subsubsection{
The cases of $N_{\rm F}>N_{\rm C}$: non-Abelian semi-local vortices
\label{sec:vtx:semi}} 
In the cases with $N_{\rm F}>N_{\rm C}$, 
there appear additional moduli for vortices, typically, 
moduli for sizes of vortices due to additional Higgs fields. 
A vortex possessing such size moduli is called a semi-local vortex
and its size is limited below by the size of ANO vortex.   
We also have non-normalizable moduli. 

As in the last subsection, 
we take elements of the moduli matrix $H_0(z)$ 
as polynomials with respect to $z$. 
This is because 
we are interested in configurations with 
boundary conditions such that 
any point of the boundary belongs to the same vacuum.
The tension of $k$ vortices in this case is given by  
\begin{eqnarray}
\hspace{-2.5cm} T = 2\pi c k 
={c\over 2} \oint dzd\bar z\ \partial \bar \partial 
{\rm log}\left({\rm det}H_0H_0^\dagger \right)
={c\over 2} \oint dzd\bar z\ \partial \bar \partial 
{\rm log}\left(\sum_{\left<\{A_r\}\right> }
|\tau ^{\left<\{A_r\}\right> }|^2\right),
  \label{eq:vtx:semilocal-tension}
\end{eqnarray}
where $\tau$ is defined similarly to equation 
(\ref{eq:wll:def-tau}) 
for the case of domain walls:
\begin{eqnarray} 
 \tau ^{\left<\{A_r\}\right> } (z) 
  \equiv {\rm det}H_0^{\left<\{A_r\}\right> } (z).
  \label{eq:vtx:tau}
\end{eqnarray}
Equation (\ref{eq:vtx:semilocal-tension})
requires that the maximal degree of a set of polynomials 
$\{\tau ^{\left<\{A_r\}\right> }\}$ 
is $k$. 

We now discuss moduli parameters of a single vortex 
in the Abelian case with $N_{\rm C}=1$ and general $N_{\rm F}$. 
The condition $k=1$ implies 
\begin{eqnarray}
 H_0(z)=\left(a_1 z+b_1, a_2z+b_2,\,\cdots, 
a_{N_{\rm F}}z+b_{N_{\rm F}}\right) , \quad
  a_A , b_A\in {\bf C}
\label{eq:vtx:semiH}
\end{eqnarray}
where $\{a_A\}$ are homogeneous coordinates of ${\bf C}P^{N_{\rm F}-1}$. 
Some of these parameters are not normalizable moduli of the vortex 
but are non-normalizable moduli which should be interpreted as  
moduli of vacua on the boundary, as shown in the following. 
In a region sufficiently far from the origin 
($z\not=0,\,|b_A/z|\ll |a_A|$),
 the moduli matrix behaves as 
\begin{eqnarray}
 H_0(z)\stackrel{z\not=0}{\simeq} 
\left(a_1 +{b_1\over z}, \,\cdots, 
a_{N_{\rm F}}+{b_{N_{\rm F}}\over z}\right)\stackrel{z\rightarrow \infty }{\rightarrow }
\left(a_1, a_2,\,\cdots,a_{N_{\rm F}}\right) 
  \label{eq:vtx:behaviour}
\end{eqnarray}
where we have used a $V$-transformation $V(z)=z^{-1}$ 
which is non-singular and is allowed 
in regions with $z\not=0$.   
We thus have found that $a_A$ parametrize the moduli space 
${\bf C}P^{N_{\rm F}-1}$ of the Higgs vacua at the boundary,  
as $H^A|_{\rm boundary}=\sqrt{c}\,a_A/\sqrt{|a|^2}$.
Therefore the set $\{a_A \}$ should be fixed 
as a boundary condition. 
Without loss of generality we can choose the boundary condition as 
${H}'|_{\rm boundary}=\sqrt{c}(1,0,\cdots,0)=H|_{\rm boundary}U$ 
with the flavor symmetry $U\in SU(N_{\rm F})$. 
Under this boundary condition the moduli matrix should be taken as
\begin{eqnarray}
 H_0'=\left(z-z_0, b'_2,b'_3,\cdots,b'_{N_{\rm F}-1 }\right)
 \simeq H_0U .  
 \label{eq:vtx:boundary} 
\end{eqnarray}
Here the parameter $z_0=-(b\cdot a^\dagger )/|a|^2 $ 
describes the position of the semi-local vortex,  
and it also has a size modulus $b'\equiv \sqrt{\sum_{A=2}^{N_{\rm F}}|b'_A|^2}
= \sqrt{|a|^2|b|^2-|b\cdot a^\dagger |^2}/|a|^2$. 
Furthermore, in the cases of $N_{\rm F}\geq   3$,  
the vortex has an internal modulus $\{b'_A/b'\}$ describing  
non-vanishing Higgs fields in its center. 
Note that even when 
the size modulus of the semi-local vortex is zero, 
$b'=0$, it becomes the ANO vortex with the size $1/g\sqrt c$.

In general cases for $N_{\rm F}$ and $N_{\rm C}$, 
the moduli matrix $H_0$ contains moduli of the vacua on the boundary 
(restricted to $H^2=0$) 
\begin{eqnarray}
 {\cal M}_{\rm boundary} 
 = G_{N_{\rm F},N_{\rm C}} 
 \simeq {SU(N_{\rm F}) \over 
  SU(N_{\rm C})\times SU(N_{\rm F}-N_{\rm C}) \times U(1)},  
  \label{eq:vtx:boundary-moduli}
\end{eqnarray}
which should be fixed.
  
Therefore, the moduli space ${\cal M}_{N_{\rm C},N_{\rm F},k}$ 
of $k$-vortices in $U(N_{\rm C})$ gauge theory 
coupled to $N_{\rm F}$ hypermultiplets 
can be formally expressed as 
the quotient of 
\begin{eqnarray} 
\hspace{-1.5cm}{\cal M}_{N_{\rm C},N_{\rm F},k}=
{\left\{H_0(z)|H_0(z)\in M_{N_{\rm C},N_{\rm F}}, 
{\rm max}\{{\rm deg}\,\tau ^{\left<\{A_r\}\right> }\} = k \right\}\over 
\left\{V(z)|V(z)\in M_{N_{\rm C},N_{\rm C}}, 
{\rm det}V(z)={\rm const.}\not=0 \right\}\times {\cal M}_{\rm boundary}} 
  \label{eq:vtx:semilocal-moduli}
\end{eqnarray}
where $M_{N,N'}$ denotes a set of 
$N\times N'$ matrices of polynomial functions of $z$. 
Let us investigate the moduli space for semi-local non-Abelian  
vortices concretely. 
Using the flavor symmetry $SU(N_{\rm F})$, 
we can choose a vacuum on the boundary as 
$\langle {\rm vac}\rangle =\langle 1,2,\cdots,N_{\rm C}\rangle $ 
without loss of generality. 
Namely we have conditions 
\begin{eqnarray}
 {\rm det}(H^{\langle {\rm vac}\rangle })
=(\sqrt{c})^{N_{\rm C}},\quad 
{\rm det}(H^{\left<\{A_r\}\right> })=0 
\quad {\rm for~} \left<\{A_r\}\right> 
\not=\langle {\rm vac}\rangle .
  \label{eq:vtx:det-H}
\end{eqnarray}
By use of the relation 
${\rm det}H^{\left<\{A_r\}\right> } 
/{\rm det}H^{\left<\{B_r\}\right> }
=\tau ^{\left<\{A_r\}\right> }/\tau ^{\left<\{B_r\}\right> } $
we find that the boundary condition with vorticity $k$ requires    
\begin{eqnarray}
{\rm deg}\, \tau ^{\langle {\rm vac}\rangle }(z)= k, \quad {\rm and ~} 
{\rm deg}\, \tau ^{\left<\{A_r\}\right> }(z)< k \quad {\rm for~} \left<\{A_r\}\right> \not=\langle {\rm vac}\rangle .
  \label{eq:vtx:deg-tau}
\end{eqnarray}
Due to the Pl\"ucker relations (\ref{eq:wll:plucker}) 
all of these conditions are 
not independent, 
but only the following two conditions turn out to be independent: 
\begin{eqnarray}
{\rm deg}\,\tau ^{\langle {\rm vac}\rangle }(z)&=&k,\nonumber \\
({\bf F}(z))^r{}_A&\equiv &\tau ^{\langle 1,\cdots,r-1,A,r+1,\cdots,N_{\rm C}\rangle  }(z)
\in {\rm Pol}(z;k), 
\label{eq:vtx:semi-cond}
\end{eqnarray}
with $N_{\rm C}<A\leq N_{\rm F}$.  

Let us decompose the moduli matrix $H_0$ to 
an $N_{\rm C}\times N_{\rm C}$ matrix ${\bf D}(z)$ and 
an $N_{\rm C}\times (N_{\rm F}-N_{\rm C})$ matrix ${\bf Q}(z)$ as
\begin{eqnarray}
 H_0(z)=\left({\bf D}(z),\,{\bf Q}(z)\right).
  \label{eq:vtx:decom-H0}
\end{eqnarray}
Then the first and the second conditions in 
equation (\ref{eq:vtx:semi-cond}) are regarded as conditions 
for the matrices ${\bf D}(z)$ and ${\bf Q}(z)$, respectively.
Under these constraints, 
we can obtain the moduli matrix for the semi-local vortices. 
For instance, in the case of $k=1$, 
the moduli matrix in a patch ${\cal U}^{(0,\cdots,0,1)}$ 
contains an ($N_{\rm C}-1$)-component vector $\vec b$ 
as orientational moduli 
and additional moduli $\{q_A\}$ as in the following:
\begin{eqnarray}
 H_0(z)\simeq \left(\begin{array}{cccccc}
 {\bf 1}_{N_{\rm C}-1}&     -\vec b &0&0&\cdots&0   \\
 {\bf 0}&z-z_0& q_{N_{\rm C}+1}&q_{N_{\rm C}+2}&\cdots&q_{N_{\rm F}}     
\end{array}\right).  
  \label{eq:vtx:semilocal-k=1}
\end{eqnarray}   
Since the monic polynomial 
$P(z)\equiv \tau ^{\langle {\rm vac}\rangle }(z)=\det {\bf D}(z)$
satisfies  
the condition (\ref{eq:vtx:semi-cond}) similar to 
equation (\ref{eq:vtx:monic}), we can use 
the same strategy as we used in the 
$N_{\rm F}=N_{\rm C}$ case
to fix the $V$-transformation. 
We thus find that 
${\bf D}(z)$ consists of $k N_{\rm C}$ complex moduli 
parameters corresponding to 
positions and orientations of $k$ vortices.
 We also 
obtain holomorphic $N_{\rm C}\times k$ matrix ${\bf \Phi}(z)$ 
whose components belong to Pol$(z;k)$ 
as a solution of the equation
\begin{eqnarray}
 {\bf D}(z){\bf \Phi }(z)={\bf J}(z)P(z).
\label{eq:vtx:DtoPhi} 
\end{eqnarray}
Next let us consider moduli described by ${\bf Q}(z)$.
Here we can easily find  the identities given by 
\begin{eqnarray}
0= H_0^{r[A_1}H_0^{1A_2}H_0^{2A_3}\cdots
H_{0}^{N_{\rm C}A_{N_{\rm C}+1}]}\propto 
H_0^{r[A_1}\tau ^{\langle A_2 \cdots A_{N_{\rm C}+1}]\rangle }
 \label{eq:vtx:identity}
\end{eqnarray}
where the square bracket means anti-symmetrization with respects to 
indices $A_s, (s=1,\cdots,N_{\rm C}+1)$. 
By setting $\{A_s\}$
to $\{1,2,\cdots,N_{\rm C},A\}$ in the above 
we obtain an identity for the matrices
\begin{eqnarray}
 {\bf D}(z){\bf F}(z)={\bf Q}(z)P(z) 
  \label{eq:vtx:F-iden}
\end{eqnarray}
By use of this identity,
we find  
the condition (\ref{eq:vtx:semi-cond}) 
requires that 
each column of ${\bf F}(z)$ should be written by a linear combination 
of column vectors of ${\bf \Phi }(z)$ in equation (\ref{eq:vtx:DtoPhi}), 
that is, ${\bf F}(z)$ should be solved with 
a constant $k\times(N_{\rm F}-N_{\rm C})$ matrix 
$\bf \widetilde \Psi $ as, 
${\bf F}(z)={\bf \Phi }(z)\widetilde {\bf \Psi }$.
Comparing equation (\ref{eq:vtx:DtoPhi}) with 
equation (\ref{eq:vtx:F-iden}) 
 we find that 
${\bf Q}(z)$ satisfying the condition must be written by
\begin{eqnarray}
  {\bf Q}(z)={\bf J}(z)\widetilde {\bf \Psi },
 \label{eq:vtx:QJPsi}
\end{eqnarray}
and conversely $\bf Q$ given in the above 
with an arbitrary $\widetilde {\bf \Psi }$ realizes the second condition.
Therefore, the matrix $\widetilde {\bf \Psi }$ describes 
the additional moduli for semi-local vortices entirely.   
As a result, we find that the dimension of the moduli space of
$k$ vortices in the cases with general $N_{\rm F}$ and $N_{\rm C}$ 
is given by
\begin{eqnarray}
 {\rm dim}{\cal M}_{N_{\rm C},N_{\rm F},k}
 =2k N_{\rm C}+2k (N_{\rm F}-N_{\rm C})=2k N_{\rm F}
 \label{eq:vtx:dim-semi}
\end{eqnarray}
in accord with the result of the index theorem \cite{Hanany:2003hp}.

In this case of semi-local vortices also 
we can extract the matrices ${\bf Z},{\bf \Psi} $ from ${\bf \Phi }(z)$ 
in equation (\ref{eq:vtx:DtoPhi}).
The $GL(k,{\bf C})$ action (\ref{eq:vtx:GLaction}) 
acts also on $\widetilde {\bf \Psi}$ as   
\begin{eqnarray}
 ({\bf Z},\,{\bf \Psi },\,\widetilde {\bf \Psi })\simeq 
(U{\bf Z}U^{-1},\,{\bf \Psi }U^{-1},\,U\widetilde {\bf \Psi })
 \label{eq:vtx:ZPsiPsi}
\end{eqnarray}
with $U\in GL(k,{\bf C})$. 
Therefore the moduli space for semi-local non-Abelian vortices 
in terms of the moduli matrix  
can be translated to that of the K\"ahler quotient as
\begin{eqnarray}
 {\cal M}_{N_{\rm C},N_{\rm F},k}\simeq 
\{{\bf Z},\,{\bf \Psi} ,\widetilde{\bf \Psi }\}/\!/GL(k,{\bf C}).
 \label{eq:vtx:kahlerquo-semi}
\end{eqnarray}
We can fix the imaginary part of 
the $GL(k,{\bf C})$ action as in equation 
(\ref{eq:vtx:Kahler}), to give 
\begin{eqnarray}
{\cal M}_{N_{\rm C},N_{\rm F},k} 
 \simeq 
 \left\{(Z,\psi,\tilde \psi) | \left[Z^\dagger,Z\right]
+\psi^\dagger \psi - \tilde \psi \tilde \psi^\dagger 
 \propto {\bf 1}_k\right\}/U(k) ,
 \label{eq:vtx:kahlerquo-semi2}
\end{eqnarray}
with $k \times (N_{\rm F} - N_{\rm C})$ matrix $\tilde \psi$. 
This again recovers the result in \cite{Hanany:2003hp}.

Finally, it may be useful to summarize
the relations between the moduli matrix $H_0(z)$ and the matrices for
 K\"ahler quotient as  
\begin{eqnarray}
\nabla ^\dagger L=\widetilde \nabla ^\dagger L=0
 \label{eq:vtx:L-semi}
\end{eqnarray}
where
an $(N_{\rm F}+k) \times N_{\rm C}$ matrix $L$,
an $(N_{\rm F}+k)\times k$ matrix $\nabla $ and
an $(N_{\rm F}+k)\times (N_{\rm F}-N_{\rm C})$ matrix $\widetilde \nabla $ 
are defined by
\begin{eqnarray}
  L^\dagger \equiv \left(H_0(z),{\bf J}(z)\right)
,\quad 
\nabla \equiv \left(
\begin{array}{c}
 -{\bf \Psi } \\ 0\\ z-{\bf Z} 
\end{array}\right)\quad {\rm and~}
\widetilde \nabla \equiv \left(
\begin{array}{c}
 0\\{\bf 1}_{N_{\rm F}-N_{\rm C}}\\
-\widetilde {\bf \Psi } 
\end{array}\right).
 \label{eq:vtx:L-def-semi}
\end{eqnarray}

\subsubsection{
Lumps as semi-local vortices in the strong coupling limit} 
Nonlinear sigma models admits lumps 
\cite{Ward:1985ij,Stokoe:1986ic} as co-dimension two 
solitons, which can be obtained from semi-local vortices 
($N_{\rm F}>N_{\rm C}$) in the limit of strong gauge 
coupling $g^2\rightarrow \infty $ 
keeping the vortex size finite. 
Thus lumps also have a size modulus. 
If we take the limit of vanishing size modulus, 
the lumps reduce to a singular configuration.
This phenomenon reflects the fact that ANO vortices 
(with size $1/g\sqrt c$) become singular 
in the strong gauge coupling limit.     
 
If we take 
the strong coupling limit keeping $\log \Omega $ smooth,
 the master equation (\ref{eq:vtx:master}) 
can be solved algebraically as
\begin{eqnarray}
 \Omega |_{g^2\rightarrow \infty }= c^{-1}H_0(z) H_0^\dagger (\bar z) .
 \label{eq:vtx:lumpsol}
\end{eqnarray}
Namely, we can obtain 
unique and exact solutions for lumps with 
given arbitrary $H_0$.
Here we should note that
if we take a solution (\ref{eq:vtx:lumpsol})
with the quantity $H_0(z)H_0(z)^\dagger $ having 
a vanishing point, 
$H_0(z_0)H_0(z_0)^\dagger =0$,
the l.h.s. of equation (\ref{eq:vtx:master}) leads to 
a singular profile implying that the strong coupling 
limit is improper in such a case. 
Therefore, the moduli matrix $H_0$ 
whose rank reduces somewhere in the $z$-plane, 
which describes an embedding of ANO vortices,  
is prohibited for lump solutions. 
To obtain the moduli space for lumps, 
subspaces with ANO vortices embedded 
should be removed from the one for the semi-local vortices.
We thus find that the total moduli space for lumps 
${\cal M}^{\rm total }_{N_{\rm C},N_{\rm F}}|_{g^2\rightarrow \infty }$
is obtained as a set of holomorphic maps from the $z$-plane to 
the Grassmann manifold $G_{N_{\rm F},N_{\rm C}}$, to give
\begin{eqnarray} 
{\cal M}^{\rm total}_{N_{\rm C},N_{\rm F}}\big|_{g^2\rightarrow \infty }&=&
{\left\{H_0(z)|H_0(z)\in M_{N_{\rm F},N_{\rm C}},{\rm rank}(H_0(z))=N_{\rm F}\right\}\over 
\left\{V(z)|V(z)\in M_{N_{\rm C},N_{\rm C}}, 
{\rm rank}(V(z))=N_{\rm C} \right\}}\nonumber \\
&=&\{H_0|{\bf C}\,
\rightarrow \,G_{N_{\rm F},N_{\rm C}},\bar \partial _zH_0=0\}.
 \label{eq:vtx:totalquotient2}
\end{eqnarray}
Due to the removal of the subspaces,
 this moduli space has singularities, which are known  
as small-lump singularities. 
In other words, 
the moduli space of semi-local vortices 
can be obtained by resolving small lump singularities 
in the lump moduli space by inserting the ANO vortices. 

\subsubsection{Vortices on cylinder}  
Interesting relation has been observed between 
domain walls and vortices \cite{Eto:2006mz}. 
In order to study the relation, it is most useful to 
consider vortices on a cylinder 
($-\infty < x^1 < \infty, 
x^2 \simeq x^2+2\pi R$) with one dimension 
compactified with the radius $R$. 
Vortices can exist when the Higgs scalars are massless. 
However, domain walls require massive Higgs scalars making 
the vacua discrete. 
It is best to introduce the mass for the Higgs scalars 
by a compactification with a twisted boundary condition, 
usually referred as a Scherk-Schwarz dimensional reduction. 
One can obtain solutions of 1/2 BPS equations for vortices 
on the cylinder 
as we have described before on a plane. 
If the vortices are placed on the cylinder and the 
twisted boundary condition is applied, the moduli 
matrix should satisfy 
\begin{eqnarray}
 H_0(z+2\pi i R)=H_0(z)e^{2 \pi i M R}, 
\end{eqnarray}
where the twisting phase $e^{2 \pi i M R}$ is related to 
the mass matrix $M$ for the hypermultiplets. 
In order to make the periodicity in $x^2$ explicit, 
one can use periodic variable $u$ instead of $z$ as 
\begin{eqnarray}
 H_0(z)=\hat H_0(u)e^{M z} ,\quad {\rm with~} u=\exp{z\over R}. 
\end{eqnarray}
The $V$-equivalence relation becomes in this case 
\begin{eqnarray}
 \hat H_0(u)\simeq \hat H_0'(u)=V(u)\hat H_0(u) 
\end{eqnarray}
with $V(u)\in {GL(N_{\rm C},{\bf C})}$ for $u\in {\bf C}-\{0\}$. 
By keeping all the Kaluza-Klein modes, we have found a duality 
between vortices and domain walls \cite{Eto:2006mz}. 

On the other hand, we should retain only the lowest mode, 
if we restrict ourselves to phenomena at energies low 
compared to the Kaluza-Klein mass scales $1/R$. 
Then we should take a particular form of moduli matrix 
given by the constant $\hat H_0$ 
\begin{eqnarray}
 H_0(z)=\hat H_0e^{M z} . 
\end{eqnarray}
This constant $\hat H_0$ is precisely the moduli matrix 
for domain walls that we have discussed. 

\subsection{Effective Lagrangians}
\label{efa}
\label{sc:efa:}

The low-energy effective Lagrangian on 
solitons is given by promoting the moduli parameters to 
fields on the worldvolume of the soliton and by assuming 
the weak dependence on the world-volume coordinates 
of the solitons \cite{Manton:1981mp}. 
In our case, we promote  the moduli parameters 
$\phi ^\alpha $ in the moduli matrix $H_0$ to fields 
on 
the worldvolume of the 
soliton, such as walls or vortices 
\begin{eqnarray}
 H_0(\phi ^\alpha )\rightarrow H_0(\phi ^\alpha (x)) , 
\label{eq:efa:moduli-fields}
\end{eqnarray}
where the coordinates on the worldvolume is denoted as $x^m$. 
We introduce ``the slow-movement parameter'' $\lambda$, 
which is assumed to be much smaller than the typical 
mass scale in the problem. 
Please note that we are using the slow-movement 
approximation to the case of nontrivial wolrdvolume 
besides the time dependence, although the original proposal 
was made for the case without the spacial woldvolume. 
It is also worth pointing out that we can obtain not only 
the effective Lagrangian for the quasi-Nambu-Goldstone 
(QNG) modes, but also for the Nambu-Goldstone (NG) 
modes, which are 
required by the spontaneously broken global symmetry. 
We will present the procedure and results in terms of 
component fields, although it is extremely useful and 
straight-forward to 
use the superfield formalism respecting the preserved 
supersymmetry especially in the case of $1/2$ BPS system 
such as walls or vortices \cite{Eto:2006uw}, 
that we are going to describe below.

\subsubsection{Effective Lagrangian on walls}
\label{eq:efa:wall}

In the case of domain walls, all the moduli parameters 
are contained in the constant moduli matrix $H_0$ 
\cite{Isozumi:2004jc,Isozumi:2004va}. 
Since the typical mass scales of the wall are $g\sqrt{c}$ 
 and the characteristic mass difference 
$\Delta m$ of hypermultiplets, we assume the slow-movement of 
moduli fields 
\begin{eqnarray}
 \lambda \ll {\rm min}(\Delta m, g\sqrt{c} ).
\label{eq:efa:slow_move}
\end{eqnarray}
The $1/2$ BPS background fields of the wall are of the order 
of $\lambda^0={\cal O}(1)$, whereas 
derivatives in terms of the worldvolume coordinates 
and induced fields by the fluctuations $\phi^\alpha$ 
are of order $\lambda$ 
\begin{eqnarray}
H^1 \sim {\cal O}(1),\quad \Sigma \sim {\cal O}(1),\quad 
\partial _m \sim {\cal O}(\lambda ). 
\label{eq:efa:h1-order}
\end{eqnarray}
\begin{eqnarray}
W_m \sim {\cal O}(\lambda ),\quad 
H^2\sim {\cal O}(\lambda ) 
\quad F_{m y}(W)\sim {\cal O}(\lambda ) .
\label{eq:efa:h2-order}
\end{eqnarray}

Decomposing the field equations in powers of 
$\lambda$, we find  the BPS equations 
(\ref{eq:wll:BPSeq-wall}) automatically 
at the order $\lambda ^0$, whereas 
we obtain all the induced fields at higher orders. 
Assuming $H^2=0$, we vary the fundamental Lagrangian to 
obtain the equations of motion (with $H^2=0$). 
The equation of motion for the gauge field 
fluctuations $W_m$ reads 
\begin{eqnarray}
0&=&{1\over g^2}{\cal D}_yF_{m y}
+{i\over g^2}[\Sigma ,\,{\cal D}_m \Sigma ]
+{i\over 2}\left(H {\cal D}_m H^{\dagger} 
-{\cal D}_m H \,H^{\dagger} \right) 
. 
\label{eq:efa:EOM-gauge-field}
\end{eqnarray}
After a long calculation we obtain the gauge field in terms 
of the matrix $S$ defined in (\ref{eq:wll:def-S}) 
and the variations $\delta_m$ with respect to 
chiral scalar fields and 
 $\delta_m^\dagger$ with respect to 
anti-chiral scalar fields 
\begin{eqnarray}
W_m = 
i\left((\delta _m S^\dagger )
S^{\dagger -1}-S^{-1}(\delta _m ^\dagger S)\right), 
\label{eq:efa:sol-gauge-field}
\end{eqnarray}
\begin{eqnarray}
\delta _m \equiv (\partial _m \phi^{\alpha})
{\partial \over \partial  \phi^{\alpha}}. 
\quad 
\delta _m^\dagger \equiv (\partial _m  \phi^{\alpha*})
{\partial \over \partial  \phi^{\alpha*}}. 
\label{eq:efa:der_chiral_moduli}
\end{eqnarray}
Similarly we obtain other fluctuations induced to order 
$\lambda$ 
\begin{eqnarray}
 {\cal D}_m H
=
S^\dagger \delta _m \left(\Omega ^{-1}H_0\right)e^{My}, 
\quad
 {\cal D}_m \Sigma +iF_{m y}
=
S^\dagger \delta _m (\Omega ^{-1}\partial _y\Omega )
S^{\dagger -1}. 
\label{eq:efa:cov-der-Sigma}
\end{eqnarray}

The effective Lagrangian is obtained 
by substituting equations (\ref{eq:efa:sol-gauge-field}) and 
(\ref{eq:efa:cov-der-Sigma}) to the fundamental 
Lagrangian 
and by integrating 
over the co-dimension $y$ of the walls 
\begin{eqnarray}
\hspace{-2cm}
&
&
{\cal L}_{\rm wall} + T_{\rm w} 
=
\int dy {\rm Tr}
\left[{\cal D}^m H{\cal D}_m H^\dagger 
+{1\over g^2}
\left({\cal D}^m \Sigma -iF^m {}_{y}\right)
\left({\cal D}_m \Sigma +iF_{m y}\right)\right]
\nonumber \\
\hspace{-2cm}
&=&\int dy{\rm Tr}\left[\Omega \delta ^m
\left(\Omega^{-1}H_0\right)e^{2My}\delta _m^\dagger 
\left(H_0^\dagger \Omega ^{-1}\right)
+{1\over g^2}\Omega \delta ^m 
\left(\Omega ^{-1}\partial _y\Omega \right)
\Omega ^{-1}\delta _m ^\dagger 
\left(\partial _y\Omega \Omega ^{-1}\right)\right]
\nonumber \\
\hspace{-2cm}
&=&
\int dy {\rm Re Tr}\left[c\delta ^m 
(\Omega ^{-1}\delta _m ^\dagger \Omega _0)+ 
{\partial _y^2\over 2g^2}
\left((\delta ^m \Omega )\Omega ^{-1}
(\delta _m ^\dagger \Omega )\Omega ^{-1}\right)\right]
\nonumber \\
\hspace{-2cm}
&=&
\int dy\delta ^m \delta _m ^\dagger 
\left[(c-{\partial ^2_y\over g^2})\log{\rm det}\Omega 
+ {1\over 2g^2}{\rm Tr}
\left(\Omega ^{-1}\partial _y\Omega \right)^2 \right]
\nonumber \\
\hspace{-2cm}
&&
-{1\over g^2}{\rm Re}{\rm Tr}
\left[\Omega ^{-1}\Omega '\delta ^m 
\left(\Omega ^{-1}\delta _m ^\dagger 
\Omega \right)\right]\Big|^{\infty }_{-\infty }
\nonumber \\
\hspace{-2cm}
&\equiv &
\delta ^m \delta _m ^\dagger K(\phi ,\phi ^*)=
K_{ij}(\phi ,\phi ^*)\partial ^m 
\phi ^i\partial _m \phi ^{j*}, 
\label{eq:efa:kahler-metric-wall}
\end{eqnarray}
where $T_{\rm w}$ is the tension of the (multi-)wall 
corresponding to the classical action of the background 
solution. 
This K\"ahler metric can be derived from 
the following K\"ahler potential for the moduli chiral 
superfields $\phi, \phi^*$ of the preserved four supersymmetry 
\begin{eqnarray}
 K(\phi ,\phi ^*)=\int dy \left[c\log{\rm det}\Omega 
+ {1\over 2g^2}{\rm Tr}
\left(\Omega ^{-1}\partial _y\Omega \right)^2 \right] . 
\label{eq:efa:kahler-potential-wall}
\end{eqnarray}
This effective Lagrangian contains both NG as well as 
QNG moduli fields. 
Equation (\ref{eq:efa:kahler-potential-wall}) 
is manifestly invariant under 
the local $U(N_{\rm C})$ gauge transformation. 
Under the $V$-equivalence transformation of $H_0$ 
with an arbitrary $N_{\rm C} \times N_{\rm C}$ matrix 
of chiral superfield $\Lambda(x,\theta,\bar\theta)$,  
given by $H_0 \to H_0{}' = e^{\Lambda} H_0$ with $V = e^{\Lambda}$, 
the K\"ahler potential receives a K\"ahler transformation from 
equation (\ref{eq:wll:omegaV}):  
\begin{eqnarray}
 \log \det \Omega \to  \log \det \Omega 
+  \det \Lambda +  \det \Lambda^\dagger. 
\label{eq:efa:Kahler-tr}
\end{eqnarray}
Since the purely chiral superfield $\log \det \Lambda$ or anti-chiral 
superfield $\log \det \Lambda^\dagger$ do not contribute 
to the effective Lagrangian. 
It is worth to point out that, 
if we regard the $\Omega$ as dynamical variables, 
the above K\"ahler potential serves as a Lagrangian from 
which the master equation (\ref{eq:wll:master-eq-wall}) 
for $\Omega$ can be derived. 
This fact can be understood most easily by means of 
superfield approach which enable us to rewrite the 
Lagrangian in terms of only $\Omega$ after solving 
the hypermultiplet part of the equations of motion in the 
slow-movement approximation.\cite{Eto:2006uw}

For the infinite gauge coupling limit $g\to \infty$, 
the effective Lagrangian for 
(multi) domain walls reduces to 
\begin{eqnarray}
\hspace{-1cm} {\cal L}_{\rm walls}^{g^2 \to \infty} 
  = c \int d^4 \theta \int dy \ \log \det \Omega_0
, 
\qquad 
 \Omega_0
= {1 \over c}H_0 e^{2 M y} H_0{}^\dagger \;. 
 \label{eq:efa:moduli_metric}
\end{eqnarray}

\subsubsection{Effective Lagrangian on vortices}
\label{eq:efa:vortex}

In the case of vortices, we have only single mass scale. 
Therefore small-movement approximation is valid when 
\begin{eqnarray}
\lambda \ll g\sqrt{c}. 
 \label{eq:efa:small_move_vortex}
\end{eqnarray}
The equation of motion for the 
gauge field fluctuations becomes 
\begin{eqnarray}
0&=&{1\over g^2}\left({\cal D}_xF_{m x}
+{\cal D}_yF_{m y}\right)
+{i\over 2}\left(H {\cal D}_m H^{\dagger} 
-{\cal D}_m H \,H^{\dagger} \right) 
. 
\label{eq:efa:vortex_EOM-gauge-field}
\end{eqnarray}
The solution $W_m$ is given by the same formula 
(\ref{eq:efa:sol-gauge-field}) as in the wall case. 
Other fluctuations induced to order $\lambda$ are similarly 
given by 
\begin{eqnarray}
 {\cal D}_m H
=
S^\dagger \delta _m \left(\Omega ^{-1}H_0\right). 
\label{eq:efa:vortex_hyper_fluc}
\end{eqnarray}

By substituting these solutions 
(\ref{eq:efa:sol-gauge-field}) and 
(\ref{eq:efa:vortex_hyper_fluc}) to the fundamental 
Lagrangian and by integrating over the co-dimension 
$x, y$ of the vortices, the effective Lagrangian on the 
vortex worldvolume is obtained in terms of 
the matrix $\Omega$ for vortices 
\begin{eqnarray}
\hspace{-1cm}
&&{\cal L}_{\rm vortex}
= \int\!\! d^2x\bigg[
\delta_m \delta_m^\dagger c\log\det\Omega\nonumber\\
\hspace{-1cm}
&+& \frac{4}{g^2}\Tr\left\{
\bar\partial_z\!
\left(\delta_m\Omega\Omega^{-1}\right)
\delta_m^\dagger\!
\left(\partial_z\Omega\Omega^{-1}\right)
- \bar\partial_z\!
\left(\partial_z\Omega\Omega^{-1}\right)
\delta_m^\dagger\!
\left(\delta_m\Omega\Omega^{-1}\right)\right\}
\bigg],
\label{eq:efa:eff-lag}
\end{eqnarray}
where $z \equiv x^1+ix^2$ and 
the variation $\delta_m$ and its conjugate 
$\delta_m^\dagger$ 
act on complex moduli fields as 
$\delta_m \Omega
= \sum_\alpha\partial_m\phi^\alpha
(\delta\Omega/\delta \phi^\alpha)$ 
and 
$\delta_m^\dagger \Omega
= \sum_\alpha\partial_m \phi^{\alpha*}
(\delta\Omega/\delta \phi^{\alpha*})$, 
respectively. 
This is a nonlinear sigma model with the K\"ahler metric 
which can be obtained from the following K\"ahler 
potential 
\begin{eqnarray}
 K 
&=&\int d^2 x 
{\rm Tr}\left[-2c V 
+e^{2 V}\Omega_0
+{16\over g^2}\int ^1_0 dt \int ^t_0 ds
\bar\partial V e^{2sL_V}\partial V\right], 
\label{eq:efa:kahler_pot_vortex}
\end{eqnarray}
where $V \equiv -{1 \over 2}\log \Omega $, 
$\Omega_0 \equiv c^{-1}H_0 H_0^\dagger$ and 
 the operation $L_V$ is defined by 
\begin{eqnarray}
L_V \times X=[V,\, X]. 
\label{eq:adjoint-operation}
\end{eqnarray}
This K\"ahler potential can be derived from the superfield 
formalism straight-forwardly \cite{Eto:2006uw} 
without going through the 
K\"ahler metric in equation (\ref{eq:efa:eff-lag}). 
The ANO case is reduced to the results in 
\cite{Samols:1991ne,Manton:2002wb,Chen:2004xu}.

In the case of a single vortex, 
the integration in the Lagrangian (\ref{eq:efa:kahler_pot_vortex}) 
can be performed explicitly to give 
\begin{eqnarray}
 K_{\rm single \; vortex} = \pi c|z_0|^2 + \frac{4\pi}{g^2}
\log\left(1 + \sum_{i=1}^{N-1}|b_i|^2\right), 
\label{eq:efa:kahler_pot_1vort}
\end{eqnarray}
in accord with the symmetry argument. 
Here $z_0$ is the position of the vortex, and $b_i$ are 
the inhomogeneous coordinates of the 
${\bf C}P^{ N_{\rm F}-1}$ as the orientational moduli.

Let us consider the limit of strong gauge coupling, 
where the gauge theory with $ N_{\rm F} >  N_{\rm C}$ 
reduces to the nonlinear sigma model on the cotangent 
bundle over the Grassmann manifold 
$T^* G_{ N_{\rm F}, N_{\rm C}}$. 
Then the semilocal vortices of 
the $N_{\rm F} >  N_{\rm C}$ case for finite gauge 
couplings become sigma-model lumps as explained in section 3.2.5. 
Since the second term in the effective Lagrangian 
(\ref{eq:efa:eff-lag}) for the vortices vanishes in 
this limit, the K\"ahler potential 
of the effective Lagrangian on the worldvolume of nonlinear 
sigma model lumps as 
\begin{eqnarray}
K_{\rm lumps} = c \int d^2x\ \log\det\Omega_0
, 
\quad 
 \Omega_0
= {1 \over c} H_0 H_0^\dagger \;.
\label{eq:efa:kahler_pot_lump}
\end{eqnarray}
This form of the K\"ahler potential has been obtained 
previously~\cite{Ward:1985ij,Stokoe:1986ic} 
in the case of the ${\bf C}P^{ N_{\rm F}-1}$ 
lumps corresponding to the case of the Abelian 
gauge theory ($N_{\rm C}=1$). 

\section{1/4 BPS solitons}
\label{qrt}
\subsection{1/4 BPS equations and their solutions}\label{4eq}
A series of 1/2 BPS equations for solitons in 
unbroken or Coulomb phases of 
non-Abelian gauge theories are well-known; 
instantons, monopoles, the Hitchin system 
which have co-dimensions 
4, 3, and 2,  respectively \cite{Tong:2005un, MantonSutcliffe}. 
\changed{
The Hitchin system is known to admit no finite energy solutions 
but, as we show in the next subsection, we can realize 
finite energy solutions in a finite region and 
so call them Hitchin vortices.
}
Monopoles and Hitchin vortices can be obtained by 
dimensional reductions of instantons.  

When these solitons are put into the Higgs phase, 
vortices and walls in sections \ref{wll} and \ref{vtx} 
are attached to these solitons. 
For instance,  magnetic flux coming out of a monopole 
must be squeezed into vortex tubes 
by the Meissner effect resulting in two vortices in opposite 
direction attached to the monopole. 
This composite soliton can be regarded as a kink on 
the worldvolume of a vortex \cite{Tong:2003pz}. 
Similarly instantons in the Higgs phase can be realized 
as vortices on a vortex \cite{Eto:2004rz}, 
and the Hitchin vortex can be 
realized at a junction of walls \cite{Eto:2005cp,Eto:2005fm}. 
These composite solitons preserve $1/4$ of supersymmetry 
and can be derived from instantons in the Higgs phase by 
either simple or  Scherk-Schwarz (SS) dimensional reduction. 

In this section we systematically derive the 1/4 BPS 
equations, the Bogomol'nyi energy bound and formal 
solutions, and describe generic structure of the moduli 
space of these composite solitons by our moduli matrix 
approach. 
The instanton-vortex-vortex (IVV) system, 
the monopole-vortex-wall (MVW) system  
and the Hitchin-wall-wall (HWW) system 
depend on coordinates along directions (co-dimensions) 
denoted by $\times$, and have the world-volume 
whose spacial part is denoted 
by ``$\bigcirc$'' as follows:
\begin{small}
\begin{eqnarray*}
\hspace{-2.5cm}
\begin{array}{c|cccc}
{\rm IVV} (d=5,6) &1 &2 &3 &4  \\
\hline
{\rm Instantons} &\times &\times &\times &\times  \\
{\rm Vortices} &\times &\times &\bigcirc &\bigcirc   \\
{\rm Vortices} &\bigcirc &\bigcirc &\times &\times   
\end{array}
\hspace{0.3cm}  
\begin{array}{c|ccc}
{\rm MVW} (d=4,5) &1 &2 &3  \\
\hline
{\rm Monopoles} &\times &\times &\times  \\
{\rm Vortices} &\times &\times &\bigcirc   \\
{\rm Walls} &\bigcirc &\bigcirc &\times  
\end{array} 
\hspace{0.3cm}  
\begin{array}{c|cc}
{\rm HWW} (d=3,4) &1 &3 \\
\hline
{\rm Hitchin\ vortices} &\times &\times  \\
{\rm Walls} &\times &\bigcirc  \\
{\rm Walls} &\bigcirc &\times   
\end{array}
\end{eqnarray*}
\end{small}
For the largest dimensions of the fundamental theory for 
each composite soliton, 
namely $d=6,5,4$ for IVV, MVW or HWW, respectively,  
another world volume direction $x^5$ is present, 
but is not written explicitly. 
Since we are interested in static solutions, we are allowed 
to choose $W_0=W_5=0$. 
The topological charges are also obtained by the 
dimensional reductions and are classified by the sign of 
their contributions to the energy density 
\changed{as summarized below. 
It has been recently found that 
Abelian gauge theories admit
negative energy objects
with the instanton charge 
localized at intersection of vortices \cite{Eto:2004rz}, 
those with the monopole charge 
at junctions of vortices and walls \cite{Isozumi:2004vg,Sakai:2005sp},  
and those with the Hitchin charge 
at junctions of walls \cite{Eto:2005cp,Eto:2005fm}.
In particular the first two are called 
{\it intersections} and {\it boojums}, respectively. 
}

\medskip
\begin{tabular}{c|c|c}
\hline
 dim $\setminus$ charge & positive & negative \\ \hline 
$d=5,6$ instantons & Instantons inside vortices & Intersectons  \\
$d=4,5$ monopole   & Monopoles attached by vortices & Boojums \\
$d=3,4$ Hitchin    & Non-Abelian wall junctions & Abelian wall junctions \\
\hline
\end{tabular} 

\subsubsection{Instanton-vortex system} 
\label{4eq:ivv}
The $1/4$ BPS equations can be derived by performing 
the Bogomol'nyi completion of the energy density as 
follows \cite{Hanany:2004ea,Eto:2004rz}
\begin{eqnarray}
{\cal E} &=& \Tr\left[
\frac{1}{2g^2}F_{mn}F_{mn} 
+ {\cal D}_mH({\cal D}_mH)^\dagger
+ \frac{1}{g^2}(Y^3)^2 \right]\nonumber\\
&=& 
\Tr\bigg[\frac{1}{g^2}\left\{
\left(F_{13} - F_{24}\right)^2
+ \left(F_{14} + F_{23}\right)^2
+ \left(F_{12} + F_{34} + Y^3\right)^2 
\right\} \nonumber\\
&& 
+({\cal D}_1 H + i  {\cal D}_2 H) 
({\cal D}_1 H + i  {\cal D}_2 H)^\dagger
+({\cal D}_3 H + i  {\cal D}_4 H) 
({\cal D}_3 H + i  {\cal D}_4 H)^\dagger
\nonumber\\
&& + \frac{1}{2g^2}F_{mn}\tilde F_{mn} 
- c(F_{12} + F_{34})
+ \partial_m J_m
\bigg]
\nonumber\\
&\ge& \Tr\left[\frac{1}{2g^2}F_{mn}\tilde F_{mn} 
-c(F_{12} + F_{34}) 
  + \partial_mJ_m\right],
 \label{eq:4eq:BPS-bound}
\end{eqnarray}
where $m,n,k,l=1,2,3,4$ and $W^0=W^5=0$ is chosen. 
The above energy density is minimized if the following set 
of the first order differential equations is satisfied: 
\begin{eqnarray}
&&
F_{13}- F_{24} =0, \ \ 
F_{14} + F_{23} =0, \ \ 
\label{eq:4eq:ivv1}\\
&& F_{12} +  F_{34} = - Y_3, \ \ 
\label{eq:4eq:ivv2}\\
&&
{\cal D}_1 H + i  {\cal D}_2 H =0, \ \ 
{\cal D}_3 H + i  {\cal D}_4 H =0. 
\label{eq:4eq:ivv3}
\end{eqnarray}
\changed{
We call a set of these as 
the self-dual Yang-Mills-Higgs (SDYM-Higgs) equation.  
This equation was also obtained by mathematicians 
\cite{Bradlow:1990ir,MundetiRiera:1999fd,CRGS} and 
is simply called the vortex equation although 
this contains instantons also. 
It has been shown recently in \cite{Popov:2005ik} 
that this set of equations can be derived 
(at least in the case of U(1) gauge group) 
from the Donaldson-Uhlenbeck-Yau equations on ${\bf C}^3$ 
\cite{DUY} by dimensional reduction on $S^2$.
}
It is easy to recognize that these equations are a 
combination of the $1/2$ BPS equations for constituent 
solitons. 
They can also be derived from the requirement of 
preserving the $1/4$ of supersymmetry defined by the 
following set of three projection operators for 
supertransformation parameters $\varepsilon^i$ 
$\gamma^{05}\varepsilon^i= - \varepsilon^i$,  
$\gamma^{12}(i\sigma_3 \varepsilon)^i= - \varepsilon^i$, 
$\gamma^{34}(i\sigma_3 \varepsilon)^i= - \varepsilon^i$, 
only two of which are independent.
The first projection corresponds to the 
supersymmetry preserved by instantons with codimensions 
in $x^{1}$-$x^{2}$-$x^{3}$-$x^{4}$, 
the second and the third projections correspond to vortices 
with codimensions $x^{1}$-$x^{2}$, and $x^{3}$-$x^{4}$ 
planes respectively. 
The Bogomol'nyi bound $T_{\rm IVV}$ for the energy density 
of the 1/4 BPS composite solitons can be rewritten as 
a sum of three topological charges 
\begin{eqnarray}
T_{\rm IVV}={\cal I}_{1234} 
           +{\cal V}_{12}  v_{34}
           +{\cal V}_{34}  v_{12}, 
\label{eq:4eq:ivv-ch}
\end{eqnarray}
where we have defined
\begin{eqnarray}
{\cal I}_{1234}
\equiv \frac{1}{4g^2}  \int d^4 x 
{\rm Tr} \left(
\epsilon_{mnkl}F_{mn}F_{kl} \right),
\label{eq:4eq:mass_i}\\
{\cal V}_{ij} \equiv -c \int dx^idx^j\ F_{ij}.
\label{eq:4eq:mass_v}
\end{eqnarray}
\changed{
Here ${\cal I}_{1234}$ is the mass of the instantons and
${\cal V}_{ij}$ is that of vortices with 
co-dimensions in the $x^i$-$x^j$ plane 
and 
$v_{kl}=\int dx^k dx^l$ is the 
$x^k$-$x^l$ world-volume 
of vortices. 
}
The world-volume integration has also the $x^5$ direction 
in the case of the fundamental theory in $d=6$.

Let us introduce the complex coordinates 
\begin{eqnarray} 
 z \equiv x^1 + ix^2 , \quad 
 w \equiv x^3 + ix^4   
 \label{eq:4eq:complex}
\end{eqnarray}
and the corresponding components for gauge 
fields 
\begin{eqnarray}
 \bar{W}_z \equiv {1 \over 2} (W_1 + iW_2) , \quad
 \bar{W}_w \equiv {1 \over 2} ({W_3 + iW_4}) .
\end{eqnarray} 
It is crucial to recognize that 
equation (\ref{eq:4eq:ivv1}) are the integrability conditions 
for the existence of the following 
solutions of equation (\ref{eq:4eq:ivv3}) 
\begin{eqnarray}
\bar{W}_z = -iS^{-1}\bar{\partial}_z S, \ \ 
\bar{W}_w = -iS^{-1}\bar{\partial}_w S, \ \ 
H = S^{-1}H_0(z,w)
\label{eq:4eq:ivv-sol}
\end{eqnarray}
with $S(z,\bar z, w,\bar w)\in GL(N_{\rm C},{\bf C})$. 
The solution is characterized by an 
$N_{\rm C} \times N_{\rm F}$ matrix function 
$H_0(z, w)$ 
whose components are holomorphic with respect to 
both $z$ and $w$. 
We call this $H_0(z,w)$ as the moduli matrix 
for the instanton-vortex-vortex system. 
By introducing a $U(N_{\rm C})$ gauge invariant matrix 
\begin{eqnarray}
 \Omega(z,\bar z, w,\bar w) \equiv SS^\dagger
  \label{eq:4eq:omega-ivv}
\end{eqnarray} 
as in the $1/2$ BPS cases, we can rewrite 
the remaining equation (\ref{eq:4eq:ivv2}) 
as \cite{Eto:2004rz} 
\begin{eqnarray}
  4 \partial_z (\Omega^{-1} \bar{\partial}_z \Omega ) 
+ 4 \partial_w (\Omega^{-1} \bar{\partial}_w \Omega )
= cg^2({\bf 1}_{N_{\rm C}} - \Omega^{-1}\Omega_0), \ \ 
c\Omega_0 \equiv H_0 H_0^\dagger, 
\label{eq:4eq:master_ivv}
\end{eqnarray}
which we call the master equation for 
the instanton-vortex-vortex system. 
\changed{
When the Higgs fields are decoupled 
by putting $c=0$ and $H_0=0$, 
this equation reduces to 
the so-called Yang's equation \cite{Yang:1977zf}, 
in the form of 
the left hand side of (\ref{eq:4eq:master_ivv}) 
being equal to zero.
The existence and uniqueness of a solution of 
the master equation (\ref{eq:4eq:master_ivv}) 
was rigorously proved in \cite{MundetiRiera:1999fd,CRGS} 
in the form of the Hitchin-Kobayashi correspondence,  
at least when the base manifold is compact K\"ahler manifold 
instead of ${\bf C}^2$ in our case. 
We simply expect that this holds for ${\bf C}^2$ 
once the moduli matrix $H_0(z,w)$ is given. 
}

Similarly to the $1/2$ BPS cases, two moduli matrices related by 
the following $V$-equivalence relation gives identical 
physics 
\begin{eqnarray}
H_0 \ \sim \ H_0 '= V(z,w) H_0, \ \ \
S \ \sim \ S '= V(z,w) S, 
\label{eq:4eq:ivv-V-equiv}
\end{eqnarray}
where $V(z,w)\in GL(N_{\rm C}, {\bf C})$ 
has components holomorphic with respect to 
both $z$ and $w$. 
Therefore the total moduli space of this system is 
the quotient divided by $\sim$ defined in the 
$V$-equivalence relation (\ref{eq:4eq:ivv-V-equiv}) 
\begin{eqnarray}
{\cal M}^{\rm total}_{\rm IVV}\equiv 
\{ H_0 \ | \  {\bf C}^2 \rightarrow 
M(N_{\rm C}\times N_{\rm F}, {\bf C}), \ 
\bar{\partial}_z H_0=0, \ 
\bar{\partial}_w H_0=0  \} / \sim .
\label{eq:4eq:moduli_sp_ivv}
\end{eqnarray}
Under the $V$-equivalence relation, $\Omega$ 
transforms as $\Omega\sim V\Omega V^\dagger$.


\subsubsection{Monopole-vortex-wall system} 
\label{4eq:mvw}

We can obtain $1/4$ BPS equations for the 
monopole-vortex-wall system by performing 
the SS reduction along the $x^4$(or $x^3$) direction 
in equations (\ref{eq:4eq:ivv1})--(\ref{eq:4eq:ivv3}):  
\begin{eqnarray}
 {\cal D}_2 \Sigma_{4} - F_{31} =0,
\quad 
 {\cal D}_1 \Sigma_4 - F_{23} =0, 
\label{eq:4eq:mvw-1}\\
 {\cal D}_3 \Sigma_4 - F_{12} 
- \frac{g^2}{2}\left(c{\bf 1}_{N_{\rm C}} - HH^\dagger\right)  =0 ,
\label{eq:4eq:mvw-2}\\
 {\cal D}_1 H + i  {\cal D}_2 H =0, 
\quad 
 {\cal D}_3 H +  \Sigma_4 H -H M_4=0,
\label{eq:4eq:mvw-3}
\end{eqnarray}
where the mass parameter $M_p$ is obtained by (SS) 
twisting the phase in compactifying along the $x^p$ 
direction. 
These equations describe composite states of monopoles with 
codimensions in $x^{1}$-$x^{2}$-$x^{3}$, 
vortices with codimensions in the $x^{1}$-$x^{2}$ plane and 
walls perpendicular to the $x^3$ direction. 
\changed{
These equations were originally found in \cite{Tong:2003pz} 
without walls and later in 
\cite{Isozumi:2004vg} with walls. 
}
The Bogomol'nyi energy bound of this system is also 
obtained by the SS reduction of equation  (\ref{eq:4eq:ivv-ch}) as 
\begin{eqnarray}
E_{\rm MVW} = {\cal M}_{123} 
             +{\cal V}_{12}   v_{3}
             +{\cal W}_{3,4}  v_{12} , 
\label{eq:4eq:mvw-ch}
\end{eqnarray}
where we have defined
\begin{eqnarray}
{\cal M}_{123} \equiv \frac{2}{g^2} \int d^3 x
{\rm Tr}\left[\epsilon_{mnk}\partial_k (F_{mn}\Sigma_4)\right],
\label{eq:4eq:mass_m}\\
{\cal W}_{3,4} \equiv c \int \partial_3 ({\rm Tr} \Sigma_4).
\label{eq:4eq:mass_w}
\end{eqnarray}
\changed{
${\cal M}_{123}$ denotes the mass of the monopoles 
and ${\cal W}_{3,4}$ denotes the mass of walls 
perpendicular to the $x^3$ direction with the mass 
matrix $M_4$ to specify the tension. 
${\cal V}_{12}$ is defined in equation~(\ref{eq:4eq:mass_v}).}
Here a check over the suffix in ${\cal I}_{123\check{4}}$ 
denotes to omit that reduced direction, 
and $v_{3}=\int dx^3 $ is the $x^3$ world-volume of vortices. 

Since equation (\ref{eq:4eq:mvw-1}) provide the integrability 
conditions, equation (\ref{eq:4eq:mvw-3}) are solved by 
\begin{eqnarray}
\bar{W}_z = -iS^{-1}\bar{\partial}_z S, \quad
{W}_3 -i\Sigma_4 = -iS^{-1}{\partial}_3 S, 
\quad
H = S^{-1}H_0(z)e^{M_4 x^3},
\label{eq:4eq:mvw-sol}
\end{eqnarray}
with $S(z,\bar z,x^3)\in GL(N_{\rm C},{\bf C})$ and $H_0(z)$ 
is the moduli matrix for the monopole-vortex-wall system, 
which is holomorphic with respect to only $z$, after the 
$e^{M_4 x^3}$ factor is extracted. 
The $V$-equivalence relation $\sim$ for this system becomes 
\begin{eqnarray}
H_0 \ \sim \ H_0 '= V(z) H_0, \ \ \
S \ \sim \ S '= V(z) S, 
\label{eq:4eq:mvw-V-equiv}
\end{eqnarray}
where $V(z)\in GL(N_{\rm C}, {\bf C})$ has components 
holomorphic with respect to only $z$. 
The total moduli space of this system is the quotient by 
this $V$-equivalence relation $\sim$ 
\begin{eqnarray}
{\cal M}^{\rm total}_{\rm MVW}\equiv 
\{ H_0 \ | \  {\bf C} \rightarrow 
M(N_{\rm C}\times N_{\rm F}, {\bf C}), \ 
\bar{\partial}_z H_0=0  \} / \sim. 
\label{eq:4eq:mod_sp_mvw}
\end{eqnarray}
In terms of a gauge invariant matrix 
\begin{eqnarray} 
  \Omega(z,\bar z,x^3) \equiv SS^\dagger, 
 \label{eq:4eq:omega-mvw}
\end{eqnarray}
equation (\ref{eq:4eq:mvw-2}) can 
be converted to the master equation 
of this system \cite{Isozumi:2004vg}
\begin{eqnarray}
\hspace{-1cm}  4 \partial_z (\Omega^{-1} \bar{\partial}_z \Omega ) 
+ {\partial}_3 (\Omega^{-1} \partial_3 \Omega ) 
= cg^2({\bf 1}_{N_{\rm C}}- \Omega^{-1} \Omega_0 ), \ \ 
c\Omega_0 \equiv H_0 e^{2M_4 x^3  }H_0^\dagger.
\label{eq:4eq:master_mvw}
\end{eqnarray}

\subsubsection{Wall webs
} 
\label{4eq:hww}
A further SS reduction 
in equations (\ref{eq:4eq:mvw-1})--(\ref{eq:4eq:mvw-3}) 
along the $x^2$(or $x^1$) direction 
gives another set of 1/4 BPS equations \cite{Eto:2005cp,Eto:2005fm}
\begin{eqnarray}
&&
F_{13} - i [\Sigma_2 ,  \Sigma_{4}] =0, \ \  
{\cal D}_1 \Sigma_4  - {\cal D}_3 \Sigma_2 =0, 
\label{eq:4eq:hww-1} \\
&& 
 {\cal D}_1 \Sigma_2 +  {\cal D}_3 \Sigma_4  =Y_3, \ \ 
\label{eq:4eq:hww-2} \\
&&
{\cal D}_1 H + \Sigma_2 H - H M_2 =0,  \ \ 
{\cal D}_3 H + \Sigma_4 H - H M_4 =0 . 
\label{eq:4eq:hww-3}
\end{eqnarray}
Note that we do not perform the SS reduction along the 
$x^3$ direction. 
These equations describe composite states of Hitchin 
vortices with codimensions in the $x^{1}$-$x^{3}$ plane 
and webs of walls as straight lines in $x^1$, 
$x^3$ plane. 
The Bogomol'nyi energy bound of this system is given by 
\begin{eqnarray}
T_{\rm HWW} =  {\cal H}_{13} 
             + {\cal W}_{1,2}  v_{3}
             + {\cal W}_{3,4}  v_{1}
\label{eq:4eq:hww-ch}
\end{eqnarray}
where we have defined
\begin{eqnarray}
{\cal H}_{12}
\equiv \frac{8}{g^2} \int d^2 x\ {\rm Tr}
\left[
\partial_{[1}\left( ({\cal D}_{3]}\Sigma_4)\Sigma_2 \right)
\right]
.
\label{eq:4eq:mass_h}
\end{eqnarray} 
\changed{
This is the mass of solitons in the Hitchin system.
It is also called $Y$-charge in literature, especially in the
context of the Abelian gauge theory.
}

Since equation (\ref{eq:4eq:hww-1}) provides the integrability 
conditions, equation (\ref{eq:4eq:hww-3}) are solved by 
\begin{eqnarray}
\hspace{-1cm} 
{W}_1 -i\Sigma_2 = -iS^{-1}{\partial}_1 S, \ \ 
{W}_3 -i\Sigma_4 = -iS^{-1}{\partial}_3 S, \ \ 
H = S^{-1}H_0 e^{M_2x^1 + M_4x^3},
\label{eq:4eq:hww-sol}
\end{eqnarray}
where $S(x^1,x^3)\in GL(N_{\rm C},{\bf C})$ and 
the mere constant complex matrix $H_0$ 
is the moduli matrix for 
the Hitchin-wall-wall system. 
Because of the $V$-equivalence relation 
\begin{eqnarray}
H_0 \ \sim \ H_0 '= V H_0, \ \ \
S \ \sim \ S '= V S, 
\quad V\in GL(N_{\rm C}, {\bf C}), 
\label{eq:4eq:hww2-V-equiv}
\end{eqnarray}
the total moduli space of this system is 
the complex Grassmann manifold 
\begin{eqnarray}
 {\cal M}^{\rm total}_{\rm HWW}\equiv 
 \{ H_0 \ | \  H_0 \ \sim \  V H_0, \ \ 
 V\in GL(N_{\rm C}, {\bf C}) \} \simeq G_{N_{\rm F},N_{\rm C}}. 
 \label{eq:4eq:mod_sp_hww}
\end{eqnarray}
This total moduli space is isomorphic to that of the 
$1/2$ BPS walls. 
In terms of a gauge invariant matrix 
\begin{eqnarray}
 \Omega(x^1,x^3) \equiv SS^\dagger, 
 \label{eq:4eq:omega-hww}
\end{eqnarray}
equation (\ref{eq:4eq:hww-2}) can 
be recast into the master equation \cite{Eto:2005cp,Eto:2005fm}
\begin{eqnarray}
\hspace{-2cm} 
  {\partial}_1 (\Omega^{-1} \partial_1 \Omega ) 
+ {\partial}_3 (\Omega^{-1} \partial_3 \Omega ) 
= cg^2({\bf 1}_{N_{\rm C}} - \Omega^{-1} \Omega_0 ), \ \ 
c\Omega_0 \equiv H_0 e^{2 M_2 x^1 + 2 M_4 x^3} H_0^\dagger.
\label{eq:4eq:master_hww}
\end{eqnarray}

\subsubsection{Summary} 
\label{4eq:sum} 
\changed{
In the strong gauge coupling limit 
in theories with $N_{\rm F}>N_{\rm C}$, 
the master equations (\ref{eq:4eq:master_ivv}), 
(\ref{eq:4eq:master_mvw}) and (\ref{eq:4eq:master_hww}) 
can be solved algebraically as 
\begin{eqnarray}
 \Omega_{g^2 \to \infty} 
 = \Omega_0 . 
 \label{eq:4eq:infity-sol}
\end{eqnarray}
We thus obtain exact solutions in all the systems. 
In this limit 
the total moduli spaces of 1/4 BPS systems are reduced, 
as in equation (\ref{eq:vtx:totalquotient2}), to
\begin{eqnarray} 
{\cal M}^{\rm total}_{\rm IVV}\big|_{g^2\rightarrow \infty }
 &\simeq&\{H_0|{\bf C}^2\,\rightarrow \,G_{N_{\rm F},N_{\rm C}},
          \bar \partial _z H_0 = \bar \partial _w H_0 = 0\} ,  \nonumber \\
{\cal M}^{\rm total}_{\rm MVW}\big|_{g^2\rightarrow \infty }
 &\simeq&\{H_0|{\bf C}\,\rightarrow \,G_{N_{\rm F},N_{\rm C}},
          \bar \partial _zH_0=0\} ,  \nonumber \\
{\cal M}^{\rm total}_{\rm HVV}\big|_{g^2\rightarrow \infty }
 &\simeq& G_{N_{\rm F},N_{\rm C}} .
 \label{eq:4eq:totalquotient}
\end{eqnarray}
The third one is isomorphic to the corresponding one at finite gauge coupling, 
while the first two develop small lump singularities 
as the case of semi-local vortices in 
section \ref{sec:vtx:semi}.
These small lump singularities are resolved in 
${\cal M}^{\rm total}_{\rm MVW}$ at finite gauge coupling 
by ANO vortices, 
but it is not the case in ${\cal M}^{\rm total}_{\rm IVV}$ 
which still contains small instanton singularities. 
}

\begin{figure}[htb]
\begin{center}
\includegraphics[width=6cm,clip]{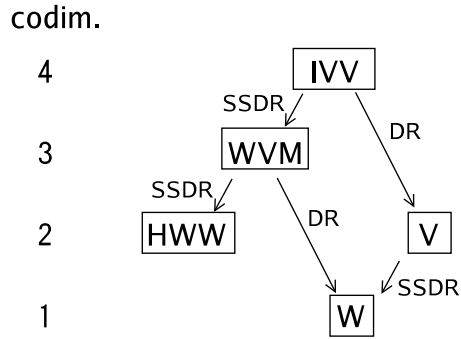}
\caption{
``SSDR'' and ``DR'' denote 
the Scherk-Schwarz and ordinary dimensional reductions, respectively.  
Systems and their equations and charges are related by 
the SS and/or ordinary dimensional reductions. 
}
\label{fig:wll:diagram}
\end{center}
\end{figure}

\changed{
We have seen that 
1/4 BPS systems and 
their BPS equations and charges 
are related by the SS dimensional reduction. 
Accordingly we obtain the relations among moduli matrices  
$H_0^{\rm IVV}(z,w)$ of the instanton-vortex-vortex system, 
$H_0^{\rm MVW}(z)$ of the monopole-vortex-wall system, 
and $H_0^{\rm HWW}$ of the Hitchin-wall-wall system. 
The construction methods 
(\ref{eq:4eq:ivv-sol}), (\ref{eq:4eq:mvw-sol}), 
and (\ref{eq:4eq:hww-sol}) in addition to the SS reductions 
reveal the following relations among moduli matrices 
\begin{eqnarray}
 H_0^{\rm IVV}(z,w)\Bigl|_{\rm MVW}=H_0^{\rm MVW}(z)e^{M_4 w}, \ \ 
 H_0^{\rm MVW}(z)\Bigl|_{\rm HWW}=H_0^{\rm HWW}e^{M_2 z}.
 \label{eq:4eq:rel-H0}
\end{eqnarray}
Namely, a particular dependence of the 
instanton-vortex-vortex moduli matrix $H_0^{\rm IVV}(z,w)$ 
on $w$ 
provides the monopole-vortex-wall moduli matrix $H_0^{\rm MVW}$, 
and a similar particular dependence of the 
monopole-vortex-wall moduli matrix $H_0^{\rm MVW}(z)$ on $z$ 
gives the Hitchin vortex-wall-wall moduli matrix $H_0^{\rm HWW}$. 
Also 1/2 BPS systems of vortices and 
domain walls can be obtained by 
ordinary dimensional reduction 
with respect to $z,\bar z$ (or $w,\bar w$) 
from the instanton-vortex system and 
the monopole-vortex-wall system, respectively.
In this case, the moduli matrices are obtained 
by just throwing away the dependence to $z$ or $w$: 
\begin{eqnarray}
 H_0^{\rm IVV}(z,w)\Bigl|_{\rm V} = H_0^{\rm V}(z), \ \ 
 H_0^{\rm MVW}(z)\Bigl|_{\rm W} = H_0^{\rm W}.
 \label{eq:4eq:rel-H0-2}
\end{eqnarray}
We summarize the relations with all 1/2 and 1/4 BPS 
systems in figure \ref{fig:wll:diagram}. 
One can easily recognizes that 
understanding the moduli matrix $H_0^{\rm IVV}(z,w)$ 
of the instanton-vortex-vortex system contains understanding 
the all the systems. 
}

\subsection{Wall junctions}\label{juc}
We will investigate the 1/4 BPS state of the domain walls 
and their junctions which are solutions of the 1/4 BPS 
equations~(\ref{eq:4eq:hww-1})--(\ref{eq:4eq:hww-3}). 
To simplify notation in this section, we choose the $x^3$ and 
  $x^4$ directions in performing the SS reduction, 
that is, we exchange the $x^3$ direction with the 
$x^2$ direction in section \ref{4eq:hww}.  
We work in the supersymmetric $U(N_{\rm C})$ gauge theory 
with $N_{\rm F}\ (>N_{\rm C})$ hypermultiplets in this 
section. 
We turn on fully non-degenerate 
complex masses 
$M =M_3+iM_4= {\rm diag}(m_1+in_1,\cdots,m_{N_{\rm F}}+in_{N_{\rm F}})$
for the hypermultiplets, and the FI parameter $c\ (> 0)$ 
in the third direction of the $SU(2)_R$ triplet. 
Most of the arguments here will follow along the lines of 
the 1/2 BPS domain walls in section~\ref{wll}. 
For instance, the total moduli space of webs of walls 
turns out to be identical to the total moduli space of 
walls: $G_{N_{\rm F}, N_{\rm C}}=SU(N_{\rm F})/[SU(N_{\rm C}) \times 
SU(N_{\rm F} - N_{\rm C}) \times U(1)]$. 
However, we will find that decomposition of 
the total moduli space into various topological sectors 
exhibits interesting differences.

\subsubsection{Webs of walls in the Abelian gauge theory}
Let us first explain the webs of walls in the Abelian gauge 
theory ($N_{\rm C}=1$) here leaving 
the non-Abelian gauge theory in the next subsection. 
As explained in section~\ref{mdl},
there exist ${N_{\rm F}}$ discrete vacua which are labeled 
by an integer $\left< A \right>$ with 
$A \in \{1,2,\cdots, N_{\rm F}\}$.
In section \ref{wll}, we have found solutions 
of the 1/2 BPS equation~(\ref{eq:wll:BPSeq-wall}) 
that interpolate between these discrete vacua and 
form stable 1/2 BPS domain walls. 
Recall that all the domain walls contained in 
the 1/2 BPS solutions are parallel and are associated with 
the relatively real masses for the hypermultiplet scalars. 
Suppose that a mass difference between non-vanishing 
hypermultiplet scalars in two vacua becomes complex. 
Even in such a situation, a $1/2$ BPS wall can be formed 
interpolating between these two vacua. 
The tension of the domain wall is determined by the 
magnitude of the mass difference. 
However, the normal vector of the wall is no longer along 
the real axis, and the wall preserves a different half of 
supercharges. 
In the case of complex masses, we can obtain walls whose 
normal vectors are in different directions in a 
two-dimensional plane. 
Such a configuration preserves a quarter of supercharges. 
Consequently, domain walls can develop webs of 
domain walls as 1/4 BPS states. 
Therefore the 1/2 BPS equations~(\ref{eq:wll:BPSeq-wall}) 
are naturally extended to the 1/4 BPS 
equations~(\ref{eq:4eq:hww-1})--(\ref{eq:4eq:hww-3}), 
once we turn on the complex masses. 

Solutions of the 1/4 BPS 
equations~(\ref{eq:4eq:hww-1})--(\ref{eq:4eq:hww-3}) 
are given in terms of the moduli matrix $H_0$ 
which is a complex $1 \times N_{\rm F}$ matrix given
in equation~(\ref{eq:4eq:hww-sol}) in the case of the 
Abelian gauge theory. 
The configurations of the webs of walls are made from 
three building blocks; the vacua, the domain walls and 
the junctions. 
Let us first explain how to understand these building 
blocks via the moduli matrix $H_0$.
The moduli matrix is represented as 
\begin{eqnarray}
H_0 = 
\left(
\tau^{\left<1\right>},
\tau^{\left<2\right>},\cdots,
\tau^{\left<N_{\rm F}\right>}\right), 
\label{eq:juc:moduli_matrix_abel}
\end{eqnarray}
with $\tau^{\left<A\right>} \in {\bf C}$.
This can be regarded as the homogeneous coordinate of 
the total moduli space ${\bf C}P^{N_{\rm F}-1}$ given 
in equation~(\ref{eq:4eq:mod_sp_hww}) by taking the 
$V$-equivalence relation (\ref{eq:4eq:hww2-V-equiv}) into 
account. 
Similarly to equation~(\ref{eq:wll:weight}) for domain 
walls, we can define a weight $\exp({\cal W}^{\left< A \right>})$ 
of a vacuum $\left< A \right>$ with a linear function 
${\cal W}^{\left< A \right>}(x^1,x^2)$ as 
\begin{eqnarray}
\exp\left(2{\cal W}^{\left< A \right>}(x^1,x^2)\right) \equiv 
\exp
2\left(
m_{A}x^1 + n_{A}x^2 + a^{\left<A\right>}
\right),
\label{eq:juc:weight_ab}
\end{eqnarray}
\begin{eqnarray}
a^{\left<A\right>} \equiv 
{\rm Re}\left(\log\tau^{\left<A\right>}\right).
\end{eqnarray}
As in the case of the parallel walls given in 
equation~(\ref{eq:wll:energy}), the energy density of the 
webs of walls 
can, then,  be nicely estimated in terms of the weights 
of vacua in equation~(\ref{eq:juc:weight_ab}) as 
\begin{eqnarray}
{\cal E} &\simeq&  
\frac{c}{2} \left(\partial_1^2+\partial_2^2\right)
\log \left(H_0e^{M_1x^1+M_2x^2}H_0^\dagger\right) 
\nonumber\\
&=&
 \frac{c}{2} \left(\partial_1^2+\partial_2^2\right)
\log \left[\sum_{\left< A \right>} 
e^{2 {\cal W}^{\left< A \right>}(x^1,x^2)}\right].
\label{eq:juc:energy_ab}
\end{eqnarray}
This approximation gives sufficiently accurate profiles 
away from the cores, although this is not good near the 
core of the domain walls. 
Furthermore this expression becomes exact at the strong 
gauge coupling limit. 
The function $\log \left[\sum_{\left< A \right>} 
e^{2 {\cal W}^{\left< A \right>}
}\right]$ to be 
differentiated in equation~(\ref{eq:juc:energy_ab}) 
is an almost piecewise linear function which is 
well-approximated by smoothly connecting linear functions 
$2{\cal W}^{\left<A\right>}(x^1,x^2)$. 
Therefore, the energy density (\ref{eq:juc:energy_ab}) 
vanishes in almost all region except near the transition 
points between $2{\cal W}^{\left<A\right>}$ and 
$2{\cal W}^{\left<B\right>}$. 
The regions where the energy density vanishes are nothing 
but the SUSY vacua and the transition lines which now 
spread on the $x^1$-$x^2$ plane correspond to the domain 
walls dividing such vacuum domains. 

We are now ready to understand the three building blocks 
via the moduli matrix $H_0$. 
Let us first note that the 1/4 BPS 
equations~(\ref{eq:4eq:hww-1})--(\ref{eq:4eq:hww-3}) 
admit solutions of  1/2 BPS equations. 
Then both the SUSY vacua and the single domain walls 
arise as solutions of the 1/4 BPS equations in terms of 
the moduli matrices with the same characteristic 
properties as in the case of the 1/2 BPS 
equation~(\ref{eq:wll:BPSeq-wall}). 
Recall that the vacuum state $\left<A\right>$ corresponds 
to the moduli matrix like equation~(\ref{eq:wll:one-tau}) 
\begin{eqnarray}
H_0 = \left(\cdots, \tau^{\left< A \right>},\cdots \right)
\sim \left( 0,\cdots,0,1,0,\cdots,0 \right),
\label{eq:juc:vac_ab}
\end{eqnarray}
and the 1/2 BPS domain wall interpolating two vacua 
$\left< A \right>$ and $\left< B \right>$ corresponds to 
the moduli matrix 
\begin{eqnarray}
H_0 = 
\left(0,\cdots,0,\tau^{\left< A \right>},0,\cdots,0,
\tau^{\left< B \right>},0,\cdots,0 \right), 
\label{eq:juc:2wall_homo_ab}
\end{eqnarray}
where there are only two non-vanishing weights of the vacua 
$\exp({\cal W}^{\left< A \right>})$ and 
$\exp({\cal W}^{\left< B \right>})$ defined 
in equation~(\ref{eq:juc:weight_ab}). 
Similarly to the 1/2 BPS case in 
equation~(\ref{eq:wll:weight_balance_ab}) for the 
domain wall positions, 
the position of the domain wall can be estimated 
as the transition line on which the two weights become 
equal 
${\cal W}^{\left< A \right>} = 
{\cal W}^{\left< B \right>}$:
\begin{eqnarray}
\left(m_{A} - m_{B}\right) x^1
+ \left(n_{A} - n_{B}\right) x^2
+ a^{\left< A \right>} 
- a^{\left< B \right>} = 0.
\label{eq:juc:weight_balance_ab}
\end{eqnarray}
When we turn off the imaginary part of the masses, 
this reduces to equation~(\ref{eq:wll:weight_balance_ab}) 
of the 1/2 BPS single wall. 
The complex masses of the hypermultiplets determine 
the angle of the domain wall in the $x^1$-$x^2$ plane 
and its position is given by difference of the parameters 
$a^{\left< A \right>} - a^{\left< B \right>}$. 
Notice that it was important to keep track of 
the ordering of the real masses in the case of 
parallel walls in section~\ref{wll}. 
Due to the ordering of the real masses, 
the single walls are classified into two types: 
elementary and non-elementary. 
However, the ordering is now meaningless in the space of 
the complex masses. 
Consequently, all the single walls become elementary 
walls in the case of the fully non-degenerate complex 
masses. 
The following tension vector is parallel to the domain wall 
and its magnitude gives the tension (per unit length) of 
the domain wall \footnote{
Here the sign of the tension vector is merely a convention. 
}
\begin{eqnarray}
\vec{T}^{\left<A\right> \left<B\right>} 
=  c \left(n_B - n_A, m_A - m_B\right). 
\label{eq:juc:wall_tension}
\end{eqnarray}
Notice that the first component of the tension is 
related to the central charge 
$Z_1$ and $Z_2$ given in equation~(\ref{eq:4eq:hww-ch}).

The difference between the 1/2 BPS 
solution~(\ref{eq:4eq:hww-sol}) and 
the 1/4 BPS solutions~(\ref{eq:4eq:hww-sol}) 
first occurs when we consider the moduli matrix 
with three non-zero elements 
\begin{eqnarray}
H_0 = 
\left(
0,\cdots,0,
\tau^{\left< A \right>},
0,\cdots,0,
\tau^{\left< B \right>},
0,\cdots,0,
\tau^{\left< C \right>},
0,\cdots,0
\right).
\label{eq:juc:junction_ab}
\end{eqnarray}
As already mentioned in section~\ref{wll}, this moduli 
matrix with the real masses describe the two parallel 
walls which divides three vacuum domains 
$\left<A\right>$, $\left<B\right>$ and $\left<C\right>$.
However, equation~(\ref{eq:juc:weight_balance_ab}) shows 
that the complex masses change the angle of the walls. 
Therefore the three walls $\left<A\right>\left<B\right>$, 
$\left<B\right>\left<C\right>$, and 
$\left<C\right>\left<A\right>$ should meet at a point to 
form a 3-pronged junction. 
Positions of component walls of the 3-pronged junction 
can be identified by the equal weight condition of two 
vacua as given in equation (\ref{eq:juc:weight_balance_ab}). 
Furthermore, the position of the domain wall junction is 
identified as a point where all the three weights become 
equal 
\begin{eqnarray}
{\cal W}^{\left<A\right>}(x^1,x^2) = 
{\cal W}^{\left<B\right>}(x^1,x^2) = 
{\cal W}^{\left<C\right>}(x^1,x^2).
\label{eq:juc:weight_balance_junc_ab}
\end{eqnarray}
One can easily show that the tension vectors of the domain walls 
given in equation~(\ref{eq:juc:wall_tension}) 
are balanced each other, so that the junction is stable:
\begin{eqnarray}
\vec{T}^{\left<A\right>\left<B\right>} + 
\vec{T}^{\left<B\right>\left<C\right>} + 
\vec{T}^{\left<C\right>\left<A\right>} 
= \vec 0. 
\label{eq:juc:tension}
\end{eqnarray}
This condition of the balance of forces is assured by 
the fact that the central charges $(Z_1,Z_2)$ of three 
constituent walls meeting at the junction 
sum up to zero. 
Besides the central charges $(Z_1,Z_2)$ associated to 
the constituent walls, junction has another characteristic 
central charge $Y$ in equation~(\ref{eq:4eq:hww-ch}). 
The $Y$-charge can be exactly calculated as
\begin{eqnarray}
Y_{\rm Abelian} = - \frac{2}{g^2} 
\left|(\vec \mu_A - \vec \mu_C) 
\times (\vec \mu_B - \vec \mu_C) \right|,
\label{eq:juc:Y_ab}
\end{eqnarray}
where the cross means the exterior product of 2-vector 
$\mu_A = (m_A,n_A)$ giving a scalar. 
Notice that the $Y$-charge of the junction in the 
Abelian gauge theory always gives negative contribution 
to the total energy, which is understood as 
the binding energy of the domain walls meeting at the 
junction point.

The set of variables $\left\{\tau^{\left< A \right>},
\tau^{\left< B \right>},\tau^{\left< C \right>}\right\}$ 
in equation~(\ref{eq:juc:junction_ab}) describes the moduli 
space of three pronged junction. 
It is just a homogeneous coordinate of ${\bf C}P^2$ 
submanifold of the total moduli space 
${\bf C}P^{N_{\rm F}-1}$. 
Let us illustrate how this ${\bf C}P^2$ manifold is 
decomposed into several topological sectors. 
To obtain the moduli space of a genuine 3-pronged junction, 
we have to remove the following subspaces: 
$\tau^{\left< A \right>}=0$, $\tau^{\left< B \right>}=0$ 
and $\tau^{\left< C \right>}=0$ from ${\bf C}P^2$, 
because such subspaces result from different boundary 
conditions due to the limits where one or two of three 
domain walls are taken to spatial infinity. 
Namely, ${\bf C}P^2$ has three ${\bf C}P^1$ submanifolds 
parametrized by 
${\bf C}P^1_{\left<\not C\right>} = \left\{\tau^{\left< A \right>},
\tau^{\left< B \right>},0\right\}$, 
${\bf C}P^1_{\left<\not A\right>} = 
\left\{0,\tau^{\left< B \right>},\tau^{\left< C \right>}\right\}$, 
and ${\bf C}P^1_{\left<\not B\right>} = 
\left\{\tau^{\left< A \right>},0,\tau^{\left< C \right>}\right\}$, 
respectively. 
Moreover, these submanifolds share the three points 
$\left\{1,0,0\right\}$, $\left\{0,1,0\right\}$ and 
$\left\{0,0,1\right\}$ corresponding to three vacua. 
Then the moduli space of the genuine 3-pronged junction is 
the open space which is given by subtracting three 
${\bf C}P^1$ subspaces from ${\bf C}P^2$ as 
\begin{eqnarray}
{\cal M}^{\rm junction} 
= {\bf C}P^2 - \bigcup_{A=1}^3 
{\bf C}P^1_{\left<\not A\right>}.
\end{eqnarray}
Each ${\bf C}P^1$ subspace consists of 
 two points corresponding to the vacuum states and an 
open space ${\bf C}^*\simeq {\bf R}\times S^1 
\simeq {\bf C}P^1 - 2\times {\bf C}P^0$ 
corresponding to the moduli space of the single domain wall, 
as was mentioned above. 
These are summarized in the following flow diagram in which 
the arrow $\to$ means $\tau^{\left<A\right>} \to 0$, 
the arrow $\nearrow$ means $\tau^{\left<C\right>} \to 0$ and 
the arrow $\searrow$ means $\tau^{\left<B\right>} \to 0$. \\
\begin{center}
\begin{tabular}{ccccc}
3-pronged junction & & single wall & & vacuum\\
\hline
&&&&\\
& & $\left\{\tau^{\left< A \right>},\tau^{\left< B \right>},0\right\}$ 
& $\to$ & $\left\{0,1,0\right\}$ \\
& $\nearrow$ & & $\nearrow\hspace*{-.425cm}\searrow$ & \\
$\left\{\tau^{\left< A \right>},\tau^{\left< B \right>},
\tau^{\left< C \right>}\right\}$
& $\to$ & $\left\{0,\tau^{\left< B \right>},\tau^{\left< C \right>}\right\}$ 
& & $\left\{1,0,0\right\}$\\
& $\searrow$ & & $\nearrow\hspace*{-.425cm}\searrow$ & \\
& & $\left\{\tau^{\left< A \right>},0,\tau^{\left< C \right>}\right\}$ 
& $\to$ & $\left\{0,0,1\right\}$\\
&&&&\\
\hline
${\bf C}P^2$ && ${\bf C}P^1$ && ${\bf CP}^0$
\end{tabular}
\end{center}

\begin{figure}[ht]
\begin{center}
\includegraphics[height=2.5cm]{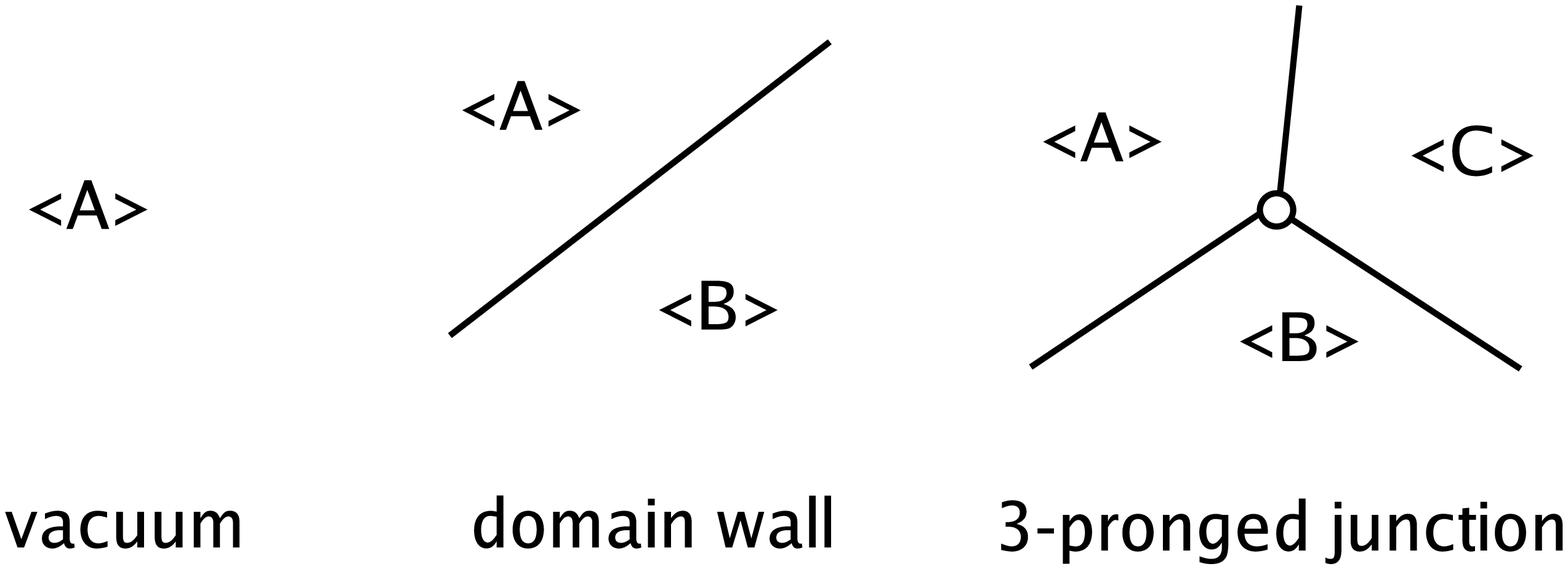}
\caption{\small{\sf 
Building blocks of the webs of walls in the Abelian gauge theory.}}
\label{fig:juc:block_ab}
\vspace*{-.3cm}
\end{center}
\end{figure}

So far, we examined the three building blocks of the webs 
of walls: the vacua, the domain walls, and the 3-pronged 
junctions, as shown in figure~\ref{fig:juc:block_ab}. 
The webs of walls are constructed by putting these building 
blocks together. 
In general, the Abelian gauge theory with $N_{\rm F}$ 
flavors admits webs of domain walls which divide 
$N_{\rm F}$ domains of vacua. 
The moduli matrix for the general configuration can be 
parametrized by the homogeneous coordinate of the total 
moduli space ${\bf C}P^{N_{\rm F}-1}$ as given in 
equation~(\ref{eq:juc:moduli_matrix_abel}) 
\begin{eqnarray}
\left\{
\tau^{\left<1\right>},
\tau^{\left<2\right>},
\cdots,
\tau^{\left<N_{\rm F}\right>} 
\right\}.
\end{eqnarray}
The area of each vacuum domain is proportional to the 
weight of that vacuum in equation~(\ref{eq:juc:weight_ab}). 
The boundary between two adjacent vacuum domains becomes 
a domain wall whose position is determined by equating the 
weights of the two vacua as in 
equation~(\ref{eq:juc:weight_balance_ab}). 
Furthermore, a junction is formed at the point 
where the vacuum weights for three vacua become equal 
as in equation~(\ref{eq:juc:weight_balance_junc_ab}). 
When we let one of the vacuum weights, for instance 
$\exp 2{\cal W}^{\left<A\right>}(x^1,x^2)$, to vanish 
(by taking the limit where $\tau^{\left<A\right>} \to 0$), 
the corresponding vacuum domain $\left<A\right>$ disappears 
in the web configuration. 
As a result, we have a smaller web configuration which 
divides $N_{\rm F}-1$ vacua and is described by the moduli 
matrix for ${\bf C}P^{N_{\rm F}-2}$: 
\begin{eqnarray}
\left\{
\tau^{\left<1\right>},
\cdots,
\tau^{\left<A-1\right>},
0,
\tau^{\left<A+1\right>},
\cdots,
\tau^{\left<N_{\rm F}\right>}
\right\}.
\label{eq:juc:homo_cp-1}
\end{eqnarray}
Since there are $N_{\rm F}$ submanifolds 
${\bf C}P^{N_{\rm F}-2}$ in the total moduli space 
${\bf C}P^{N_{\rm F}-1}$, the moduli space of the 
maximal webs of walls is given by 
\begin{eqnarray}
{\cal M}^{\rm max} = {\bf C}P^{N_{\rm F}-1} - 
\bigcup^{N_{\rm F}}_{A=1} 
{\bf C}P^{N_{\rm F}-2}_{\left<\not A\right>}
\end{eqnarray}
where ${\bf C}P^{N_{\rm F}-2}_{\left<\not A\right>}$ 
is a submanifold parametrized by the homogeneous 
coordinate (\ref{eq:juc:homo_cp-1}).

\begin{figure}[ht]
\begin{center}
\includegraphics[height=4cm]{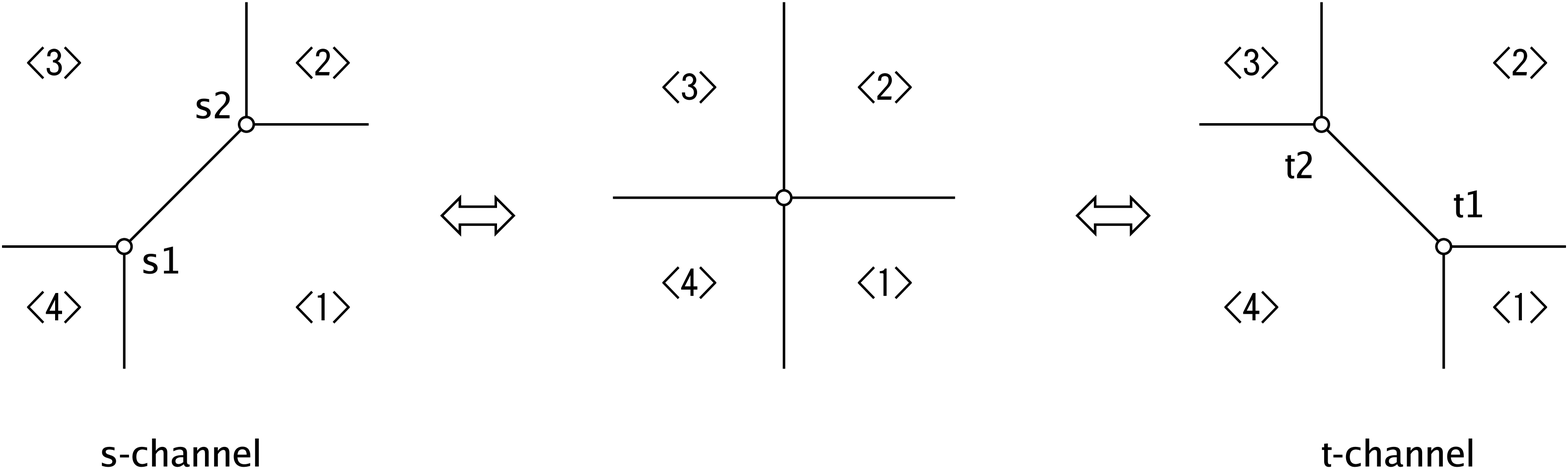}
\caption{\small{\sf 
The s-channel and the t-channel of webs of walls. 
The s-channel appears when 
$a^{\left<1\right>}+a^{\left<3\right>} > 
a^{\left<2\right>}+a^{\left<4\right>}$ 
while the t-channel appears when 
$a^{\left<1\right>}+a^{\left<3\right>} < 
a^{\left<2\right>}+a^{\left<4\right>}.$ 
}}
\label{fig:juc:juc_st}
\vspace*{-.3cm}
\end{center}
\end{figure}

Let us give an example of webs of walls in $N_{\rm F}=4$ 
model with masses $(m_A,n_A)=\{(1,0),(1,1),(0,1),(0,0)\}$. 
Then we have four linear functions 
${\cal W}^{\left<1\right>} = x^1 + a^{\left<1\right>}$, 
${\cal W}^{\left<2\right>} = x^1 + x^2 + a^{\left<2\right>}$, 
${\cal W}^{\left<3\right>} = x^2 + a^{\left<3\right>}$ and 
${\cal W}^{\left<4\right>} = a^{\left<4\right>}$. 
The shape of the web varies as we change the moduli 
parameters $a^{\left<A\right>}$ ($A=1,2,3,4$). 
The configuration has two branches which we called 
s-channel and t-channel in Ref.\cite{Eto:2005cp}. 
The s-channel has an internal wall dividing the vacua 
$\left<1\right>$ and $\left<3\right>$ while the t-channel 
has the other internal wall which divides the vacua 
$\left<2\right>$ and $\left<4\right>$, 
as shown in figure~\ref{fig:juc:juc_st}. 
The s-channel has two junctions is denoted as s1 and s2. 
The s1 separates three vacua 
$\left<1\right>$, $\left<3\right>$ and $\left<4\right>$ at 
$(x^1,x^2)=\left(a^{\left<4\right>}-a^{\left<1\right>}, 
a^{\left<4\right>}-a^{\left<3\right>}\right)$. 
The s2 separates three vacua 
$\left<1\right>$, $\left<2\right>$ and $\left<3\right>$ at 
$(x^1,x^2)=\left(a^{\left<3\right>}-a^{\left<2\right>},
a^{\left<1\right>}-a^{\left<2\right>}\right)$. 
These junctions consistently appears in the parameter 
region where 
$a^{\left<1\right>}+a^{\left<3\right>} 
> a^{\left<2\right>}+a^{\left<4\right>}$. 
The two junctions s1 and s2 approach each other when we 
let 
$\left(a^{\left<1\right>} + a^{\left<3\right>}\right) -
\left(a^{\left<2\right>} + a^{\left<4\right>}\right) \to 0$ 
as shown in the middle of figure~\ref{fig:juc:juc_st}. 
When $a^{\left<2\right>} + a^{\left<4\right>}$ grows 
over $a^{\left<1\right>} + a^{\left<3\right>}$, the 
configuration makes a transition from the s-channel 
to the t-channel which has another two junctions t1 
and t2. 
The t1 separates three vacua 
$\left<1\right>$, $\left<2\right>$ and $\left<4\right>$ at 
$(x^1,x^2)=\left(a^{\left<4\right>}-a^{\left<1\right>}, 
a^{\left<1\right>}-a^{\left<2\right>}\right)$. 
The t2 separates three vacua 
$\left<2\right>$, $\left<3\right>$ and $\left<4\right>$ at 
$(x^1,x^2)=\left(a^{\left<3\right>}-a^{\left<2\right>},
a^{\left<4\right>}-a^{\left<3\right>}\right)$. 
Other examples of the webs of walls are shown in 
Ref.~\cite{Eto:2005cp}. 

\subsubsection{Webs of walls in the non-Abelian gauge theory}
In this subsection, we will study the webs of domain walls 
in the non-Abelian gauge theory ($N_{\rm C}>1$), which has 
$_{N_{\rm F}}C_{N_{\rm C}}$ discrete vacua 
labeled by a set of $N_{\rm C}$ different integers
$\left< A_1 \cdots A_r \cdots A_{N_{\rm C}} \right>$, 
as given in section \ref{sc:mdl:vacua-UN}. 
Similarly to the Abelian case, all the 1/4 BPS solutions 
are given by the moduli matrix $H_0$ which is, now, a 
complex $N_{\rm C}\times N_{\rm F}$ matrix 
given in equation~(\ref{eq:4eq:hww-sol}). 
We have found the total moduli space parametrized by 
$H_0$ to be the complex Grassmaniann 
$G_{N_{\rm F},N_{\rm C}} \simeq \left\{H_0 \sim VH_0\right\}$ 
with $V \in GL(N_{\rm C},{\bf C})$ in 
equation~(\ref{eq:4eq:hww2-V-equiv}). 
In section~\ref{wll} we have seen that it is also useful to 
introduce the Pl\"ucker coordinate instead of the moduli 
matrix $H_0$ itself: 
\begin{eqnarray}
\tau^{\left<\{A_r\}\right>} 
\equiv \det H_0^{\left<\{A_r\}\right>}. 
\label{eq:juc:plucker-coordinate1}
\end{eqnarray}
Here $\left< A_1 \cdots A_r \cdots A_{N_{\rm C}} \right>$ 
is abbreviated as $\left<\{A_r\}\right>$, and the minor 
matrix $H_0^{\left<\{A_r\}\right>}$ is constructed by 
picking up the $\{A_r\}$-th row from the moduli matrix 
$H_0$. 
Since the number of determinants 
$\tau^{\left<\{A_r\}\right>}$ 
are $_{N_{\rm F}}C_{N_{\rm C}}$, which is the same as 
the number of SUSY vacua, 
we assemble the Pl\"ucker coordinate in a 
$_{N_{\rm F}}C_{N_{\rm C}}$ component vector:  
\begin{eqnarray}
\left\{\cdots,\tau^{\left<\{A_r\}\right>},\cdots,
\tau^{\left<\{B_r\}\right>} ,\cdots\right\}.
\label{eq:juc:plucker-coordinate}
\end{eqnarray}

Similarly to the weight of the vacuum 
(\ref{eq:juc:moduli_matrix_abel}) 
in the Abelian gauge theory, 
we can define the weight of the vacuum in the non-Abelian 
gauge theory by 
\begin{eqnarray}
\exp 2{\cal W}^{\left<\{A_r\}\right>} 
\equiv 
\exp 2 \left( \sum_{r=1}^{N_{\rm C}}
\left(
m_{A_r}x^1 + n_{A_r}x^2\right) + a^{\left<\{A_r\}\right>}
\right),
\label{eq:juc:weight}\\
a^{\left<\{A_r\}\right>} + i b^{\left<\{A_r\}\right>} 
\equiv \log \tau^{\left<\{A_r\}\right>},
\end{eqnarray}
which is a natural extension 
from equation~(\ref{eq:wll:weight}) for 
the case of parallel walls. 
The energy density of the webs of walls 
can be estimated in terms of these weights of vacua as 
\begin{eqnarray}
{\cal E} 
&\simeq& \frac{c}{2} \left(\partial_1^2+\partial_2^2\right)
\log\det \left( H_0e^{M_1x^1+M_2x^2} H_0^\dagger \right)
\nonumber\\
&=& \frac{c}{2} \left(\partial_1^2+\partial_2^2\right)
\log \sum_{\left<\{A_r\}\right>} 
e^{2 {\cal W}^{\left<\{A_r\}\right>}}.
\label{eq:juc:energy}
\end{eqnarray}

We can now find out the structure of the webs in the 
non-Abelian gauge theory similarly to the Abelian gauge 
theory. 
The webs are also made of the three building blocks:
vacua, single domain walls and their junctions.
Similarly to the Abelian case in 
equation~(\ref{eq:juc:vac_ab}), vacua are represented 
by a single non-vanishing component 
$\tau^{\left<\{A_r\}\right>}$ 
in the Pl\"ucker coordinate 
(\ref{eq:juc:plucker-coordinate}):
$
\left\{
0,\cdots,0, \tau^{\left<\{A_r\}\right>}, 0,\cdots,0
\right\}
\sim 
\left\{
0,\cdots,0, 1, 0,\cdots,0
\right\}.
$
The domain walls are described by 
equation~(\ref{eq:juc:2wall_homo_ab})
in the Abelian gauge theory, whereas 
the domain walls in the non-Abelian gauge theory 
are represented by only two non-vanishing 
components in the Pl\"ucker coordinate. 
Namely, the domain wall interpolating the vacuum 
$\left<\{A_r\}\right>$and $\left<\{B_r\}\right>$ is 
given by 
$
\left\{
0,\cdots,0, \tau^{\left<\{A_r\}\right>}, 0,\cdots,0, 
\tau^{\left<\{B_r\}\right>}, 0,\cdots,0
\right\}
$. 
However, not all the Pl\"ucker coordinates are independent 
as we have already seen in section~\ref{wll}. 
The Pl\"ucker coordinates are constrained  by the Pl\"ucker 
 relations~(\ref{eq:wll:plucker}) in order for them to 
describe the Grassmaniann. 
For instance, the  Pl\"ucker 
 relations does not allow moduli matrix with only 
two non-vanishing components $\tau^{\left<A_r\right>}$ 
whose labels differ in only one element 
such as $\left<\ \underline{\cdots}\ A \right>$ 
and $\left<\ \underline{\cdots}\ B \right>$.
It follows that no single domain wall exists interpolating 
two such vacua. 

Locations of domain walls are estimated by comparing 
weights of the two adjacent vacua as in 
equation~(\ref{eq:juc:weight_balance_ab}): 
${\cal W}^{\left< \underline{\cdots}A \right>} = 
{\cal W}^{\left< \underline{\cdots}B \right>}$ gives 
the domain wall interpolating 
$\left<\ \underline{\cdots}\ A \right>$ 
and $\left<\ \underline{\cdots}\ B \right>$ to lie 
\begin{eqnarray}
\left(m_{A} - m_{B}\right) x^1
+ \left(n_{A} - n_{B}\right) x^2
+ a^{\left<\underline{\cdots}A \right>} 
- a^{\left<\underline{\cdots}B \right>} = 0.
\label{eq:juc:weight_balance}
\end{eqnarray}

The 3-pronged junctions of the domain walls 
in the non-Abelian gauge theory 
are described by the Pl\"ucker coordinate which has only 
three non-vanishing components:
\begin{eqnarray}
\left\{
0,\cdots,0,
\tau^{\left<\{A_r\}\right>},
0,\cdots,0,
\tau^{\left<\{B_r\}\right>},
0,\cdots,0,
\tau^{\left<\{C_r\}\right>}
0,\cdots,0
\right\}. 
\end{eqnarray}
The position of the 3-pronged junction can be estimated 
by equating the three vacuum weights 
$
{\cal W}^{\left<\{A_r\}\right>} =
{\cal W}^{\left<\{B_r\}\right>} =
{\cal W}^{\left<\{C_r\}\right>}
$ as equation~(\ref{eq:juc:weight_balance_junc_ab}). 
In the previous subsection, we have found that junctions 
in the Abelian gauge theory dividing vacua 
$\left<A\right>$, $\left<B\right>$ and $\left<C\right>$ 
are always characterized by the negative contribution 
to the energy from the topological charge $Y$ 
(\ref{eq:juc:Y_ab}). 
On the other hand, there are two kinds of domain wall 
junctions in the non-Abelian gauge theory. 
Junctions of walls are specified 
by choosing three different vacua 
$\left<\{A_r\}\right>$, $\left<\{B_r\}\right>$ and 
$\left<\{C_r\}\right>$. 
In the non-Abelian gauge theory, 
the Pl\"ucker relation (\ref{eq:wll:plucker}) requires 
any pairs of those three vacua to have flavor labels 
which are different only in one color component. 
Then there are two possibilities to choose three different 
vacua. 
One possibility is the junction which separates three 
vacua with the same flavors except one color component: 
$\left<\ \underline{\cdots}\ A \right>$, 
$\left<\ \underline{\cdots}\ B \right>$ and 
$\left<\ \underline{\cdots}\ C \right>$. 
The other is that dividing a set of vacua 
with the same flavor labels except two color component: 
$\left<\ \underline{\cdots}\ AB \right>$, 
$\left<\ \underline{\cdots}\ BC \right>$ and 
$\left<\ \underline{\cdots}\ CA \right>$. 
The former (latter) is called the Abelian (non-Abelian) 
junction. 
The Abelian junction dividing the vacua 
$\left<\ \underline{\cdots}\ A \right>$, 
$\left<\ \underline{\cdots}\ B \right>$ and 
$\left<\ \underline{\cdots}\ C \right>$ 
is essentially the same as the junction in the Abelian 
gauge theory dividing the vacua 
$\left< A \right>$, 
$\left< B \right>$ and 
$\left< C \right>$. 
Actually the topological charge is the same as that given 
in equation~(\ref{eq:juc:Y_ab}), namely it is negative and 
should be interpreted as the binding energy of the domain 
walls. 
On the contrary, the non-Abelian junction separating the 
three vacuum domains 
$\left<\ \underline{\cdots}\ AB \right>$, 
$\left<\ \underline{\cdots}\ BC \right>$ and 
$\left<\ \underline{\cdots}\ CA \right>$ 
is essentially the same as the junction dividing 
three vacua 
$\left< AB \right>$, 
$\left< BC \right>$ and 
$\left< CA \right>$ 
in the $U(2)$ gauge theory.
Notice that the non-Abelian junctions do not exist in the 
Abelian gauge theory. 
The remarkable property of the non-Abelian junction is 
that the topological $Y$-charge given in 
equation~(\ref{eq:4eq:hww-ch}) always contribute positively 
to the energy density, so that it can not be regarded as 
the binding energy, in contrast to the Abelian $Y$-charge 
\begin{eqnarray}
Y_{\rm non-Abelian} = \frac{2}{g^2}
\left|(\vec \mu_A - \vec \mu_C) 
\times (\vec \mu_B - \vec \mu_C) \right| > 0.
\label{eq:juc:Y_na}
\end{eqnarray}

\begin{figure}[ht]
\begin{center}
\begin{tabular}{ccc}
\includegraphics[height=3cm]{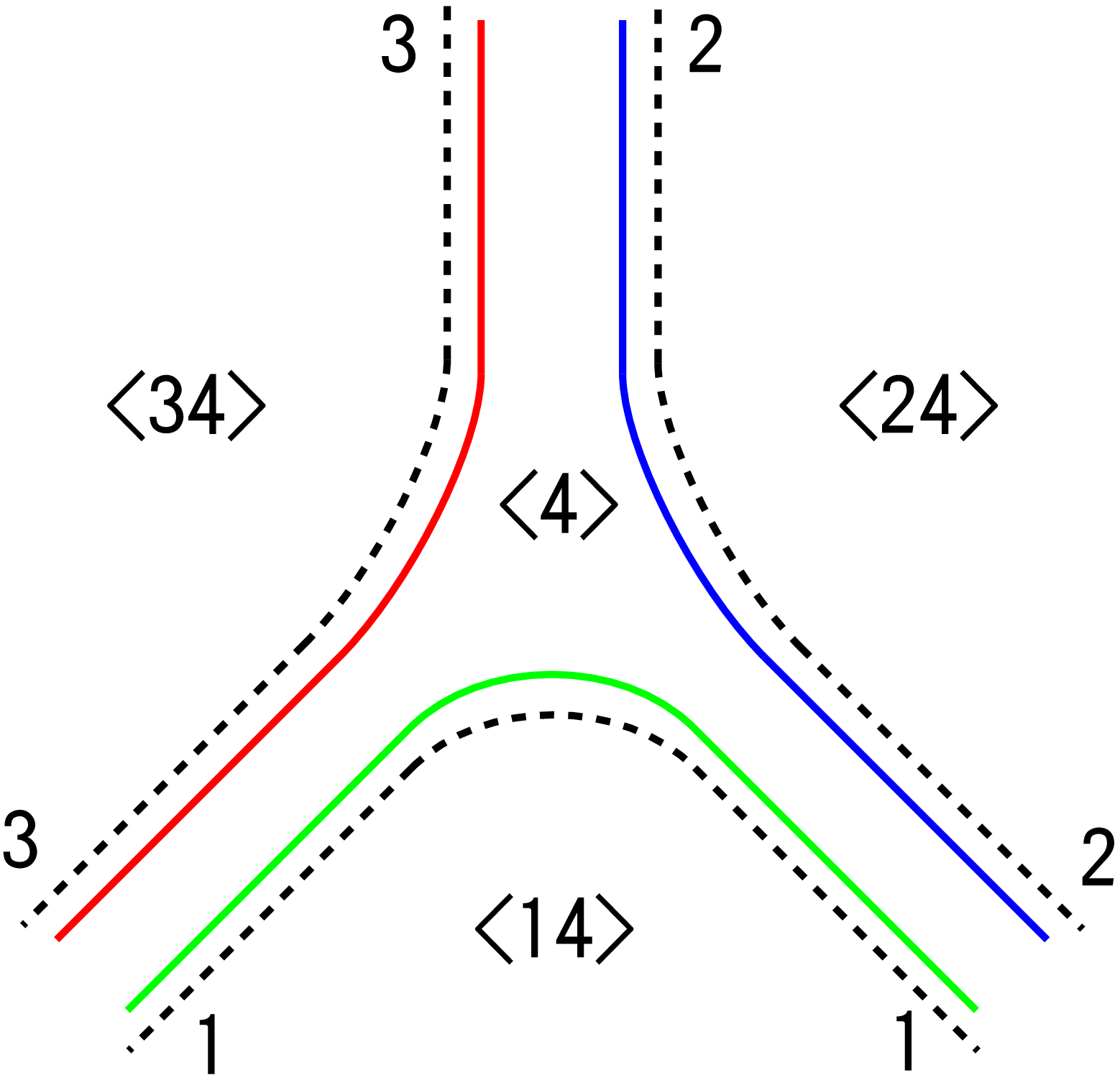}
&\qquad\qquad&
\includegraphics[height=3cm]{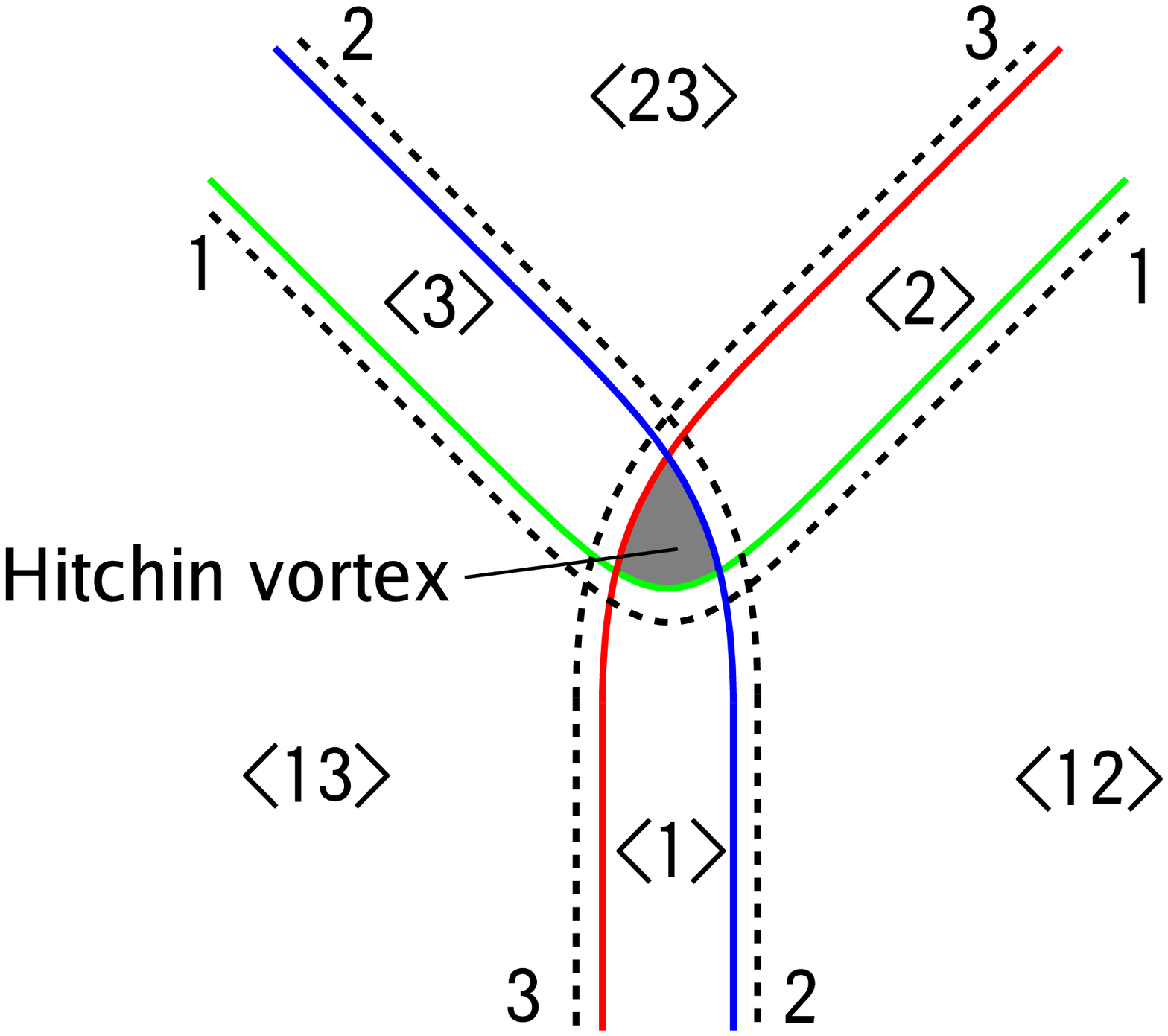}
\end{tabular}
\caption{\small{\sf Internal structures of the junctions
 with $g \sqrt c \ll |\Delta m+i\Delta n|$}}
\label{fig:juc:hitchin}
\vspace*{-.3cm}
\end{center}
\end{figure}
In order to understand the origin of negative and positive 
$Y$-charges, we find it useful to pay attention to the 
internal structures of the junction points of the domain 
walls. 
To this end, let us consider
the model with $N_{\rm F}=4$ and $N_{\rm C}=2$ which has $_4C_2=6$ discrete vacua 
$\langle 12 \rangle$, $\langle 23 \rangle$, 
$\langle 13 \rangle$,
$\langle 14 \rangle$, $\langle 24 \rangle$ and 
$\langle 34 \rangle$. 
The 1/4 BPS wall junction interpolating 
the three vacua $\langle 14 \rangle$, 
$\langle 24 \rangle$ and $\langle 34 \rangle$ is 
the Abelian junction 
while that interpolating 
$\langle 12 \rangle$, $\langle 23 \rangle$ and 
$\langle 13 \rangle$
is the non-Abelian junction.
Internal structures of these junctions are schematically 
shown in figure~\ref{fig:juc:hitchin}.
The left of the figure~\ref{fig:juc:hitchin} shows
the Abelian junction and the right shows
the non-Abelian junction. As explained in section~\ref{wll},
each component domain wall
of the junction has three-layer structure
in the weak gauge coupling region 
$(g\sqrt c \ll |\Delta m+i\Delta n|)$, 
see figure~\ref{fig:wll:wall}. 
The same $U(1)$ subgroup is recovered in all 
three middle layers of the Abelian junction, as 
denoted by $\langle 4\rangle$. 
They are connected at the junction point so that the 
middle layer of the wall junction is also in the same 
phase $\langle 4 \rangle$ 
as can be seen in the left of figure~\ref{fig:juc:hitchin}.
On the other hand, the non-Abelian junction has a complicated
internal structure as shown in the right of 
figure~\ref{fig:juc:hitchin}.
Although it also separates three different vacua 
$\langle 12 \rangle$, $\langle 23 \rangle$ and 
$\langle 13 \rangle$,
their middle layers preserve different $U(1)$ subgroups, 
$\langle1\rangle$, $\langle2\rangle$ and $\langle3\rangle$ as 
in the right of figure~\ref{fig:juc:hitchin}. 
However, all the hypermultiplet scalars $H$ vanish when 
all three middle layers overlap near the junction point, 
so that only the $U(2)$ vector multiplet scalar $\Sigma$ is 
active there. 
The key observation is that the 
1/4 BPS equations given 
in equations~(\ref{eq:4eq:hww-1})--(\ref{eq:4eq:hww-3})
reduce to the 1/2 BPS Hitchin equations 
\begin{eqnarray}
 F_{12}=i\left[\Sigma_3,\Sigma_4\right],\quad 
 {\cal D}_1\Sigma_4-{\cal D}_2\Sigma_3=0,\quad 
 {\cal D}_1\Sigma_3+{\cal D}_2\Sigma_4=0,
\label{eq:juc:hitchin}
\end{eqnarray} 
if we pick up the traceless part of 
equations~(\ref{eq:4eq:hww-1})--(\ref{eq:4eq:hww-3}) and 
discard the hypermultiplet scalars $H$. 
This reduction occurs at the core of the 
non-Abelian junction, since hypermultiplet scalars 
vanish as we mentioned above. 
Therefore the system reduces to the Hitchin system of 
$SU(2)$ subgroup in the middle of the non-Abelian junction. 
Furthermore, the charge of the non-Abelian junction
given in equation~(\ref{eq:4eq:hww-ch}) 
completely agrees with the charge of the Hitchin system 
 \cite{Eto:2005sw}. 
Thus we conclude that the positive $Y$-charges of the 
non-Abelian junctions given in equation~(\ref{eq:juc:Y_na})
are the charges of the Hitchin system.

\begin{figure}[ht]
\begin{center}
\includegraphics[height=2.5cm]{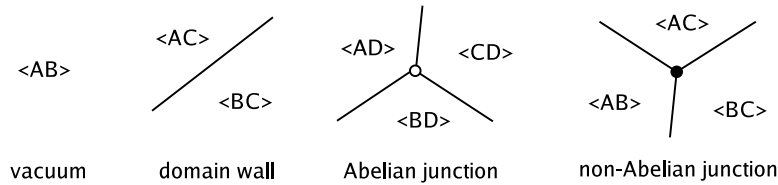}
\caption{\small{\sf 
Building blocks of the webs of walls in the non-Abelian gauge theory.}}
\label{fig:juc:block_na}
\vspace*{-.3cm}
\end{center}
\end{figure}

Now, we have found four kinds of building blocks for the webs
of walls in the non-Abelian gauge theory.
We have the SUSY vacua, the domain walls interpolating 
these discrete vacua 
and the Abelian and the non-Abelian junctions.
These are shown in figure~\ref{fig:juc:block_na}.
There is one more fundamental object in the non-Abelian 
gauge theory. 
It is the trivial intersection, namely the 4-pronged 
junction without a junction charge, of the domain walls. 
Such 4-pronged junction accidentally, of course, appears 
in the Abelian gauge theory as a special configuration in 
which two different 3-pronged junctions get together. 
However, these are decomposed to two 3-pronged 
junctions by varying moduli parameters, so we should not 
regard it as the building 
block of the webs in the Abelian gauge theory. 
We have already met an example in figure~\ref{fig:juc:juc_st}.
There the s-channel and t-channel are interchanged 
when the moduli parameters accidentally satisfy 
$a^{\left<1\right>}+a^{\left<3\right>} =
a^{\left<2\right>}+a^{\left<4\right>}$.
\begin{figure}[ht]
\begin{center}
\includegraphics[height=3.5cm]{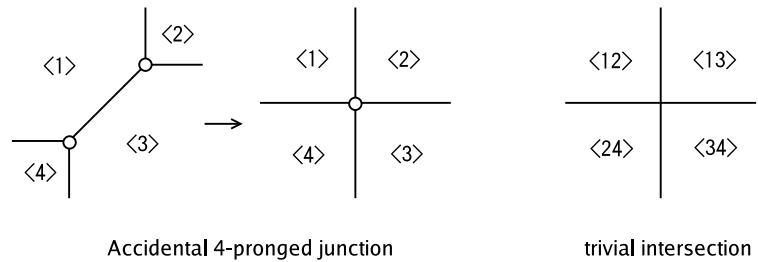}
\caption{\small{\sf 
Accidental 4-pronged junction and a trivial intersection of walls.}}
\label{fig:juc:intersection}
\vspace*{-.3cm}
\end{center}
\end{figure}  
On the other hand, the configuration dividing a set of 
four vacua, for instance 
$\left<12\right>$, $\left<34\right>$, 
$\left<13\right>$ and $\left<24\right>$, 
has to obey the Pl\"ucker relation~(\ref{eq:wll:plucker}).
The Pl\"ucker relation restrict the moduli parameters 
under the condition 
$a^{\left<12\right>}+a^{\left<34\right>} = 
a^{\left<13\right>}+a^{\left<24\right>}$. 
So all the domain walls of the configuration 
certainly get together at a point, as shown 
in the right of figure~\ref{fig:juc:intersection}.
One can easily show that the $Y$-charge in 
equation~(\ref{eq:4eq:hww-ch}) around this 4-pronged 
junction always vanishes, so this is the trivial 
intersection of the two domain walls without any 
kinds of additional junction charge apart from the wall 
tension. 

\subsubsection{Rules of construction}
\begin{figure}[ht]
\begin{center}
\includegraphics[height=3cm]{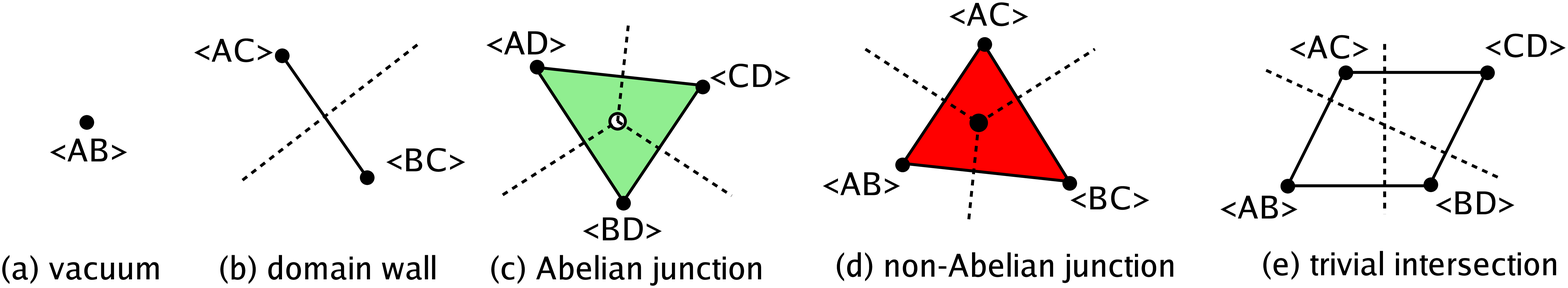}
\caption{\small{\sf Building blocks of the grid diagrams}}
\label{fig:juc:grid}
\vspace*{-.3cm}
\end{center}
\end{figure}

Now we are ready to construct the webs of walls both in 
the Abelian and the non-Abelian gauge theories. 
As is clear from equations~(\ref{eq:juc:weight_balance_ab}) 
and (\ref{eq:juc:weight_balance}), slopes of walls are 
determined by the mass parameters $(m_A,n_A)$ for the 
hypermultiplet scalars. 
In general great pains are needed to clarify shapes of the 
webs corresponding to every points on the moduli space 
$G_{N_{\rm F},N_{\rm C}}$ as the number of flavors 
$N_{\rm F}$ increases. 
A nice tool to overcome this complication is provided by 
the grid diagram \cite{Eto:2005cp,Eto:2005fm}, 
where informations are discarded about actual positions of 
each walls and their junctions. 
Namely, we try to capture the webs in the complex 
$\Tr \Sigma = \Tr (\Sigma_3+i\Sigma_4)$ plane, 
instead of considering them in the real space. 
Let us start with the SUSY vacua. 
The VEV of the adjoint scalar in the 
$\left<\{A_r\}\right>$ vacuum is given by 
$\Sigma = {\rm diag}\left(m_{A_1}+in_{A_1},m_{A_2}+in_{A_2},
\cdots,m_{A_{N_{\rm C}}}+in_{A_{N_{\rm C}}}\right)$. 
Hence the vacuum $\left<\{A_r\}\right>$ is located at the 
point 
$\sum_{r=1}^{N_{\rm C}} \left(m_{A_r}+in_{A_r}\right)$ 
in the $\Tr\Sigma$ plane, as shown in 
figure~\ref{fig:juc:grid}(a). 
Next, domain walls can be regarded as segments between 
possible pairs of these points (labeled by 
$\left<\ \underline{\cdots}\ A\right>$ and 
$\left<\ \underline{\cdots}\ B\right>$) in the 
$\Tr\Sigma$ plane. 
A nice feature of this representation of walls is that 
the magnitude of the tension given in the 
equation~(\ref{eq:juc:wall_tension}) is proportional to 
the length of the segment. 
Moreover, the dual line which is the vector normal to the 
segment is parallel to 
the corresponding domain wall in the real space and to 
the tension vector for the wall in 
equation~(\ref{eq:juc:wall_tension}). 
In the figure~\ref{fig:juc:grid}(b) we denote a solid line 
for a segment connecting two vacua, and denote a broken 
line for its dual line parallel to the domain wall in 
the real space. 
The Abelian 3-pronged junctions 
of domain walls are realized as triangles whose 
vertices are labeled by 
$\left<\ \underline{\cdots}\ A\right>$, 
$\left<\ \underline{\cdots}\ B\right>$ and
$\left<\ \underline{\cdots}\ C\right>$ 
in the $\Tr\Sigma$ plane as shown 
in figure~\ref{fig:juc:grid}(c). 
The non-Abelian 3-pronged junctions are similarly 
realized as triangles whose vertices are labeled by 
$\left<\ \underline{\cdots}\ AB\right>$, 
$\left<\ \underline{\cdots}\ BC\right>$ and
$\left<\ \underline{\cdots}\ CA\right>$ as shown 
in figure~\ref{fig:juc:grid} (d). 
Their topological $Y$-charges given in 
equations (\ref{eq:juc:Y_ab}) and (\ref{eq:juc:Y_na}) are 
proportional to areas of the triangles. 
Furthermore, the equilibrium condition of the tension 
vectors of the 3 components walls given in 
equation~(\ref{eq:juc:tension}) is obvious 
because the tension vector of the wall connecting 
two vacua is dual ($\pi/2$ rotation) to the vector 
connecting these two vacua which forms a closed triangle. 
Lastly we identify the trivial intersection dividing the 
vacua 
$\left<\ \underline{\cdots}\ AB\right>$, 
$\left<\ \underline{\cdots}\ CD\right>$, 
$\left<\ \underline{\cdots}\ AC\right>$ and 
$\left<\ \underline{\cdots}\ BD\right>$ as a parallelogram 
like in the figure~\ref{fig:juc:grid}(e).
All together they constitute the building blocks for the 
webs of walls. 

The followings are the rules to construct grid diagrams for 
the webs of walls to assemble the building blocks: 
\begin{enumerate}[i)]

\item \label{rule1}
Determine mass arrangement $m_A+in_A$ and
plot $_{N_{\rm F}}C_{N_{\rm C}}$ vacuum points 
$\langle A_r\rangle$ at 
$\sum_{r=1}^{N_{\rm C}}(m_{A_r}+in_{A_r})$ in the complex 
$\Tr\Sigma$ plane.

\item \label{rule2}
Draw a convex polygon by choosing a set of vacuum points, which
determines the boundary condition of a BPS solution.
Here each edge of the convex polygon must be a 1/2 BPS single wall
between pairs of the vacuum points
$\langle\ \underline{\cdots}\ A\rangle$ and 
$\langle\ \underline{\cdots}\ B\rangle$.

\item \label{rule3}
Draw all possible internal segments within the convex 
polygon describing
1/2 BPS single walls forbidding any segments to cross.

\item \label{rule4}
 Identify Abelian triangles with vertices  
$\langle\ \underline{\cdots}\ A\rangle$, $\langle\ 
\underline{\cdots}\ B\rangle$
and $\langle\ \underline{\cdots}\ C\rangle$ to
Abelian 3-pronged junctions.
Identify non-Abelian triangles with vertices  
$\langle\ \underline{\cdots}\ AB\rangle$, $\langle\ 
\underline{\cdots}\ BC\rangle$
and $\langle\ \underline{\cdots}\ CA\rangle$ 
to non-Abelian 3-pronged junctions.
Identify parallelograms with vertices
$\langle\ \underline{\cdots}\ AB\rangle$, 
$\langle\ \underline{\cdots}\ CD\rangle$,
$\langle\ \underline{\cdots}\ AC\rangle$ and 
$\langle\ \underline{\cdots}\ BD\rangle$
to intersections with vanishing 
$Y$-charges.

\end{enumerate}
Shapes of the web diagrams in the configuration space can 
be obtained by drawing a dual diagram by exchanging 
points and faces of the grid diagram \cite{Eto:2005cp}.

\begin{figure}[ht]
\begin{center}
\includegraphics[height=5cm]{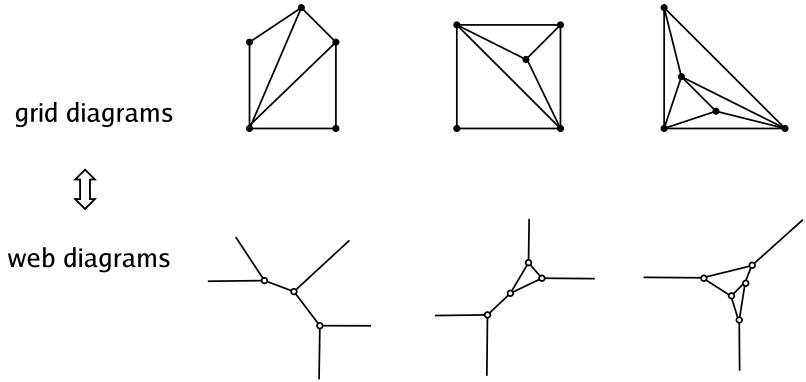}
\caption{\small{\sf 
Examples of the grid and the web diagrams for the Abelian 
webs in the model with 5 flavors.}}
\label{fig:juc:juc_exam}
\vspace*{-.3cm}
\end{center}
\end{figure}

In the figure~\ref{fig:juc:juc_exam} we show several 
examples of the webs in the Abelian gauge theory with 
$N_{\rm F}=5$ flavors.
In general, the grid diagrams have $N_{\rm F}=5$  vertices.
Some of these are the vertices of the convex polygon and 
the others are the internal points inside the polygon.
The shape of the grid diagram is determined by the mass 
arrangement and there are $N_{\rm F}-2$ $(= 3)$ kinds of 
the convex polygons according to 
the number of the internal points inside the convex polygons, 
see figure~\ref{fig:juc:juc_exam}. 
The number of the internal points of the grid diagram is 
the same as the number of the loops of the web 
diagrams in the real space. 
Then one can easily read the graphical relation
of the configuration as 
\begin{eqnarray}
{\rm dim}_{\bf C} {\cal M} = N_{\rm F} = 
E + L,
\end{eqnarray}
where $E$ is the number of the external legs and $L$ is 
the number of the loops in the web diagram.

The webs of walls develop richer species of configurations 
in the non-Abelian gauge theories.
The number of different kinds of webs is the same as 
that in the Abelian gauge theories, namely there are 
$N_{\rm F}-2$ kinds of the webs.
\begin{figure}[ht]
\begin{center}
\includegraphics[height=6cm]{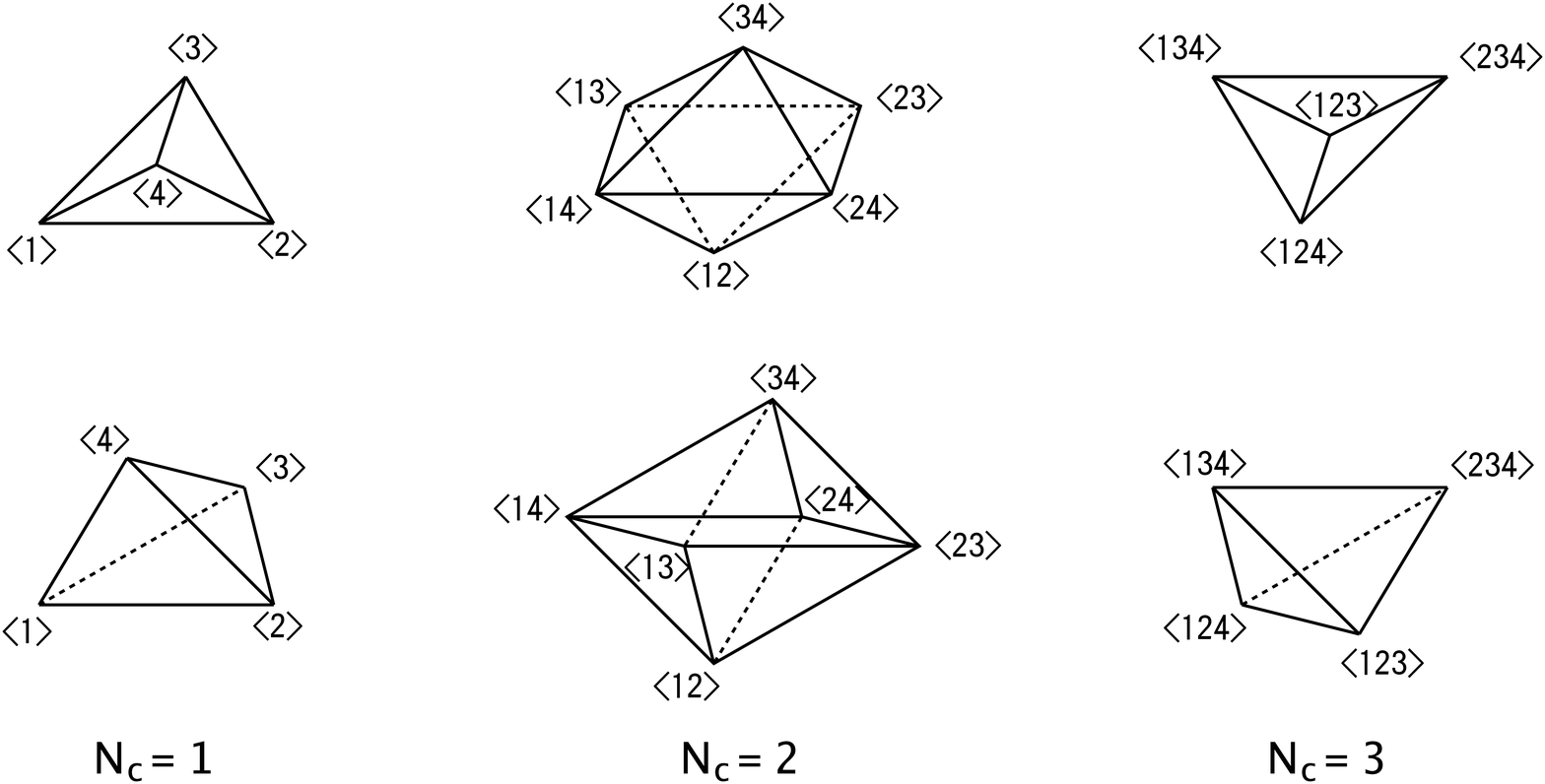}
\caption{\small{\sf 
Examples of the grid diagrams in the model with 4 flavors. 
Each row uses the same hypermultiplet 
mass arrangement which is shown in the $N_{\rm C}=1$ case. 
}}
\label{fig:juc:juc_grid_42}
\vspace*{-.3cm}
\end{center}
\end{figure}
We show several examples of the grid diagrams for the 
model with $N_{\rm F}=4$ flavors and various numbers 
of colors $N_{\rm C}=1,2,3$ in 
figure~\ref{fig:juc:juc_grid_42}. 
The vacuum points of the grid diagrams of the 
$N_{\rm C}=1$ case shows the values of hypermultiplet 
masses directly: 
the upper one has an internal point, whereas the lower one 
does not. 
The same mass arrangement for each flavor of 
hypermultiplets is used to draw the grid diagrams of 
$N_{\rm C}=2, 3$ cases in the same 
row of figure~\ref{fig:juc:juc_grid_42}. 
Let us finally illustrate an interesting phenomenon of 
a transition between webs with different types of junctions 
such as Abelian and non-Abelian by changing 
moduli parameters. 
We draw two grid diagrams in the model with $N_{\rm C}=2$ 
and $N_{\rm F}=4$ corresponding to two different 
mass arrangements in figure~\ref{fig:juc:example_na}.
One is the hexagon type with no internal points, 
which is a tree type web diagram. 
The other is the parallelogram type diagram with two 
internal points, which has a loop. 
By varying the moduli parameters of the grid 
diagrams, we find that the internal structure of web 
diagrams changes as shown in 
figure~\ref{fig:juc:example_na}(a2) and (b2): 
webs with Abelian junctions makes a 
transition to webs with non-Abelian junctions 
and vice versa. 
The changes of the shape of the webs have been 
explicitly worked 
out in terms of the Pl\"ucker coordinates 
$\tau^{\left<\{A_r\}\right>}$ in 
equation~(\ref{eq:juc:plucker-coordinate}) \cite{Eto:2005fm}.
\begin{figure}[ht]
\begin{center}
\begin{tabular}{ccc}
\includegraphics[height=2.5cm]{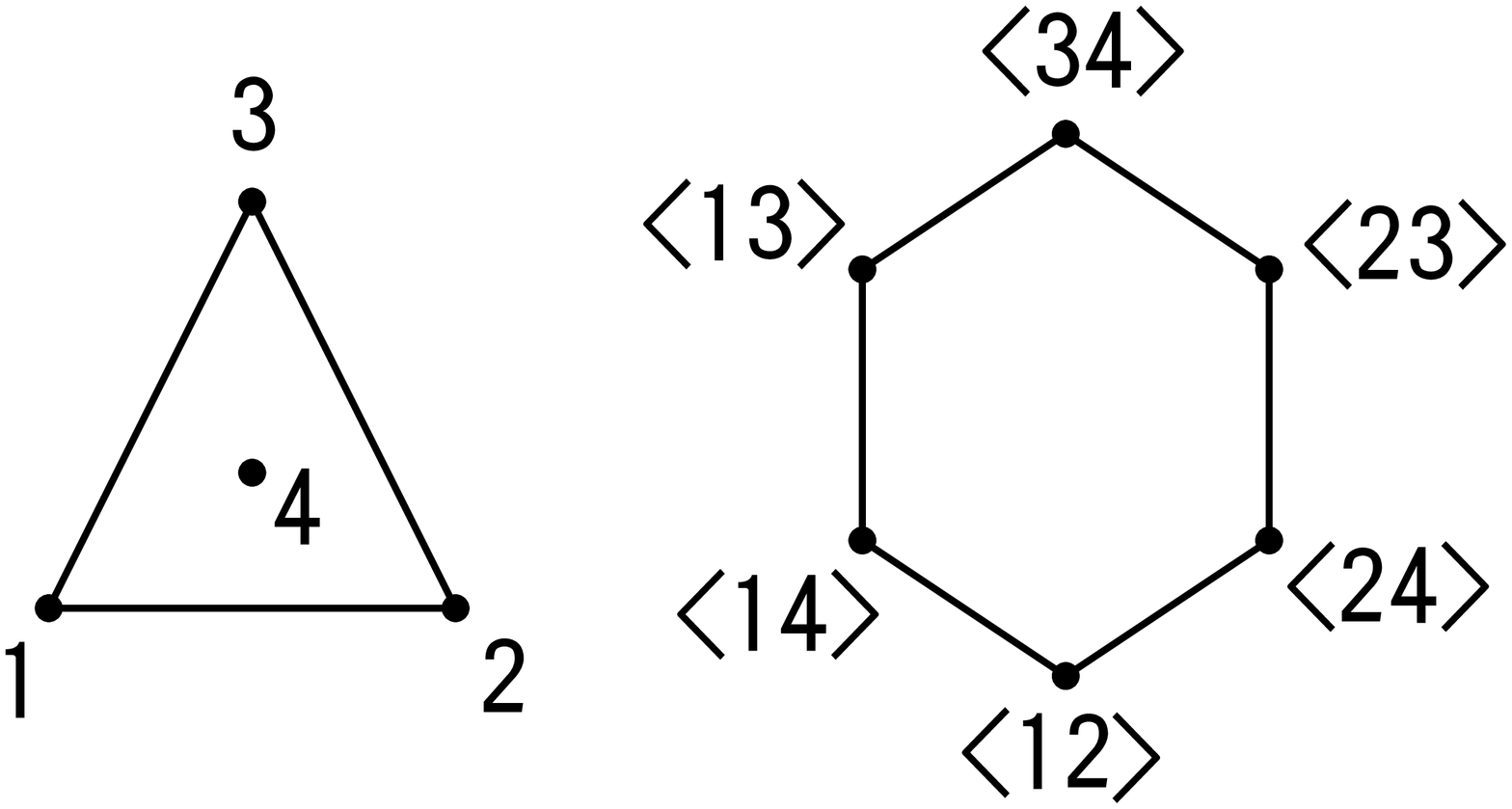}
&\qquad&
\includegraphics[height=2.3cm]{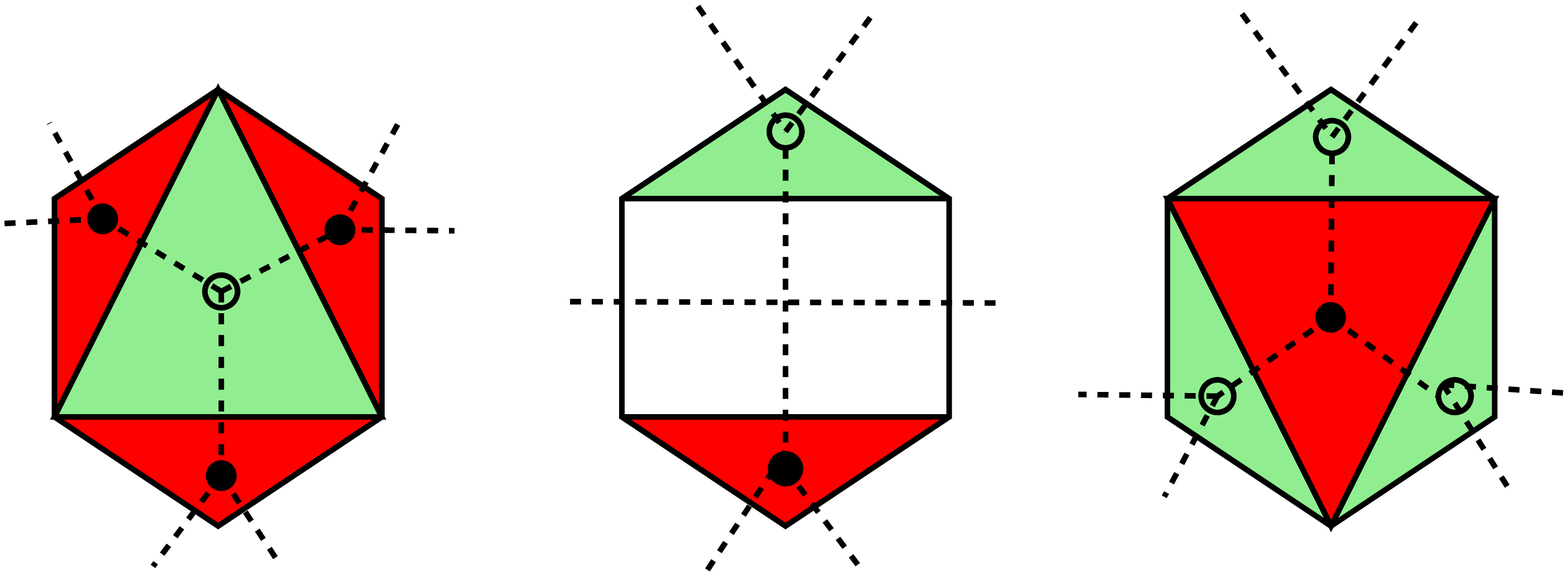}\\
\small{\sf (a1) mass arrangement and vacua} & & 
\small{\sf (a2) grid diagrams and web configurations}\\
\includegraphics[height=2cm]{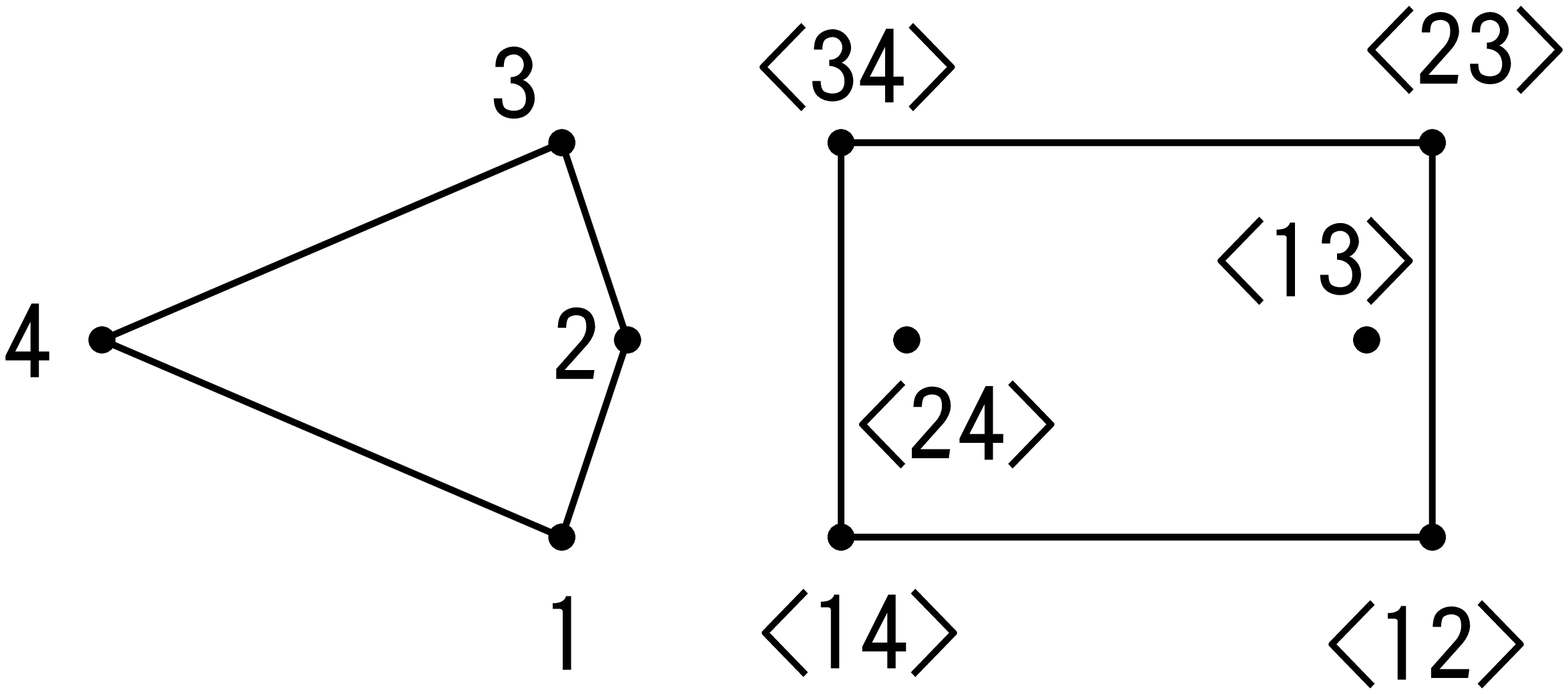}
&\qquad&
\includegraphics[height=3cm]{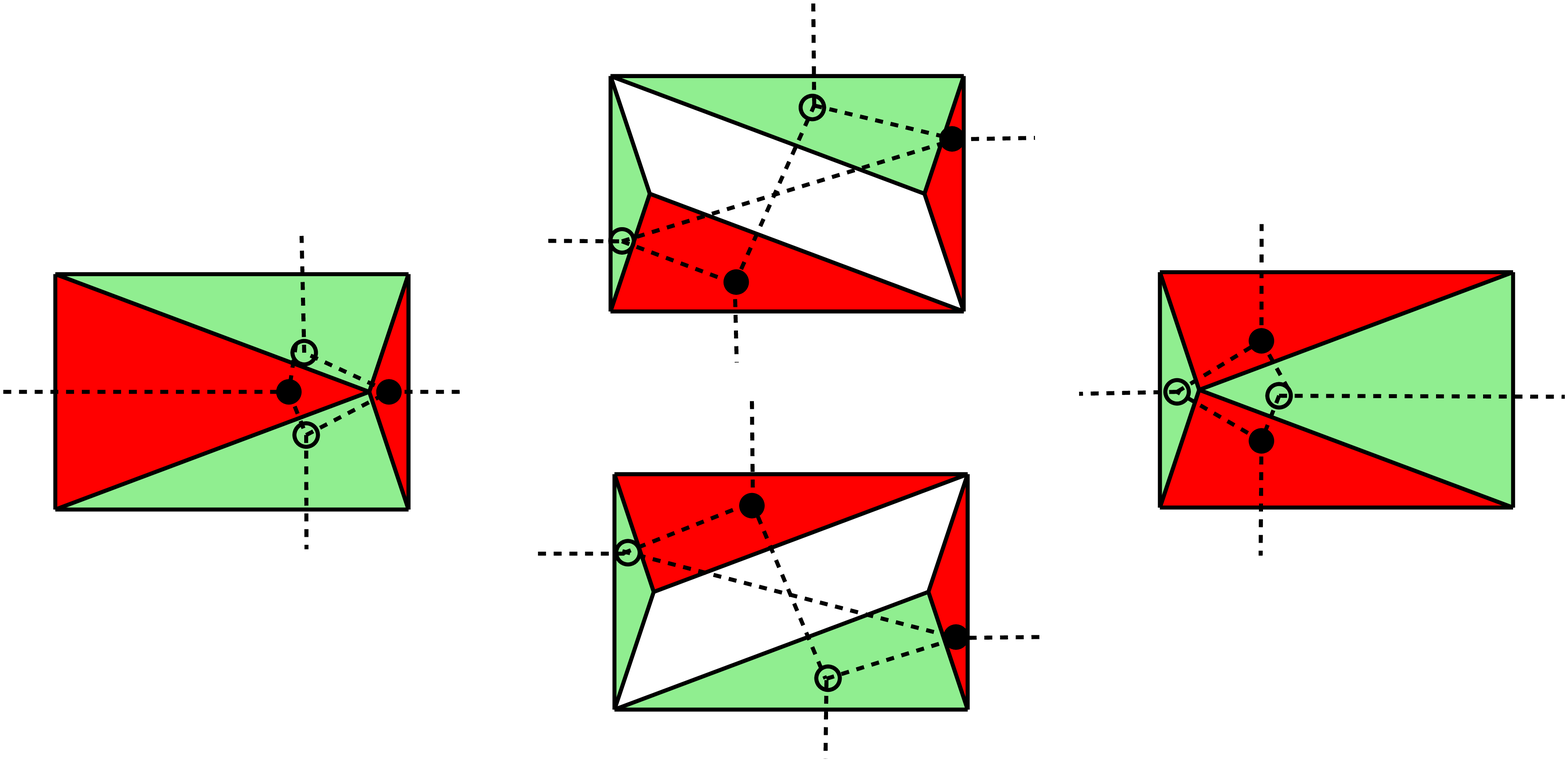}\\
\small{\sf (b1) mass arrangement and vacua} & & 
\small{\sf (b2) grid diagrams and web configurations}
\end{tabular}
\caption{\small{\sf (a1) and (a2) for the hexagon-type and
(b1) and (b2) for the parallelogram-type web.}}
\label{fig:juc:example_na}
\vspace*{-.3cm}
\end{center}
\end{figure}

\begin{figure}[ht]
\begin{center}
\includegraphics[height=5.5cm]{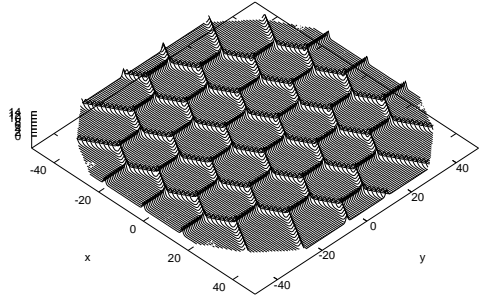}
\caption{\small{\sf Honeycomb lattice of webs of walls.}}
\label{fig:juc:honey}
\vspace*{-.3cm}
\end{center}
\end{figure}

\changed{
At the end of this subsection, we give an exact solution
in the strong gauge coupling limit $g^2\to\infty$. 
As explained in section~\ref{4eq}, 
the master equation reduces 
to just an algebraic equation. Then we can exactly solve them.
In figure~\ref{fig:juc:honey} we give a complicated configuration 
of webs of walls.
}

Interestingly, solutions of the KP and coupled KP 
equations were found to contain very similar web 
structure to our solutions of webs of walls 
\cite{junction-cKP}.

\subsection{Composite of walls, vortices and monopoles}\label{mon}
We will focus on another 1/4 BPS composite state of 
monopoles, vortices and domain walls in this subsection.
The 1/4 BPS equations of this system has been already 
derived in 
equations~(\ref{eq:4eq:mvw-1})--(\ref{eq:4eq:mvw-3}).
\changed{This 1/4 BPS system was studied qualitatively in
\cite{Sakai:2005sp} and lots of interesting features were found.}
\changed{We found that solutions of them}
can be described by the moduli matrix $H_0(z)$ which 
is an $N_{\rm C}\times N_{\rm F}$ holomorphic matrix 
with respect to $z=x^1+ix^2$. 
Similarly to the webs of walls dealt with in the previous 
subsection, the moduli matrix is also a powerful tool 
to clarify properties of this 1/4 BPS system 
\cite{Isozumi:2004vg}.

First of all, it should be stressed that the 1/4 BPS 
equations (\ref{eq:4eq:mvw-1})--(\ref{eq:4eq:mvw-3}) 
are composite of the three types of 1/2 BPS solitons:
$B_i = {\cal D}_i\Sigma$ for monopoles, 
$({\cal D}_1+i{\cal D}_2)H=0$,
$B_3 + \frac{g^2}{2}(c{\bf 1}_{N_{\rm C}} - HH^\dagger) = 0$
for vortices and 
${\cal D}_3 H + \Sigma H - HM = 0$,
${\cal D}_3\Sigma - \frac{g^2}{2}
(c{\bf 1}_{N_{\rm C}} - HH^\dagger) = 0$
for domain walls. 
In this section we omit the 
subscript ``4'' of $\Sigma_4$ and $M_4$. 
Solutions of these 1/2 BPS equations, of course, 
are also solutions of the 1/4 BPS
equations (\ref{eq:4eq:mvw-1})--(\ref{eq:4eq:mvw-3}).
When different types of 1/2 BPS solitons coexist, the 
configuration becomes a 1/4 BPS state as we will show. 
There are two kinds of solutions of the 1/4 BPS equations. 
One is a junction of two vortices living in the same Higgs 
vacuum, but with different orientations in the internal 
symmetry. 
This kind of composite soliton does not exist in Abelian 
gauge theory and is intrinsically non-Abelian. 
Since there is a unique vacuum in this case, domain walls 
do not appear. 
So the topological charges characterizing this system are 
the vortex charge and the monopole charge 
\begin{eqnarray}
{\cal E} = - c\Tr B_3 
+ \frac{2}{g^2}\Tr \partial_m\left(B_m\Sigma\right),
\label{eq:wvm:charge_mv}
\end{eqnarray}
where the first one is the charge of the non-Abelian 
vortex and the second one is that of the ordinary 
t' Hooft-Polyakov type monopole in the $SU(N_C)$ gauge 
theory. 
The monopole appears at the junction point of two vortices. 
This configuration is called the monopole in the Higgs 
phase which was found in Ref.~\cite{Tong:2003pz}. 
The other 1/4 BPS state is also a composite state of 
vortices, but is now the junction of vortices living 
in different vacua. 
So the domain wall interpolating these vacua is formed. 
This kind of 1/4 BPS state can exist in Abelian gauge 
theory with $N_{\rm F}\ge 2$ flavors and is essentially 
a composite soliton in Abelian gauge theory. 
The topological charges characterizing this type of 
soliton in Abelian gauge theory is given by 
\begin{eqnarray}
{\cal E} = c\partial_3\Sigma - c B_3 
+ \frac{2}{g^2} \partial_m\left(B_m\Sigma\right),
\label{eq:wvm:charge_vw}
\end{eqnarray}
where the first one is the charge of the domain wall, 
the second one is the charge of the ANO vortex and 
the third one is a somewhat strange charge which has a 
form very similar to the monopole charge in the 
$SU(N_{\rm C})$ 
gauge theory. 
This monopole-like charge gives a negative contribution 
to the energy density and will be understood as the 
binding energy (boojum) of the domain wall and the 
ANO vortex \cite{Isozumi:2004vg}, \cite{Sakai:2005sp}.

\subsubsection{Vortices in the massive theories
\label{sec:mvw:vortex}}
In this subsection, vortices will play a prominent role. 
We have clarified 1/2 BPS vortices in the massless 
theory in section~\ref{vtx}. 
Here we deal with 1/2 BPS vortices in the massive theory with 
$M={\rm diag}\left(m_1,m_2,\cdots,m_{N_{\rm F}}\right),\ (m_A > m_{A+1})$. 
Let us start with the vortices in the 
$N \equiv N_{\rm C}=N_{\rm F}$ model. 
First recall that the massless (fully degenerate masses) 
model has the unique color-flavor locking vacuum given by 
the condition 
\begin{eqnarray}
HH^\dagger = c{\bf 1}_N,\quad \Sigma H = 0.
\label{eq:wvm:vac_massless}
\end{eqnarray}
The vacuum has the diagonal $SU(N)_{\rm G+F}$ symmetry 
as explained in section~\ref{mdl}. 
This system admits the 1/2 BPS vortices which are solutions 
of the 1/2 BPS equations~(\ref{eq:vtx:BPSeq1}) and (\ref{eq:vtx:BPSeq2}). 
Solutions of the 1/2 BPS equations are described by the 
moduli matrix $H_0(z)$ which is a $N$ by $N$ matrix 
holomorphic with respect to $z$ as shown in 
equation~(\ref{eq:vtx:solution}). 
Notice that we need to keep the additional condition 
$\Sigma H=0$ to get regular solutions in the massless 
theory, although this condition is trivially satisfied 
by setting $\Sigma = 0$.

Roughly speaking, the vortex solutions in the non-Abelian 
gauge theory are obtained by embedding the Abelian 
ANO vortex solutions to the moduli matrices in the 
non-Abelian case. 
Then a single vortex solution breaks the 
$SU(N)_{\rm G+F}$ vacuum symmetry into 
$U(1)\times SU(N-1)$, so that the Nambu-Goldstone modes 
taking values on 
${\bf C}P^{N-1} \simeq SU(N)_{\rm G+F}/[U(1)\times SU(N-1)]$ 
arise as orientational moduli. 
In terms of the $N\times N$ moduli matrix $H_0(z)$ 
given in equation~(\ref{eq:vtx:solution}), 
$k$-vortex solutions are generated by the matrix whose 
determinant is of order $z^k$ as 
\begin{eqnarray}
\tau \equiv \det H_0(z) =   \prod_{i=1}^k
\left(z - z_i\right)
\label{eq:wvm:k-ano_n}
\end{eqnarray}
and their orientational moduli which is the homogeneous 
coordinate of ${\bf C}P^{N-1}$ is defined by 
$H_0(z_i)\vec\phi_i = \vec 0$:
\begin{eqnarray}
\vec\phi_i{}^T 
= \left(\phi_i^1,\phi_i^2,\cdots,\phi_i^N\right) 
\in {\bf C}P^{N-1}.
\end{eqnarray}

\begin{figure}[ht]
\begin{center}
\includegraphics[height=3.5cm]{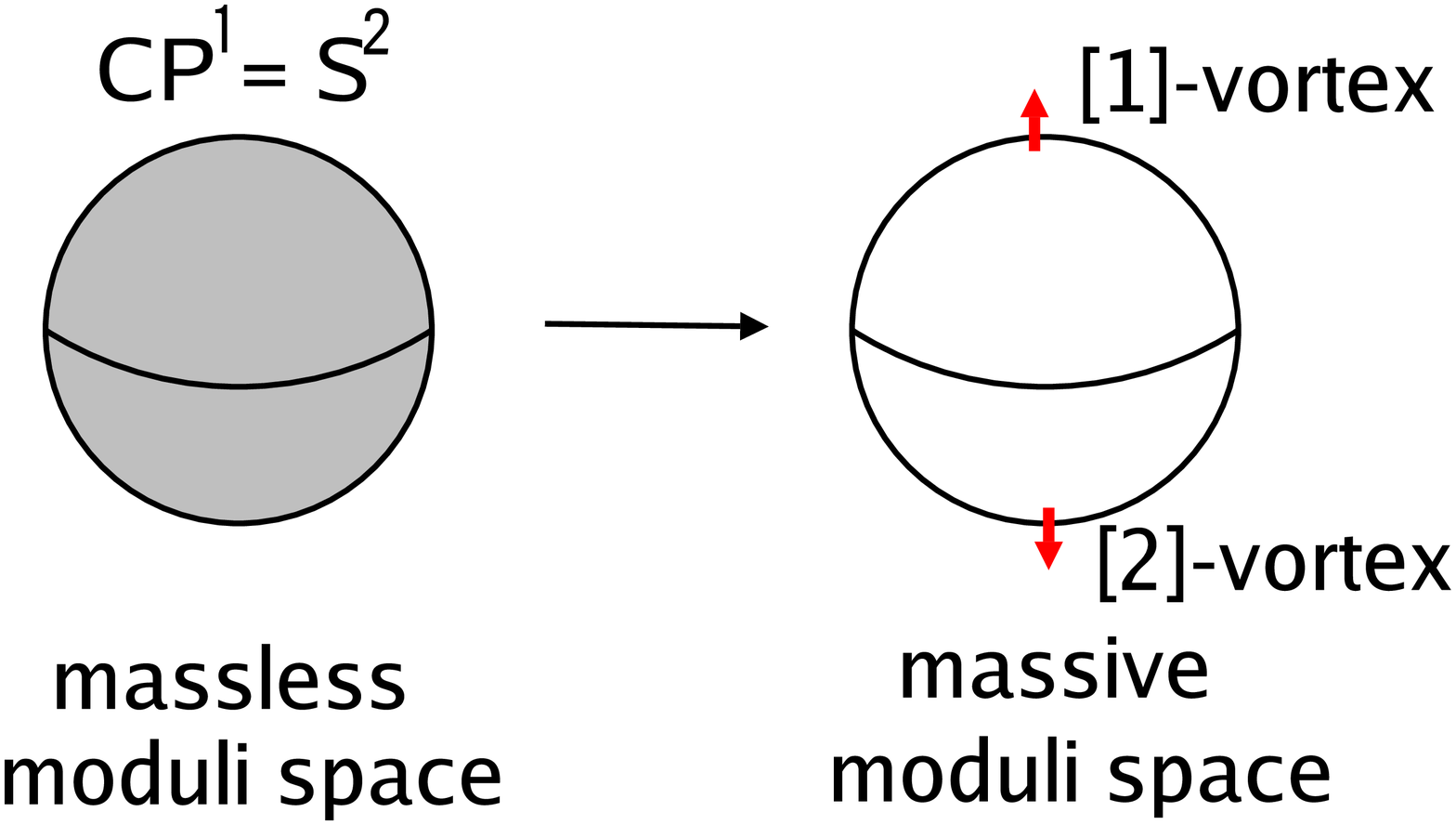}
\caption{\small{\sf Orientation of non-Abelian vortices}}
\label{fig:wvm:wvm_orientaion}
\vspace*{-.3cm}
\end{center}
\end{figure}

When we turn on the non-degenerate masses $M$ for the 
hypermultiplet scalars, the vacuum is still unique 
but the vacuum condition (\ref{eq:wvm:vac_massless}) is 
changed to 
\begin{eqnarray}
HH^\dagger = c{\bf 1}_N,\quad
\Sigma H - HM = 0.
\label{eq:wvm:vac_massive}
\end{eqnarray}
The second equation requires 
$\Sigma = {\rm diag}(m_1,m_2,\cdots,m_N)$, so that the 
flavor symmetry reduces from $SU(N)$ to $U(1)^{N-1}$. 
Furthermore $U(N)$ gauge symmetry also reduces 
to $U(1)^{N}$ by the VEV of the adjoint scalar 
(of course, these gauge symmetries are completely broken 
in the true vacuum by the VEV of hypermultiplet scalars). 
This means that there no longer exist the orientational 
moduli for the non-Abelian vortices and the system reduces 
to just $N$ vortices of the ANO type. 
In other words, the non-degenerate masses $M$ lift almost 
all the points of the massless orientational moduli space 
${\bf C}P^{N-1}$ except for $N$ fixed points of the $U(1)^{N-1}$ 
isometry which remain as solutions of the massive theory. 
Actually, the 1/2 BPS equations are not changed from the 
massless one given in equations~(\ref{eq:vtx:BPSeq1}) and (\ref{eq:vtx:BPSeq2}), and 
their solutions are also given by the same form as 
equation~(\ref{eq:vtx:solution}). 
However, we have to impose the additional condition 
$\Sigma H - HM = 0$ instead of $\Sigma H=0$, then the 
moduli matrix can have non-trivial elements only in their 
diagonal elements. 
As a result, the $N$ different ANO vortices live in the 
diagonal elements of the moduli matrix as 
$H_0(z) = 
{\rm diag}\left(H_{0\star}^{1}(z),H_{0\star}^{2}(z),
\cdots, H_{0\star}^{N}(z)\right)$ with 
$H_{0\star}^{A}(z) = a_A\prod_{i=1}^{k_A}(z-z_i)$, 
$(A=1,2,\cdots,N)$. 
The orientational moduli in the massive theory reduces 
to just $N$ different vectors as 
\begin{eqnarray}
\vec\phi_i{}^T = 
\left(\phi_i^1,\phi_i^2,\cdots,\phi_i^N\right)
\rightarrow
\left\{
\begin{array}{ccl}
\left(1,0,0,\cdots,0,0\right)&:& \left[1\right]-{\rm vortex},\\
\left(0,1,0,\cdots,0,0\right)&:& \left[2\right]-{\rm vortex},\\
\vdots\\
\left(0,0,0,\cdots,0,1\right)&:& \left[N\right]-{\rm vortex}.
\end{array}
\right.
\end{eqnarray}
Thus we find that $N$ different species of vortices, 
which we call $\left[ A \right]$-vortex, can live in the 
unique vacuum of the massive $N=N_{\rm C}=N_{\rm F}$ 
system. 
The $\left[A\right]$-vortex is associated with the $U(1)$ 
phase of the $A$-th flavor element $H_{0\star}^{A}(z)$. 
In figure \ref{fig:wvm:wvm_orientaion} we show an example 
of $N=2$ model. 
The massless moduli space is ${\bf C}P^1 \simeq S^2$. 
Almost all of them are lifted by the non-degenerate 
masses except for the north and the south poles 
($\left[1\right]$- and $\left[2\right]$-vortex) which 
remain as solutions of the massive 1/2 BPS equation. 

\begin{figure}[ht]
\begin{center}
\includegraphics[height=3.5cm]{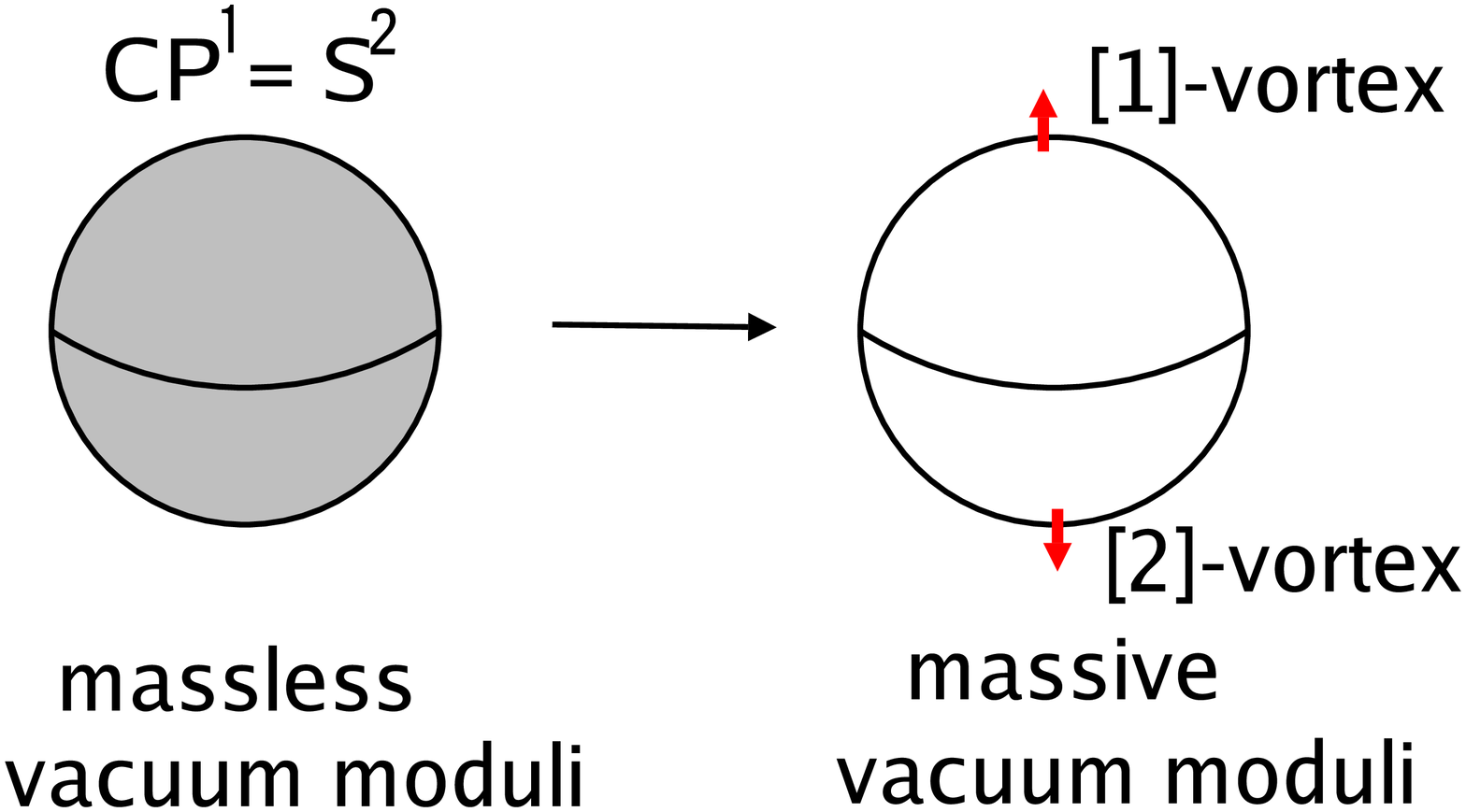}
\caption{\small{\sf Semi-local vortices in the massive 
Abelian gauge theory.}}
\label{fig:wvm:wvm_massive_semilocal}
\vspace*{-.3cm}
\end{center}
\end{figure}

Let us next consider the non-Abelian semi-local vortices 
which arise in the massive theory with the $N_{\rm F}$ 
flavors being greater than the $N_{\rm C}$ colors.
We have investigated the semi-local vortices in the 
massless theory in section~\ref{sec:vtx:semi} 
and found that they have additional moduli parameters 
(size moduli) compared to the ANO vortices in 
the model with $N_{\rm C}=N_{\rm F}$ theory. 
These additional moduli parameters are lifted 
and only several different species of ANO vortices 
remain, because of the same reason as 
the $N_{\rm F}=N_{\rm C}$ case mentioned above. 
In fact, the massless vacuum manifold 
$Gr_{N_{\rm F},N_{\rm C}}$ shrinks to 
the $_{N_{\rm F}}C_{N_{\rm C}}$ discrete vacua labeled by 
$\left<A_1A_2\cdots A_{N_{\rm C}}\right> = \left<\{A_r\}\right>$
by the non-degenerate masses, as explained in section~\ref{wll}. 
The $N_{\rm C}$ different species of vortices, namely 
$\left[A_r\right]$-vortex ($r=1,2,\cdots,N_{\rm C}$), can 
live in each vacuum. 
To clear matters, let us consider the semi-local vortex 
in the Abelian gauge theory. 
In the case of the massless theory the moduli matrix for 
the semi-local vortices are written as 
$H_0(z) = \left(H_{0\star}^1(z),\cdots,H_{0\star}^{N_{\rm F}}(z)\right)$
with $H_{0\star}^A(z)\equiv a_A\prod_{i=1}^{k_A}(z-z_i)$ 
and the moduli parameters $z_i$ and $a_A$ can be understood 
as positions and sizes of the vortices, respectively. 
Additionally there exist non-normalizable moduli 
parameters which specify the position in the vacuum 
manifold as the boundary condition, see 
section~\ref{sec:vtx:semi}.
When we turn on the non-degenerate masses, the connection 
between different flavors is turned off: 
the size moduli and the non-normalizable moduli are 
frozen in solutions of the 1/2 BPS semi-local vortices 
in the massive theory. 
Therefore, the allowed moduli matrix is 
just $N_{\rm F}$ species of the ANO vortices labeled as 
$\left[A\right]$-vortex $(A=1,2,\cdots,N_{\rm F})$:
\begin{eqnarray}
H_0(z) = 
\left(H_{0\star}^1(z),\cdots,H_{0\star}^{N_{\rm F}}(z)\right)
\rightarrow
\left\{
\begin{array}{c}
\left(H_{0\star}^1(z),0,0,\cdots,0,0\right),\\
\left(0,H_{0\star}^2(z),0,\cdots,0,0\right),\\
\vdots\\
\left(0,0,0,\cdots,0,H_{0\star}^{N_{\rm F}}(z)\right).
\end{array}
\right.
\label{eq:wvm:semilocal_massive}
\end{eqnarray}
For example, the massless vacuum manifold of 
$N_{\rm C}=1$ and $N_{\rm F}=2$ model is 
${\bf C}P^1 \simeq S^2$ and we can choose each point 
on the ${\bf C}P^1$ as the boundary condition of the 
semi-local vortex. 
However, when the masses are turned on, there are only 
two choice for the boundary condition, either the north 
pole or the south pole, as illustrated in 
figure~\ref{fig:wvm:wvm_massive_semilocal}.
The figure~\ref{fig:wvm:wvm_orientaion} for 
the vortices in the theory with $N_{\rm C} = 2$ and 
$N_{\rm F}=2$ appears similar to the 
figure~\ref{fig:wvm:wvm_massive_semilocal} 
for vortices in the theory with $N_{\rm C} = 1$ 
and $N_{\rm F}=2$. 
However, their properties are very different. 
The former has the unique vacuum $\left<12\right>$ 
and there exist $\left[1\right]$-vortices and 
$\left[2\right]$-vortices simultaneously in the vacuum. 
In contrast, the latter has two discrete vacua 
$\left<1\right>$ and $\left<2\right>$, and the 
$\left[1\right]$-vortices can live only in the vacuum 
$\left<1\right>$ while the $\left[2\right]$-vortices can 
in the vacuum $\left<2\right>$. 

In the case of the massive model with 
$N_{\rm F}>N_{\rm C}$, there are 
$_{N_{\rm F}}C_{N_{\rm C}}$ discrete vacua. 
Although there exist $N_{\rm F}$ species of ANO vortices 
labeled by $\left[A\right]$-vortex 
($A=1,2,\cdots,N_{\rm F}$), 
only $N_{\rm C}$ of them can live in each vacuum.
The vacuum $\left<A_1A_2\cdots A_{N_{\rm C}}\right>$ 
allows the $\left[A_r\right]$-vortex 
($r=1,2,\cdots,N_{\rm C}$) to live therein. 
With respect to the moduli matrix, the determinants of 
the minor matrices $H_0^{\left<\{A_r\}\right>}$ 
characterize the configuration. 
Namely, $k_{A_r}$ $\left[A_r\right]$-vortices in the 
$\left<\{A_r\}\right>$ vacuum is 
generated by the determinant of the moduli matrix 
\begin{eqnarray}
\tau^{\left<\{A_r\}\right>}
= \det H_0^{\left<\{A_r\}\right>}
=   \prod_{i=1}^{k_{A_r}}(z-z_i),\quad
{\rm others}=0,
\label{eq:wvm:plucker}
\end{eqnarray}
like the case of $N= N_{\rm C}=N_{\rm F}$ given in 
equation~(\ref{eq:wvm:k-ano_n}). 
These variables provide convenient coordinates of the 
moduli space and are called the Pl\"ucker coordinates 
as given in equation (\ref{eq:wll:def-tau}) in section 
\ref{wll}. 
Thus we conclude that $N_{\rm C}$ different 
species of ANO vortices can exist in every one of the 
$_{N_{\rm F}}C_{N_{\rm C}}$ vacua in the 
$U(N_{\rm C})$ gauge theory with the massive 
$N_{\rm F}$ flavors.
\begin{figure}[ht]
\begin{center}
\includegraphics[height=3.5cm]{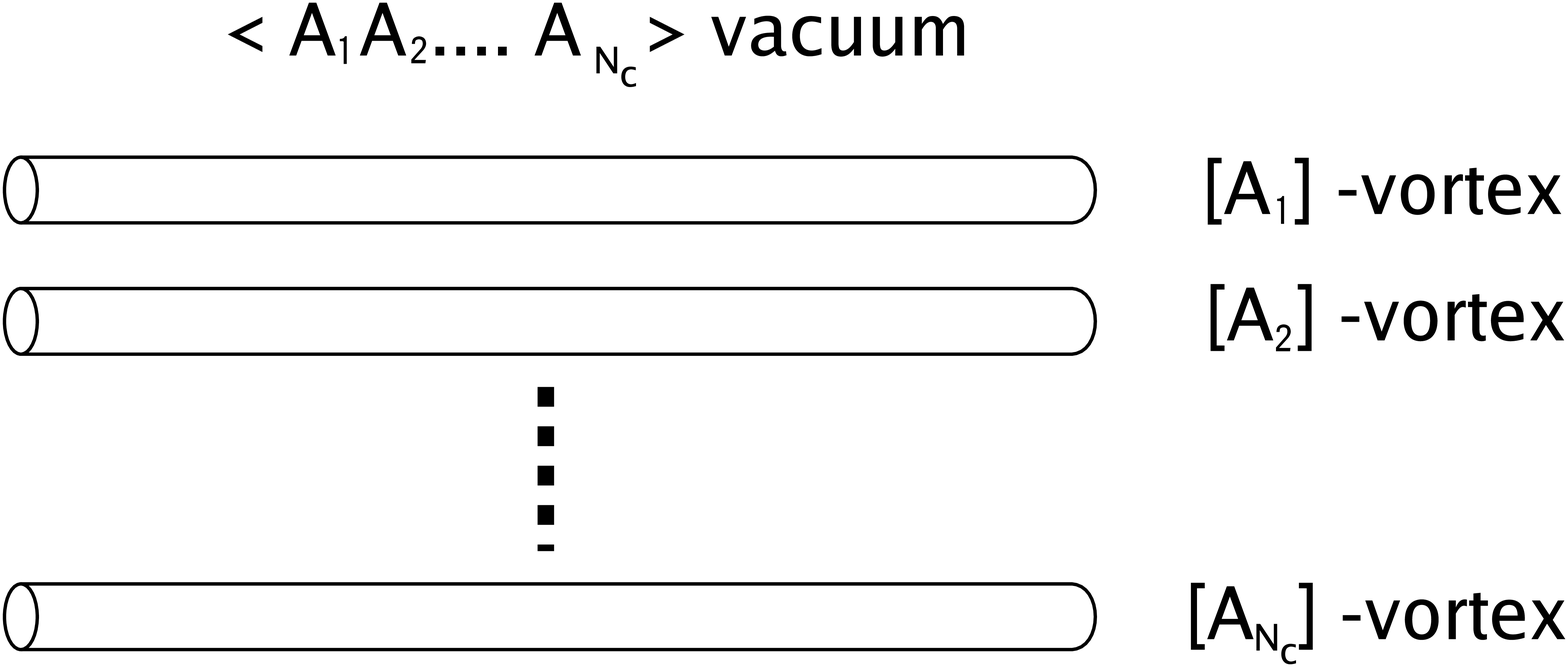}
\caption{\small{\sf Orientation of a non-Abelian vortex}}
\label{fig:wvm:n_vortex}
\vspace*{-.3cm}
\end{center}
\end{figure}

\subsubsection{Monopoles in the Higgs phase
\label{sec:mvw:monopole}}

We begin with the 1/4 BPS states of vortex junctions in a 
single vacuum which accompany monopoles as given in 
equation (\ref{eq:wvm:charge_mv}). 
We shall work in $U(2)$ gauge theory with $N_{\rm F}=2$ 
massive flavors with 
$M = {\rm diag}(m_1,m_2),\ (m_1>m_2)$.
As we saw above, $\left[1\right]$- and 
$\left[2\right]$-vortices with orientations 
$\vec\phi{\ }^T = (1,0)$ and $(0,1)$ can exist 
in the vacuum $\left<12\right>$. 
The 1/2 BPS vortices containing $k_1$ 
$\left[1\right]$-vortices and 
$k_2$ $\left[2\right]$-vortices are given by the diagonal 
moduli matrix $H_0(z)e^{Mx^3}$ in the 1/4 BPS solution 
(\ref{eq:4eq:mvw-sol}) as 
\begin{eqnarray}
H_0 (z) e^{Mx^3}
= {\rm diag}
\left(
H_{0\star}^{1}(z), H_{0\star}^{2}(z)
\right)e^{Mx^3}, 
\label{eq:wvm:mm_1/2_vortex1}\\
S(x^1,x^2,x^3) = {\rm diag}
\left(
S_\star^{1}(x^1,x^2), S_\star^{2}(x^1,x^2)
\right)e^{Mx^3},
\label{eq:wvm:mm_1/2_vortex2}
\end{eqnarray}
with $H_{0\star}^{A} = \prod_{i=1}^{k_A}(z-z_i)$.
In fact, plugging this solution into the 1/4 BPS equations 
(\ref{eq:4eq:mvw-1})--(\ref{eq:4eq:mvw-3}), we find that 
the equations reduce to the 1/2 BPS equations for the 
ANO vortices: 
$\partial_z\bar\partial_z \log \Omega^{A}_\star = g^2
\left(c-(\Omega^{A}_\star)^{-1}|H_{0\star}^{A}|^2\right)$
with $\Omega_\star^{A} \equiv |S_\star^{A}|^2$.
Furthermore, the solutions 
given in equation~(\ref{eq:4eq:mvw-sol}) 
reduce to the 1/2 BPS solutions given in 
equation~(\ref{eq:vtx:solution}). 
Especially the additional condition $\Sigma H - HM= 0$ 
is automatically satisfied as 
\begin{eqnarray}
W_3 - i \Sigma = -i {\rm diag}
\left(
m_1 , m_2
\right).
\end{eqnarray}
When we turn on the off-diagonal elements of the moduli 
matrix $H_0(z)e^{Mx^3}$ given in 
equation~(\ref{eq:wvm:mm_1/2_vortex1}), the moduli matrix 
does not give a 1/2 BPS solution because 
$\Sigma H - HM = 0$ is no longer satisfied. 
Instead, it gives a 1/4 BPS solution satisfying 
${\cal D}_3H + \Sigma H - H M = 0$, which is one of 
the 1/4 BPS equations (\ref{eq:4eq:mvw-1}).

\begin{figure}[ht]
\begin{center}
\includegraphics[height=4cm]{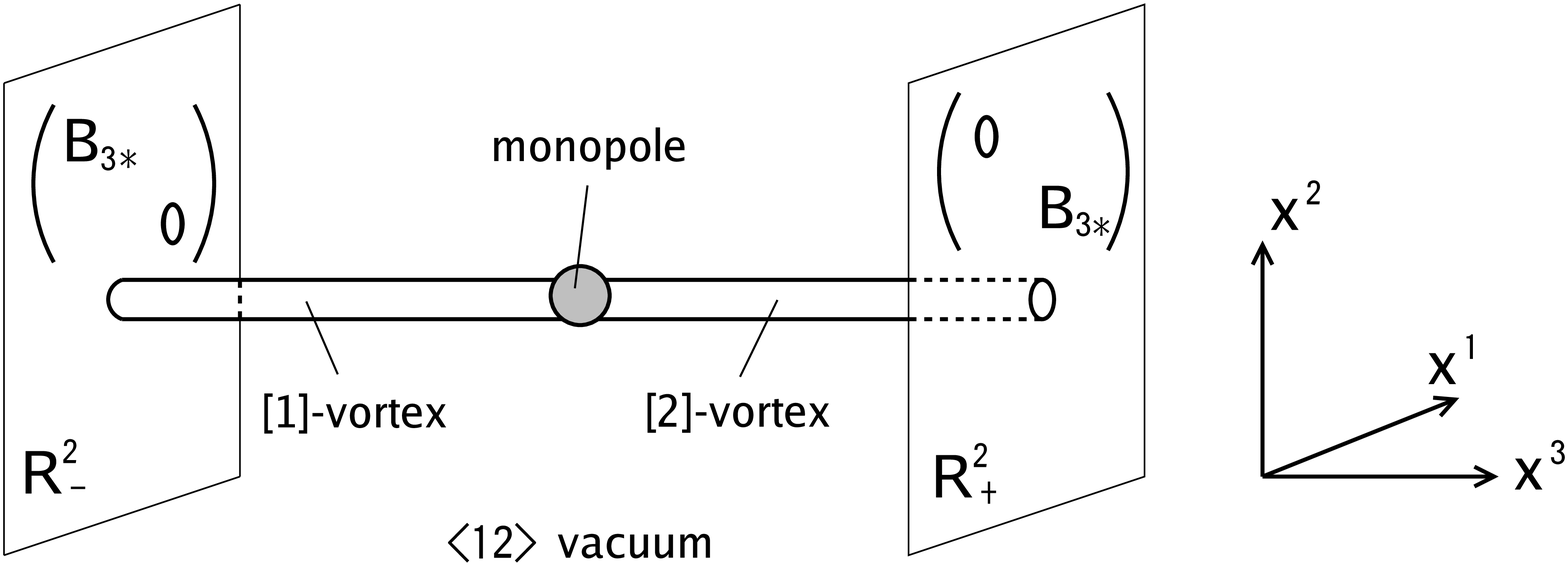}
\caption{\small{\sf Monopole in the Higgs phase.}}
\label{fig:wvm:charge_vm}
\vspace*{-.3cm}
\end{center}
\end{figure}

To clear matters, let us consider a single vortex 
configuration $k = k_1 + k_2 = 1$.
The general moduli matrix with a unit vorticity $k=1$ 
($\det H_0(z) = {\cal O}(z)$) is of the form
\begin{eqnarray}
H_0(z)e^{Mx^3} = 
\left(
\begin{array}{cc}
1 & b \\
0 & z-z_1
\end{array}
\right)e^{Mx^3}
\sim
\left(
\begin{array}{cc}
z-z_1 & 0 \\
b' & 1
\end{array}
\right)e^{Mx^3},
\label{eq:wvm:mm_mv}
\end{eqnarray}
where $b,b' \in {\bf C}$ and they are related by 
$b = 1/b'$ except for $b=0$ or $b'=0$. 
Notice that this moduli matrix is quite similar to that 
for the single non-Abelian vortex given in 
equation~(\ref{eq:vtx:N=2}) in the massless 
theory where the moduli parameter $b \in {\bf C}P^1$ 
is the orientational moduli of the vortex. 
In the massive theory as mentioned above, 
the parameter $b$ is no longer an orientational moduli 
(only $b=0$ and $b=\infty$ give the ANO vortex solution). 
Nevertheless, treating $b$ as the orientation in analogy 
with the massless model gives us a powerful insight in 
understanding the 1/4 BPS solution. 
To this end, it is useful to rewrite the above moduli 
matrix (\ref{eq:wvm:mm_mv}) as
\begin{eqnarray}
H_0(z)e^{Mx^3} = 
e^{Mx^3}
\left(
\begin{array}{cc}
1 & \tilde b(x^3)\\
0 & z-z_1
\end{array}
\right)
\sim
e^{Mx^3}
\left(
\begin{array}{cc}
z-z_1 & 0 \\
\tilde b'(x^3)& 1
\end{array}
\right),
\label{eq:wvm:mm_mv2}
\end{eqnarray}
with $\tilde b(x^3)\equiv b e^{-(m_1-m_2)x^3}$ 
and $\tilde b'(x^3)\equiv b' e^{(m_1-m_2)x^3}$. 
By absorbing the prefactor $e^{Mx^3}$ by the 
$V$-equivalence relation (\ref{eq:4eq:mvw-V-equiv}) 
at every slice at $x^3={\rm const.}$, 
we can regard this moduli matrix as the moduli matrix 
of 1/2 BPS vortices given in equation~(\ref{eq:vtx:solution}). 
Then the orientational moduli 
$\vec\phi{\ }^T = (\tilde b(x^3),1) = (1,\tilde b'(x^3))$ 
depends on the $x^3$ coordinate. 
This means that the orientation of the vortex changes 
along the $x^3$ axis. 
Since we have chosen $m_1 - m_2 >0$, $\tilde b(x^3) \to 0$ 
as $x^3 \to + \infty$ and $\tilde b'(x^3) \to 0$ as 
$x^3 \to - \infty$. 
Therefore, we find that the moduli matrix 
(\ref{eq:wvm:mm_mv}) gives a composite soliton which 
reduces to $[2]$-vortex ($\vec\phi{\ }^T = (0,1)$) at 
$x^3 = +\infty$ and $[1]$-vortex 
($\vec\phi{\ }^T = (1,0)$) at $x^3 = -\infty$. 
Let us next focus on the transition between the 
$[1]$-vortex and $[2]$-vortex. 
As we mentioned above, the transition is accompanied by 
the monopole charge, see equation~(\ref{eq:wvm:charge_mv}).
We can directly calculate the monopole charge by taking 
account of limits where $B_3 = {\rm diag}(B_{3\star},0)$ 
at $x^3\to -\infty$ and 
$B_3 = {\rm diag}(0,B_{3\star})$ at $x^3\to +\infty$ 
while $B_1=B_2=0$ and $\Sigma = {\rm diag}(m_1,m_2)$ 
both at $x^3=\pm\infty$. 
Here $B_{3\star}$ is the flux of the single ANO vortex 
defined by 
$B_{3\star} = - \frac{1}{2}(\partial_1^2+\partial_2^2)\Omega_\star$
and it satisfies $\int B_{3\star}\ dx^1dx^2 = - 2\pi$. 
Then the monopole charge $M_+$ is calculated as
\begin{eqnarray}
M_+
&=& \frac{2}{g^2}\int d^3x\ 
\partial_m\Tr\left(B_m\Sigma\right)
\nonumber\\
&=& \frac{2}{g^2}\Tr\left[
\left\{
\int_{{\bf R}^2_+}\!\!\! d^2x
\left(
\begin{array}{cc}
0 & 0 \\
0 & B_{3\star}
\end{array}
\right)
- 
\int_{{\bf R}^2_-}\!\!\! d^2x
\left(
\begin{array}{cc}
B_{3\star} & 0 \\
0 & 0
\end{array}
\right)
\right\}
\left(
\begin{array}{cc}
m_1 & 0 \\
0 & m_2
\end{array}
\right)
\right]
\nonumber\\
&=& \frac{4\pi}{g^2}\left(m_1 - m_2\right) > 0
\label{eq:wvm:calc_mv}
\end{eqnarray}
where ${\bf R}^2_{\pm}$ are the boundary surface at 
$x^3=\pm \infty$, see figure \ref{fig:wvm:charge_vm}.
Notice that the ordering of the mass parameters 
$m_1 > m_2$ is not important here. 
In fact, if we reconsider this system with the opposite 
ordering $m_2 > m_1$, we again obtain the monopole with 
positive definite mass $\frac{4\pi}{g^2}(m_2 - m_1) >0$ 
since the orientation also changes as 
$\vec\phi{\ }^T \to (1,0)$ at $x^3\to+\infty$ 
and $(0,1)$ at $x^3\to-\infty$. 

\begin{figure}[ht]
\begin{center}
\includegraphics[height=3.5cm]{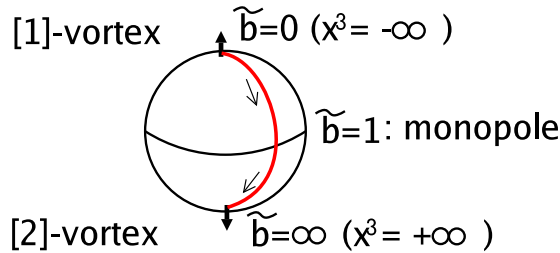}
\caption{\small{\sf Orientation of the vortex and 
position of the monopole.}}
\label{fig:wvm:wvm_orientation_mv}
\vspace*{-.3cm}
\end{center}
\end{figure}

Let us next consider the physical meaning of the moduli 
parameter $b$ in the moduli matrix (\ref{eq:wvm:mm_mv}) 
in the massive theory. 
At this stage, $\tilde b (x^3) = be^{-(m_1-m_2)x^3}$ can 
be thought of a quantity representing how close to the 
$[1]$- or $[2]$-vortex the configuration is. 
Especially, we found that $\tilde b = 0,\infty$ 
correspond to $[1]$-vortex and $[2]$-vortex, respectively. 
The transition point $\tilde b = 1$ which is the middle 
point between the north pole and the south pole 
of ${\bf C}P^1$, can be regarded as the monopole from the 
view point both of the moduli space and the real space, 
see figure \ref{fig:wvm:wvm_orientation_mv}. 
So we conclude that $b$ is related to the position of the 
monopole in the real space as 
\begin{eqnarray}
\left|\tilde b(x^3)\right| \approx 1 \quad \Leftrightarrow 
\quad
x^3 \approx \frac{1}{m_1 - m_2}\log \left|b\right|. 
\end{eqnarray}
Notice that the monopole goes away to the spacial infinity 
when we take the limit of $b \to 0,\infty$. 
This observation agrees with the previous 
argument that the moduli matrix (\ref{eq:wvm:mm_mv}) 
reduces to those for the 1/2 BPS vortex given in 
equation~(\ref{eq:wvm:mm_1/2_vortex1}). 

A natural extension of this is multiple monopoles in the 
Higgs phase which is sometimes called beads of monopoles. 
To get such a configuration, we need to consider $U(N)$ 
gauge theory with $N$ massive flavors.
The vortices in the massless theory has orientational 
moduli 
$\vec\phi{\ }^T = (\vec b{\ }^T, 1)\in {\bf C}P^{N-1}$, 
which reduce to $N$ fixed points when we turn on the 
nondegenerate masses. 
\begin{figure}[ht]
\begin{center}
\includegraphics[height=2.3cm]{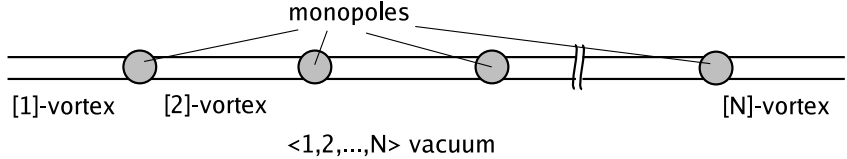}
\caption{\small{\sf Beads of monopoles.}}
\label{fig:wvm:beads}
\vspace*{-.3cm}
\end{center}
\end{figure}
Similarly to the case of $N=2$, the orientational moduli 
$\vec b$ can be understood as the positions of $N-1$ 
monopoles connecting $N$ different species of the ANO 
vortices. 
The moduli matrix describing the beads of monopoles 
penetrated by vortices are of the form 
\begin{eqnarray}
H_0 e^{Mx^3} = 
\left(
\begin{array}{cc}
{\bf 1}_{N-1} & \vec b\\
0 & z - z_1
\end{array}
\right)
e^{Mx^3}
= e^{Mx^3}
\left(
\begin{array}{cc}
{\bf 1}_{N-1} & \vec{\tilde b}(x^3)\\
0 & z - z_1
\end{array}
\right),
\label{eq:mvw:beads}
\end{eqnarray}
where we have defined the orientational vector
$\vec\phi{\ }^T = \left(\vec{\tilde b}(x^3)^T, 1\right)$ 
with 
$\vec{\tilde b}(x^3)^T \equiv 
\left(
b_1 e^{-(m_1-m_N)x^3},
\cdots,
b_{N-1} e^{-(m_{N-1}-m_N)x^3}
\right)
$.
Positions of the monopoles are estimated by
$|b_A| e^{-(m_A-m_N)x^3} \approx |b_{A+1}| e^{-(m_{A+1}-m_N)x^3}$:
\begin{eqnarray}
x^3 \approx 
\frac{1}{m_A - m_{A+1}}\log \left|\frac{b_A}{b_{A+1}}\right|,
\end{eqnarray}
where $A = 1,2,\cdots,N-1\ (b_N = 1)$.

\subsubsection{Boojums: Junctions of Walls and Vortices
\label{sec:mvw:boojum}}

Let us next investigate the other composite 1/4 BPS state 
made of the ANO vortices and the domain walls whose 
topological charges were given in 
equation~(\ref{eq:wvm:charge_vw}). 
As already mentioned above, this composite soliton is 
essentially a 1/4 BPS state in the Abelian gauge theory. 
The moduli matrix is just an $N_{\rm F}$ component complex 
vector in the Abelian gauge theory. 
Although the moduli matrix is completely the same as 
that for the 1/2 BPS semi-local vortex, it is very 
important to realize that the moduli matrix is accompanied 
by the factor $e^{Mx^3}$ as given in 
equation~(\ref{eq:4eq:mvw-1}). 
The 1/4 BPS equations 
(\ref{eq:4eq:mvw-1})--(\ref{eq:4eq:mvw-3}) admit 
the 1/2 BPS solutions also. 
Namely, the 1/2 BPS vortices in the massive theory 
dealt in the section~\ref{sec:mvw:vortex} are solutions 
of the 1/4 BPS equations. 
As already explained, a part of moduli of the semi-local 
vortices in the massless theory are lost when they are 
put into the massive theory. 
Then the moduli space of semi-local vortices are split 
into those of the ANO vortices in the massive theory. 
Indeed, the moduli matrix for the 1/2 BPS solutions turns 
into that for the $N_{\rm F}$ ANO vortices as given in 
equation (\ref{eq:wvm:semilocal_massive}). 
However, due to the additional factor $e^{Mx^3}$, the 
general moduli matrix which has two or more non zero 
components can give solutions of the BPS equations. 
But it no longer gives 1/2 BPS solutions but 1/4 BPS 
solutions:
\begin{eqnarray}
H_0 e^{Mx^3} = 
\left(
H_{0\star}^{1}(z) e^{m_1x^3},
H_{0\star}^{2}(z) e^{m_2x^3},
\cdots,
H_{0\star}^{N_{\rm F}}(z) e^{m_{N_{\rm F}}x^3}
\right)
\label{eq:wvm:mm_vw}
\end{eqnarray}
where $H_{0\star}^{A}(z)$ is again the moduli matrix for 
the ANO vortices defined by 
$H_{0\star}^{A}(z) \equiv a_A\prod_{i_A=1}^{k_A}(z-z_{i_A})$. 
We will show that the moduli parameters contained in the 
general moduli matrix in equation (\ref{eq:wvm:mm_vw}) can 
be reinterpreted as the moduli of the ANO vortices and the 
domain walls in the massive theory instead of those of 
the semi-local vortices in the massless theory. 
Notice that the set of $H_0(z)e^{Mx^3}$ can be thought 
of as the moduli matrix for the domain walls if we fix 
the coordinate $z$, whereas it can be regarded as that 
for the semi-local vortices if we fix the coordinate 
$x^3$. 
Thus, the moduli parameters in the moduli matrix 
(\ref{eq:wvm:mm_vw}) will be reinterpreted in terms of 
both the semilocal vortices and the domain walls in the 
following. 

Let us first consider the simplest example of a junction 
of a vortex and a domain wall in the $N_{\rm F}=2$ theory 
with the nondegenerate mass $M={\rm diag}(m_1,m_2)$ 
ordered as $m_1>m_2$. 
We focus on a single semilocal vortex given by the 
moduli matrix 
$H_0(z) = \left(a_1(z-z_1),\ a_2(z-z_2)\right)$ 
where the moduli parameters $z_1$, $z_2$ are positions 
and the size of the semilocal vortex, 
and the ratio $a \equiv a_1/a_2 \in {\bf C}P^1$ 
correspond to the position of the wall. 
In the massive theory this moduli matrix is multiplied 
by the additional factor $e^{Mx^3}$ and can be 
reinterpreted as follows
\begin{eqnarray}
H_0(z)e^{Mx^3} 
&\sim& e^{m_2x^3} 
\left(\tilde a(x^3)(z-z_1),\ z-z_2\right)\nonumber\\
&\sim& e^{m_1x^3}  
\left(z-z_1,\ \tilde a'(x^3)(z-z_2)\right) 
\label{eq:wvm:mm_single_vw}
\end{eqnarray}
with $\tilde a(x^3) \equiv e^{(m_1-m_2)x^3}a_1/a_2$ 
and $\tilde a'(x^3) \equiv \tilde a(x^3)^{-1}$. 
The new parameter $\tilde a(x^3)$ again gives us the 
boundary condition at the spatial infinity $|z|\to \infty$ 
but now it varies along $x^3$ axis in the massless vacuum 
manifold ${\bf C}P^1$. 
Since $\tilde a \to 0$ at $x^3 \to -\infty$, the moduli 
matrix (\ref{eq:wvm:mm_single_vw}) reduces to that for 
the 1/2 BPS $\left[2\right]$-vortex sitting on $z = z_2$ 
at $x^3 = - \infty$. 
On the other hand, $\tilde a' \to 0$ at $x^3 \to +\infty$, 
then the moduli matrix (\ref{eq:wvm:mm_single_vw}) 
reduces that for the 1/2 BPS $\left[1\right]$-vortex 
sitting on $z=z_1$ at $x^3 = + \infty$. 
Thus the parameter $\tilde a(x^3)$ gives the flow 
connecting $\left[1\right]$-vortex and 
$\left[2\right]$-vortex. 
\begin{figure}[ht]
\begin{center}
\includegraphics[height=3.5cm]{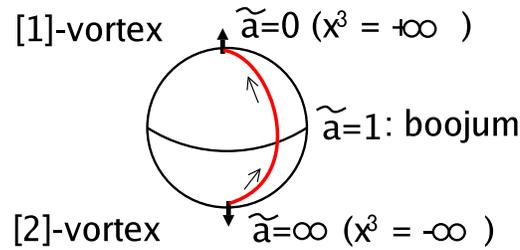}
\caption{\small{\sf Boundary condition of the vortex and 
position of the domain wall.}}
\label{fig:wvm:wvm_orientation_vw}
\vspace*{-.3cm}
\end{center}
\end{figure}
One can easily recognize a similarity between the 
monopoles in the Higgs phase in the previous section 
and this system: 
the orientational moduli $b$ of the non-Abelian vortex 
is promoted to a function 
$\tilde b = b e^{-(m_1-m_2)x^3}$ which is 
reinterpreted as the flow in the massless moduli 
space ${\bf C}P^1$ while the boundary moduli $a$ of the 
semi-local vortex is also promoted to a function 
$\tilde a = a^{(m_1-m_2)x^3}$ which is reinterpreted as the 
anti-flow of the massless vacuum manifold ${\bf C}P^1$, 
see figure~\ref{fig:wvm:wvm_orientation_vw}. 

\begin{figure}[ht]
\begin{center}
\includegraphics[height=5cm]{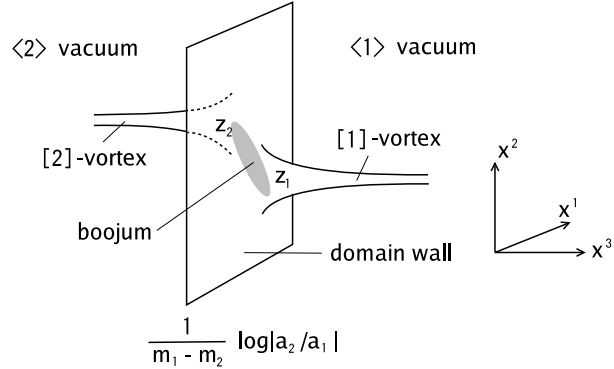}
\caption{\small{\sf Two vortices ending on the domain 
wall from both sides.}}
\label{fig:wvm:wvm_vw_simple}
\vspace*{-.3cm}
\end{center}
\end{figure}

As studied in section~\ref{sec:mvw:vortex}, the 
$\left[A\right]$-vortex can live only in the vacuum 
$\left<A\right>$. 
So the transition between $\left[1\right]$-vortex and 
$\left[2\right]$-vortex necessarily accompany a 
transition between $\left<1\right>$ and $\left<2\right>$ 
vacua, namely a domain wall. 
The position of the domain wall can be estimated by 
comparing the weights of vacua 
$|{\cal W}^{\left<1\right>}(z)| 
\approx |{\cal W}^{\left<2\right>}(z)|$ 
with ${\cal W}^{\left<A\right>}(z) \equiv a_A(z-z_A)e^{m_Ax^3}$ 
as we explained in section \ref{wll}. 
Then we get the position of the domain wall by 
\begin{eqnarray}
x^3(z) \approx \frac{1}{m_1 - m_2} 
\log \left|\frac{a_2(z-z_2)}{a_1(z-z_1)}\right|
\to \frac{1}{m_1 - m_2} 
\log \left|\frac{a_2}{a_1}\right|
\label{eq:wvm:wall_position}
\end{eqnarray}
as $|z|\to \infty$.
Thus, the moduli parameters $z_1,z_2$ for the positions 
of the semilocal vortices are reinterpreted as the positions 
of the ANO vortices ending on both sides of domain walls, 
whereas the size of the semilocal vortex $a = a_1/a_2$ is 
reinterpreted as the position of the domain wall. 
We give a schematic figure of this system in 
figure~\ref{fig:wvm:wvm_vw_simple}.

Let us next study the junction charge given in the last 
term of equation~(\ref{eq:wvm:charge_vw}). 
Notice that we now consider the Abelian gauge theory and 
then the charge is not the usual monopole charge in the 
non-Abelian gauge theories. 
It is called the boojum, see Ref.\cite{Sakai:2005sp}. 
We can explicitly calculate the boojum charge $M_-$ similarly 
to the monopole charge in equation~(\ref{eq:wvm:calc_mv}).
The domain wall sits on 
$x^3={1 \over m_1-m_2} \log |a_2/a_1|$ 
and the vortex 
in the left vacuum $\left<2\right>$ resides at 
$z = z_2$, and the vortex in the right vacuum 
$\left<1\right>$ at $z=z_1$. 
At the both infinities $x^3 = \pm \infty$ the magnetic 
flux reduces to that of the ANO vortex 
$\vec B = (0,0,B_{3\star})$. 
On the other hand, the VEV of the scalar $\Sigma$ in the 
vector multiplet approaches $\Sigma = m_1$ 
at $x^3\to+\infty$ and $\Sigma =m_2$ at $x^3\to-\infty$. 
Then we obtain 
\begin{eqnarray}
M_{-} &=& \frac{2}{g^2}\int d^3x\ 
\partial_m\left(B_m\Sigma\right)
\nonumber\\
&=& \frac{2}{g^2}\left(
\int_{{\bf R}^2_+}d^2x\ B_{3\star} m_1
- 
\int_{{\bf R}^2_-}d^2x\ B_{3\star} m_2
\right) 
\nonumber\\
&=& - \frac{4\pi}{g^2}(m_1 - m_2) < 0
\label{eq:wvm:calc_vw}
\end{eqnarray}
where we have used $\int d^2x B_{3\star} = -2\pi$. 
Here the ordering of mass parameters $m_1 > m_2$ 
is not essential because the opposite ordering 
$m_2 > m_1$ requires us to set $\Sigma = m_2$ at 
$x^3 \to + \infty$ and $\Sigma = m_1$ at 
$x^3 \to -\infty$ to get a consistent domain wall 
configuration, see section~\ref{wll}.
So the boojum charge always gives a negative contribution 
to the energy although their magnitude is the same as the 
monopole charge given in equation~(\ref{eq:wvm:calc_mv}). 
This negative charge of the boojum is understood as the 
binding energy of the domain wall and the ANO vortex 
\cite{Sakai:2005sp}. 
The sign difference between the monopole energy in 
equation~(\ref{eq:wvm:calc_mv}) and the boojum energy in 
equation~(\ref{eq:wvm:calc_vw}) comes from a different 
mechanisms of picking up the mass difference at the 
boundary. 
Different orientations of left and right vortices gives 
the mass difference for the monopole, whereas different 
VEV of $\Sigma$ of left and right vacua provides it for 
the boojum. 
This distinction is also reflected in the opposite 
direction of the flows $\tilde b(x^3) = b e^{-(m_1-m_2)x^3}$ 
for the monopole and $\tilde a(x^3) = a e^{(m_1-m_2)x^3}$ 
for the boojum in the ${\bf C}P^1$ manifold of the vortex 
orientation, as shown in 
figure~\ref{fig:wvm:wvm_orientation_mv} and 
figure~\ref{fig:wvm:wvm_orientation_vw}. 
Notice that the position of the boojum can be estimated 
similarly to the position of the monopole in the 
previous section: 
the north pole ($\tilde a = 0$) corresponds to the 
$\left[1\right]$-vortex while the south pole 
($\tilde a = \infty$) to the $\left[2\right]$-vortex. 
Then the middle point of ${\bf C}P^1$, namely 
$|\tilde a(x^3)| =1$ can be 
regarded as the position of the boojum 
\begin{eqnarray}
|\tilde a(x^3)| =1
\quad\Rightarrow\quad
x^3 = \frac{1}{m_1-m_2}\log\left|\frac{a_2}{a_1}\right|.
\label{eq:wvm:boojum_position}
\end{eqnarray}
Notice that the position of the boojum given above agrees 
with the position of the domain wall in equation 
(\ref{eq:wvm:wall_position}). 
The boojum energy is spread inside the domain wall on 
which the two vortices end from both sides, as shown in 
figure~\ref{fig:wvm:wvm_vw_simple}.

\begin{figure}[ht]
\begin{center}
\includegraphics[width=13cm]{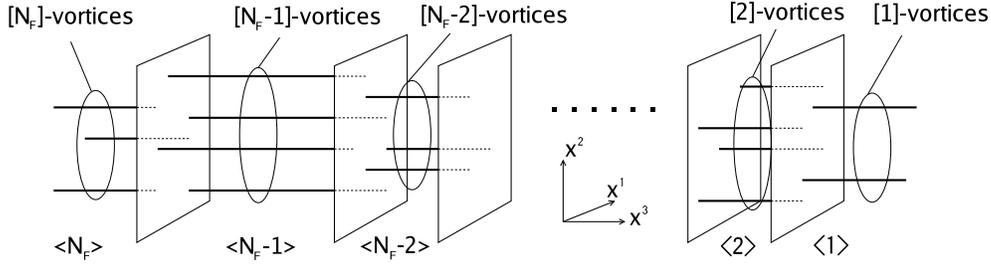}
\caption{\small{\sf A generic configuration of the ANO 
vortices ending on domain walls.}}
\label{fig:wvm:wvm_vw_general}
\vspace*{-.3cm}
\end{center}
\end{figure}

Let us consider more general configurations with 
multiple domain walls and multiple vortices ending 
on the domain walls. 
In Abelian gauge theory, the moduli matrix has been 
given in equation (\ref{eq:wvm:mm_vw}). 
From the view point of the domain wall, we can regard 
${\cal W}^{\left<A\right>}(z) = H_{0\star}^{A}(z)e^{m_Ax^3}$ 
as the weight of the $\left<A\right>$ vacuum. 
Then the position of the domain wall interpolating two 
vacua $\left<A\right>$ and $\left<A+1\right>$ can be 
estimated by equating weights of the the adjacent vacua 
$|H_{0\star}^{A}(z) e^{m_Ax^3}| 
\approx |H_{0\star}^{A+1}(z) e^{m_{A+1}x^3}|$:
\begin{eqnarray}
x^3(z) \approx \frac{1}{m_A - m_{A+1}}
\log\left|\frac{H_{0\star}^{A+1}(z)}{H_{0\star}^{A}(z)}\right|.
\label{eq:wvm:posi_vw}
\end{eqnarray}
Let us next change the view from the domain wall to the 
vortex. 
Each component $H_{0\star}^{A}(z)$ of the moduli matrix 
(\ref{eq:wvm:mm_vw}) is then thought of as the moduli 
matrix for the $\left[A\right]$-vortex in the 
$\left<A\right>$ vacuum. 
Therefore, 
$H_{0\star}^{A}(z) = a_A \prod_{i_A=1}^{k_A}(z-z_{i_A})$ 
gives the $k_A$ vortices sitting on $z=z_{i_A}$ in the 
vacuum $\left<A\right>$. 
Thus we conclude that the moduli matrix 
(\ref{eq:wvm:mm_vw}) gives us the domain walls 
interpolating $N_{\rm F}$ vacua as 
$\left<N_{\rm F}\right> \leftrightarrow
\left<N_{\rm F}-1\right> \leftrightarrow
\cdots \leftrightarrow
\left<2\right> \leftrightarrow
\left<1\right>$ and each vacuum holds the $k_A$ 
ANO vortices at $z_{i_A}$ which end on the domain walls 
sandwiching the vacuum $\left<A\right>$. 
A generic configuration of the vortices ending on domain 
walls is depicted in figure \ref{fig:wvm:wvm_vw_general}.

\begin{figure}[ht]
\begin{center}
\includegraphics[height=3.5cm]{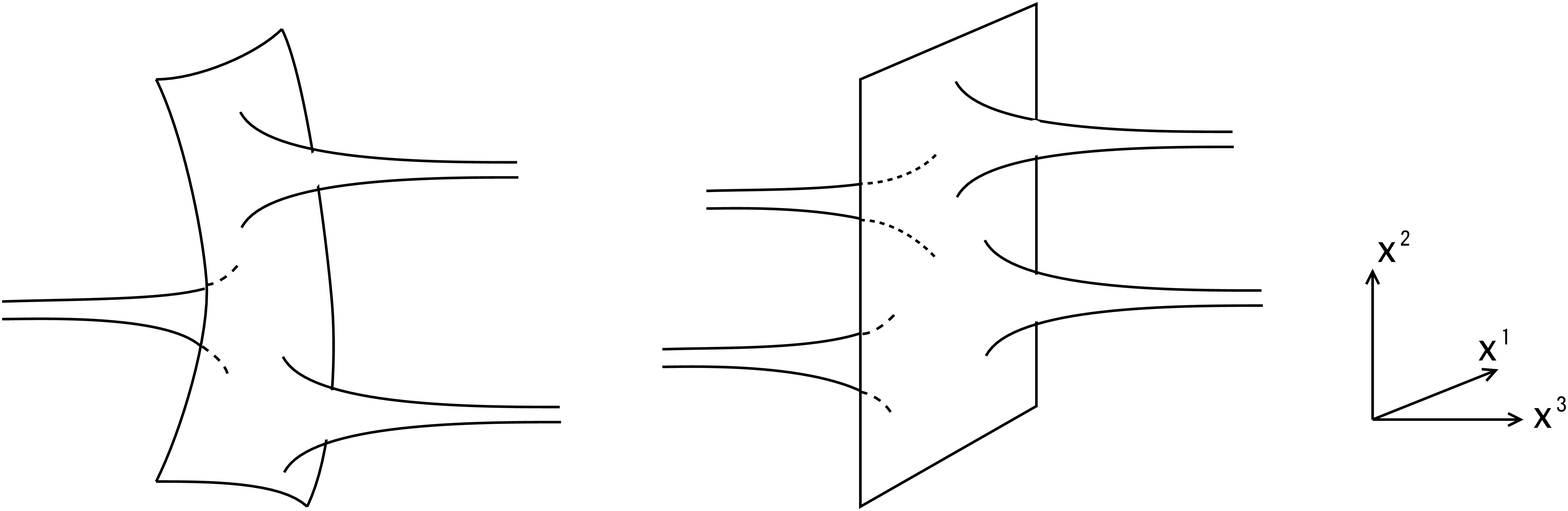}
\caption{\small{\sf Logarithmic bending and flat domain walls.}}
\label{fig:wvm:wvm_log}
\vspace*{-.3cm}
\end{center}
\end{figure}

It is worth commenting on the expression of wall 
positions given in equation~(\ref{eq:wvm:posi_vw}). 
When the number $k_{A+1}$ of the vortices ending on 
the wall from the left ($\left<A+1\right>$ vacuum) and 
the number $k_A$ of that ending on the wall from the right 
($\left<A\right>$ vacuum) is different from each other, 
the domain wall bends logarithmically. 
This is because the vortices pull the domain wall and 
the domain wall needs to pull the vortices back to keep 
themselves static and stable. 
The logarithmic bending always appears when $p$-brane 
ends on $(p+2)$-brane at a point like $D$-branes in the 
string theory. 
Notice that completely the same result as 
equation~(\ref{eq:wvm:posi_vw}) has been obtained from the 
view point of the low energy effective theory on the 
world volume of the domain wall in Ref.~\cite{Sakai:2005sp}.
When $k_A = k_{A+1}$, the ratio 
$\frac{H_{0\star}^{A+1}(z)}{H_{0\star}^{A}(z)}$ 
becomes constant at the spatial infinity on the $z$ 
plane ($|z|\to \infty$) 
\begin{eqnarray}
x^3(z) 
\to  \frac{1}{m_A - m_{A+1}}
\log\left|\frac{a_{A+1}}{a_{A}}\right|
\label{eq:wvm:posi_vw_flat}
\end{eqnarray}
as $|z|\to\infty$.
So the domain wall with the same number of vortices at 
left and right vacua becomes asymptotically flat, see 
figure \ref{fig:wvm:wvm_log}.

\subsubsection{General configurations}

So far we have dealt with the minimal models: One is the 
$N_{\rm C}=N_{\rm F}=2$ for the monopoles in the Higgs 
phase and the other is the $N_{\rm C}=1$, $N_{\rm F}\ge2$ 
for the composite of vortices and domain walls (the boojums). 
Let us next investigate a more general configuration 
which includes both the monopole and the boojum as the 
junction charges of vortices and domain walls. 
For that purpose, we shall consider $U(2)$ gauge
theory with $N_{\rm F}=3$ flavors with 
$M = {\rm diag}(m_1,m_2,m_3)$ ordered as $m_1>m_2>m_3$. 
This model has three discrete vacua $\left<12\right>$, 
$\left<23\right>$ and $\left<13\right>$. Let us focus on 
the configuration which has the single domain wall 
interpolating the $\left<13\right>$ vacuum at 
$x^3=+\infty$ and the $\left<23\right>$ vacuum at 
$x^3 = -\infty$. 
Furthermore, we put a vortex in both sides of the 
domain wall. 
In terms of the Pl\"ucker coordinates in equation 
(\ref{eq:wvm:plucker}), this configuration is represented 
by $\tau^{\left<12\right>}=0$, 
$\tau^{\left<13\right>}= {\cal O}(z)$ and 
$\tau^{\left<23\right>}= {\cal O}(z)$. 
Exploiting the $V$-equivalence relation 
(\ref{eq:4eq:mvw-V-equiv}), we can reduce all 
possible moduli matrix satisfying these conditions to 
either one of the following three different kinds of the 
moduli matrices $H_0e^{Mx^3}$: 
\begin{eqnarray}
\left(
\begin{array}{ccc}
1 & a_1 & b \\
0 & 0 & z-z_1
\end{array}
\right)e^{Mx^3} \sim
\left(
\begin{array}{ccc}
\frac{1}{a_1}(z-z_1) & z-z_1 & 0 \\
\frac{1}{b} & \frac{a_1}{b} & 1
\end{array}
\right)e^{Mx^3},
\label{eq:wvm:mm_mb1}\\
\left(
\begin{array}{ccc}
z-z_2 & a_2(z-z_3) & 0\\
0 & 0 & 1
\end{array}
\right)e^{Mx^3},
\label{eq:wvm:mm_mb2}\\
\left(
\begin{array}{ccc}
1 & a_3 & 0 \\
0 & 0 & z-z_4
\end{array}
\right)e^{Mx^3},
\label{eq:wvm:mm_mb3}
\end{eqnarray}
with $z_1,z_2,z_3,z_4 \in {\bf C}$ and 
$a_1,a_2,a_3,b \in {\bf C}^*$.

\begin{figure}[ht]
\begin{center}
\includegraphics[height=5cm]{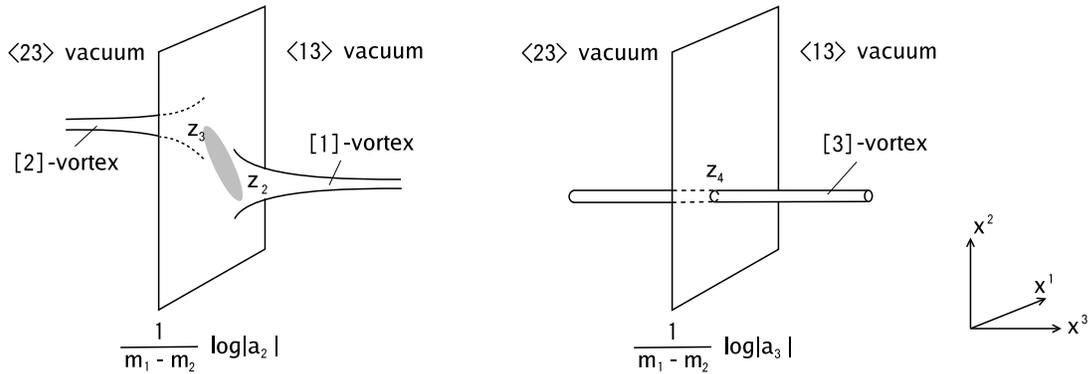}
\caption{\small{\sf The junction and intersection 
between the vortices and the domain wall.}}
\label{fig:wvm:wvm_bm23}
\vspace*{-.3cm}
\end{center}
\end{figure}

Let us begin by investigating the moduli matrices 
(\ref{eq:wvm:mm_mb2}) and (\ref{eq:wvm:mm_mb3}). 
These are simple in the sense that the 
moduli matrices do not have off-diagonal elements which mix 
the first color and the second color components.
In fact, $\Omega_0 = H_0(z) e^{2Mx^3}H_0^\dagger$ which
is the source of the master equation (\ref{eq:4eq:master_mvw}) 
becomes diagonal. 
Then we can deal with the first row and the 
second row independently. 
In the case of the moduli matrix (\ref{eq:wvm:mm_mb2}) 
the second row gives no physical effects, so the moduli 
matrix is essentially that for the Abelian gauge theory. 
In fact, the first row 
$\left((z-z_2)e^{m_1x^3}, a_2(z-z_3)e^{m_2x^3},0\right)$
is nothing but the moduli matrix for the vortices ending 
on the domain wall 
with boojum studied in section \ref{sec:mvw:boojum}.
Then the position of the domain wall interpolating 
$\left<13\right>$ vacuum and $\left<23\right>$ vacuum 
can be estimated by equation (\ref{eq:wvm:wall_position}), 
so we have $x^3 \approx \frac{1}{m_1-m_2}\log\left|a_2\right|$.
A $\left[1\right]$-vortex ends on the domain wall 
from the right (vacuum $\left<13\right>$) at $z=z_2$
while a $\left[2\right]$-vortex ends on 
from the left (vacuum $\left<23\right>$) at $z=z_3$.
They accompany the boojum whose $x^3$ position is 
the same as that of the domain wall, see
equation (\ref{eq:wvm:boojum_position}).
The configuration is depicted in the left of 
figure~\ref{fig:wvm:wvm_bm23}.
On the other hand, the moduli matrix 
(\ref{eq:wvm:mm_mb3}) 
has the domain wall in the first row 
$\left(e^{m_1x^3},a_3e^{m_2x^3},0\right)$ 
and the $\left[3\right]$-vortex 
$\left(0,0,(z-z_4)e^{m_3x^3}\right)$ in the 
second row. 
Since these two rows contribute to $\Omega_0$ as an 
incoherent sum, the moduli matrix (\ref{eq:wvm:mm_mb3}) can 
be regarded as that for the direct product of two decoupled 
$U(1)$ gauge theories rather than the $U(2)$ gauge theory. 
The first row gives the domain wall interpolating 
$\left<13\right>$ and $\left<23\right>$ vacua and 
the second row gives the $\left[3\right]$-vortex. 
The domain wall and the vortex do not interact, 
so the composite soliton represents just an intersection 
without the monopole and/or the boojum. 
The position of the domain wall is 
$x^3 \approx \frac{1}{m_1-m_2}\log|a_3|$ 
and that of the $\left[3\right]$-vortex is $z_4$. 
This situation is depicted in the right of 
figure~\ref{fig:wvm:wvm_bm23}.

Let us next investigate a more interesting moduli 
matrix (\ref{eq:wvm:mm_mb1}) which has an off-diagonal 
element $b$ and is intrinsically non-Abelian. 
It turns out that the configuration has
the domain wall, the vortices, the monopoles and the boojum.
We have
$
\tau^{\left<12\right>}(z) = 0, 
\tau^{\left<23\right>}(z) = a_1(z-z_1)
$ and $
\tau^{\left<13\right>}(z) = z-z_1
$ and the weights of vacua ${\cal W}^{\left<AB\right>}(z) 
= \tau^{\left<AB\right>}(z)e^{(m_A+m_B)x^3}$:
\begin{eqnarray}
{\cal W}^{\left<12\right>}(z) = 0,\\ 
{\cal W}^{\left<23\right>}(z) = a_1(z-z_1)e^{(m_2+m_3)x^3},\\ 
{\cal W}^{\left<13\right>}(z) = (z-z_1)e^{(m_1+m_3)x^3}.
\end{eqnarray}
The position of the domain wall dividing those two vacua 
can be estimated by the same method as 
equation~(\ref{eq:wvm:posi_vw}):
\begin{eqnarray}
x^3\bigg|_{\rm wall} \approx \frac{1}{(m_1+m_3) - (m_2+m_3)}\log 
\left|\frac{\tau^{\left<23\right>}}{\tau^{\left<13\right>}}\right|
= \frac{1}{m_1-m_2}\log\left|a_1\right|.
\label{eq:wvm:wall_postion_bm1}
\end{eqnarray}
The $z-z_1$ in $\tau^{\left<23\right>}$ and 
$\tau^{\left<13\right>}$ gives us the single ANO vortex 
both in the vacua $\left<23\right>$ and $\left<13\right>$ 
sit on $z=z_1$. As shown previously, 
$\tau^{\left<23\right>} \propto z-z_1$ means a $[2]$- or 
$[3]$-vortex in the vacuum $\left<23\right>$ while 
$\tau^{\left<13\right>} \propto z-z_1$ means a $[1]$- or 
$[3]$-vortex in the vacuum $\left<13\right>$. 
In order to identify which vortex arises in the vacua, 
we rewrite the moduli matrix (\ref{eq:wvm:mm_mb1}) in the 
following form and understand it from the view point of 
the vortex moduli matrix 
\begin{eqnarray}
e^{M^{\left<13\right>}x^3}
\left(
\begin{array}{ccc}
1 & \tilde a_1(x^3) & \tilde b(x^3) \\
0 & 0 & z-z_1
\end{array}
\right)
\sim
e^{M^{\left<23\right>}x^3}
\left(
\begin{array}{ccc}
\frac{z-z_1}{\tilde a_1(x^3)} & z-z_1 & 0 \\
\frac{1}{\tilde b(x^3)} & 
\frac{\tilde a_1(x^3)}{\tilde b(x^3)} & 1
\end{array}
\right)
\end{eqnarray}
with $M^{\left<AB\right>} \equiv {\rm diag}(m_A,m_B)$,
$\tilde a_1(x^3) \equiv e^{-(m_1-m_2)x^3}a_1$ and
$\tilde b(x^3) \equiv e^{-(m_1-m_3)x^3}b$. 
Taking the mass ordering $m_1 > m_2 > m_3$ into account,
we easily find that the configuration at both the boundary 
$x^3 \to\pm \infty$ from this expression:
\begin{eqnarray}
\left(
\begin{array}{ccc}
1 & \tilde a_1(x^3) & \tilde b(x^3) \\
0 & 0 & z-z_1
\end{array}
\right)
\to 
\left(
\begin{array}{ccc}
1 & 0 & 0 \\
0 & 0 & z-z_1
\end{array}
\right)
\end{eqnarray}
as $x^3 \to \infty$ and 
\begin{eqnarray}
\left(
\begin{array}{ccc}
\frac{z-z_1}{\tilde a_1(x^3)} & z-z_1 & 0 \\
\frac{1}{\tilde b(x^3)} & \frac{\tilde a_1(x^3)}{\tilde b(x^3)} & 1
\end{array}
\right)
\to 
\left(
\begin{array}{ccc}
0 & z-z_1 & 0 \\
0 & 0 & 1
\end{array}
\right)
\end{eqnarray}
as $x^3\to-\infty$.
Then the moduli matrix (\ref{eq:wvm:mm_mb1}) has 
the $[3]$-vortex in the $\left<13\right>$ vacuum at 
$x^3\to+\infty$ and 
the $\left[2\right]$-vortex
in the $\left<23\right>$ vacuum at $x^3 \to -\infty$.
When going along the $x^3$ axis, the $[2]$-vortex makes 
a transition to the $[3]$-vortex with the monopole 
and/or the boojum charge. 

\begin{figure}[ht]
\begin{center}
\includegraphics[height=5cm]{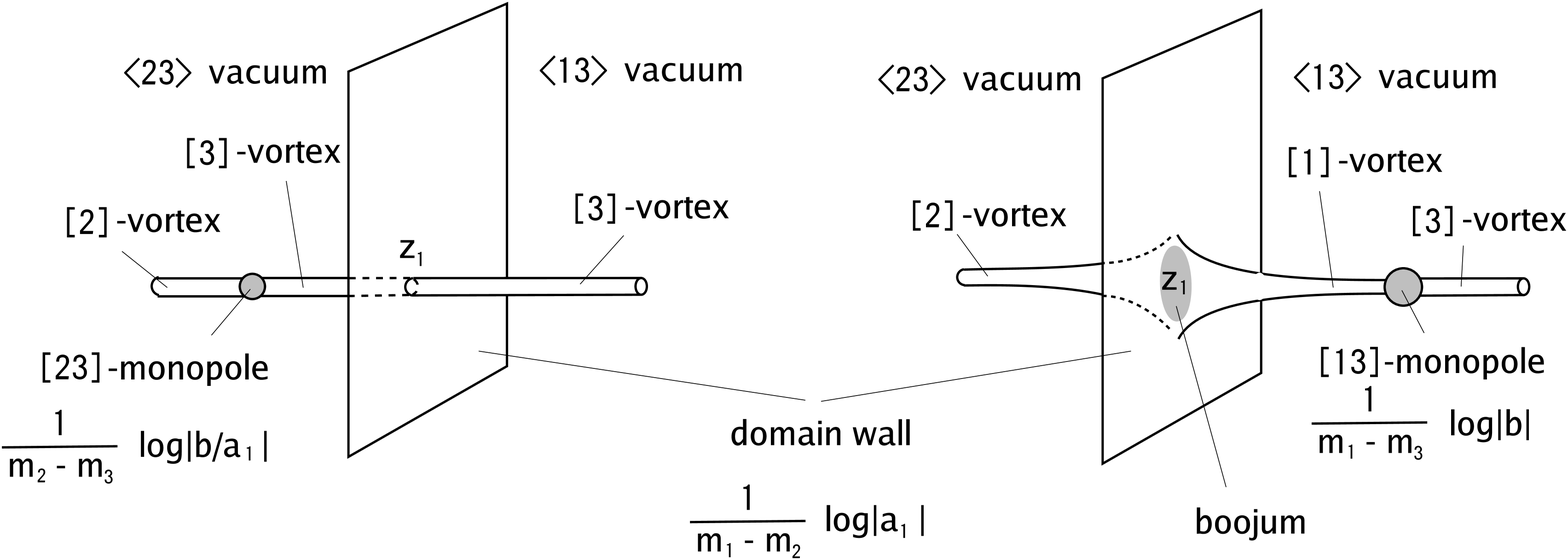}
\caption{\small{\sf The composite states of the vortices, 
the domain wall, the monopoles and the boojum. 
The left figure is for the parameter region 
$|a_1| \gg |b|$ while the right is for $|a_1| \ll |b|$.}}
\label{fig:wvm:wvm_bm1}
\vspace*{-.3cm}
\end{center}
\end{figure}

Let us first consider a parameter region where 
$|b|\gg |a_1|,1$. 
Then there exists a region where 
$\left|\tilde a_1(x^3)\right| \ll \left|\tilde b(x^3)\right| \ll 1$
and the moduli matrix reduces to
\begin{eqnarray}
e^{M^{\left<13\right>}x^3}
\left(
\begin{array}{ccc}
1 & 0 & \tilde b(x^3) \\
0 & 0 & z-z_1
\end{array}
\right).
\end{eqnarray}
Since the second column has no contribution, this moduli 
matrix has the same form as the middle moduli matrix in 
equation (\ref{eq:wvm:mm_mv}) and gives the monopole 
attached by 
the $[1]$-vortex from the left and the $[3]$-vortex 
from the right in the vacuum $\left<13\right>$. 
The mass of the $[13]$-monopole is given by 
$
M_+^{[13]} = \frac{4\pi}{g^2}(m_1-m_3),
$
and its position is given by the condition 
$\left|\tilde b(x^3)
\right| \approx 1$ as 
$
x^3\big|_{[13]-{\rm monopole}} 
\approx \frac{1}{m_1 - m_3} \log\left|b\right|
$.
There is another region 
where $\left|\frac{1}{\tilde b(x^3)}\right|, 
\left|\frac{\tilde a_1(x^3)}{\tilde b(x^3)}\right| \ll 1$
and there the moduli matrix reduces to the following form
\begin{eqnarray}
e^{M^{\left<23\right>}x^3}
\left(
\begin{array}{ccc}
\frac{z-z_1}{\tilde a_1(x^3)} & z-z_1 & 0 \\
0 & 0 & 1
\end{array}
\right)
\end{eqnarray}
This is nothing but the moduli matrix for the domain wall 
interpolating $\left<13\right>$ and $\left<23\right>$ 
vacua on which the $[1]$-vortex ends from the right and 
the $[2]$-vortex from the left with the boojum. 
The mass of the boojum is given by 
$
M_-^{[12]} = - \frac{4\pi}{g^2}(m_1-m_2).
$
and its position is the same as that of the domain wall 
given by equation (\ref{eq:wvm:wall_postion_bm1}). 
The configuration is depicted in the right of 
figure~\ref{fig:wvm:wvm_bm1}.

Let us next consider the parameter region where 
$|b|\ll |a_1|$. 
Then the moduli matrix reduces to the following form 
in the region where 
$1,\left|\tilde a_1(x^3)\right| \gg  \left|\tilde b(x^3)\right|$
\begin{eqnarray}
e^{M^{\left<13\right>}x^3}
\left(
\begin{array}{ccc}
1 & \tilde a_1(x^3) & 0 \\
0 & 0 & z-z_1
\end{array}
\right).
\end{eqnarray}
This moduli matrix gives us the trivial intersection of 
the domain wall and the $[3]$-vortex as explained for 
the moduli matrix (\ref{eq:wvm:mm_mb3}). 
There also exists a region where 
$1,\left|\frac{1}{\tilde b(x^3)}\right|
\ll \left|\frac{\tilde a_1(x^3)}{\tilde b(x^3)}\right|$.
There the moduli matrix reduces to 
\begin{eqnarray}
e^{M^{\left<23\right>}x^3}
\left(
\begin{array}{ccc}
0 & z-z_1 & 0 \\
0 & \frac{\tilde a_1(x^3)}{\tilde b(x^3)} & 1
\end{array}
\right).
\end{eqnarray}
Since the first column has no contribution, this moduli 
matrix has the same form as the right hand side of equation 
(\ref{eq:wvm:mm_mv}) and gives the $[23]$-monopole 
sandwiched by $[2]$-vortex from the left and $[3]$-vortex 
from the right. 
The mass of the monopole is given by 
$
M_+^{[23]} = \frac{4\pi}{g^2}(m_2-m_3),
$
and its position is given by 
$\left|\frac{\tilde a_1(x^3)}{\tilde b(x^3)}\right| \approx 1$:
$
x^3\big|_{[23]-{\rm monopole}} 
\approx \frac{1}{m_2 - m_3} \log\left|\frac{b}{a_1}\right|.
$
The configuration is depicted in the left of 
figure~\ref{fig:wvm:wvm_bm1}.

Let us summarize the configuration given by the moduli 
matrix (\ref{eq:wvm:mm_mb1}). 
There are two types of infinitely heavy objects: one is the 
domain wall sitting at $x^3= \frac{1}{m_1-m_2}\log|a_1|$, 
and the other type is the ($[1]$-, $[2]$- and $[3]$-)vortices 
penetrating the domain wall at $z=z_1$. 
For the time being, let us fix the parameters $z_1$ and $a_1$. 
Then the only the parameter $b$ remains as a free 
parameter corresponding to the position of the monopole 
which has a finite mass. 
When $|b| \ll |a_1|$, the $[23]$-monopole sandwiched 
by the $[2]$-vortex from the left and $[3]$-vortex from 
the right in the vacuum $\left<23\right>$, namely in the 
left of the domain wall. 
As the parameter $b$ grows, the monopole moves to the left 
along the $x^3$ axis. 
Around the region where $|b| \sim |a_1|$, the monopole, 
the vortices and the domain wall merge. 
After the $[23]$-monopole passes through the domain wall, 
namely the region where $|b| \gg |a_1|$, the $[2]$-vortex 
ends on the domain wall from the left and the $[1]$-vortex 
appears from the domain wall with the [13]-boojum left in 
the wall. 
Furthermore, the $[1]$-vortex makes a transition to 
the $[3]$-vortex at the $[13]$-monopole. 
Of course, the masses of the monopole and the boojum are 
preserved before and after the monopole passes the 
domain wall:
$
M_+^{[13]} + M_-^{[12]} = M_+^{[23]}
$. 
Interestingly, the center of mass of the $[13]$-monopole 
and the $[12]$-boojum for $|b| \gg |a_1|$ becomes the 
position of the $[23]$-monopole for $|b| \ll |a_1|$.

Before closing this section, we give a comment on the 
relation between the three moduli matrices 
(\ref{eq:wvm:mm_mb1})--(\ref{eq:wvm:mm_mb3}). 
When we take $b \to 0$ in the moduli matrix 
(\ref{eq:wvm:mm_mb1}), it reduces to the same form as 
the matrix (\ref{eq:wvm:mm_mb3}). 
This fact implies that the $[23]$-monopole can be sent 
to the minus infinity of $x^3$ axis, so that the 
only trivial intersection between the domain wall and 
the $[3]$-vortex remains. 
On the other hand, when we take $b \to \infty$ in the 
moduli matrix (\ref{eq:wvm:mm_mb1}), namely the 
$[13]$-monopole is sent to the plus infinity of the 
$x^3$ axis, the moduli matrix reduces to almost the same 
form as the matrix (\ref{eq:wvm:mm_mb2}). 
However, there is a small but crucial difference between 
them: the positions of two vortices can take different 
values $(z_2 \neq z_3)$ in the matrix (\ref{eq:wvm:mm_mb2}) 
while they must coincide in the limit ($b\to\infty$) of 
the moduli matrix (\ref{eq:wvm:mm_mb1}). 
Therefore we conclude that the positions of the vortices 
have to coincide when monopoles attach anywhere on the 
vortices. 
Only after removing the monopoles by sending them to the 
spatial infinity, the vortex in the left of the domain 
wall and that in the right can separate on the domain wall. 

\begin{figure}[ht]
\begin{center}
\includegraphics[height=5cm]{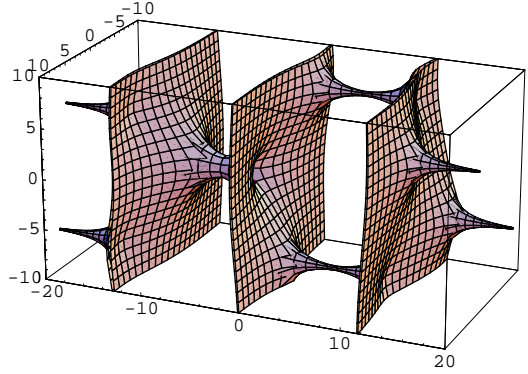}
\caption{\small{\sf Composite soliton of vortices (lumps) and domain walls.}}
\label{fig:wvm:amida}
\vspace*{-.3cm}
\end{center}
\end{figure}

\changed{
At the end of this subsection, we give an exact solution
in the strong gauge coupling limit $g^2\to\infty$. 
As explained in equation~(\ref{eq:4eq:infity-sol}), 
the master equation reduces
to just an algebraic equations. Then we can exactly solve them.
In figure~\ref{fig:wvm:amida} we give a configuration which is 
composite of vortices (lumps) and domain walls.
}

\subsection{Composite of vortices and instantons}\label{ist}
Although the total moduli space 
${\cal M}^{\rm total}_{\rm IVV}$ in equation 
(\ref{eq:4eq:moduli_sp_ivv}) 
of the $1/4$ BPS composite system of instantons \changed{and vortices }
is a parent of other $1/4$ BPS composite 
solitons, we leave its full analysis to future works. 
Instead we will here restrict ourselves to two cases of 
physical interest: one is 
$1/4$ BPS solutions interpretable as $1/2$ BPS lumps on 
the world volume of $1/2$ BPS vortices in the 1-3 plane, 
and the other is the intersection of two vortices. 
We wish to stress, however, that these solutions 
are genuine solutions of the 1/4 BPS 
equations, rather than solutions of the effective 
theory on host vortices.

\subsubsection{Instantons as lumps on vortices}

We first consider $1/2$ BPS lumps in 
the effective theory on the world volume 
of a vortex (\ref{eq:efa:kahler_pot_1vort}) before constructing a genuine solution of 
the 1/4 BPS equations. 
For simplicity, let us take the 
$N_{\rm C}=N_{\rm F}\equiv N=2$ case. 
Defining $z=x^1+ix^2$, the BPS equations 
(\ref{eq:vtx:BPSeq1}) and (\ref{eq:vtx:BPSeq2}) for vortices 
give a solution (\ref{eq:vtx:solution}). 
\changed{
For a single vortex 
the moduli matrix in a patch ${\cal U}^{(1,0)}$\, (\ref{eq:vtx:single})
is given by
\begin{equation}
H_{{\rm v}0}^{\rm single}(z;z_0,b) \equiv  \left(
\begin{array}{cc}
z-z_0 & 0\\
b & 1
\end{array}
\right),
\label{eq:ist:MM_vortex1}
\end{equation}
where $b\in {\bf C} $ is an orientational modulus of 
a vortex and an inhomogeneous coordinate of 
${\bf C}P^1$ as we explained.}
By promoting the moduli parameter $b$ to a field on the 
world volume of the vortex, we obtain the effective 
Lagrangian on the vortex (\ref{eq:efa:eff-lag}). 
By an almost identical argument to obtain (\ref{eq:vtx:BPSeq1}) and 
(\ref{eq:vtx:BPSeq2}) 
in section \ref{vtx}, 
we obtain a 1/2 BPS equation for lumps 
\cite{Polyakov:1975yp} on the vortex 
\begin{equation}
 \bar\partial_w b(w, \bar w) = 0, 
\qquad w\equiv x^3+ix^4 . 
\label{eq:ist:BPSeq-lump}
\end{equation}
This equation gives a $k$-lump solution which can be expressed 
in terms of rational functions of degree 
$k$~\cite{Polyakov:1975yp,Ward:1985ij} 
\begin{eqnarray}
b(w) = \frac{P_k(w)}{\alpha P_k(w) + aQ_{k-1}(w)},
\label{eq:ist:k-lumps}
\end{eqnarray}
\begin{eqnarray}
P_k(w) \equiv \prod_{i=1}^{k}(w-p_i), \qquad
Q_{k-1}(w) 
\equiv
\prod_{j=1}^{k-1} (w- q_j). 
\end{eqnarray}
Among the moduli parameters, $\{p_1,p_2,\cdots,p_k\}$ 
correspond to the positions of the $k$-lumps on 
the host vortex, and $a$ 
to the total size of the configurations, 
and $\{q_1,q_2,\cdots,q_{k-1}\}$ to the relative 
sizes of the $k$-lumps. 
The remaining modulus $\alpha$ specifies the boundary 
condition at $|w|\rightarrow \infty$ and 
parameterizes a point in the vacuum manifold 
${\bf C}P^1$, 
since $b(w) \to 1/\alpha$ as $|w| \to \infty$. 
When $\alpha=0$,  $\{p_i,a,q_j\}$ can be 
identified with positions and sizes of $k$-lumps precisely. 
The zeros of the denominator in equation (\ref{eq:ist:k-lumps}) 
are mere coordinate singularities caused by the use of an 
inhomogeneous coordinate $b$ of the ${\bf C}P^1$ manifold. 
The configurations are smooth and continuous at these 
coordinate singularities. 
However, the point $a=0$ and the points $p_i =q_j$ 
are true singularities of the moduli space of 
the lumps and are called small lump singularities.

For a more general case of $N_{\rm }=N_{\rm F}=N$, 
the orientational moduli space for the non-Abelian vortex is 
$SU(N)/[SU(N-1)\times U(1)] \simeq {\bf C}P^{N-1}$
~\cite{Hanany:2003hp,Auzzi:2003fs,Eto:2004ii}. 
The multi-lump solutions on the vortex in this case 
are obtained as lump solutions 
for the ${\bf C}P^{N-1}$ nonlinear sigma model, 
which are also known \cite{Stokoe:1986ic}.

\subsubsection{
1/4 BPS solutions of the instantons in the Higgs phase}

With the aid of the lump solution in the effective theory on 
the world volume of vortex, we can now obtain the genuine 
solutions of the 1/4 BPS equations 
(\ref{eq:4eq:ivv1})--(\ref{eq:4eq:ivv3}) 
for instantons in the Higgs phase. 
Our idea is to start replacing the moduli parameter 
$b$ in the moduli matrix $H_{{\rm v}0}^{\rm single}(z;z_0,b)$ 
in equation (\ref{eq:ist:MM_vortex1}) for a single vortex 
by the lump solution $b(w)$ in equation (\ref{eq:ist:k-lumps}) 
\begin{eqnarray}
H_0(z,w) \sim H_{{\rm v}0}^{\rm single}(z;z_0,b(w))
=  
\left(
\begin{array}{cc}
z - z_0 & 0\\
\frac{P_k}{\alpha P_k + aQ_{k-1}} & 1
\end{array}
\right). 
\label{eq:ist:1/2_to_1/4}
\end{eqnarray}
This moduli matrix is very close to the solution, 
except for the following deficiency: 
$b(w)=\frac{P_k}{\alpha P_k + aQ_{k-1}}$ is 
not holomorphic at some points in $w$ where $b(w)$ 
diverges. 
As stated in equation (\ref{eq:4eq:moduli_sp_ivv}), 
all components in the moduli matrix $H_0(z,w)$ 
should be holomorphic with respect to both $z$ and $w$ 
at any point $(z,w)\in{\bf C}^2$. 
We can overcome this problem by noting that the lump 
solution $b(w)$ is given in terms of 
an inhomogeneous coordinate $b$ on ${\mathbf C}P^1$. 
We now transform the moduli matrix 
$H_{{\rm v}0}^{\rm single}(z;z_0,b(w))$
written in the inhomogeneous coordinate $b$ 
into the one in homogeneous coordinates. 
The correct moduli matrix should then be 
\begin{eqnarray}
\hspace{-1cm}
H_0(z,w)
=  \left(
\begin{array}{cc}
(z-z_0)A_{k-1}(w) & (z-z_0)(\alpha A_{k-1}(w) + aB_{k-2}(w))\\
P_k(w) & \alpha P_k(w) + aQ_{k-1}(w)
\end{array}
\right) , \label{eq:ist:k-instantons}
\end{eqnarray}
with $A_{k-1}$ and $B_{k-2}$ as 
polynomial functions of order $k-1$ and $k-2$ in $w$, given by 
\begin{eqnarray}
A_{k-1}(w) &=& \sum_{i=1}^{k} \frac{1}{Q_{k-1}(p_i)}
\prod_{i' (\neq i)=1}^{k}\left(\frac{w-p_{i'}}{p_i-p_{i'}}\right),
\label{eq:ist:A}
\\ 
B_{k-2}(w) &=& \sum_{j=1}^{k-1} \frac{-1}{P_k(q_j)}
\prod_{j' (\neq j)=1}^{k-1}\left(
\frac{w-q_{j'}}{q_j-q_{j'}}\right).
\label{eq:ist:B}
\end{eqnarray}
We can determine these $A_{k-1}(w)$ and $B_{k-2}(w)$ 
uniquely by the following condition 
\begin{eqnarray}
 A_{k-1}Q_{k-1} - B_{k-2}P_k = 1, 
\label{eq:ist:aq-bp}
\end{eqnarray}
which requires the vorticity of the solution to agree with 
the one in equation (\ref{eq:ist:1/2_to_1/4}): 
the solution should have a single vortex in the 1-3 plane 
and no vortices in the 2-4 plane. 
We see that 
the right hand side of 
equation (\ref{eq:ist:1/2_to_1/4}) 
and equation (\ref{eq:ist:k-instantons}) 
are related by 
\begin{eqnarray}
H_0(z,w)
= V(P_k(w),Q_{k-1}(w))
\ H_{{\rm v}0}^{\rm single}(z;z_0,b(w)), 
\label{eq:ist:relation}
\end{eqnarray}
with the matrix $V(P_k,Q_{k-1})$ defined by 
\begin{eqnarray}
V(P_k,Q_{k-1}) \equiv \left(
\begin{array}{cc}
\frac{a}{\alpha P_k + aQ_{k-1}} & (z-z_0)(\alpha A_{k-1} + aB_{k-2})\\
0 & \alpha P_k + aQ_{k-1}
\end{array}
\right).
\end{eqnarray}
In a particular region of $w$ with non-vanishing 
$\alpha P_k + aQ_{k-1}$, this matrix $V(P_k,Q_{k-1})$ 
is a $V$-equivalence transformation 
(\ref{eq:4eq:ivv-V-equiv}). 
However, it cannot be a legitimate $V$-equivalence 
transformation, since it has a singularity in $w$. 
Although $V(P_k,Q_{k-1})$ is not a valid 
$V$-equivalence transformation because of these 
singularities in $w$, it is needed precisely to 
compensate singularities of 
$H_{{\rm v}0}^{\rm single}(z;z_0,b(w))$ in equation 
(\ref{eq:ist:1/2_to_1/4}), if we wish 
to obtain the regular moduli matrix 
(\ref{eq:ist:k-instantons}).

We now examine the moduli parameters of the $k$-instantons 
in the Higgs phase in detail. 
Since no new parameters appear in $A_{k-1}$ and $B_{k-2}$, 
the configuration of $k$-instantons in the Higgs phase 
has the $2k+2$ complex moduli parameters 
$(z_0,\{p_i\},\{q_j\},a,\alpha)$. 
The position of the single vortex on the 1-3 plane is 
given by the moduli parameter $z_0$, which decouples 
from other moduli parameters 
and has a flat metric. 
This decoupling of $z_0$ can be recognized also from the 
K\"ahler potential (\ref{eq:efa:kahler_pot_1vort}). 
Therefore the moduli space of instantons in the 
Higgs phase can be written as 
\begin{eqnarray}
\hspace{-1cm}
{\cal M}^{k{\rm -instantons}} \simeq
{\bf C} \times {\cal M}^{k{\rm -lumps}}
\simeq {\bf C} \times 
   \{\varphi|{\bf C} \to \hat{\cal M}^{1{\rm -vortex}}, 
             \bar \partial_w \varphi = 0\}.
\end{eqnarray}
We easily realize that $\{p_i\}$ correspond to 
the positions of $k$-instantons inside the vortex, 
$a$ to the total size and the orientation of the 
configurations 
and $\{q_j\}$ to the relative sizes and the 
orientations of the instantons. 
In the limit of vanishing $a$, 
the rank of the moduli matrix (\ref{eq:ist:k-instantons}) 
reduces by one and its determinant vanishes. 
Then the point $a \to 0$ is singular in the moduli space.
On the other hand, the small lump singularities coming
from $p_i = q_j$ in equation (\ref{eq:ist:k-lumps}) 
arise as divergences of $1/P_k$ and $1/Q_{k-1}$ in 
$A_{k-1}$ and $B_{k-2}$ in equations 
(\ref{eq:ist:A}) and (\ref{eq:ist:B}). 
Therefore we observe 
that the small lump singularities 
with $a=0$ or $p_i = q_j$ 
in equation (\ref{eq:ist:k-lumps}) are now interpreted as 
the small instanton singularities in the Higgs phase. 
We can easily confirm that the points $p_i = p_{i'}$ 
for $i\neq i'$ and $q_j = q_{j'}$ for $j\neq j'$, 
respectively, are not singularities of equations 
(\ref{eq:ist:A}) and (\ref{eq:ist:B}). 
The remaining parameter $\alpha$ parameterizes ${\bf C}P^1$ 
similarly to the lump solutions. 
In the case of $1/2$ BPS vortex, 
this $\alpha \in {\bf C}P^1$ 
is a normalizable moduli, whereas it becomes a 
nonnormalizable moduli in the case of the $1/4$ BPS lumps. 
This sort of phenomenon occurs often: normalizable moduli 
of the host soliton can become a non-normalizable moduli 
of a soliton on the host soliton. 
In summary we find 
$z_0 \in {\bf C}$, $p_i \in {\bf C}$, 
$a \in {\bf C}^* \equiv {\bf C} - \{0\} \simeq 
{\bf R}\times S^1$,  
$q_j \in 
{\bf C} - \{p_1,p_2,\cdots,p_k\}$ 
and $\alpha \in {\bf C}P^1$.

Let us consider the simplest case of 
a single instanton $(k=1)$ with $A_0 = 1$ and $B_{-1}=0$ 
in some detail. The lump solution 
$b(w) = \frac{w-p_1}{\alpha(w-p_1) + a}$ in the effective theory 
on a vortex suggests a solution of the 1/4 BPS equations 
with the moduli matrix 
\begin{eqnarray}
 H_0^{1-{\rm instanton}} =  
 \left(
  \begin{array}{cc}
   z-z_0 & \alpha(z-z_0)\\
   w-p_1 & \alpha(w-p_1) + a
  \end{array}
 \right).
\label{eq:ist:1vortex-moduli-matrix}
\end{eqnarray}
We can clarify the physical significance of these 
four complex moduli parameters $z_0,p_1,a,\alpha$, 
by transforming the moduli matrix in 
equation (\ref{eq:ist:1vortex-moduli-matrix})
into that with $\alpha = 0$ 
by an $SU(2)_{\rm F}$ rotation $U$ combined 
with a $V$-equivalence transformation:
\begin{equation}
H_{{\rm v}0}^{1-{\rm instanton}}(z;a,p_1,\alpha)
\sim H_{{\rm v}0}^{1-{\rm instanton}}(z;a_0,p_0,\alpha=0) U,
\label{eq:ist:alpha_tr}
\end{equation}
Then we obtain the physical
position $p_0$ and the size 
$|a_0|$ of the instanton in the vortex as 
\begin{eqnarray}
 p_0 = p_1-\frac{\alpha^*}{1+|\alpha|^2}a, 
  \quad
 |a_0| = 
 \frac{|a|}{1+|\alpha|^2} , 
\label{eq:ist:pos-size}
\end{eqnarray}
which are invariant under the $SU(2)_{\rm F}$ rotation. 

Let us now consider the topology 
of the the moduli space of one instanton in the Higgs phase. 
The moduli matrix $H_0^{1-{\rm instanton}}$ can be 
transformed into $H_0^{'1-{\rm instanton}}$ in another 
patch of the moduli space by a $V$-equivalence 
transformation in (\ref{eq:4eq:ivv-V-equiv}) 
\begin{eqnarray}
 H_0^{'1-{\rm instanton}}
& \equiv & 
 \left(
  \begin{array}{cc}
   \alpha'(w-p'_1) + a' & w-p'_1\\
   \alpha'(z-z_0) & z-z_0
  \end{array}
 \right)
\sim H_0^{1-{\rm instanton}} ,
\label{eq:ist:rel-vortex-moduli-matrix}
\end{eqnarray}
with the following relation between coordinates 
in two patches
\begin{eqnarray}
 \alpha' = \frac{1}{\alpha},
  \quad
 a' = -\frac{a}{\alpha^2},
\quad
 p'_1 = p_1 -\frac{a}{\alpha}.
\end{eqnarray}
Both $\alpha$ and $\alpha'$ are the standard inhomogeneous 
coordinates of the ${\mathbf C}P^1$ 
in different patches, which are enough to
cover the whole manifold. 
Since we find that $a$ requires a nontrivial transition 
function $-1/\alpha^2$ between two patches, 
it is a tangent vector as a fiber on the ${\mathbf C}P^1$.
On the other hand, we can use an invariant global 
coordinate for two patches, $p_0$, instead of $p_1$. 
This implies that the space ${\bf C}$ parameterized by 
$p_0$ is a direct product to the ${\mathbf C}P^1$. 
Therefore the topology of the moduli space of one 
$U(2)$ instanton in the Higgs phase is given by 
\begin{eqnarray}
 (z_0, p_0, a, \alpha) \in
{\bf C} \times
{\bf C} \times ({\bf C}^* \times^* {\bf C}P^1)
  \simeq {\cal M}^{1-{\rm instanton}}, 
\label{eq:ist:single-inst}
\end{eqnarray}
where $({\bf C}^* \times^* {\bf C}P^1)$ 
is the tangent bundle with a base space ${\bf C}P^1$ and 
a fiber ${\bf C}^*$.

For $N>2$, we can specify the moduli matrix 
$H_0 (z,w)$ for a particular class of $1/4$ BPS solutions 
which can be interpreted as $1/2$ BPS states in the 
vortex theory, similarly to the case of $N=2$. 
Thus we can obtain the $1/4$ BPS states 
corresponding to the $U(N)$ instantons 
in the Higgs phase by repeating the same discussion.

\subsubsection{Intersection of vortices}

We can obtain more varieties of solutions, 
if we do not restrict ourselves to solutions interpretable 
as solitons in the effective theory on a host vortex. 
For instance, intersection of two or more vortices 
cannot be understood as solitons on vortices, 
since the energy of such composite solitons diverges 
in the effective theory. 
The moduli matrix approach allows us to construct 
such solitons 
of intersecting vortices directly.

In the theory with $N_{\rm C}=N_{\rm F}\equiv N=2$, 
the following two moduli matrices give configurations with 
$\nu_{\rm v}=k_z (\ge 0)$ vortices in the 1-3 plane and 
$\nu_{\rm v'}=k_w (\ge 0)$ vortices 
in the 2-4 plane 
\begin{eqnarray}
H_0 &=&
\left(
\begin{array}{cc}
z^{k_z} & 0\\
0 & w^{k_w}
\end{array}
\right), 
\label{eq:ist:trivial_int}
\\ 
H_0 &=&
\left(
\begin{array}{cc}
z^{k_z} w^{k_w} & 0 \\
              0 & 1
\end{array}
\right).
\label{eq:ist:intersecton}
\end{eqnarray}
The two vortices intersect at a point $z=w=0$ in 
both cases. 
The moduli matrix in equation (\ref{eq:ist:trivial_int}) 
gives a trivial intersection carrying no instanton charge. 
On the other hand, the moduli matrix in equation 
(\ref{eq:ist:intersecton}) gives two vortices with a 
nontrivial intersection carrying the instanton charge 
$\nu_{\rm i} = - k_z k_w$ at the intersection point. 
In this case the instanton charge contributes negatively 
to the energy of the composite soliton. 
This negative contribution can be interpreted as a binding 
energy of two vortices at the intersection, similarly 
to the case of an Abelian junction of domain walls 
in equation~(\ref{eq:juc:Y_ab}). 
The (infinitely) large energy coming from vortices is 
slightly canceled by the negative contribution from 
the intersection giving a positive energy as a whole. 
We can call this composite soliton as an ``intersecton", 
since the instanton charge is stuck at the intersecting point 
of vortices. 
It cannot move once the vortices are fixed.

Let us consider the case of $N_{\rm C} < N_{\rm F}$ 
theory where semi-local vortices are available. 
In this case, we can take a strong coupling limit 
$g^2\rightarrow\infty$ to obtain 
an exact solution. 
In this strong coupling limit we obtain 
a nonlinear sigma model 
whose target space is the cotangent bundle over the 
complex Grassmann manifold, 
$T^* (G_{N_{\rm F},N_{\rm C}})$, as explained 
in section~\ref{sc:mdl:BPSWSIGC}. 
Then the master equation (\ref{eq:4eq:master_ivv}) 
can be solved algebraically as 
$\Omega^{g^2\rightarrow \infty}= \Omega_0 =c^{-1}H_0H_0^\dagger$. 
For simplicity we take the $U(1)$ gauge theory 
with four flavors: $N_{\rm C}=1$, $N_{\rm F}=4$. 
We obtain 
non-trivially intersecting vortices with 
$\nu_{\rm v}=k_z,\ \nu_{\rm v'}=k_w$
by considering the following moduli matrix 
\begin{eqnarray}
 H_0
 = \left(
  z^{k_z}w^{k_w},\ z^{k_z},\ w^{k_w},\ 1
 \right) . 
\label{eq:ist:U(1)instanton}
\end{eqnarray}
We find the exact solution 
$H = (1/ \sqrt{ \Omega^{g \to \infty}} ) H_0$ with 
\begin{eqnarray}
  \Omega^{g \to \infty}  
  = \Omega_0 = (|z|^2 + 1)^{k_z}(|w|^2 + 1)^{k_w}. 
\end{eqnarray}
The instanton charge is found to be the product of 
vorticities, namely $\nu_{\rm i} = - k_z k_w$. 
This solution explicitly shows that 
the $U(1)$ instantons are 
stuck at the intersection of vortices. 
This instanton charge also contributes negatively 
to energy, in agreement with our observation that 
the instanton charge in Abelian gauge theories can be 
interpreted as a binding energy of vortices. 
However, we have observed that the instanton charge 
$\nu_{\rm i}$ changes its sign under the duality 
transformation 
$N_{\rm C} \leftrightarrow N_{\rm F} - N_{\rm C}$ 
(with fixed $N_{\rm F}$) 
of nonlinear sigma models in equation (\ref{eq:mdl:duality}). 
By using this duality transformation, we can also obtain 
intersections of vortices in nonlinear sigma models 
coming from non-Abelian gauge theories. 
As a result, we find 
intersections of vortices that contribute positively to 
energy of the composite soliton, 
similarly to the non-Abelian junction contributing 
positively to the energy.

Let us summarize this section by observing that there exist 
three types of instantons. 
The first type is interpretable as lumps living inside a 
vortex. 
The second type is an instanton stuck at the intersection 
point of vortices. 
The third type is the intersection of vortices without 
any interaction. 
This last type has been observed in equation 
(\ref{eq:ist:trivial_int}). 
We expect that the most general solution is given by 
the mixture of these configurations, similarly to the 
webs of walls in section \ref{juc}.


\subsection{Solitons in world volume of solitons}\label{sos}
In this section 
we have  discussed 1/4 BPS states of 
composite solitons. 
To do that we have worked out  
the moduli parameters 
in the moduli matrix 
for 1/4 BPS systems. 
If we could solve the master equations 
analytically or numerically 
we would obtain the full solutions. 
As discussed below 
one big advantage of this method is that 
the moduli matrix contains non-normalizable modes also.

In order to discuss the composite solitons 
there exists another method, 
the moduli approximation.  
In this method one first constructs the effective action 
of a host 1/2 BPS soliton by using 
the Manton's method \cite{Manton:1981mp} as discussed 
in section \ref{efa}. 
Then one constructs 1/2 BPS solitons in this effective theory. 
In the end one has to check matching of topological charges 
in the original theory and in the effective theory. 

It was this method to find a confined monopole 
in the Higgs phase \cite{Tong:2003pz}.
Namely one constructs effective theory on a single non-Abelian vortex 
in the model with $N_{\rm F}=N_{\rm C}=2$ with massive hypermultiplets. 
It is the ${\bf C}P^1$ model with a potential, 
which contains two vacua. 
Its K\"ahler potential is 
given in equation (\ref{eq:efa:kahler_pot_1vort}). 
Here note the coefficient in the second term 
(the K\"ahler class of ${\bf C}P^1$) 
is given by $4\pi / g^2$. 
Then one constructs a kink interpolating between these two vacua. 
The energy of the kink can be calculated from 
equation (\ref{eq:wll:tension}) 
as the product of the K\"ahler class $4\pi / g^2$ 
and the mass difference $\Delta m$, to give $4\pi \Delta m/ g^2$. 
This coincides with the energy (\ref{eq:wvm:calc_mv}) of a monopole, 
and the topological charges match 
in the original theory and in the effective theory. 
Therefore the kink in the vortex can be interpreted 
as a monopole in the original (bulk) theory.
In fact it was this argument for the authors in \cite{Shifman:2004dr} 
to determine the K\"ahler class from only symmetry argument,  
while we have derived it in equation (\ref{eq:efa:kahler_pot_1vort}) 
from direct calculation. 

In the same way, one can construct instantons. 
First one constructs a single vortex 
in the model with $N_{\rm F}=N_{\rm C}=2$ with massless hypermultiplets. 
The effective action on it is the ${\bf C}P^1$ model without potential. 
Then one can construct the ${\bf C}P^1$-lumps. 
The energy of lumps can be calculated by the product of 
the K\"ahler class $4\pi / g^2$ and 
the lump number $n \in {\bf Z} \simeq \pi_2({\bf C}P^1)$, 
to give $4\pi n/ g^2$.  
This coincides with the instanton number, and so 
lumps in the vortex can be regarded as instantons 
in the original theory \cite{Eto:2004rz}.

The wall-vortex system can be constructed from the wall point of view. 
When a vortex ends on a domain wall, 
it can be understood as a BIon in 
the effective theory on the wall \cite{Gauntlett:2000de}. 
The negative energy of the boojum can also be obtained 
from the effective theory on a double wall \cite{Eto:2006uw}.

These configuration can be classified into two cases.
The first is the case that the soliton is made of 
the internal (orientational) moduli of the host soliton.
The second is the case that it is made of 
space-time moduli. 
The first case contains monopoles (instantons) on a vortex while 
the second case contains 
vortices ending on a domain wall.

Clearly this moduli approximation has limitations.  
First the slow movement approximation is used to construct 
effective action. 
Therefore this method cannot be applied 
to a region with rapidly varying fields. 
For instance, 
the center of vortex of wall-vortex composite system 
cannot be described accurately by 
the wall effective action. 
Second it cannot be applied to non-normalizable modes 
because one cannot construct an effective action for those 
modes. 
Our method of the moduli matrix overcomes both of these 
problems. 
We do not use the slow movement approximation at all.
Moreover the moduli matrix contains non-normalizable modes 
also. 
When we wish to view 1/4 BPS system of monopole, vortex 
and walls  from the vortex point of view, for instance, 
the incorporation of non-normalizable modes provides the 
following advantage. 
Semi-local vortices (lumps) have non-normalizable modes 
as well as normalizable modes, as seen in section \ref{vtx}. 
If 
\changed{a modus corresponding to 
normalizable mode in the moduli matrix for 1/2 BPS vortex  
is promoted to 
a ``field" depending on the vortex world volume 
in the sense of (\ref{eq:mvw:beads})}, 
it can produce a kink 
inside the vortex, namely a confined monopole attached by 
vortices.  
On the other hand, if non-normalizable mode depends on the 
coordinates corresponding to the vortex world volume, 
domain walls appear in the bulk instead of inside the vortex. 
This is because non-normalizable moduli of a vortex are 
bulk modes living at spatial infinities of the vortex. 
This point of view is missing in the moduli approximation. 

However the moduli approximation 
is powerful for non-BPS composite solitons, 
for which the method of the moduli matrix cannot be used. 
For instance, domain walls and instantons cannot coexist 
as a BPS state \cite{Eto:2005sw}. 
If we consider them together, the configuration breaks 
all SUSY and therefore it is non-BPS. 
Nevertheless we can construct 
the effective action on walls as usual, 
and can construct a soliton on it. 
We have found in \cite{Eto:2005cc}
that effective action on walls is precisely the Skyrme model 
including the four derivative term. 
Skyrmions are non-BPS and so we cannot compare energy in effective 
theory and the original theory.  
Instead we should calculate topological charges directly. 
Since we can show that 
the baryon (Skyrmion) number in the effective theory coincides 
with the instanton number, 
the Skyrmions in the effective theory 
can be regarded as instantons 
in the original theory. 

\section{Discussion}\label{cad}
We have seen that the SDYM-Higgs equation 
describing the instanton-vortex system 
is the most general in the sense that it 
gives all other equations by 
dimensional reductions. 
Accordingly, the moduli matrix $H_0(z,w)$ 
of this system 
contains the moduli matrices of all other systems 
as special cases. 
However a complete understanding of $H_0(z,w)$ is still 
missing at present while the moduli matrices have been 
more throughly studied in this paper for walls, vortices, 
domain wall webs and wall-vortex-monopole systems. 

In this paper we have considered only static solutions. 
Time dependent stationary solitons, like Q-kinks 
\cite{Abraham:1992vb} and Q-lumps \cite{Leese:1991hr}, 
are also BPS states. 
Some of these dyonic objects have been discussed in 
\cite{Lee:2005sv,Eto:2005sw}. 
These solitons are worth studying in more detail. 

We have considered effective Lagrangian on only 
elementary solitons, namely vortices and walls. 
Construction of effective Lagrangian of composite solitons 
is also possible, if they have normalizable modes. 
For instance, effective Lagrangian of monopoles (instantons) 
in a vortex may be obtainable, but essentially 
the same dynamics should already be contained in the 
kink (lump) solutions using the effective Lagrangian on 
vortex. 
Some nontrivial examples of effective Lagrangian on composite 
solitons are:
\begin{itemize} 
\item
loops in domain wall webs,  
\item
vortices stretched between domain walls 
in the wall-vortex-monopole system. 
\end{itemize}
We know from 
the discussion of conserved SUSY \cite{Eto:2005sw} 
that they are described by a $(2,0)$ sigma model
in $d=1+1$ dimensions or its dimensional reduction. 
Construction of these effective Lagrangians is very 
interesting because they resemble the $(p,q)$ 5-brane 
webs \cite{Aharony:1997ju} and the Hanany-Witten brane 
configuration \cite{Hanany:1996ie}, respectively.

We should note that another set of 1/4 BPS equations and 
the unique set of 1/8 BPS equations have 
also been found \cite{Lee:2005sv,Eto:2005sw}. 
Let us list their co-dimensions in $d=5+1$ dimensions 
\cite{Eto:2005sw}. 
Another set of 1/4 BPS equations contains triply 
intersecting vortices \cite{Naganuma:2001pu} whose 
co-dimensions are listed by $\times$ (world volume 
is denoted by $\bigcirc$) as 
\begin{eqnarray*}
\begin{array}{c|cccccc}
1/4 \;\; {\rm VVV} &0 &1 &2 &3 &4 &5 \\
\hline
{\rm Vortex}&\bigcirc &\bigcirc &\times &\times &\bigcirc &\bigcirc \\
{\rm Vortex}&\bigcirc &\times &\bigcirc &\times &\bigcirc &\bigcirc \\
{\rm Vortex}&\bigcirc &\times &\times &\bigcirc &\bigcirc &\bigcirc
\end{array}
\end{eqnarray*}
The unique set of 1/8 BPS equations can contain various 
solitons with co-dimensions denoted by $\times$ (world 
volume is denoted by $\bigcirc$) as 
\begin{eqnarray*}
\begin{array}{c|cccccc}
1/8 \;\; {\rm IV}^6 &0 &1 &2 &3 &4 &5 \\ \hline
{\rm Instanton}&\bigcirc &\times &\times &\times &\times &\bigcirc  \\
{\rm Vortex}&\bigcirc &\bigcirc &\times &\times &\bigcirc &\bigcirc \\
{\rm Vortex}&\bigcirc &\times &\bigcirc &\times &\bigcirc &\bigcirc \\
{\rm Vortex}&\bigcirc &\times &\times &\bigcirc &\bigcirc &\bigcirc \\
{\rm Vortex}&\bigcirc &\times &\bigcirc &\bigcirc &\times &\bigcirc \\
{\rm Vortex}&\bigcirc &\bigcirc &\times &\bigcirc &\times &\bigcirc \\
{\rm Vortex}&\bigcirc &\bigcirc &\bigcirc &\times &\times &\bigcirc
\end{array}
\end{eqnarray*}  
Systematically solving these equations is much more 
difficult than so far studied and remains as a future 
problem.  

Let us finally list some of other interesting future 
directions: 
quantum effects of solitons, 
non-perturbative dynamics of field theories 
as well as string theory,
and applications to particle physics, 
cosmology, condensed matter physics 
and nuclear physics. 
We hope this review article to be useful to 
explore these and other problems. 

\section*{Acknowledgements}
We would like to thank Masato Arai, Toshiaki Fujimori, 
Kazuya Kakimoto, Nobuhito Maru, Masashi Naganuma, 
Kazutoshi Ohta, Yuji Tachikawa, David Tong, and Yisong Yang 
for collaborations in various stages. 
We also thank Joao Baptista, Jarah Evslin, Koji Hashimoto,  
Kimyeong Lee, Nick Manton, and Ho-Ung Yee for a useful discussion. 
This work is supported in part by Grant-in-Aid for 
Scientific Research from the Ministry of Education, 
Culture, Sports, Science and Technology, Japan No.17540237 
(N.~S.) and 16028203 for the priority area ``origin of 
mass'' (N.~S.). 
The work of M.~N.~and K.~O.~(M.~E.~and Y.~I.) is 
supported by Japan Society for the Promotion 
of Science under the Post-doctoral (Pre-doctoral) Research 
Program. 


\vspace{1cm}

\noindent
{\large {\bf References}}

\end{document}